\documentclass [a4paper,12pt,twoside]{report}
\usepackage{amsmath,amssymb,graphicx,cite,delarray}
\usepackage[a4paper]{geometry}
\usepackage[bf]{caption2}
\renewcommand{\captionlabeldelim}{.}
\begin{document}

\begin{center}M. P. KASHCHENKO\end{center}
\vspace{20mm}
\begin{center}\begin{Large}THE WAVE MODEL OF MARTENSITE GROWTH \\
FOR THE $\gamma-\alpha$ TRANSFORMATION \\
\vspace{2mm}
OF IRON-BASED ALLOYS
\end{Large}\end{center}
\vspace{15mm}
\begin{figure}[htb]
\centering
\includegraphics[clip=true,width=.8\textwidth]{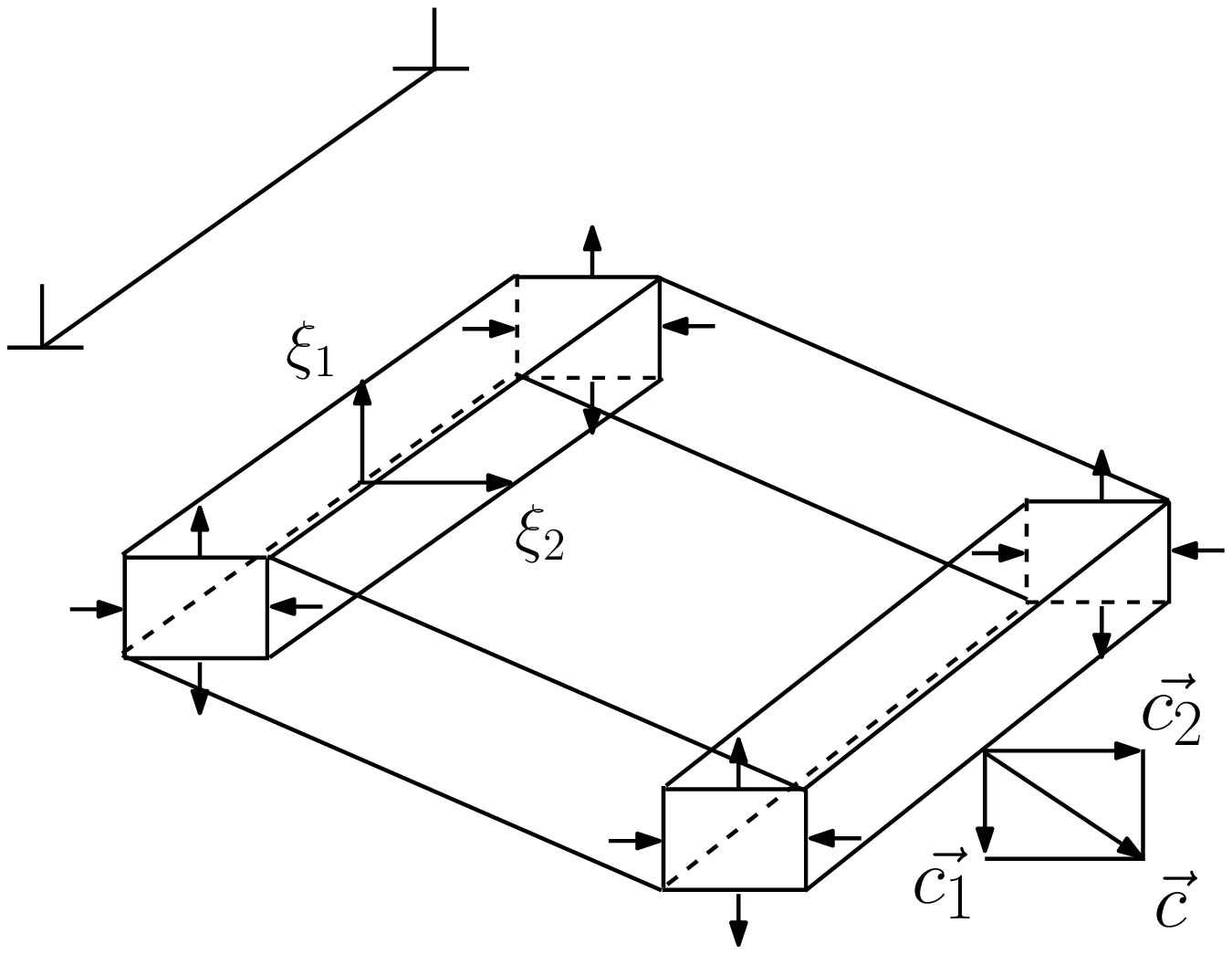}
\end{figure}
\vspace{30mm}
\begin{center}EKATERINBURG\quad\large{2005}\end{center}

\newpage
\rule{0pt}{10cm}
\pagebreak

\rule{0pt}{3cm}
\begin{center}M.P.KASHCHENKO\end{center}
\begin{center}THE WAVE MODEL OF MARTENSITE GROWTH \\ 
FOR THE $\gamma-\alpha$ TRANSFORMATION  \\
OF IRON-BASED ALLOYS\end{center}

\begin{center}Prof. Michael  P. Kashchenko\\
Ural State Forest  Engineering University \\
Physics Chair \\
Sybirskyi Tract 37 \\
620100 Ekaterinburg \\
RUSSIA \\
mpk@usfeu.ru or  mpkathome@usfeu.ru\end{center}
\rule{0pt}{7cm}

\begin{center}Ekaterinburg \\ 
UIF Publishing-House "Nauka" \\
	Original Russian Edition 1993\\
\vspace{5mm}
English Edition 2005.\\  
Translation from Russian by J. Gerlts and U. Kayser-Herold
\end{center}

\newpage
\rule{0pt}{10cm}
\pagebreak

\pagestyle{plain}
\tableofcontents

\sloppy

\chapter*{Preface}
\addcontentsline{toc}{chapter}{Preface}

Since the early seventies, the Author and his Team have been 
engaged in their still ongoing task and vision of developing a comprehensive 
physical notion of the $\gamma -\alpha $ Martensitic Transformation ($\gamma 
-\alpha $ MT). The main results of their research work are published in 
this monograph. Shortly after an initial review and analysis of existing 
literature on the subject, the following initial conclusions could be drawn:

\begin{itemize}
\item[-] The $\gamma -\alpha $ MT is a cooperative lattice 
transformation proceeding under highly pronounced deviations from local 
equilibrium conditions;

\item[-] The velocity of rapid controlled growth of $\alpha $-Martensite 
crystals is so large that only a wave-model has the physical potential for a 
conclusive detailed and comprehensive explanation of the transformation and 
growth ;
 
\item[-] A microscopic theory of the transformation is missing. 
\end{itemize}

From the very beginning, the development of a new and comprehensive growth 
model of a single martensite crystal was mainly guided by the following two 
factors: On the one hand, the realization that a substantial fraction of 
energy released during the martensitic transformation process is immediately 
being converted into the energy of cooperative lattice displacement waves, 
by the mechanism of stimulated emission of coherent phonons, becoming most 
effective in a subsystem of non-equilibrium 3-d electrons, which "jump" down 
to a lower energetic level at the very instant and location of the 
martensitic lattice transformation, and on the other hand, the formulation 
of a simple two-wave scheme of the crystal growth control mechanism. Even 
though these two factors were known to the author since 1976, considerable 
efforts were necessary to figure out in detail the kind and sequence of 
events during lattice wave generation, and to consistently interpret in 
their entirety the observed morphological and kinetic particularities of the 
$\gamma -\alpha $ MT, including those of some binary alloys in a wide 
range of concentrations of their second component, as well as to prove 
experimentally a variety of postulates and predictions related to the 
problem. The matter described in this monograph nearly reproduces the contents 
of the habilitation paper of the Author at the Department of Radiating Materials
Science of the Ural Polytechnical Institute (nowadays the Ural State Technical 
University) and at the Cathedra of Physics of the Ural Forest Engineering 
Institute (nowadays Ural Forest Engineering State University) completed 
in 1986.

The decision to publish this monograph was motivated, on the one hand, by 
several wishes expressed during the many and extensive discussions with 
specialists in the disciplines of Solid State Physics and Phase-Transitions, 
as well as with Metal- and Materials scientists, and on the other hand, by 
the presently achieved success in the realization of theoretical and 
experimental research programs on $\alpha $- Martensite nucleation focusing 
on the specific aspects of the Wave-Growth-Model. The results obtained from 1989 
to 1991 are supposed to be worth a presentation in a single special 
monograph. The short survey in the final part of this monograph is to complete 
the notion of the $\gamma -\alpha $ MT. Here, let us only mention the most 
important: 
\begin{itemize}
\item[-] It has been found that 60\r{ } straight line dislocations with 
$\langle110\rangle$  lines could act as dislocation nucleation centers 
(DNC) for $\alpha $-Martensite crystals with habit planes close to \{557\} - 
\{225\}, whereas 30\r{ } dislocations with $\langle112\rangle$  lines were 
acceptable candidates for the role of DNC for crystals with habits close to 
\{259\} - \{3 10 15\}. 

\item[-] Physical models have been developed for the excited 
pre-martensitic lattice condition, being typical of the state of nucleation. It 
was also possible to reproduce the physical conditions prevailing in the 
transformation wave-front, by means of picosecond-laser irradiation directed 
onto the surface of a Fe - 31,5 Ni single crystal. The results of these 
investigations can be interpreted as supportive and further detailing the 
qualitative notion of nucleation, which had been presupposed during the 
development of the growth model.
\end{itemize}

This is the first time that the mechanism of phonon generation during a solid 
state transformation has been analyzed and identified as a fundamental aspect 
of the $\gamma-\alpha$ transformation.
Obviously, the newly developed model of spontaneous martensitic nucleation 
and growth is also suitable for a physical description of the reverse 
martensitic $\alpha -\gamma $-transformation, as well as for other 
transformations of the type (i.e. cubic-bcc $ \leftrightarrow$ cubic-fcc 
lattice of Cu-Zn alloys). Moreover, the example of the $\gamma -\alpha$ MT 
shows that the key for comprehensive understanding of the mechanism controlling 
the growth of a new phase during reconstructive cooperative transformations is 
to reveal the dynamic characteristics of an excited lattice in the 
non-equilibrium system conditions developing in the frontal zone of a non-linear 
transformation wave. This latter conclusion is fundamental as it includes the 
essence of the paradigm complementing the traditional conceptual notion of 
equilibrium thermodynamics.

The Author hopes that this monograph will close (to some degree) the gap 
which previously had existed in the theory of reconstructive martensitic 
transformations, and will attract the attention of other scientists to the 
problem, and be useful to a wider audience of readers.

The Author wants to express his thanks and acknowledgments to all his 
Colleagues and Specialists who discussed this matter with him, thereby 
promoting the progress in research, as well as to the cooperative 
"Alvo-Materialovedenie", who sponsored the publication of this monograph.

\chapter*{Introduction}
\addcontentsline{toc}{chapter}{Introduction}

Since many epochs, the production and use of steel and ferrous alloys have 
been of high and ever increasing importance, for almost any branch of 
industry and economy. Thus the development of new types of steel with 
prescribed combinations of the desired mechanical and physical properties, 
as well as economical benefits, have been the final aim of numerous and still 
ongoing metallurgical research projects. On the one hand, such research 
traditionally comprises the empirical development of highly sophisticated 
metallurgical production methods, like new thermo-mechanical methods of 
treatment, by making use of the most suitable phase- and structural 
transformations and, on the other hand, has the final aim of comprehensive 
and fundamental theoretical understanding of the underlying phase 
transitions and related physical processes. 

In the metallurgy of steel, the spontaneous martensitic $\gamma -\alpha$ 
-transformation plays a key role among structural transformations, as it 
constitutes the most important hardening process ending up in the formation of 
martensite. Traditionally, the spontaneous martensitic transformation (MT) is 
produced by quenching of the high-temperature (austenitic) phase. The 
spontaneous MT of ferrous alloys is intriguing and challenging from a scientific 
point of view. One of the most characteristic features of MT is their 
diffusionless transformation mechanism, by means of rapid cooperative 
rearrangement of the face-centered cubic (fcc) high-temperature $\gamma$-phase 
(austenite) into the body- centered cubic (bcc) or body-centered tetragonal 
(bct) low temperature  $\alpha$-phase (martensite). The results of specific 
research focusing on martensitic transformations substantially contributed to 
the increase of understanding of the many relevant details of diffusionless 
structural transformations. However until recently, some key features of 
martensitic transformations, like the growth process and lattice-dynamics of the 
spontaneous transformation, remained as one of metal physics greatest 
intractable problems. 
The disclosure of the dynamical mechanisms and principles underlying the 
process of martensitic transformations would enlighten their intrinsic 
characteristics and physical nature, and can thus be regarded as one of the 
fundamental problems of metal physics. Till now, dynamic approach has only 
been (partially) developed for transformations of the distortional type, in 
which the lattice symmetry groups of the final and of the initial phases are 
mutually subordinated. In principle, the transformations of the distortional 
type are close to a second-order phase transition (as a rule they exhibit 
weakly pronounced characteristics of first-order transitions). Such 
transformations are almost satisfactorily explained within the "Soft-Mode" 
theory, which, for the most part, focuses around the Shape-Memory-Effect 
(SME). In contrast, the reconstructive phase transitions, to which the 
spontaneous $\gamma -\alpha$  MT is related, demonstrate pronounced features of 
the first-order transitions, namely, considerable temperature hysteresis 
(between the direct and reverse transformations) and thermal and volume 
effects. Therefore, the "Soft-Mode" theory cannot be directly applied to 
them. Theoretical research on the $\gamma -\alpha$  MT is mainly characterized 
by the parallel development of the lattice-geometrical, thermodynamic and wave 
approaches. The few attempts to describe the martensite growth within the 
wave theory remained uninformative, mainly due to the poorly understood 
mechanisms of wave excitation and stabilization, as well as of the types of 
waves and their directions of propagation. On the other hand, the results of 
such investigations remained in the shadow of an impressive success in the 
interpretation of the morphology of the martensite crystals, which had 
mainly been achieved by means of lattice-geometrical and thermodynamic 
investigations. At the same time it turned out more and more obvious that 
only the wave approach would have the full potential for a comprehensive 
description of the dynamical aspects of the transformation process. This 
circumstance determines the immediateness of development of the wave models 
of martensite crystal growth. 

The main problems of the wave theory of the $\alpha$ -phase growth are 
the type of the wave process and the mechanism of energy supply that ensures 
their steady-state propagation. Experimental results indicate that the frontal 
velocity of martensitic lattice growth only weakly depends on temperature. 
Its speed of propagation of about $\sim  10^{3}$ m/s is of the same 
order of magnitude as the sound velocity, presumably exceeding the velocity 
of longitudinal elastic waves. Supersonic growth speed, on the one hand, is 
a clear indication of a complicated, non-linear character of a wave-process 
and on the other hand, of its adiabatic nature, which, in combination with 
supercooling by about $\Delta T \approx 200$K below phase equilibrium 
temperature T$_{0}$, together with considerable heat- and volume-transformation 
effects, must inevitably lead to a strongly pronounced non-equilibrium condition 
with large temperature- and chemical potential gradients, in the area between 
the two phases and, finally, to intensive electronic flows. In accordance with 
the general laws of non-equilibrium thermodynamics, specific self-organizing 
processes and structures can in principle emerge within any kind of 
pronounced non-equilibrium system. In the particular case of interactions 
between  radiation fields comprising photons or phonons with inversely 
occupied radiative (electronic) sub-systems, such conditions may strongly 
support the generation of coherent waves. To achieve our final aim of 
determination of the mechanism controlling the cooperative atom 
displacements (in a microscopic frame of description) associated with the 
$\gamma -\alpha$  MT, it is necessary to analyze the dynamical properties of a 
non-equilibrium system comprising essential electron-phonon interactions. 
The direction of research initiated by this work, being mainly related to 
the $\gamma -\alpha$  martensitic transformation of ferrous alloys, can 
essentially be formulated as follows:
\vspace{-1.5mm}
\begin{itemize}
\item[]{A theoretical description of rapid growth of a martensite crystal as 
a self-organizing process being inherently associated with the propagation 
of displacement waves generated and amplified by non-equilibrium 3-d 
electrons.}
\end{itemize}
\vspace{-1.5mm}
Thus the aim of this work will focus on the substantiation of a wave-model 
of martensite growth, which in principle should enable us to determine the 
basic relationships between the microscopic particularities of the 
electronic structure and the macroscopic (morphologic) features of the 
transforming systems. 

To achieve this goal, the following basic prerequisites will have to be met:
\begin{enumerate}
\item{The development of a methodology for specific search and the 
identification of pairs of electronic states, that are potentially active for the 
generation of displacement waves in a local non-equilibrium medium, and an 
assessment of the amount of such pairs in the electronic spectrum of 
fcc-iron during the $\gamma -\alpha $  MT.}
\item{Definition of the threshold conditions for the generation of elastic 
displacement -waves by nonequilibrium electrons, including an estimation of 
maximum wave amplitudes.}
\item{Substantiation of the possibility to realize the conditions required for 
wave generation within a wide range of temperatures and concentrations of 
the second component in ferrous alloys.}
\item{Creation of a model associating a variety of morphologic features of 
martensite crystals with the coordinated propagation of lattice displacement 
waves.}
\item{Investigation of the motion of the martensitic crystal interface within 
the concept of a coordinated propagation of a "switching-wave" for relative 
volume deformation and displacement waves ensuring the onset of threshold 
deformation.}
\end{enumerate}

A qualitative notion of the physical state of rapid frontal growth of 
martensitic lattice, as obtained from theoretical analysis, thus must at 
least comprise the following key features: 

\begin{enumerate}
\item{Frontal growth of martensite is controlled by quasi-longitudinal lattice 
distortion waves fastly propagating through a $\gamma$- phase; }

\item{The characteristics of such waves define the basic spatial and temporal 
scales of martensitic growth, and as a consequence the gradients of the 
chemical potential of electrons (and thereby also the magnitudes of the 
electrical field strength and current density) and of temperature in the 
transformation zone;}

\item{During martensitic lattice growth, the physical parameters of the 
interphase region comprise key characteristics of a highly pronounced 
non-equilibrium condition;}

\item{The amplitudes of lattice distortion waves are maintained owing to the 
process of stimulated emission of phonons by non-equilibrium 3-d electrons 
located in the transformation zone, thereby revealing the physical nature of 
the real process controlling a spontaneous martensitic reaction.}
\end{enumerate}

Detailed physical information on the various field characteristics (i.e. 
displacements, temperature, electrical and magnetic fields) within the 
non-equilibrium area in the vicinity of a growing lattice will be of 
fundamental importance for the design of experiments aiming at a simulation 
of the process of lattice nucleation of the $\alpha$ -phase, as well as 
for the design of experimental devices for dynamical excitation of a 
pre-martensitic lattice condition by hypersonic frequencies in a range of $\sim 
10^{10} s^{-1}$, which would have to coincide with the range of 
frequencies of the distortion waves controlling martensite lattice growth. 
There exists a variety of possibilities for controlled oriented lattice 
growth during the $\gamma -\alpha$  - transformation of austenitic single 
crystals, for example by means of the combined influence of hypersound, 
mechanical stress and strong external magnetic fields. In particular, the effect 
of controlled orientation of martensitic crystal growth by means of a strong 
magnetic field, as predicted by the author, has experimentally been confirmed at 
the "Institute of Physics of Metals " of the Academy of Sciences of the USSR.

The first chapter of this monograph includes a short review of the most 
important results of theoretical and experimental research on the subject of 
 $\gamma -\alpha $ martensitic transformations. The theoretical difficulties 
arising at the description of a martensite nucleation and rapid growth of a 
martensite crystal will be discussed (in particular, the deficiencies of 
existing theories dealing with the wave model), also the discussion of the 
physical problem will be presented in detail.

Contents of Chapters 2 to 6 is the absolutely original and corresponds to 
the objectives of the aforementioned points 1 to 5. Briefly it is presented 
in a synopsis. In this synopsis, the conclusions playing a major role in 
understanding of the dynamical sequence of processes during the  $\gamma -\alpha 
$ martensitic transformation with a survey of results relating to a 
description of the martensitic nucleation process are presented. In 
addition, some objectives of proposed further research are outlined.

In order to understand this monograph, it is essential to read Chapter 1. 
However, some alternatives exist for the further sequence of reading. For 
instance, those readers only being interested in an explanation of the 
morphology of the products of the $\gamma -\alpha $  MT, can immediately skip 
over to Chapter 5. Chapter 2 is of interest for the experts in kinetic 
properties of metals and alloys. In Chapter 4, specialists in phase-diagrams and 
electronic structures of alloys can find new information related to the 
dependence between concentration and optimum temperature for the onset of 
the $\gamma -\alpha $  MT, from the point of view of the growth theory. None the 
less, the author recommends to read this monograph in sequence to those readers 
who want to thoroughly understand such a complicated process as the $\gamma 
-\alpha $ MT.

\chapter{Basic notions on the $\gamma -\alpha $ martensitic 
transformation in iron based alloys}

\section{Position of the $\gamma -\alpha$ martensitic 
transformation in relation to other solid state structural transformations} 

The proposition that the causal dependence between a structural state and 
physical properties of a material is fundamental for solid state physics and 
defines thus raising permanent interest and intensive research efforts on 
structural transformations (ST). From the point of view of symmetry-theory, 
the following two types of ST may be emphasized \cite{Iziumov1984}. The transformations are 
related to the first type (ST1) if the spatial group of one of the phases (as a 
rule, the low-temperature phase) is a subgroup of the initial phase symmetry 
group. They are called distortional transitions. The success of 
lattice-dynamical theories can mainly be attributed to the notion of 
"Soft-Phonons". Accordingly, the frequencies of certain phonons exhibit a 
strong temperature-dependence and can even vanish at temperature $T_{0}$ 
(Ms) owing to interactions in the system. At the critical point $T_{0}$, a 
macroscopic amount of zero-frequency phonons emerges, and the associated 
static rearrangements of atoms can be interpreted and visualized by the 
concept of Soft-Modes "frozen" in initial crystal. At present, this approach 
for ST-I is well elaborated and described in detail in the literature (i.e. 
\cite{Iziumov1984, Vaks1973, Gufan1982, Brus1984}). For this reason, we shall neither dwell on examples of ST-1 nor 
on details of the mechanism of the phonons softening, only remarking here that 
such structural transformations may proceed as second-order phase 
transitions (as a rule, the features of the first-order phase transitions 
are only weakly expressed). 

In contrast, for the structural transformations of the second type (ST-2) 
there is no subordination between the symmetries of the initial phase and 
the final phase. Phase transitions from the fcc to a bcc or to a close 
packed hexagonal phase are the examples of the ST-2 transitions. ST-2 exhibits 
the pronounced characteristics of the first-order phase transitions, namely, the 
significant thermal and volume effects and a temperature hysteresis (between the 
direct and reverse transformations). This type of transition is called 
"reconstructive".

A separate class (ST-3) of structural transitions belongs to the type 
"order-disorder", during which the atoms occupy mainly one of the 
quasi-equilibrium positions on temperatures below a transition temperature 
$T_{0}$ whereas at temperatures above $T_{0}$ 
different (not less than two) positions are occupied with the same 
probability. In addition to \cite{Iziumov1984, Vaks1973, Gufan1982, Brus1984}, an extensive survey of ST-3 
transformations is given in \cite{Vaks1983}.

As to the fourth type (ST-4) of structural transitions, it is possible to 
assign them as "symmetric-noncommensurate-commensurate" phase transition, in 
which an intermediate phase features a "super-structure" with a lattice period 
that is non-commensurate with the period of the parent structure (see more in 
detail \cite{Iziumov1984,Brus1984,Strukov1983}). 

The term "Martensitic Transformation" is used as a common attribute for 
diffusionless structural transitions proceeding by a cooperative 
rearrangement of the lattice structure. With such an extensive definition, the 
term MT would equally apply to the above defined ST-1, ST-2 and ST-4 
transitions, with specific exclusion of such ST-3 transformation, for which 
atomic displacements over distances comparable to the inter-atomic distances 
would be an indispensable prerequisite, typical of this transformation process. 
It should however be noted that the term "Martensite" originally was used for 
designation of the lamellar component in the microstructure of hardened steel 
\cite{Kaufman61}. If the extensive definition of MT encompasses the below mentioned totality 
of transformation characteristics, then we would rather prefer to use term MT 
in a more restricted sense, i.e. only for the sub-set of MT relating to 
diffusionless transformations of the ST-2 type. As mentioned in articles
\cite{Warlimont76,Warlimont77} it is necessary to refine
the terminology. Among the reasons for the 
insufficient present classification of MT are, on the one hand, its 
incompleteness, and on the other hand, the dissimilarities between the 
microscopic mechanisms of structural transformations of various types, whose 
clarification would require an utmost differentiation of MT from the outset. 
And vice-versa, in a lattice-geometrical or thermodynamic model of MT, it 
would suggest itself to strive for a standardized interpretation of all the 
numerous classes of MT.
 
If kinetic characteristics \cite{Kristian78, Umanskii78} were used for the classification of 
diffusionless spontaneous phase transitions, and if their speed of 
transformation was mainly determined by the velocity of their mobile phase 
boundaries, then the distinguishing features of MT would be their cooperative 
way of lattice transformation and their large velocity of growth (with a magnitude 
of about the sound velocity), being practically independent on temperature. In 
contrast, "normal" or polymorphic transformations are characterized by an 
alternative group of distinguishing features (like non-correlated migration of 
atoms over the phase boundaries, as well as relatively slow, thermally activated 
motion of phase boundaries).
 
Due to a missing of the standardized classification it is obvious to regard 
the existing approaches as complementary, each of them is focused on a 
different set of features and aspects of a MT.

Within the frame of this theoretical research work on the $\gamma -\alpha 
$ MT (related to fcc, bcc or bct lattice, respectively), particular 
attention will be paid to steel and iron-based (ferrous) alloys. Thus in 
this paper, the term "Martensitic Transformation" (MT) will only be used in 
its more restricted meaning, if not otherwise explicitly noted. In addition 
to its great practical importance, the choice of the $\gamma -\alpha$ 
MT as a subject of research has also been guided by the large amount of 
already available experimental data. As contemporary well matured 
lattice-geometrical and thermodynamical approaches are not able to provide a 
comprehensive notion of MT, there is an increasing need to extend them by a 
microscopic description, with the final aim to develop a dynamical model of 
the $\gamma -\alpha$ MT. Of course, the objectives of this work are 
more modest than the development of a complete microscopic theory of the 
$\gamma -\alpha$ MT. We shall formulate our objectives after a brief 
review of the main and well established notions on the $\gamma -\alpha$ 
MT in steel and ferrous alloys.
\vspace{\stretch{1}} 


\section{Characteristic features of the spontaneous $\gamma -\alpha$ 
  martensitic transformation}

There exist many monographs
\cite{Kristian78,Umanskii78,Kurdjumov60,Kurdjumov77,Lysak75,Krivoglaz77,Bernshtein83} and
survey papers (i.e. \cite{Kaufman61,Roitburd72,Roitburd1968,Roitburd78}) 
dealing with the history of research, as well as with basic aspects of the 
$\gamma -\alpha$ MT. For this reason there is no need here to dwell 
with history. We only would like to remark that the basis for quantitative 
research on the subject of $\gamma -\alpha$ MT has been founded in the 
twenties. In those days it had already been determined by means of X-ray 
investigations \cite{Seljakow27} that the low-temperature phase in steel (martensite) 
features a body-centered tetragonal (bct) lattice, whereas the 
high-temperature phase (austenite) features a fcc-lattice. A bct lattice is 
typical of interstitial alloys (Fe-C, Fe-N), in which tetragonality 
increases with increasing concentration of the interstitial component, 
whereas a bcc-lattice is more typical of substitutional alloys (Fe-Ni, 
Fe-Mn). Martensitic transformations of both kinds of alloys feature similar 
characteristics. As for their classification, we shall preferentially follow 
\cite{Kurdjumov77} and choose the systems Fe-C and Fe-Ni for a quantitative 
presentation. 

The definition of the martensitic transformation (MT) mechanism by G.W. 
Kurdjumov essentially postulates that a MT is a spontaneous process of 
lattice transformation following certain rules, one of them being that the 
relative migration distance of neighbor atoms does not exceed the 
interatomic distance of the parent lattice. An immediate conclusion from 
this definition is that a MT is as a phase transition without change of 
composition, analog to phase transitions in a one-component system. For 
1$^{st}$ order transitions, the equilibrium-temperature $T_{0}$ of phases 
\cite{Fridkin79} is determined by the Helmholtz-Equation of free energy F: $F_{\gamma} 
(T_{0}) = F_{\alpha} (T_{0})$. Thus the energy to be invested 
for creation of a new phase-boundary always requires a given amount of 
supercooling below $T_{0}$ (or, conversely, of superheating above $T_{0}$, 
during the reverse transition). Typical measured values of the 
temperature-hysteresis between these two transformations are quite 
substantial and amount to ($M_{S}  - A_{S})  \simeq 400$ K, 
with similar magnitudes of supercooling and super-heating, respectively: 
$T_{0} - M_{S} \approx A_{S} -  T_{0} \approx 200$K. 
Here $M_{S}$ is the starting temperature of the $\gamma -\alpha $ MT and 
$A_{S}$ the starting temperature of the reverse $\alpha -\gamma $ MT . 
The change of specific volume also is considerable and has been assessed by 
different authors up to values within a range of 2 to 5{\%}, where the 
larger figures of the specific volume effect (per unit weight) are more 
characteristic of the $\alpha$ -phase. The specific heating effect \cite{Vinnikov74} 
during a MT amounts to some hundreds of calories per mol, resulting in an 
increase of the temperature of a sample by some dozens of centigrades 
\cite{Vinnikov74,Robin1977}. 

Let us now enumerate some of the indisputable structural (morphologic) 
characteristics of the $\gamma -\alpha$ MT.
\begin{enumerate}
\item{Martensite emerges in the shape of lamellae featuring a low ratio of 
thickness to the other of their characteristic linear dimensions, or in the 
shape of lenticular crystals, in the middle of which a lamella (midrib) sets 
off in the first stage of a MT. The thickness of such lamellae varies 
between 10$^{-7}$ and 10$^{-6}$ m (i.e. from 0.1 to a few $\mu$m).}

\item{The habit-plane of a lamellae (i.e. the phase-boundary or flat 
midrib-boundaries) has some stable orientations (depending on constitution), 
in relation to the crystallographic axis of the $\gamma$- and  $\alpha 
$-phases. In the systems Fe-C, Fe-Ni, there have been observed habit-planes 
coming close to the \{557\} $\div$ \{111\} (up to 0,6 weight{\%} 
C, up to 29 {\%} Ni), \{2 2 5\} - (0,6 $\div$ 1,4 weight.{\%} C), 
\{2 5 9\} $\div$ \{3 10 15\} - (1,4 $\div$ 1,8 weight.{\%} C, 
29 + 34 {\%} Ni ). \footnote{In the following, the crystallographic terms 
without notation will relate to the axis of the $\gamma $-parent-phase (i.e. 
three right-hand vectors along the axis of symmetry of the 4$^{th}$ order 
$\langle 001 \rangle$).} }
\item{The shape of the transformed areas changes with the emergence of a 
surface-relief, the shape of which being characterized by distinct 
macroscopic shear-parameters. This latter feature suggested the commonly 
used designation "Shear-Transformation" as a synonym for MT.}

\item{There exists an orientational relationship among the $\gamma$- and the 
$\alpha$-phase, indicating parallelism (or approximate parallelism) among 
the densely packed phase-planes: $\{111\}_{\gamma} \; \Vert \;
\{110\}_{\alpha}$ and among angles of rotation in relation to the 
orientation of the densely packed parallel planes.}

\item{Martensite crystals exhibit an ordered internal structure, and in many 
cases a more or less ordered relative arrangement. This way, crystals with 
habit-planes \{557\} $\div $ \{111\} are characterized by a rather 
complicated lattice displacement structure. They form colonies (packets) of 
crystals with almost equal orientation, whereas for crystals with habit-planes 
\{225\}, \{259\} $\div $ \{3 10 15\}, the formation of internal 
transformation twins and other crystal arrangements are more typical (more 
detailed in \cite{Izotov1972}).}
\end{enumerate}

An important conclusion raised in \cite{Bernshtein83} states that " \ldots In a single crystal 
of martensite, its habit-plane incorporates a substantial aspect of an 
orientational relation, being closely linked to the macroscopic shear 
parameters or, in other words, only a single way of martensitic 
transformation corresponds to a given habit-plane." 

Any classification of the kinetic characteristics of MT should therefore 
clearly distinguish between micro-kinetics, which characterize the growth of 
a single lamella, and macro-kinetics, which characterize the increase of the 
bulk amount of martensite in a sample. 

In terms of micro-kinetics, a MT can be classified as athermal, because 
firstly, the speed of transformation and growth of a single lamella is very 
large (in the order of magnitude of the speed of sound) and independent on 
temperature, even though the MT may occur within a wide range of temperature 
T - $(0 \div  10^3)$ K, and secondly, because the transformed 
volume is determined by the emergence of discrete new crystals and not by 
the steady growth of previously existing ones. Thus the growth proceeds with 
almost non-existent thermal activation, giving reason for the attribute 
"athermal". 

In terms of macro-kinetics, it can be distinguished between isothermal and 
athermal MT. Isothermal MT can proceed at constant temperature, attainable 
by an isothermal high-capacity heat receptacle (in relation to the transforming 
volume), while there exists a characteristic temperature (depending on the 
composition of an alloy) at which the rate of increase of the macroscopic amount 
of martensite is utmost. If significant deviations from the optimum temperature 
are forced in (i.e. by rapid quenching), then isothermal MT can be suppressed. 
This category of macro-kinetics is typical for the formation of 
packet-martensite. 

If athermal macro-kinetics predominate, then the amount of martensite is 
determined by the degree of supercooling below $M_{S}$, while isothermal 
stops of the quenching process would not cause any increase in the 
transformed volume. These characteristics are typical of the growth of 
crystals with habit planes \{259\} $\div$ \{3 10 15\}. The extreme 
case of athermal growth is the "explosive" MT, during which an appreciable 
fraction (some dozens of percents) of the total transformed volume is 
produced almost instantaneously within a short "explosive" period. 

Remarkably, there have also been discovered alloys featuring both types of 
reaction kinetics \cite{Georgieva1969}. Though, as a rule, the athermal martensite 
formation normally proceeds after isothermal (during the cooling process), the 
observation of the reverse sequence has also been reported in \cite{Zambrzhitzkii1983}. 

In spite of the crucial importance of data required for a precise 
determination of the speed of growth of martensite crystals, as well as of 
the transformation mechanism, only few experimental data with quantitative 
details like those shown in Table \ref{tab1} have been published.

\begin{table}[htbp]
\renewcommand{\captionlabeldelim}{.}
\caption{Experimental data on the speed of growth of martensite}
\begin{center}
\begin{tabular}{|c|c|c|}
\hline

Material  & Speed of growth $10^{3}$ m/s & Source  \\ 
 \hline

Fe - 29,5 {\%} Ni  & 1 & \cite{Bunshah1953}  \\ 
\hline
Fe - 30 {\%} Ni  &  (1,8 $\div $ 2)  &  \cite{Mukherjee1968}  \\
\hline 
& 0.1  & \cite{Robin1977} \\
Fe - 32 {\%} Ni  &  1.1 & \cite{Robin1982} \\
& 0.2  &  \cite{Robin1982}
\\ \hline                
Fe - 0,35 {\%} C - 8 {\%} Mn  &  6,6  &  \cite{Lokshin1968,Lokshin1967}  \\
 \hline 
Steel Type 18-8  &  (0,1 $\div $ 0,2)  & \cite{Takashima84}      \\
\hline
\end{tabular}
\end{center}
\label{tab1}
\end{table}

Let us now comment the data shown in Table \ref{tab1}. Obviously, the 
emergence of a single martensite crystal immediately changes the electrical 
resistance of a sample and, in turn, triggers an electrical signal which can be 
registered by means of a simple oscilloscope, as reported in \cite{Bunshah1953}. The speed v = 
$l  \cdot \tau^{-1}$ has been determined from previously known values of 
the signal period $\tau$ and the grain size $l$. Basically in the same way, the 
speed v has been estimated in \cite{Mukherjee1968}. The different order of magnitude of the data 
presented in \cite{Robin1977} can mainly be deduced from the different experimental method 
of determination of the signal period $\tau $, which in this case was obtained 
by observation of the change in reflectivity of the sample, being caused by a 
dull surface relief showing up simultaneously with the MT and remaining engraved 
in the aftermath. Instead of a characteristic timescale $\tau \approx 10^{-7}$ 
s, as used in \cite{Bunshah1953,Mukherjee1968}, a timescale of $\tau \approx 5 \cdot 10 ^{-6}$ s 
was used in \cite{Robin1977}, which however is not directly related to the growth of a 
single crystal, but to the "explosive" bulk growth of the crystals as a whole. 
It has been clarified in \cite{Robin1982} that the speed of $2 \cdot 10^{2}$ m/s only 
relates to the lateral speed of crystal growth, while a radial (frontal) speed 
of growth of an order of magnitude of 10$^{3}$ m/s could be confirmed. In \cite{Bunshah1953}, 
the potential coexistence of two different, superimposed stages of growth of 
martensite crystals has been postulated for the first time, based on signal-form 
analysis. In another analysis of \cite{Bunshah1953} by Arskij, published in \cite{Arskii66}, it was 
further asserted that the initially upshooting edge of the signal is caused by 
the rapid formation of a two-dimensional crystal nucleus, in conjunction with 
radial growth. We note that the surface of such a nucleus represents an 
inhomogeneity, immediately prompting an increase in electrical resistance. Based 
on this assumption, it can be inferred that the observed declining signal edge, 
having a period 5 to 6 times as large as that of the increasing edge, can be 
used as a natural time gauge for lateral growth of the nucleus (bearing in mind 
that the specific resistance of martensite is less than that of austenite). 
Based on such differentiation of growth stages and on data published in \cite{Bunshah1953}, 
Lokschin \cite{Lokshin1968,Lokshin1967} obtained a value of $7 \cdot 10 ^{3}$ m/s for radial growth 
velocity, clearly exceeding the velocity of longitudinal sound waves in 
austenite. This evaluation was substantiated in \cite{Lokshin1968,Lokshin1967} by measurements of the 
propagation of detonation waves in steel. In the case where the MT was initiated 
by an external detonation, the speed of wave propagation was $(6,5 \div  6,6) 
\cdot 10 ^{3}$ m/s. By contrast, in cases where a MT in steel was induced by a 
precooling, the speed of the detonation wave was remarkably lower: $(4,8 \div  
4,9) \cdot 10^{3}$ m/s. Assuming a measurement error of 4 {\%} in \cite{Lokshin1967}, this 
difference can be interpreted as an indication of a large proper speed of the 
transformation. Even though in \cite{Mukherjee1968} no reason is given for a differentiation of 
the growth mechanism into two different stages, the observations reported in 
\cite{Sadovskii78,Schastlivtsev81,Schastlivtsev83}, relating to the nucleation of 
thin martensite lamellae being triggered 
by a strong magnetic field, where an initially rapid nucleation process was 
followed up by a significantly slower growth stage, convincingly support the 
assumption that a differentiation or classification of the growth mechanism is 
realistic. Without doubt from a methodological point of view, such nucleation 
and growth phenomena would be most promising for further experimental research 
aiming at the measurement of frontal velocity of growth, by systematic 
measurement and analysis of signals. And finally, it should be noted that the 
data reported in \cite{Takashima84}, which were obtained by evaluation of acoustic signals, 
only pertain to the average speed of growth of martensite crystals and not to 
that of a single lamellae, as would be required. Thus, on the basis of 
experimental data, an order of magnitude of $10^{3}$ m/s for radial (frontal) 
 growth of a martensite crystal during the $\gamma -\alpha $ MT presumably 
would be realistic. The results of direct experiments reported in \cite{Lokshin1968,Lokshin1967} 
further support the assumption that the frontal velocity of growth is greater 
than the velocity of longitudinal elastic waves. 


\section{Lattice-stability near the $M_{S}$-temperature and the 
problem of nucleation during the $\gamma -\alpha $ martensitic 
transformation}

The pronounced characteristics of a first-order transition of the $\gamma 
 -\alpha $ MT suggest to assume the separate existence 
of a nucleation and of a growth stage of the new phase. Thus an important 
question is related to the stability of the austenitic lattice at 
$M_{S}$ - temperature, being closely related to the question of the possible 
existence of a temperature $T_{C}$ at which the stability of the austenitic 
lattice would completely vanish. In fact, the existence of an absolute 
temperature $T_{C } > 0$, at which a supposedly perfect lattice 
becomes unstable with respect to infinitesimally small fluctuations, will lead 
us to the conclusion that a MT is almost inevitable during the cooling process. 
Due to an obvious disparity indicated by the inequality $ M_{S} > T_{C}$, it can be 
inferred for alloys with isothermal reaction kinetics, which 
can normally be suppressed by sufficiently rapid quenching below $M_{S}$, that a 
distinguished $T_{C}$ - temperature is missing. This in turn is a clear 
indication of metastability (relative stability) of the parent lattice at 
temperatures comprising not only the $T_{0} > T \geqslant 
M_{S}$ region, but also the $T < M_{S}$ region. An important 
characteristic of alloys with athermal transformation kinetics, being mentioned 
in most relevant publications, is the independence of $M_{S}$ from the quenching 
rate, which can be interpreted as a clear indication of the impossibility of 
supercooling below $M_{S}$. However, according to experimental results 
reported in \cite{Serebriakov77}, it has also been possible to supercool an Fe -33,7{\%} Ni 
alloy down to 4.2 K and to trigger an "explosive" MT only in a subsequent 
reheating process, which also included some isothermal holding stages. 
Further indication of the non-existence of an unequivocal $T_{C}$ for 
athermally transforming alloys may also be inferred from the character of 
incompleteness, being typical of them: The transformation process ceases at 
a temperature $M_{f}$, with $0 < M_{f} < M_{S}$, whilst 
subsequent cooling will not lead to any further growth of martensite, in spite 
of the existence of a substantial volumetric fraction ( up to some dozens of 
percent) of residual, non-transformed parent phase. 

Investigations on the temperature-dependence of elastic moduli published in 
\cite{Haush73,Delaey79}, as well as of dispersion curves of phonon-spectra 
\cite{Haush73,Delaey79,Hallman69,Endoh79} show 
that, in most cases, not even a faint tendency towards loss of lattice 
stability is evident. The observed abnormal course of the 
temperature-dependency of elastic moduli \cite{Haush73} of alloys of the 
Fe-Ni-systems ($> 30${\%} Ni), which already attain a ferromagnetic 
state of order for $T > M_{S}$, can easily be attributed to the 
effect of ferromagnetic ordering \cite{Zverev72,Zverev84}, instead of lattice instability. 
This means that in general, a normal temperature characteristic of the elastic 
moduli of ferromagnetic systems is the case. In other words, an austenitic 
lattice essentially maintains its inherent stability against certain 
phonon-modes of small amplitude. 

The results of experiments with small particles (diameter $ > 10^{-6}$ m),
reported in \cite{Cech56}, convincingly demonstrated that when particles of 
similar size are cooled down, a MT only occurs in a certain fraction of the 
involved particles, this being further evidence of the heterogeneous nature of 
martensite nucleation in certain preferenced centers. The same conclusion 
can also be drawn from observations of thermo-cycling, during which repeated 
(cyclic) growth of parent crystals and of martensite crystals takes place at 
the same location. A variety of electron-microscopical observations 
documented in \cite{Vintaikin83} give evidence of an intimate relationship between 
dislocations and the process of martensitic nucleation. 

Even though there is no doubt in general on the heterogeneous character of 
nucleation, there exists a variety of experiments related to the microscopic 
mechanism of nucleation \cite{Iurchikov71}, like diffuse electron- and X-ray scattering 
\cite{Tiapkin76} (see also \cite{Vintaikin83,Kondrat'ev82}), and measurements of reactive forces against 
microscopic deformations at temperatures near $M_{S}$ \cite{Tiapkin76}, all of them 
providing evidence of a special state of the austenitic lattice prior to its 
transformation, manifesting itself by an anisotropic increase of the squared 
mean of the oscillatory amplitudes of lattice modes with wave-vectors of 
selected orientation (and of selected polarization), as well as in a 
reduction of the mechanical resistance to small deformation \cite{Sarrak82}. In the 
case of a perfect lattice being free of phonon-mode softening, this 
particularity can be interpreted as the result of non-linear (anharmonic) 
mode-interaction, associated with the emergence of isolated, near-order 
regions of displacements (NORD), as proposed in \cite{Kondrat'ev79}. 

The investigations performed in \cite{Kondrat'ev79} mainly deal with constitutive 
thermodynamic aspects of MT (Investigation of lattice stability against 
quasi-static lattice deformations, as modeled by a package of "frozen-in" 
waves). Obviously, the more normal situation would be characterized by a 
transformed spectrum of elementary lattice oscillations, leading to the 
emergence of movable near-order domains of the soliton type \cite{Falk84}, their 
stability being endorsed by the accumulation (localization) of electrons in 
the vicinity of isolated NORD (in a similar way as in the case of the 
fluctuational model in \cite{Krivoglaz73}). From our point of view, a reasonable 
qualitative notion of martensite formation, comprising both heterogeneous 
and homogeneous nucleation, would have to be a nucleation model encompassing 
the basic process of NORD -localization in the vicinity of inhomogeneities 
produced by single or grouped dislocations. Obviously this point of view is 
both in accordance with pure thermodynamical concepts \cite{Kristian78,Roitburd81}, because 
any weakening of the binding energy in the proximity of a lattice defect 
would compensate for the increase in energy related to the creation of a 
boundary area, as well as with the Pinning-effect \cite{Gurevich84} in non-linear lattice 
dynamics. This model thus is closely related to the "reactive path" model by 
Cohen, Machlin and Paranyan \cite{Kaufman61} in that: 
\begin{enumerate}
\item{A coordinated motion of atoms proceeds by means of "rapid sequential 
change of intermediate structures in a given range\ldots" growing "\ldots in a 
similar way as the propagation of an elastic wave".} 

\item{"The activation occurs by means of fluctuating rearrangements of the 
atomic configuration within the nucleus, and not through changes of its 
size"}

\item{Nucleation occurs in the centers of lattice distortion, including 
dislocation rows.}
\end{enumerate}

An important feature reported in \cite{Kondrat'ev79} is the possible loss of stability of 
NORD, which could be related to their increase in size in conjunction with 
partial lattice-softening. However, as local softening is always existent in 
the vicinity of a dislocation, the lateral dimension $2 \cdot r_{0}$ of a 
settling NORD must be larger than the dimension of mobile NORD without defect. 

In pure dislocation nucleation models (e.g. in \cite{Petrov78}), the $M_{S}$ 
-temperature is put into relation with the emergence of the first 
dislocation loop, whereas the creation of follow-up dislocation loops is 
supposed to occur spontaneously. Subsequently, a rapid nucleation process up 
to macroscopic dimensions would be possible by means of expansion of such 
dislocation loops, at a speed close to the speed of sound. With the 
assumption that a nucleus would have the following main linear dimensions: 
Length as determined by the straight section of the dislocation line, 
lateral diameter $2 \cdot r$, for $r\approx(10 ^{-6}\div 10 ^{-7})$m as a 
realistic gauge for the thickness of a typical lamella, it is possible to 
assess the period of macro-nucleation $t_{N}$ by $t_{N} \approx r\cdot  c ^{-1}
\approx (10^{-9}\div  10^{-10})$ s, where $c$ is a speed close to the speed of 
sound. A model based on the notion of localized NORD thus would have to presuppose 
an initial quantity $r_{0}$ of a NORD, being significantly smaller than $r$ (\cite{Kondrat'ev79} 
assumes $r_{0} \approx 10^{-9} $ m), whilst the event of NORD-localization would 
simultaneously stimulate the creation of the first dislocation loop. The follow
up expansion of these loops would then be characterized by synchronized growth 
from $r_{0}$ to a macroscopic size $r$. It should however be considered that the 
NORD-model as defined in \cite{Kondrat'ev79}, is inherently confined to short wave length displacements.
For this reason, the existence of soliton solutions with dimensions $ r_{0}$ of about
the order of magnitude of $r$ cannot generally be excluded. The transformation 
of NORD from their initial size $r_{0} \approx 10^{-7}m$ into a macro-nucleus takes 
place within a period of $ t_{N} \approx 10 ^{-11}$ s, as the participating atoms would
only have to overcome a distance of about $\varepsilon \cdot $ r, within a 
period $t_{N}$, where the quantity $\varepsilon \approx 0,1$ suggests itself to 
be taken as a natural gauge of relative deformation during the $\gamma 
-\alpha $ MT \cite{Kashchenko84}. We have to remark that the usually assumed shape of 
the nucleus as a thin, flattened spheroid was mainly determined, on the one 
hand, by thermodynamical reasons, and on the other hand, by a presupposed 
inherent similarity between the initial shape of the nucleus and the 
lenticularly shaped final crystal. The last circumstance led to a nucleation 
growth model in the pattern of expanding dislocation loops, where the edges of 
the growing crystal would feature mono-atomic thickness in radial orientation. 
The model of a slowly thickening spheroid for lenticularly shaped crystals 
requires some further explanation related to its central lamellar zone - the 
midribs - whose constant thickness can, for good reasons, be regarded as a 
characterizing gauge of the bulk structure. In terms of thermodynamics, this 
would be analog to a gauge for critical crystal thickness \cite{Roitburd68}, being 
related to phenomena of transition between different mechanisms of 
additional deformation (i.e. low thickness by twinning vs. larger thickness 
by slipping). Thus, in principle, the creation of midribs and their 
expansion from mono-atomic size to their final dimensions of $r \approx 
10 ^{-6} \div 10 ^{-7}$ m would be 
feasible. 

Another alternative might be the above mentioned possibility for rapid 
emergence of a macro-nucleus, whose dimension $r$ determined the length of 
lattice-displacement waves excited under non-equilibrium conditions. Thus, 
if a mechanism for effective maintenance of the wave amplitude would exist 
in such a way that the amplitude $\varepsilon$ of deformation 
associated with these waves exceeded the threshold-value required for 
initiation of the martesitic transformation, then the growth had to proceed in a 
wave-mode, with characteristic structural gauges of the macro-nuclei 
corresponding to certain wave-lengths. Obviously, the notion of 
non-equilibrium conditions, being typical of the nucleation process, will 
constitute the focus of such a model. It is fairly obvious that the highly 
pronounced non-equilibrium conditions prevailing in the vicinity of a 
rapidly growing phase can only be maintained if the process of heat 
liberation is adiabatic \cite{Krizement61,Mogutnov72}. Then, the temperature difference between 
the $\gamma $ - and the $\alpha $ - phase $\Delta T  \approx Q \cdot C_{sp}^{-1}$ 
(where Q - heat energy density, $C_{sp}$ - specific heat), would be comparable 
with the magnitude of supercooling $T_{0} - M_{S}$. This in turn allows to
refine the formulation of our basic task: 

\textbf{Investigation of a non-equilibrium electron-phonon system with the 
final aim of clarification and substantiation of the principles and 
conditions for efficient generation of lattice-displacement waves}. 

In the above outlined qualitative pattern, the transition from nucleation to 
the growth stage features some characteristics of a continuous transition, 
thus being an indication of a potential instability with respect to rapid 
growth of nuclei emerging under non-equilibrium conditions. Such a 
conclusion would however have to presuppose the existence of effective rapid 
growth mechanisms. This latter circumstance may thus play the crucial role 
in the transition process of the initial system into a new state. In fact, 
the following basic rule of non-stationary thermodynamics applies: If a 
given system can potentially exist in different states, each of them having 
a relative minimum of free energy, then " \ldots those states emerging at the 
greater speed will be realized first". Just this way, the kinetic factor 
comes to the fore, which is quite natural in a growth stage. Even though the 
kinetic aspect of the transformation does not belong to the classic realm of 
thermodynamics, there exists a link between the thermodynamical and the 
kinetic conditions of such phase generation. In our case however, we shall 
confine the role of thermodynamics to the determination of the thermodynamic 
stimulus of the transformation, and of the activation energies required for 
nucleation and growth.

The missing quantity $T_{0}$, and therefore the asserted relative stability 
of the fcc-lattice at $M_{S}$ -temperature, implies a question related to the 
magnitude of the energy barrier, which has to be overcome by the system 
during the processes of nucleation and growth. Obviously, this question is 
closely related to the problem dealing with the large observed magnitude of 
supercooling $T_{0} - M_{S} \approx  200$ K, which has 
been clearly formulated in \cite{Krizement61,Mogutnov72}. This specific problem arises from a 
consideration of the remarkably small temperature difference $\Delta 
T = T_{0} - M_{S} \approx 45 $K, being required 
to satisfy the energy requirement of $E_{e} \approx 50$ kal/mol to 
carry out the exterior mechanical work during a MT, and of the great 
observed magnitude of the supercooling. One of the acceptable explanations 
stated in \cite{Kaufman61} is that "\ldots the kinetics of the process is determined by the 
speed at which the transient system overcomes the energy barrier associated 
with the formation of the nucleation center of the new phase, as further 
growth of martensite crystals proceeds almost instantly." Thus, any 
explanation of the magnitude of supercooling $T_{0} - M_{S}$ must 
inevitably consider the key thermodynamic and kinetic process parameters. 

In the above discussed model of adiabatic generation of a macro-nucleus, the 
heating effect Q associated with the transformation, treated in \cite{Kashchenko84}, can be 
interpreted as a result of the distribution of atomic oscillation energy E 
in the vicinity of new equilibrium states, among all available $3 \cdot  N_{N}$
oscillatory modes of the nucleus ($N_{N}$ - number of atoms in the nucleus), 
with only a small number of polarization orientations 
excluded (for anisotropic deformation). We evaluate E by the assumption that 
during the process of nucleation, the excitation of longitudinal 
oscillations being polarized in two orientations (called x and y), will be 
predominant. Let an oscillatory mode be enumerated by a wave-vector of 
components $\textbf{q}_{i}$ ($i=x,y$). Then, their contribution to E will be

\begin{equation}
\label{eq1}
E_{\mathbf{q}_{i}} = \frac{1}{2}\, M \,\omega_{\mathbf{q}_{i}}^{2} 
\arrowvert {\mathbf{u}_{\textbf{q}_{i}}} \arrowvert^{2} N_{N},
\end{equation}
where M - atomic mass, $\omega_{\textbf{q}_{i}}$ - cyclic frequency, $\textbf{u}_{\textbf{q}_{i}}$ - oscillatory 
amplitude. 

Further, the relative deformation $\varepsilon$ associated 
with such oscillations can be determined by the extreme condition

\begin{equation}
\label{eq2}
\varepsilon_{m} = \frac{2 \,u}{\lambda/2} 
= \frac{2 \,u\, q}{\pi},
\end{equation}
where $\lambda$/2 is the distance between two adjacent planes, 
supposed to oscillate in mutually opposed phases ($\lambda $  --
wave-length). We further assume the same $\varepsilon$ for  all $q_{i}$,
replacing $\omega_{\textbf{q}_{i} } $ by $\omega_{\textbf{q}_{i }} = c\, q_{i} $,  where $c$ -
speed of sound, and consider that $q_{x}q_y $ attain the value  $N_{Nx}
\approx N_{N y} \approx 2 \,r_{N} /a$ ($r_{N}$ - radius of the
macro-nucleus, $a$ - lattice  parameter). Then from \eqref{eq1} and
\eqref{eq2}:

\begin{equation}
\label{eq3}
E = \sum_{i,\textbf{q}_{i}} {E_{\textbf{q}_{i}}} \approx \frac{\pi^{2}}{2a}
r_{N} 
N_{N} M c^{2} \varepsilon_{m}^{2}.
\end{equation}

At the instant of nucleation $t = t_{N}$ we get

\begin{equation}
\label{eq4}
\frac{E}{N_{N} M} \approx \frac{\pi^{2}}{2}\,r_{N} \,c^{2}\, \varepsilon 
_{m^{2}}  \approx Q,
\end{equation}
 where Q is the specific transformation heating effect [J/kg]. Using
$Q \approx 4\cdot 10 ^{4}$ J/kg (for a Fe-30 Ni - alloy 
according to \cite{Vinnikov74}) we get $r_{N} \approx 10^{-7} m$, $a  \approx 3 \cdot
10^{-10}$ m, $c = 5 \cdot 10 ^{3}$ m/s and from (\ref{eq4}): $\varepsilon _{m} \approx 10^{-3}$.

This way we can reasonably infer a threshold-value for the deformation 
$\varepsilon_{th}$ of about a magnitude $\varepsilon_{m} \approx 
10 ^{-3}$, at least for long-living modes with quasi-momenta 
 $q \approx q_{min} \approx  \pi \cdot r_{N}^{-1}$ at 
the instant $t=t_{N}$, coinciding with the final instant of the nucleation stage 
and with the moment of growth initiation. With respect to the effect of external 
stress on MT, it would be useful to compare our estimate of 
$\varepsilon_{m}$ with data published in \cite{Kurdjumov77,Sarrak82}. 

The reasons for such comparison - for instance with data of the strain 
effect - are obvious, if we consider that the term $\lambda$/2 in 
(\ref{eq2}) plays the role of the initial length $l$ of the sample, while
the  amplitude $u$ corresponds to the incremental length $\Delta l$,  and
that the quantity $\lambda_{max} = 2 \cdot \pi /q_{min} \approx (10^{-7} 
\div 10 ^{-6})$ m is a multiple of the lattice parameter $a$. In 
conclusion, the order of magnitude of these quantities appears to be fairly 
adequate for a description by continuum mechanics. Useful information can
further  be inferred from the mere existence of a temperature $M_{elast}$ 
(in the notion of \cite{Sarrak82} $M_{elast}$ is close to MS, i.e. $M_{elast} - M_{S}
\approx  35 \div  40$ K), the quantity $M_{elast}$ being such that if a MT 
proceeded within the temperature range $M_{S} < T < M_{elast}$ it could be  
induced by elastic stress $\sigma_{m}$, thus satisfying the inequality 
$\sigma_{m} < \sigma_{y}$ ($\sigma_{y}$ -- yield stress). Remarkably,  
an evaluation of the stress-strain curve shows that the values of $\sigma   
\le \sigma_{y}$ correspond to strains $\varepsilon_{th} \le  10 ^{-3}$.  
This observation may further support our estimate of $\varepsilon_{th}  
\approx  10^{-3}$ for threshold deformation. Thus in the notion of a developing  
macro-nucleus, supposed to be inherently unstable in the growth process, a  
strongly pronounced supercooling $T_{0} - M_{S}$ is regarded as an  imperative 
condition for the excitation of oscillations with final (peak)  amplitudes 
(and frequencies) satisfying the condition $\varepsilon >  \varepsilon_{th} 
\approx  10^{-3}$. Moreover, the same condition could be used as a qualitative 
criterion for lattice stability (or instability) of the  $\gamma$ - phase at the
$M_{S}$ - point.

 
\section{Solved and unresolved problems on the theoretical description of the 
$\gamma -\alpha $- martensitic transformation}

Any comprehensive reasoning on the theory of phase transitions will inevitably 
lead to the following two fundamental questions, as noted in \cite{Kristian78}: 
\begin{enumerate}
\item{Why does a given phase transition occur?}

\item{Which is its transformation mechanism?}
\end{enumerate}
The first question will normally be approached by investigations on the 
relative stability of different possible phases. In this case, the phase 
with lowest value of free Gibbs-Energy G(P,T) will usually be in the focus 
of interest. A generalized expression for G is given by the constitutive 
expression in \cite{Landau76}:
\begin{equation}
\label{eq5}
G(P,T) = H - TS = U + PV - TS                       
\end{equation}
where H, U, S, V are the commonly used specific values of enthalpy H, internal 
energy U, entropy S, volume V and pressure P (relating to 1 mol). 

A calculus of the (macroscopic) thermodynamic quantities in \eqref{eq5} within the 
frame of a strict (quantum-statistical) approach would be an extraordinarily 
complicated task, as the G-values of different phases differ almost 
inappreciably, and to the worse, any such calculus cannot be effectively 
simplified. Nonetheless, today there exist sufficiently clear and 
substantiated notions on the causes and mechanisms of polymorphic 
transitions in pure iron. 

Presumably, the high-temperature transition fcc-bcc ($\gamma -\delta$) at 
$T_{\gamma - \delta}$ = 1665 K is determined by the entropy term in \eqref{eq5}, 
as the contribution to S from atomic oscillations in the more weakly bound 
bcc lattice, as well as from disordered (large) magnetic moments, is 
relatively larger for the $\delta$-phase than for the $\gamma$-phase \cite{Ziner70}. 
The low-temperature transition fcc-bcc ($\gamma -\alpha$) at $T_{\gamma - 
\alpha }$ = 1183 K, which would appear paradox from the aspect of the 
oscillatory contributions of the atoms to S, can be put into relation with 
the reduction of internal energy of the $\alpha$-phase in the ferromagnetic 
state of order \cite{Dovgopol82,Miodownik82,Hasegawa83}. Taking for granted that the answers are already 
given, we shall refrain from dealing here any more in detail with questions 
related to phase stability. 

Of course, a constitutive solution to our initial question "Why"  related 
to the $\gamma -\alpha $ MT  has not yet been found, as the second part 
of our question, being related to a "given phase transition", belongs to the 
particularities of the $\gamma -\alpha $ MT and is intimately linked up 
with the question for the mechanism of this particular transition. 

\begin{figure}[htb]
\centering
\includegraphics[clip=true,width=.8\textwidth]{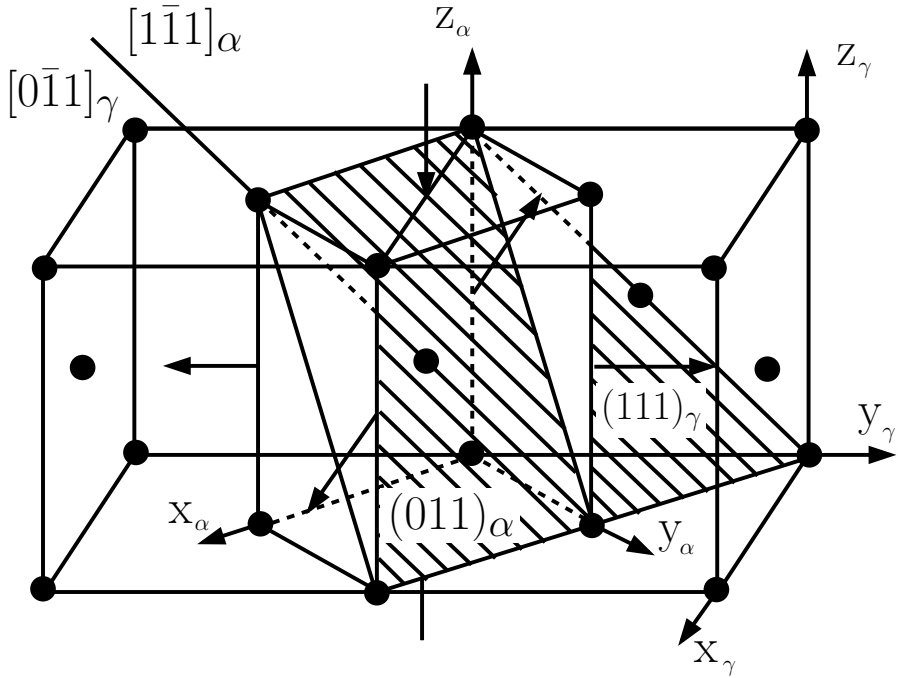}
\renewcommand{\captionlabeldelim}{.}
\caption{Conformance of austenitic and martensitic planes during a 
fcc $\to$ bcc Bain- transformation [13]. Planes (111)$_{\gamma } 
\|(011)_{\alpha }$ are hatched. Bold arrows point to orientations of 
compression [001]$_{\gamma }$ and tension [100]$_{\gamma }$, [010]$_{\gamma 
}$, respectively.}
\label{fig1}
\end{figure}

\begin{figure}[htb]
\centering
\includegraphics[clip=true,clip=true,width=.8\textwidth]{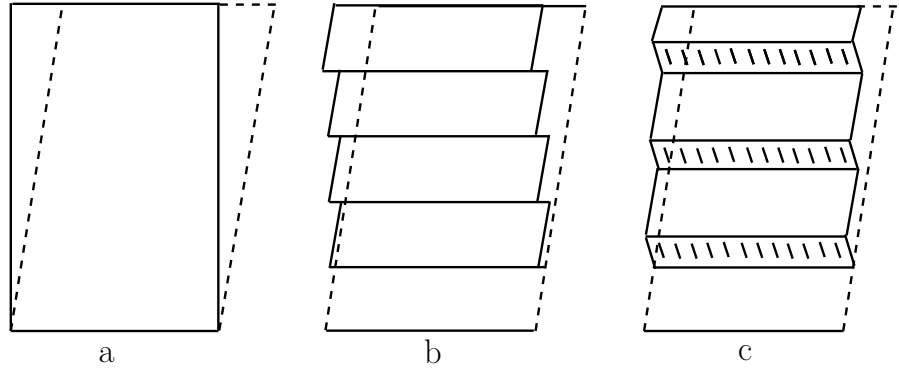}
\renewcommand{\captionlabeldelim}{.}
\caption{Reconstruction of the shape of a transformed region by 
additional deformation [13]: (a) = Mode of transformation; (b) = 
Reconstruction of the shape by means of slipping; (c) = Reconstruction of 
the shape by twinning.}
\label{fig2}
\end{figure}

The morphological (phenomenological) features of low-temperature martensite 
with habit planes $\{2\ 5\ 9\} \div \{3\ 10\ 15\}$ (see 1.2) have 
successfully been interpreted by the crystallographic theory. A detailed 
description of its most important results is published in
\cite{Kurdjumov77,Roitburd1968}. From a 
crystallographic point of view, MT are regarded as a lattice deformation 
process, characterized by a macroscopically invariant plane, the habit 
plane. This deformation includes pure lattice-distortion (PLD), as well as 
lattice invariant deformation (LID). PLD can either be described within a 
pattern including two (or more) shear motions, onto which a dilatation 
mechanism is superimposed, or within the pattern of Bain-Distortion (BD) 
with consecutive rotations satisfying the constitutive orientational 
relationships. The BD-pattern is shown in Fig. \ref{fig1}, which has been 
extracted from \cite{Kurdjumov77}. Imagine an elementary bct-lattice cell (with a 
tetragonality of $\sqrt 2$) being partially configured by two adjacent 
fcc-cells. Then, in order to transform that bct-cell into a cubic bcc-cell, the 
bct-cell has to be subjected to a compression by 20 {\%} along its [001] axis 
and to a dilatation by about 13 {\%} along its [100] and $[0 \bar{1} 0]$ axis 
(or along its [110] and $[1\bar {1}0]$ axis). In Fig.\ref{fig1}, the 
principal orientations of compression and dilatation are marked by bold arrows. 
The densely packed $(111)_{\gamma }\Vert (011)_{\alpha }$ plane is hatched. 
Fig.\ref{fig2} shows an inhomogeneous lattice resulting from LID in the 
case of shearing Fig.\ref{fig2}a and twinning Fig.\ref{fig2}b. It 
is mandatory to consider LID, because pure BD does not comprise a 
deformation with an invariant plane (DIP) (during DIP, one of the main 
components vanishes, whereas the two remaining deformation components 
feature opposite sign). Thus, if LID is used as a method to introduce 
auxiliary deformation, it is possible in some cases to attribute the bulk 
macro-deformation of the transformed region to a deformation with an 
invariant plane. Among other aspects, the twinning mechanism of 
low-temperature martensite was used to prove LID, in that different 
volumetric fractions of twins were predicted, as confirmed later on by 
experimental evidence. Among the drawbacks of this approach are those 
mentioned in \cite{Barret84}, as there are: The lack of physical justification for the 
selection of a specific mechanism of deformation, as well as " \ldots of some of 
the most essential principles like: attainment of minimum surface energy, 
minimum active shear parameter, or minimum of atomic motion." In addition, 
the aforementioned phenomenological theories are not suitable to reasonably 
explain "any change of a crystallographic feature of martensitic 
transformations in relation to changes of their composition or of 
transformation temperature." 

The thermodynamical approach in \cite{Roitburd72,Roitburd78,Roitburd74}, 
which takes into consideration 
the energy and enthalpy associated with internal stress, within a given 
system of coexisting $\gamma $- and $\alpha $-phases, is mainly based on the 
assumption that the formation of a final structure reflects the tendency of 
the system to minimize its free energy (i.e. a minimum of elastic energy 
caused by internal stress), thus giving reasonable justification for 
regarding the crystallographic pattern of martensite as an isolated case of 
phases being in immediate contact in an invariant plane, without 
consideration of the appearance of mechanical stress. In addition, it 
explains the lamellar equilibrium shape of the crystals, particularities of 
their internal structure (like autocatalytic twinning) and other general 
features related to the emergence of crystal groups, under the conditions of 
a sample with free external boundaries as well as under the conditions 
imposed by an externally applied field of mechanical tension
\cite{Patel63,Kosenko78,Pankova84}. 
An interpretation of the kinetic particularities must be based on an evaluation 
of: a) the level of the energy-barriers which have to be surmounted during this 
type of growth. b) The thermodynamic driving force of the transformation. c) The 
role of relaxation processes in this environment. To take an example, the 
observed differences between lath-shaped and lamellar martensite might be 
deducible to differences between the dominating relaxation mechanisms (i.e. 
slipping processes in austenite and twinning in martensite). 

\begin{figure}[htb]
\centering
\includegraphics[clip=true,width=.8\textwidth]{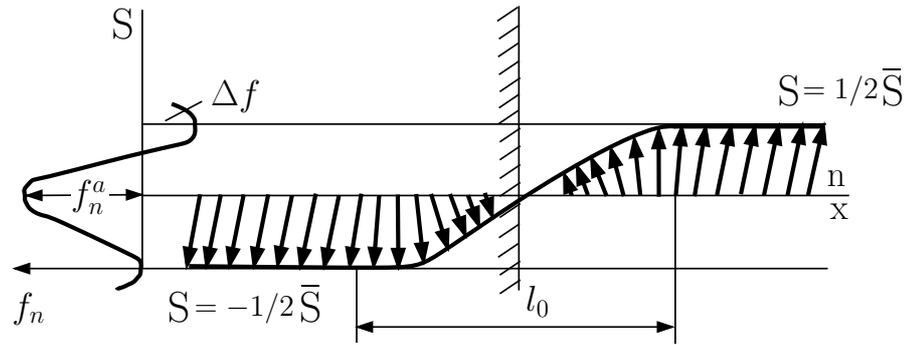}
\renewcommand{\captionlabeldelim}{.}
\caption{Structural pattern of a phase boundary \cite{Roitburd78}}
\label{fig3}
\end{figure}

The question about the growth mechanism will lead us into a consideration of 
the model of a phase-boundary and into an evaluation of its mobility. For 
example, in \cite{Roitburd72,Roitburd78,Roitburd62} there has been investigated a model of a phase 
boundary arising during a MT, with a transformation pattern characterized by 
an invariant plane. In Fig. \ref{fig3}, which has been taken from \cite{Roitburd78}, 
the phase boundary is represented by a transition layer together with a 
transition vector \textbf{S} of variable size and orientation, indicating the 
locations of atoms ($l_{0}$ - thickness of the boundary layer, being about the 
size of several lined up lattice parameters). The term $f_{n}$ relates to free 
energy density, x - coordinate axis oriented parallel to the normal \textbf{n} 
of the invariant plane of the boundary layer, $f_{n}^{a}$ and $\Delta f $ 
indicate the magnitude of the activation energy threshold, or of the 
thermodynamic driving force, respectively. This representation of a phase 
boundary thus is akin to that of a ferromagnetic domain \cite{Landau82}, which also 
includes a term related to the free energy gradient. 

\begin{figure}[htb]
\centering
\includegraphics[clip=true,width=.8\textwidth]{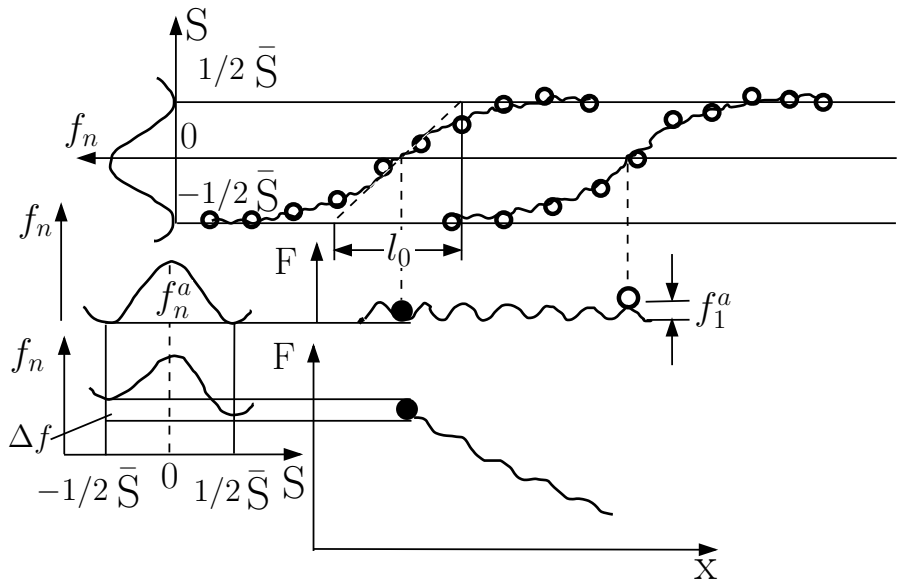}
\renewcommand{\captionlabeldelim}{.}
\caption{Determination of the $f_{1}^{a}$-threshold at the phase 
boundary, analogous to the Peierls-threshold of a dislocation \cite{Roitburd78}}
\label{fig4}
\end{figure}

As for the MT of alloys of the Fe - Ni, Fe- C systems, a realistic 
representation of a phase boundary would, at the first sight, have to 
satisfy the following two, obviously contradicting requirements: the pronounced 
persistence of metastable austenite at the $M_{S}$ temperature (see also the 
explanation in 1.3) implies the existence of a finite energy threshold 
$f_{n}^{a}$. The almost temperature-independent speed of growth of an individual 
martensitic crystal (at least within a wide range of temperature) is a 
typical feature of a thresholdless type of growth or motion. In 
\cite{Roitburd72,Roitburd78,Roitburd62} it is 
proposed to consider the mobility of a phase boundary analogous to the mobility 
of a dislocation through a periodic potential as determined by the lattice with 
its structural periodicity in \textbf{n} - orientation, in a similar way as that 
described by the Frenkel-Kontorova-Model. As a result, the $f_{1}^{a}$ threshold 
is considered instead of the $f_{n}^{a}$ threshold, (see Fig.\ref{fig4} as 
taken from \cite{Roitburd78}), which separates the relatively unstable states in the 
boundary area (i.e. where the location of individual atoms deviates 
significantly from the stable location with minimum potential energy in relation 
to the peaks of the potential "comb") from relatively stable ones (i.e. the 
atoms are located near the bottom of the periodic lattice potential), if put 
into relation with the driving force $\Delta f$. Thus for $\Delta f 
 > f_{1}^{a} < f_{n}^{a}$ a boundary would move 
(jump) nearly unhindered over a threshold, at a speed comparable to the velocity 
of (transversal) sound waves, nearly independent of temperature. However it has 
to be considered that a calculus of the above postulated speed of propagation 
has not yet been performed (as this would not be a typical problem of 
thermodynamic analysis). Thus the postulate of a large velocity of boundary 
propagation still needs to be proven. Moreover, according to data published in 
\cite{Lokshin1968,Lokshin1967}, the velocity  of growth of martensite crystals not only exceeds the velocity 
$c_{t}$ of transversal, but also the velocity $c_{l}$ of longitudinal elastic 
waves. Thus any concept based on an upper velocity limit $c_{t} < 
c_{l}$ would be fundamentally inadequate for the description of radial 
(frontal) growth of a lamellar martensitic crystal, even though an 
interpretation of lateral growth might be conceivable within such a concept. 

What's more, the assumption of supersonic velocity of martensitic crystal 
growth will inevitably lead into another problem: The usually observed 
$\gamma -\alpha $-phase boundaries belong to the semi-coherent type \cite{Kurdjumov60}. 
High mobility of semi-coherent boundaries usually is put into relation with 
the existence of parallel rows of slipping dislocations making up a boundary 
\cite{Roitburd75}. Under the assumption that the motion of a boundary, including that of a 
martensite crystal, is produced by slipping dislocations, then we must 
inevitably conclude that the motion speed $v$ of such 
dislocations will also take place at the of sound velocity $ c_{t}$. We remember 
that the supposed motion of dislocations at a velocity exceeding $c_{t}$ is 
prohibited within the frame of a continuous model \cite{Khirt72}. In principle however, 
this could yet be possible in crystals of a discrete atomic structure \cite{Kottrel69}. 
Moreover, we must consider that a dislocation in supersonic motion $v  
>  c_{t}$ would inevitably and permanently emit acoustic radiation 
energy, due to an effect analogous to the emission of Tscherenkov-radiation. In 
accordance with \cite{Kottrel69}, it is possible to calculate the stress $\sigma$ needed to 
maintain a dislocation speed $v > c_{t}$, by the relation

\begin{equation}
\label{eq6}
\sigma \approx [ \frac{\upsilon^{2}}{c_{t}^{2}} - 1 ] 
\cdot \frac{\mu \;b}{2\,\pi \,a}
\end{equation}
where $\mu $ - shear-modulus; $b$ - Burgers-Vector, $a$ - distance between the 
planes of shear. 

For example, if $v \approx  2 \,c_{t}$, and $b \approx  a$ were inserted 
in (\ref{eq6}), then $\sigma$ would exceed the 
theoretical yield stress $\approx  0,1 \cdot \mu $, a result 
which makes no physical sense. The same kind of problem would basically also 
appear in any investigation of shock-wave propagation in solid state matter. 
Some alternative opinions about the role of dislocations in shock-wave 
propagation are outlined in the papers by Wirthman, Meyers and Moor in \cite{redaktsia84}. 
According to Wirthman, supersonic propagation of dislocations 
(Smith-dislocations) should theoretically be possible in conjunction with the 
propagation of a shock-wave at a maximum stress level being equivalent to the 
yield strength. Meyers and Moor, whose position is based on experimental 
evidence however reject this point of view, arguing that dislocations generated 
in conjunction with a shock wave would lag behind the shock wave and soon come 
to a stop. Rejecting the thesis that the level of stress during the $\gamma 
-\alpha$ MT is close to theoretical yield strength, the motion of the 
phase boundary should rather be regarded as some kind of non-linear wave 
process producing dislocations which at the same time set up a boundary 
between the phases arising immediately behind the propagating wave front. 

Machlin and Cohen in \cite{Machlin51} were first to clearly formulate a wave model as a 
conceptual approach for resolving the martensitic growth problem. Without 
getting into details on the meaning of velocities, the authors of \cite{Machlin51} 
propose a pattern of two sequential deformation waves. The first of the 
waves spreads out into radial orientations starting from a minute 
platelet-shaped martensitic nucleus, thereby causing a homogenous 
deformation of a lamellarly shaped area, the surface of which (i.e. the 
habit plane) matches with the invariant plane. As soon as the deformation 
induced stress delivered by the first wave attains certain amount 
(presumably the yield limit), the second wave starts to spread into a 
direction perpendicular to the habit plane, thereby causing shear 
deformation (being homogenous only at a macroscopic scale). The residual 
inhomogeneous shear deformation completes the growth process of a martensite 
lamella. In principle, such wave pattern would make possible a direct 
interpretation of the crystallographic (geometrical) theory: The first wave 
produces pure lattice deformation, and the second one the required lattice 
invariant deformation. But in contrast to the geometrical pattern, in which 
the invariant (habit) plane can only arise during a combined lattice-variant 
and lattice-invariant deformation, \cite{Machlin51} requires that the development of a 
habit plane must already occur in the first step of pure deformation. 
Obviously, there do not exist any reasons for assuming that inhomogeneous 
deformation (i.e. by twinning) takes place only after formation of the 
lamellar region. At present, there is no doubt about inhomogeneous deformation 
occurring during a martensitic transformation (transformation twinning according 
to \cite{Kurdjumov77}). Nonetheless, the notion of a clear distinction between two stages of 
crystal growth still appears to be current. 

\begin{figure}[htb]
\centering
\includegraphics[clip=true,width=.4\textwidth]{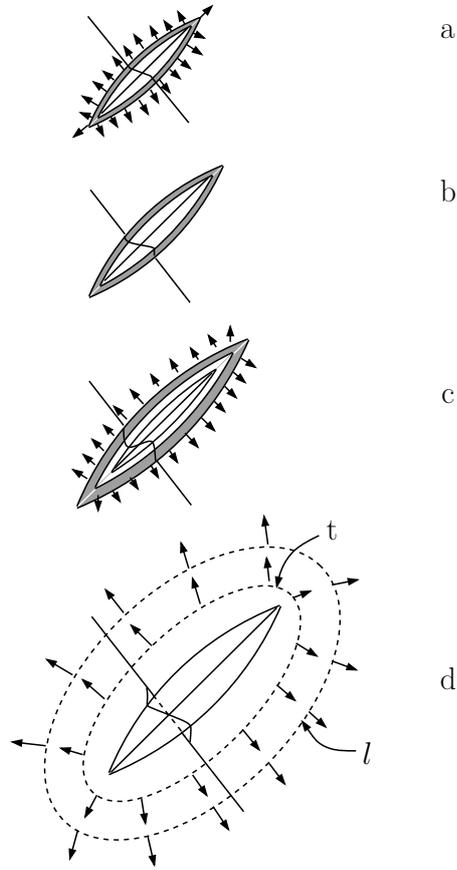}
\renewcommand{\captionlabeldelim}{.}
\caption{Growth pattern of a lenticularly shaped martensite 
crystal \cite{Meyers90}:
(a): Plastic shear and compression waves are propagating ahead of the 
transformation front; (b) and (c): Crystal growth process has ceased, but 
abating waves of plastic deformation are still propagating in the aftermath; 
(d): Waves of plastic deformation faded away, after having excited elastic 
transverse-(t) and longitudinal $(l)$ waves.}
\label{fig5}
\end{figure}

Essentially, the aforementioned basic ideas are also proposed in \cite{Meyers90}. In 
addition, it is postulated therein that the radially propagating wave is of 
the longitudinal type, leading to a MT in the central area (midrib) of a 
martensitic crystal. The transformed midrib-regions thus are about to play 
the role of second-order nuclei for excitation of a transversal 
transformation wave, which would propagate into an orientation perpendicular 
to the habit plane. In conclusion, it can only be the transversal wave which 
initiates the growth process in the region adjacent to a midrib. The most 
important single factor ensuring growth initiation is the high pressure 
(about 7,4$ \cdot $10$^{9}$ Pa) being produced in the immediate vicinity of 
the midrib, due to its increase in width by about 5 {\%}. This way, right in 
front of the transformation wave, short compressive and plastic shear wave 
pulses are surmised to move about, slowly widening until the growth process 
has ceased, after which the wave-pulses fade down to the level of ordinary 
elastic waves. The whole scenario is shown in Fig.\ref{fig5}, extracted 
from \cite{Meyers90}. Fig.\ref{fig5} shows a cross-section of a growing lenticular 
crystal (the drawing plane is assumed to run perpendicular to the habit-plane).

The midrib is represented by a straight line, splitting the lenticular 
region into two symmetrical halves, while the arrows are pointing into the 
orientation of wave propagation. The hatched region represents the region of 
compressive and shear wave pulses, and the broken line intersecting the 
crystal indicates how an originally straight line of reference (dotted line 
in Fig.\ref{fig5}d) is distorted during and after a MT. Using this pattern of 
transformation, together with the assumption that the longitudinal and the 
transversal transformation waves propagate at $v_{l} = c_{l}$, 
and $v_{t} = c_{t}$, respectively, there has been reported in 
\cite{Meyers90} an equation for the shape of the surface of a lenticular crystal, which has 
been the final aim of that work. We would further like to remark that the notion 
of dislocation generation under the condition of a high level of mechanical 
stress being built up in the waves, as postulated in \cite{Meyers90}, is justified, 
whilst the motion of the boundary is not put into clear relation with the 
motion of the dislocations. Even though the values of $v_{l}$ 
in \cite{Meyers90} appear to be estimated too low, and too high for $v_{t}$ 
(see also the interpretation of the data related to the speed of growth 
in 1.2), the qualitative model looks quite promising. However, what is obviously 
missing in \cite{Machlin51,Meyers90} is a description of the development of a midrib with an 
internal twinning structure, which is supposed to take up a leading role during 
the formation of lenticular crystals. 

In \cite{Lokshin1967,Lokshin69,Crussard}, the growth of martensite has been treated from the point of 
view of the shock wave theory. Lokschin also investigated the case of strong 
shock waves, opening up a possible explanation for supersonic speed of 
growth. According to estimates in \cite{Lokshin1967,Lokshin69}, the compressive stress or 
tension being developed within a shock wave amounts to about $(1,4 \div  
1,6)  \cdot 10 ^{10}$ Pa, being about twice as large as the value used in 
\cite{Meyers90}, thus attaining the same order of magnitude as theoretical yield 
stress. Based on data published in \cite{Bunshah1953}, Crussard analyzed the case of weak 
shock waves with a speed of propagation lower than sonic speed. The common 
weakness of these treatises is their lack of interpretations of the 
extensive amount of morphological transformation characteristics. 

If the role of dislocations during formation and propagation of the phase 
boundary was not properly considered in the aforementioned models, then the 
motion of a phase boundary could simply be regarded as the propagation of a 
lonely front (called "switching" wave in \cite{Rabinovich84}), akin to a step like 
excitation (as shown in Fig. \ref{fig3}. To the left and to the right of 
such a "switching" wave, different values of deformation $\varepsilon 
_{\alpha ,\gamma }$ would be produced, being typical of the $\gamma $- and 
$\alpha $ - phases. Such a process can usually be described on the basis of a 
quasi-linear parabolic equation (non-linear diffusion equation), to be dealt 
with in Chapter 6. 

Let us now briefly discuss the theories focusing on the role of hyperbolic 
waves during a MT. Generally, these waves correspond to solutions of 
non-linear equations arising from a basic linear hyperbolic equation (i.e. a 
classical wave equation featuring spatial and temporal derivatives of the 
second order) for which (harmonic) sinusoidal waves are fundamental 
solutions. In our opinion, the general interest on waves of this category is 
due to their potential for simple representations by (quantisized) 
totalities of phonons. Our next logical step will thus be to pass over to 
phonon theory, which will finally enable us to take advantage of the well 
developed and efficient microscopic models of solid state physics, and to 
establish later on an important link between microscopic and macroscopic 
descriptions of MT. 

The assertions in \cite{Wasilewski75} advocate for a description of MT in the wave model. 
Remarkably in \cite{Wasilewski75}, a decidedly passive role is assigned to dislocations, 
including the proposal to reject a differentiation between the nucleation 
and growth stages, apparently justified by the lack of observable static 
nuclei. Thus the kinetics of any kind of MT should in general be allocated 
within the explosive type. (Even the case of slow macroscopic growth of a 
martensite crystal can easily be explained within the general notion of 
explosive transformation, by assuming small micro-explosions occurring in 
sequence, with fairly long interruptions in between). In addition to the 
above thesis we want to point to another assertion stated in \cite{Wasilewski75}: It is 
being proposed to interpret the particularities of the martensitic structure 
as a result of the interaction of a combination of one (or more) lattice 
oscillation modes, propagating at the velocities of elastic waves).

\begin{figure}[htb]
\centering
\includegraphics[clip=true, width=.4\textwidth]{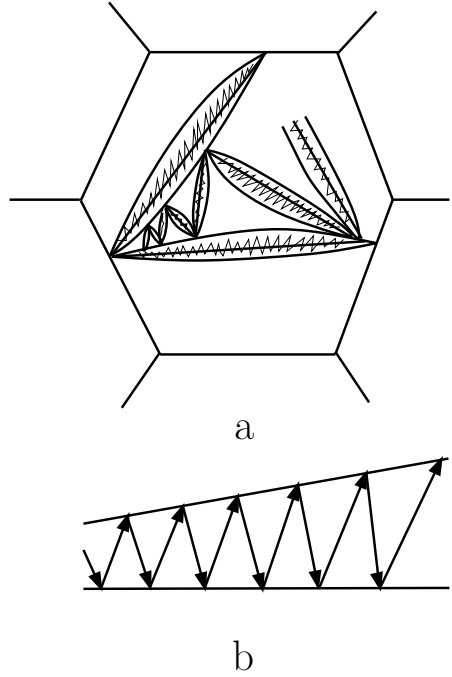}
\renewcommand{\captionlabeldelim}{.}
\caption{Schematic structural arrangement of typical martensite 
crystals: 
(a): Totality of the lenticularly shaped twinned crystals in a realistic 
pattern \cite{Izotov1972}; 
(b): Idealized pattern of multiple reflections of excited phonons \cite{Kayser72}. }
\label{fig6}
\end{figure}

Kayser \cite{Kayser72} was first in trying to associate the athermal activation of 
explosive MT in ferrous alloys with the principle of stimulated emission of 
phonons, which led to the conceptual notion of a 
phonon-maser  \footnote{Editorial note: Phonon-maser and laser 
research and development proceeded almost in parallel in their initial 
theoretical and experimental stages in the sixties, but phonon-maser 
development was abandoned later on due to lack of practical need for 
phonon-masers. See e.g. Phys. Rev. Lett. vol. 12, pp. 592-595, May 1964.} 
effect coming into being at a given amount of supercooling below the 
equilibrium temperature $T_{0}$ of both phases. The optimum frequency assumed in 
\cite{Kayser72} is the Debye-frequency $\nu_{AM} \approx  10^{13}$ 
Hz, while the associated phonon energy $h \nu_{AM}$ would equate with the free 
energy difference $G_{A} - G_{M}$ (related to one atom) between 
metastable (pre-martensitic) austenite $G_{A}$ and stable martensite $G_{M}$. 
The radiation system is supposed to be given by the lattice atoms, which, during 
the MT, perform a coordinated (cooperative) jump from an energy level $G_{A}$ 
down to a level $G_{M}$, simultaneously with a structural transformation. As the 
initial occupation of the $G_{M}$ - level is assumed to be nil (i.e. when the 
atoms make up the austenitic lattice) and if the condition $G_{A} > 
G_{M}$ was given, then the starting scenario would correspond to the utmost 
possible inversion between the occupations $G_{A}$ and $G_{M}$. The propagation 
of the emitted phonons through the transforming lattice is supposed to proceed 
in such a way that the amount of phonons would steadily increase, resulting in a 
steadily increasing amplitude of the acoustic wave, up to a magnitude 
enabling the creation of lattice defects. Kayser also proposed to interpret the 
zigzag structures, comprising a macroscopic group of relatively small martensite 
crystals, grown and confined within the space enclosed by two larger, 
pre-existing lamellae, as a sequence of multiple reflections of stimulated 
phonon radiation between the lamellae. Fig. \ref{fig6}a shows a schematic 
illustration of the observed zigzag structure (being more commonly known in 
metallurgy under the term ("lightning" or "lace" structure), whereas 
Fig. \ref{fig6}b shows the Kayser-pattern (arrow lines). Typical of this 
structure are two groups of lamellae: Any two lamellae selected from the same 
group are parallel, whilst lamellae belonging to different groups adjoin at an 
acute angle. We would however like to remark that any two of such totalities of 
lamellae cannot be built up only on the basis of reflections between 
non-parallel larger lamellae. Thus the interpretation by Kayser cannot be 
regarded as consistent. Obviously the structure shown in Fig. \ref{fig6}b 
represents nothing more than an observed phenomenological rule, which however 
does not reconcile with the classical laws of reflection. The general definition 
of the radiating system as well as of the radiation frequency also appear to be 
contestable. In fact, in a well developed two-level maser (i.e. see \cite{Zvelto79}), 
pairs of inversely occupied energy levels would simultaneously exist at 
certain points of space, and the active radiating sub-system normally is 
subjected to the Fermi-Dirac-Statistik. However in the Kayser-pattern, the 
physical state corresponding to $G_{M}$-energy would only potentially be 
possible, as the atoms are in a state with $G_{A}$-energy. Furthermore, 
short-wave phonons generally have a highly pronounced rate of decay, 
corresponding to short average lifetimes and ranges, thus the argument for 
the possibility of being generated by the proposed mechanism appears fairly 
improbable.

A more promising description of the emission of radiation in conjunction 
with atomic jumps is obviously set up by the concept of vacant atomic 
states, being closely linked with the physical state of quenched austenite 
(i.e. see \cite{Panin72}). 

An explanation of the mechanism of phonon generation has 
also been tried by Zhang in \cite{Zhang84}. Refraining here from further and detailed 
enumeration or discussion of the various theoretical possibilities, we can 
conclude that, in the aforementioned papers, a more or less reasonable 
concept of the microscopic mechanism of phonon generation during MT is 
missing. This can be easily realized by comparing points 2 and 3 of 
Table \ref{table12} (next page) which has been borrowed from \cite{Zhang84}.
\begin{table}[tbp]
\renewcommand{\captionlabeldelim}{.}
\caption{Analogies between the Laser-effect and martensitic transformations}
\begin{center} 
\begin{tabular}{|p{6.3cm}|p{7.5cm}|}
\hline
Laser & Martensitic Transformation \\ 
\hline
1. Coherent photons & 1. Coherent phonons \\
\hline
2. Occupational inversion &  2. Mechanism ensuring a (potential) possibility for 
stimulated emission of phonons\\
\hline
3. Under the conditions of an occupational 
inversion, coherent photons are emitted during the transition of electrons to 
the lower energy level. & 3. Under given 
conditions (critical rate of cooling, supercooling down to a certain 
temperature), the transition of atoms occurs from their initial lattice 
position of the high-temperature-phase into lattice positions of the low 
temperature equilibrium phase. Simultaneously with this phase transition, 
coherent phonons are emitted. \\
\hline

4. Resonator\footnotemark 
& 4. Lattice boundary, plane lattice defects, grouped point defects, 
dislocations etc. \\
\hline
5. Van-der-Pol-Equation  & 5. Van-der-Pol- and Duffing-Equation \\
\hline
\end{tabular}
\end{center} 
\label{table12}
\end{table}
              
In Table \ref{table12}, it has been tried to draw some general analogies 
(theoretical and phenomenological) between the laser or maser effect and the 
MT. Moreover, we assume that the requirement of a critical rate of cooling 
down to $M_{S}$ (pt. 3 in Table \ref{table12}) is not mandatory.

In order to suppress diffusion, it would suffice to rapidly pass through the 
high-temperature region, where diffusion processes predominate, whilst further 
cooling can succeed at a rather slow rate. For a description of the lateral 
displacement wave, Zhang proposes a wave equation borrowed from 
Laser-Theory:
\begin{equation}
\label{eq7}
\nabla^{2} \vec{\xi}+ 
\frac{1}{c^{2}}\frac{\partial^{2}\vec{\xi}}{\partial t^{2}} = 
- \frac{\partial^{2}\textbf{p}}{\partial t^{2}},
\end{equation}
where: $c$ - speed of sound; $\vec{\xi}= [\nabla , \textbf{u}]$ 
- axial vector of rotation is correlated with the antisymmetric rotation 
tensor \cite{Nai67}; $\textbf{u}$ - vector of translational displacement; $\nabla$ - gradient 
operator in a spatial frame of reference; \textbf{p} - analog to the 
polarization-vector of a lasing medium as used in laser-theory, given by 
$\textbf{p} = c^{-2} \hat{\varepsilon} \vec{\xi}$, 
where $\hat{\varepsilon}$ - matrix of lattice distortion during a 
lattice-transformation. As reported in \cite{Zhang84}, there have been performed investigations on the 
characteristics of stationary wave amplitudes, using a phenomenological 
approach by the commonly known Van-der-Pol and Duffing equations, under the 
assumption that, in a harmonic approximation, the square of oscillatory 
frequencies exhibit a critical dependence on temperature, thus resembling a 
typical characteristic of "Soft-Modes", i.e.: $\omega_{0}^{2} \sim (T - T_{0})$.

A discussion of reports related to investigations on the effects of 
externally applied magnetic fields and of magnetic ordering in austenite on 
MT will be subject of Chapter 5. 

\footnotetext{{A laser-resonator, being a characteristic part of most  lasers,
is not an absolute requirement for realization of laser/maser radiation,  as
in some radiating systems comprising a sufficiently high population inversion 
density (threshold-inversion), a "single-shot" or "superradiance" (Dicke) 
laser/maser-effect can emerge. (Superradiance however mostly is an undesired, 
potentially destructive and hardly controllable side-effect in many 
high-intensity lasers.) Moreover, there have also been discovered innumerable 
cosmic sources of (continuous) laser/maser radiation,  clearly demonstrating 
that laser/maser activity is a natural effect, not requiring sophisticated 
laboratory equipment like a resonator at all.}}


\section{General physical tasks and objectives}

A brief review of existing publications dealing with the spontaneous MT of 
paramagnetic austenite lead us to the conclusion that the wave-approach, 
though being regarded as most promising for further development of a 
microscopic dynamical MT-theory and description of the $\gamma -\alpha $- 
MT, presently is only weakly developed.

At first, the mechanism of wave-generation (and -amplification, 
respectively) in the stage of martensite growth has to be determined, and 
secondly, the substantial variability of crystallographic characteristics of 
Martensite must be interpreted in the wave-model, as without such an 
interpretation, the wave-approach would be largely invalidated with respect 
to materials science. 

\begin{enumerate} 
\item{ Let us assume that the process of fast development of a 
macro-nucleus of a lateral dimension of about 2 r $ \approx  (10^{-6 
}\div 10^{-7})$ m (see 1.3) occurs within a period of $t_{N}  
\approx (10^{-9}\div  10^{-10})$ s. Then it is fairly easy to 
realize that this postulate represents some kind of specification going 
beyond the normally used model of fluctuation nucleation. Let us further 
assume that certain spatial scale $l$ can be defined along an orientation 
perpendicular to the axis of nucleation, in such a way that $l$ would be 
complementary to r, as shown in Fig. \ref{fig7}. 
\begin{figure}[htb]
\centering
\includegraphics[clip=true, width=.8\textwidth]{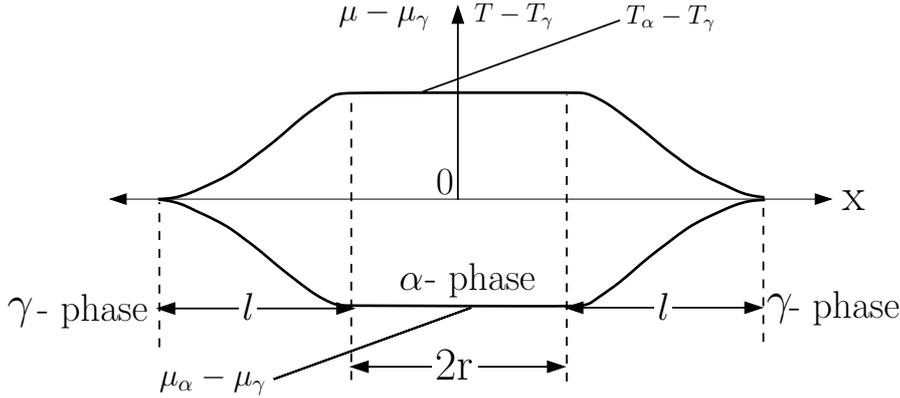}
\caption{Assumed distribution of temperature T and chemical 
potential $\mu $ at the instant of macro-nucleation of the $\alpha $-phase. 
The x-axis is oriented perpendicular to the axis of nucleation.}
\label{fig7}
\end{figure}
Then $l$ indicates the width of a specific region (we call it 
$B_{\gamma - \alpha }$) characterized by significant inhomogeneities of 
such fundamental macro-parameters like temperature T and chemical potential 
$\mu$ \footnote{The designation $\mu $, having been used in 1.4 as the designation of 
the shear-modulus, will from now on exclusively be used as the designation 
of chemical potential.}. In other words, the magnitudes of the spatial gradients 
$\nabla T$ and $\nabla \mu$ in the a.m. 
$B_{\gamma - \alpha }$-region are relatively large. In order to evaluate the 
size of $l$, we start from the assumption that heat liberation as well as a 
remarkable volume change simultaneously occur in the region penetrated by an 
elasto-plastic wave-front (being closely linked up with the 
$\gamma -\alpha $-transformation). Let us further assume that the 
wave-front propagates at a speed of $ c  \sim 10^{3}$ m/s, through a 
region transforming within a period $t_{N}$, rendering: $l  \sim  c 
\cdot  t_{N}  \sim  (10^{-6}  \div  10^{-8})$ m (see also 
an evaluation of $l$ under 6.3.3). The order of magnitude of the quantities 
$\nabla T$, $\nabla \mu$ can easily be assessed from the 
relations:

\begin{equation}
\label{eq8}
\nabla T \sim \frac{\Delta T}{l} \approx \frac{T_{\alpha} - T_{\gamma} 
}{l},\quad \nabla \mu \sim \frac{\Delta \mu }{l} \approx \frac{\mu_{\alpha }
- \mu_{\gamma} }{l}.
\end{equation}

The quantity $\Delta T$ is comparable with the degree of supercooling 
$T_{0} - M_{S}$, i.e., $\Delta T \sim  100$ K (see also the beginning 
of 1.2) and, as a result, the temperature gradient would become 
$\nabla T  \sim (10^{8} \div  10^{10})$ K/m. In assessing 
$\nabla \mu$, we take into consideration that the specific volume of 
the $\alpha$-phase increases due to the volume-effect, and that simultaneously 
the concentration of electrons n, and thus also $\mu$, decrease. Let $\mu  
 \sim  n^{2/3}$ be the generally used relation between $\mu$ and 
$n$ in the free-electron model (see for example \cite{Zaiman74}). 

Then we get 

\begin{equation}
\label{1.9}
\frac{\Delta \mu}{\mu } = -\frac{2}{3}\frac{\Delta V}{V}.
\end{equation}

For an assumed ratio of volume change $\Delta V / V  \approx  2,4 
\cdot 10^{-2}$, being typical of the Bain-deformation, and with $\mu \sim $ 
10 eV, we get from \eqref{1.9}: $\Delta \mu  = 0,16 $eV, being equivalent to 1860 K at 
a temperature scale. This way, the quantity $\nabla \mu / k_{B}$ 
($k_{B}$ - Boltzmann-Factor) can easily exceed the value of $\nabla T$ by one 
order of magnitude, for the same value of $l$. It should be noted that our 
estimate of $\Delta \mu $ with adequate consideration of the volume-effect 
should be essentially correct, due to the use of the relation $\mu \sim 
n^{\frac{2}{3}}$  for a sub-system of s-electrons.  Of course the same results 
can be obtained analytically for a d-electron subsystem (see 4.5). 

The remarkable gradients $\nabla T$ and $\nabla \mu $ however require 
further analysis under non-equilibrium conditions of the electron- and 
phonon-subsystems of the $ B_{\gamma - \alpha }$-region. A substantiation of such 
non-equilibrium conditions will then enable us to confine our further task to 
the determination of the mechanism of generation (or selective amplification) of 
phonons by non-equilibrium electrons.}

\item{We recall that the physical basis of the maser-effect is that 
the processes of stimulated absorption are outnumbered by simultaneously 
occurring processes of stimulated emission in a given region of a system, 
provided an inverted occupation is given in the radiating system (i.e. in a 
lasing system the occupation of higher energy levels must substantially 
exceed that of the lower energy levels). If we choose a group of electrons as a 
radiating system and further assume that the band model is correct, then it will 
be straight forward to conclude that inversely occupied energy levels are 
present wherever electronic flows exist, provided their assumed energy levels 
are correct. Let $f_{j\textbf{k}}$ denote the non-equilibrium distribution function of 
electrons in the state (j,\textbf{k}), where j is a band number, $\textbf{k}$ is 
the wave vector and $ \hbar\,\textbf{k}$  the quasi-momentum ( $\hbar$ - Planck's 
constant) associated to a given electron of energy $\varepsilon_{j\textbf{k}}$. Let us 
now regard the graphs pertaining to a non-equilibrium distribution of electrons 
(ref. i.e. \cite{Zaiman74}) being already known from the theories of heat ( for 
$\nabla T \ne  0$) and of electrical conductivity (for 
$\nabla \mu \ne $ 0). In Fig.\ref{fig8}, an equilibrium 
function $f_{\textbf{k}}^{0}$ (normal line), and a non-equilibrium function $f_{\textbf{k}}$ 
(dotted line) are plotted, the degree of non-equilibrium being determined by the 
quantities $\nabla T$ Fig.\ref{fig8}a and $\nabla \mu$ Fig.\ref{fig8}b. 
In case (a), it is clearly obvious that two oppositely 
oriented electronic flows arise (namely, electrons with energy $\varepsilon  
> \mu $ predominantly move from the warmer to the cooler regions, 
i.e., against the temperature-gradient $\nabla T$, whereas electrons 
with $\varepsilon < \mu $ predominantly move parallel to the 
direction being set by $\nabla T$).

Thus, if the quasi-momentum $\hbar\,\textbf{k}$ of electrons is oriented 
anti-parallel to the temperature-gradient $\nabla T$, a 
non-equilibrium term $\Delta f_{\textbf{k}} > 0$ for $\varepsilon_{\textbf{k}} 
> \mu $, with $\Delta f_{\textbf{k}} <  0$ for $\varepsilon 
_{\textbf{k}} < \mu$, has to be considered. And vice-versa, if $\hbar\,\textbf{k}$ 
is oriented parallel to $\nabla T$, then $\Delta f_{\textbf{k}} > 
0$ for $\varepsilon_{\textbf{k}} < \mu$  and  $\Delta f_{\textbf{k}} < 0$ 
for $\varepsilon_{\textbf{k}} > \mu$. Thus the states with quasi-momentum 
$\hbar\,\textbf{k}$ oriented anti-parallel or parallel to $\nabla T$ would be 
inversely occupied above or below the Fermi-level $\mu$. In Fig.\ref{fig8}a, 
two such inversely occupied energy levels are plotted by thin horizontal lines, 
in relation to the Fermi level $\mu$. In Fig.\ref{fig8}b, the particular 
case of parallel electron motion towards decreasing chemical potential, i.e. 
against $\nabla \mu$, is illustrated, where the sign of the 
non-equilibrium addend to the equilibrium distribution function $f_{\textbf{k}}^{0}$ is 
independent of the relation between $\varepsilon_{\textbf{k}}$ and $\mu $, i.e.: $\Delta 
f_{\textbf{k}}> 0$ for \textbf{k} $ \uparrow  \downarrow \nabla \mu$, 
and $\Delta f_{\textbf{k}} < 0$ for $\textbf{k}\uparrow \uparrow \nabla 
\mu$. The above concept of formation of inversely occupied states related to 
the non-equilibrium conditions during the stage of growth of the $\alpha$-phase 
has been proposed in \cite{Kashchenko79,Kashchenko77,Kashchenko82}.}

\item{Our investigation of the generation of elastic waves by 
non-equilibrium electrons should most reasonably start with a definition of 
the conditions required for the excitation (or amplification) of a plane 
wave of atomic displacement, in the ideal case of a crystal of infinite 
dimensions, with homogenous and stationary temperature- and chemical 
potential gradients.

From the point of view of quantum-mechanics, a plane wave of relative 
\begin{figure}[htb]
\centering
\includegraphics[clip=true, width=.8\textwidth]{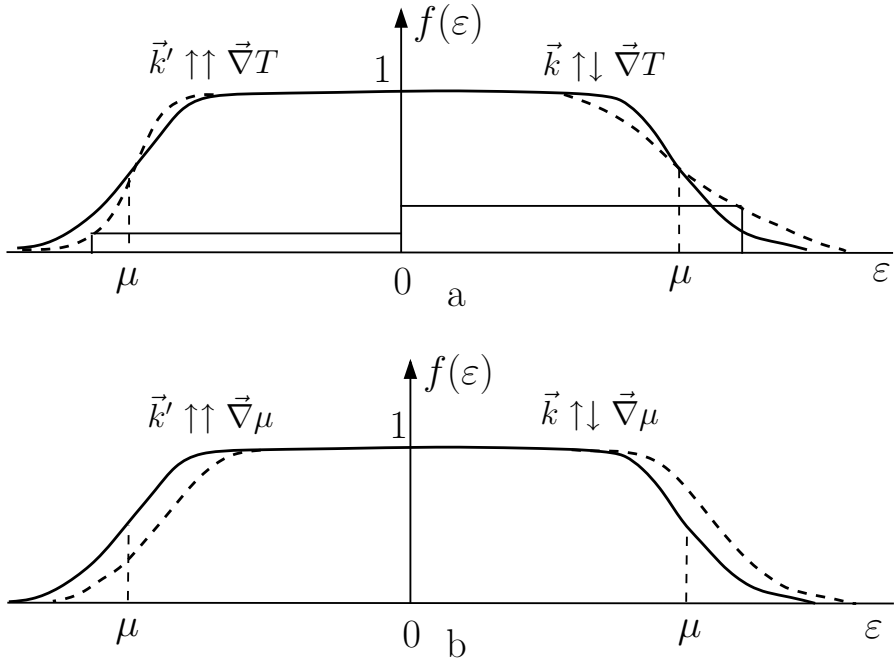}
\renewcommand{\captionlabeldelim}{.}
\caption{Equilibrium (normal line) and non-equilibrium (dashed 
line) electron energy distributions: (a): $\nabla T \ne 0$; 
(b) $\nabla \mu \ne 0$}
\label{fig8}
\end{figure}
atomic displacement: $\textbf{u}(\textbf{r},t) = \textbf{u}_{0}cos(\omega_{\textbf{q}}t - \textbf{q r}$) 
represents the 
macroscopic effect of a totality of coherent phonons of energy $\hbar\omega 
_{\textbf{q}}$, with wave-vectors \textbf{q}. This means that the generation of waves by a 
maser-mechanism would become feasible whenever a macroscopic number of 
inversely occupied electronic states ($\textbf{k}, \textbf{k}^{\prime}$) was existent, and transitions 
between them would predominantly result in the emission of phonons, in 
accordance with the laws of energy- and quasi-momentum conservation: 

\begin{equation}
\label{eq10a}
\varepsilon _{i \textbf{k}} - \varepsilon _{i^{\prime} \textbf{k}^{\prime}} =
\omega _{j \textbf{q}}, 
\end{equation}

\begin{equation}
\label{eq10b}
\textbf{k} -  \textbf{k}^{\prime} - \textbf{q} = 0, \textbf{Q}. 
\end{equation}

In \eqref{eq10a}, $j$ is the index of the phonon-branch, whereas the equality of 
\eqref{eq10b} depends on whether the electronic transitions belong to a normal 
"N-process" or to a "U-process" \cite{Zaiman74} (see Fig.\ref{fig9}). 
\begin{figure}[htb]
\centering
\includegraphics[clip=true, width=.6\textwidth]{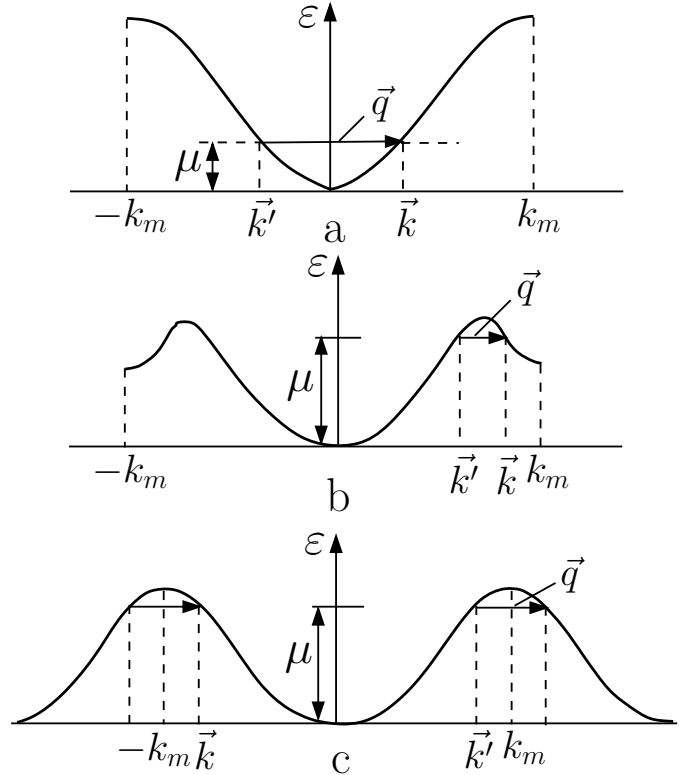}
\renewcommand{\captionlabeldelim}{.}
\caption{One-dimensional pattern of N-processes (cases a and b) and 
U-process (case c) $k_{m}$ -- upper momentum limit in the 1$^{st}$ BZ,
\textbf{q} -- phonon-momentum. The energy difference $\varepsilon_{\textbf{k}}-\varepsilon 
_{\textbf{k}^{\prime}}$ is not considered, $\mu$ -- Fermi-level.}
\label{fig9}
\end{figure}
For an N-process, the right side of \eqref{eq10b} vanishes, while for a U-process, the 
right side of \eqref{eq10b} is the vector \textbf{Q} of the reciprocal lattice. 
This way, our task is equivalent to finding macroscopic sets of pairs of 
inversely occupied electronic states in \textbf{k}-space, which would have 
to be equidistant to satisfy the conditions of (\ref{eq10a},\ref{eq10b}). The main objective 
in processing with this task will thus be the identification and 
classification of such pairs of electronic states, starting with the 
definition of points located in a surface separating inversely occupied 
states in \textbf{k}-space. Thus it will be important from the outset of our 
analysis, to identify and pre-select such lattice oscillations being 
characterized by an implicit tendency to become unstable under 
non-equilibrium conditions. This analysis will be based on the assumption 
that the structure of the electronic spectrum in the transforming $B_{\gamma 
- \alpha}$-region maintains all relevant properties of the electronic 
spectrum of the fcc-phase of iron, this being a significant prerequisite 
from the viewpoint of the phonon-generation mechanism. 

If we think of a MT as the manifestation of a particular kind of 
lattice-instability, it would be reasonable to identify some key features of 
solid-state instabilities in general. Firstly, it can be stated that, except 
for some specific aspects of non-equilibrium states, basically the same - 
finally stated - particularity of an electronic spectrum is used to explain 
lattice-instabilities like the Peierls-instability of a three-dimensional 
lattice, as well as during a variety of electronic and magnetic phase 
transitions, within the frame of a wave concept for charge- and 
spin-densities \cite{Brus1984,Afanasev62,Bulaevskii75,Egorushkin83,Kulikov84}: 
The key point thus is: The macroscopicity and 
collectivity of a number of electronic states, corresponding to 
(\ref{eq10a}, \ref{eq10b}). 
In fact: For the presented examples, there is a requirement for the existence of 
more or less large regions on the Fermi-surface which must nest during a 
translation by a given vector \textbf{q}. This very vector \textbf{q} thus 
characterizes the resulting structure. The existence of such particularities 
must manifest itself through the appearance of density-peaks of the electronic 
states in the proximity of the Fermi-energy $\mu $
\cite{Gor'kov74,Gor'kov76,Vonsovskii77}. 
Of course, the pairs of electrons that will attract our particular interest 
later on, must also have energies near $\mu$.}

\item{ By analogy with a photon-maser \cite{Leks75}, it can be expected that, in 
order to realize a phonon-maser, mere satisfaction of the 
threshold-condition

\begin{equation}
\label{eq11}
\sigma_{0} > \sigma_{th} = \frac{\Gamma \,\varkappa}{\left| W 
\right|^2 \,R},
\end{equation}
would be sufficient, which already considers that the inverted initial 
difference of the $\sigma_{0}$-populations has to exceed a given 
threshold $\sigma_{th}$. This threshold value thus is proportional to the 
inherent attenuation $\Gamma $ of radiating electrons, as well as to the rate 
of damping $\varkappa$ of generated phonons, and inversely proportional to the 
square of the matrix-element W of electron-phonon interaction, as well as to 
the number of pairs of equidistant electronic states R. It should be noted that 
Debye-Phonons turn out to be incompetitive, due to their high attenuation in 
relation to the attenuation of phonons with longer waves, as indicated by 
\eqref{eq11}.}

\item{ It has to be clearly distinguished between a dynamical 
instability, as manifested by the excitation of elastic waves (in accordance 
with \eqref{eq11}) and the lattice-instability arising as soon as the waves attain 
a threshold amplitude of $\varepsilon_{th} \sim 10^{-3}$, as 
required to initiate plastic deformation (see 1.3). Thus, in order to 
materialize the lattice instability to be dealt with here, a more demanding 
condition than that of \eqref{eq11} has to be satisfied.}

\item{ As the $\gamma -\alpha $ MT of ferrous alloys occurs in a wide 
range of concentrations of its other component (-s), causing significant 
variations of the $M_{S}$-temperatures (from $M_{S} < 10^{3}$ K, for 
low Ni-concentrations, down to $M_{S} <  4,2 $ K, for 34{\%} Ni in a Fe 
- Ni - System), we have to identify a specific criterion for an effective 
working principle of phonon-maser-excitation, which must be valid for the 
observed wide range of temperatures and concentrations. The specific 
difficulties arising from this task can most easily be overcome by a two-band 
model of the electronic spectrum of a metal, comprising s and d electrons (a 
more comprehensive definition of this task will be given under 4.1).}

\item{ After having established a conceptual notion of a phonon-maser 
for single wave generation, let us now get into an interpretation of some 
morphological characteristics of martensite. It would be most convenient to 
select the habit plane as a basic characteristic of a MT, as the habit plane 
is unequivocally linked up with a MT, as already mentioned under 1.2. For 
further interpretation of MT, it will be convenient to start with an outline 
of the following geometrical representation, which will play a key role in 
our further analysis: Imagine a straight line of fixed (non-rotating) 
orientation, moving with a constant velocity relative to an inertial 
reference frame (associated, for example, with crystal axes  $\langle 100 
\rangle$). As a flat surface can geometrically be defined as the location 
of all points in space through which such a line has passed, then this line in 
motion will generate a plane. (Analogously, a line can be generated by the trail 
of a moving point imprinted in a given plane). In our next step, we assume that 
the moving line itself is defined and generated by the geometric locus of a 
moving intersection of two flat (non-parallel) wave fronts (i.e. with 
non-collinear wave-vectors). (The idea of using a combination of waves has 
first been expressed in \cite{Wasilewski75}). Further substantiation of the wave 
characteristics should aim at defining the class of waves (i.e. longitudinal 
or transversal), their direction(s) of propagation, and their 
phase-relationships. It can however be anticipated that the combined action 
of such waves might suffice to build up favorable conditions for a 
structural transition. With respect to the $\gamma -\alpha 
$-transformation, longitudinal (or quasi-longitudinal) waves, supposed to 
propagate near to the orientations  $\langle 001 \rangle$,  $\langle 110 
\rangle$ will be selected. In addition and in accordance with the pattern of 
Bain-deformation (see Fig.\ref{fig1}), let us further suppose that the growth 
of a martensite lamella proceeds by progressive unification with its adjacent 
areas, which are subjected to the synchronized dilatational and compressive 
action of the waves intersecting them. In Fig.\ref{fig10}, two selected 
perpendicular directions of wave propagation are marked by indices 1 and 2, 
while the velocities and lengths of the two waves are marked by 
$\textbf{c}_{1}$, $\textbf{c}_{2}$, $\lambda_{1}$, $\lambda_{2}$, 
respectively. The hatched cross-sections represent the areas of favorable 
transformation conditions with respect to the orientations of compressive- 
(acting in direction 1) and tensile stress (acting in direction 2). Each of the 
waves is shown at a given starting time $t_{0}$ and at a somewhat later instant 
$t$. Obviously in Fig.\ref{fig10}, the area enclosed by the bold lines 
represents an idealized cross-section of a lamella with an approximate width of 
$(1/2) \lambda_{1,2}$, thus resembling a prototype image of a 
growing martensite lamella.

\begin{figure}[htb]
\centering
\includegraphics[clip=true, width=.8\textwidth]{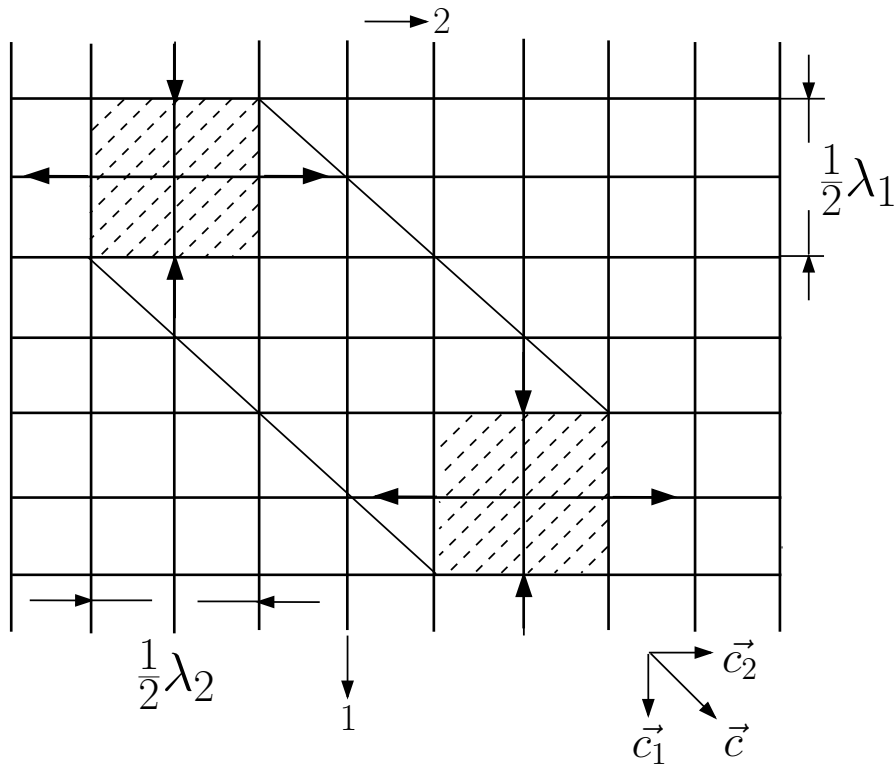}
\renewcommand{\captionlabeldelim}{.}
\caption{Fundamental growth pattern of a martensite lamella in 
the notion of two flat longitudinal waves propagating perpendicular to each 
other: $c_{1}$, $c_{2}$ -- wave - velocity, $\lambda_{1}$, 
$\lambda_{2}$ -- wavelength.}
\label{fig10}
\end{figure}

It is easy to conceive from Fig.\ref{fig10} that the hatched area of 
wave-superposition, as well as the intersecting area of the wave-fronts, are 
simultaneously propagating at the velocity equivalent to the vector-sum 
(i.e. geometrical sum) of their individual velocities $\textbf{c}_{1}, 
\textbf{c}_{2}$\footnote{In case of non-orthogonality between 
$c_{1}$, $c_{2}$, the velocity c would deviate from \eqref{eq12}}, i.e.
\begin{equation}
\label{eq12}
\textbf{c} = \textbf{c}_{1} + \textbf{c}_{2},\quad \quad c = \arrowvert \textbf{c} \arrowvert = \sqrt 
{c_{1}^{2} + c_{2}^{2}}.
\end{equation}

As the value of $\arrowvert \textbf{c} \arrowvert$, on the one hand, characterizes the frontal 
speed of growth of a martensite lamella, and on the other hand can exceed 
the longitudinal velocity of sound in direction of \textbf{c} in 
Fig.\ref{fig10}, it is possible in principle to explain this way the 
aligned supersonic growth of martensite crystals (being controlled by a pair 
of ordinary longitudinal waves), thus inherently representing an important 
kinetic particularity of the growth stage. This interpretation was proposed 
for the first time in \cite{Mints77}. We should also keep in mind that the dominating 
role of longitudinal waves in the formation stage of midribs was postulated 
in \cite{Meyers90} too. If, on the one hand, the types of generated waves were put in 
relation with the particularities of the electronic spectrum and on the other 
hand, with the morphology of the end product, then we could expect new 
explanations of structural and kinetic particularities of the MT (including the 
effects of changes in concentration of the second component), which might be 
complementary to some important conclusions drawn from thermodynamic 
analysis of MT.}

\item{After having convinced ourselves of the "functionality" of a 
two-wave pattern, we shall deal in our next step with the various 
implications related to the coordinated propagation of displacement waves in 
the model of "switching waves", as we will call the isolated, step-shaped 
fronts of temperature, chemical potential and distortion. This might finally 
consolidate our understanding of the key phenomena associated with the 
motion of the interface of a martensite crystal in the growing stage.

The following chapters will mainly be devoted to the solution of the tasks 
defined in above sub-points 3 to 8. In brief, the ultimate objective of this 
work is to analyze and understand the growth stage of a martensitic crystal 
during the $\gamma -\alpha$-transformation of ferrous alloys, within the 
conceptual frame of a self-organizing process, during which the mechanism of 
generation (and amplification) of lattice-displacement-waves by means of 
non-equilibrium electrons will play a dominant role.

To avoid any misunderstandings when reading this monograph, the author 
recommends to bear in mind that different physical quantities may sometimes 
be denoted by one and the same symbol. For this reason, the indices appended 
to symbols, as well as the "local" definition of abbreviations, will 
normally be introduced in the "local" text and should be duly considered. 
Thus in one instance, $\varepsilon_{k}$ may represent the energy of an 
electron of wave-vector \textbf{k}, whereas $\varepsilon$ may represent 
linear deformation, and $\tilde{\varepsilon}$ represents relative volume 
deformation. Vectors will be represented either by bold letters or by an 
arrow placed above a letter.}
\end{enumerate}

\chapter{Particularities of an electronic energy spectrum as required for 
realization of a Phonon-Maser}

\section{The Hamiltonian Problem}

In accordance with No. 1 and 2 of point 1.5 of this monograph we associate
to each state of an electron a band-index j and a wave-vector \textbf{k}.
Let us further assume that the basic characteristics of the band-structure
of $\gamma$ - iron will essentially also apply to other ferrous alloys with
fcc-lattice, at least in a coherent potential approximation
\cite{Erenreikh76,Vediaev77}
(also see \cite{Egorushkin82}). Thus our consistent first step will be to analyze the
dynamical characteristics of the electron-phonon system under
non-equilibrium conditions, within the frame of energy of the band spectra
of the involved electrons and phonons.

We choose the standard formulation of the Hamiltonians for non-interacting
electrons $H_e$ and Phonons $H_p$, respectively

\begin{equation}
\label{2.1}
 H_e=\sum_{j\textbf{k}} \varepsilon_{j\textbf{k}} a_{j\textbf{k}}^+  a_{j\textbf{k}},
\end{equation}

\begin{equation}
\label{2.2}
H_p=\sum_{i\textbf{q}} \hbar\omega_{i\textbf{q}}
(b_{i\textbf{q}}^{+}b_{i\textbf{q}}+\frac{1}{2}),
\end{equation}
where $a_{j\textbf{k}}^{+}$, $a_{j\textbf{k}}$, $b_{i\textbf{q}}^{ + }$,
$b_{i\textbf{q}}$ - are the creation and annihilation operators for
electrons and phonons, respectively.  Accordingly, the acoustic branches of
the phonon spectrum with frequencies $\omega $ and wave-vectors \textbf{q}
will be enumerated by i = 1,2,3. In consideration of the predominance of
single-phonon processes during electron-phonon interactions, it is justified
to use the Fr\"{o}hlich-Hamiltonian $H_{ep}$ as interaction operator, being
linear in  the operators $b^{+}$ and $b$:

\begin{equation}
H_{ep}= \sum_{\textbf{q}\textbf{k}ijj^{\prime}} W^{*}_{i\textbf{q}jj^{\prime}} b_{i\textbf{q}}
a^{+}_{j\textbf{k}}a_{j^{\prime}\textbf{k}^{\prime}}+ 
W_{i\textbf{q}jj^{\prime}} b^{+}_{i\textbf{q}} a^{+}_{j^{\prime}\textbf{k}^{\prime}}
a_{j\textbf{k}}.
\label{2.3}
\end{equation}

In this case the quasi-momenta obey to the conservation-law (\ref{1.9}b), and
W$_{\textbf{q}}$ is the matrix-element pertaining to electron-phonon interaction 
(W$_{\textbf{q}}$* is complexly coupled to W$_{\textbf{q}})$. In the tight binding
approximation it is possible to get an estimated value for W$_{\textbf{q}}$ by 
limiting the calculus to a linear reduction of the resonance integral G,
with respect to atomic displacements, where the magnitude of G determines 
the width of the electron band \cite{Bulaevskii75}. For $q < q_{max}$ with $q_{max} \sim
\pi $/$a$, near the boundary of the 1$^{st}$ Brillouin-Zone (BZ), we get for
Normal (N-processes) as well as for Umklapp (U-processes): 

\begin{equation}
W_{\textbf{q}} \approx  i\left[ \frac{\hbar}{2MN\omega_{\textbf{q}} }
\right]^{\frac{1}{2}}G(\textbf{e}_{\textbf{q}},\textbf{q}),
\label{2.4}
\end{equation}
where M - atomic mass, N - number of atoms, $\textbf{e}_{\textbf{q}}$- phonon
polarization vector, $i$ - imaginary unit in (\ref{2.4}) (we omitted the band and
branch indices). 

Apart from the Hamiltonians (\ref{2.1})-(\ref{2.3}), the complete Hamiltonian of system
H would also have to take into account electron-electron interactions
$H_{ee}$ as well as anharmonic phonon-phonon interactions $H_{pp}$, the 
explicit form of which will however not be used further.


\section{Form of a non-equilibrium addend to the electronic distribution
function and definition of points separating inversely occupied states in a
one-dimensional electronic spectrum}

The task related to the identification of pairs of potentially active
electronic states (ES) being involved in the generation of phonons (see
sub-point 3 under point 1.5) requires that some additional conditions have
to be satisfied, besides those already defined in \eqref{1.9}. Above all we have
to remark that in the context of this monograph, the notion "potentially
active" addresses the population inversion of pairs of ES under
non-equilibrium conditions in a transforming lattice. As a population
inversion of the kind to be discussed here always coexists with electronic
flows, it can be anticipated that the uppermost inversion between pairs of
ES will only be realized for such orientations of the group-velocities
$\textbf{v}_{\textbf{k}} \uparrow \downarrow  \textbf{v}_{\textbf{k}^{\prime}}$ of electrons 
(with wave-vectors $\textbf{k}$, $\textbf{k}^{\prime}$, respectively) being mutually
opposed to the relevant gradient of a locally existing spatial inhomogenity. 

It is easy to convince ourselves of the correctness of this latter statement
by writing out a stationary non-equilibrium addend $f_{\textbf{k}} - 
f_{\textbf{k}}^{0}$ (for stationary and homogenous temperature and chemical
potential gradients), as determined by the standard kinetic equations of
electronic populations $f_{\textbf{k}}$ and using the relaxation time approximations
presented in \cite{Zaiman74}. Our determination of relaxation time $\tau$ will be
based on the consideration that the impact-integral can be replaced by
expression $(f - f^{0}) \tau^{-1}$ , where $\tau$ is the
average relaxation time to equilibrium distribution $f^{0}$.

\begin{equation}
f_{\textbf{k}} - f_{\textbf{k}}^{0} \approx  \frac{\partial
f_{\textbf{k}}^{0}}{\partial y_{\textbf{k}}} \frac{y_{\textbf{k}}\tau}{T}
(\textbf{v}_{\textbf{k}},\vec{\nabla} T),
\label{2.5}
\end{equation}

\begin{equation}
f_{\textbf{k}} - f_{\textbf{k}}^{0} \approx \frac{\partial
f_{\textbf{k}}^{0}}{\partial y_{\textbf{k}}} \frac{\tau}{k_{B} T}
(\textbf{v}_{\textbf{k}},\vec{\nabla} \mu).
\label{2.6}
\end{equation}

In (\ref{2.5}) and (\ref{2.6}) $f_{\textbf{k}}^{0}$ - Fermi-Distribution

\begin{equation}
f_{\textbf{k}}^{0} =\frac{1}{e^{y_{\textbf{k}}}+1},\quad y_{\textbf{k}}
= \frac{\varepsilon_{\textbf{k}}-\mu}{k_{B}T},
\label{2.7}
\end{equation}

$k_{B}$ - Boltzmann-Factor.

It immediately follows from (\ref{2.5}) and (\ref{2.6}) that the
non-equilibrium addends in brackets simultaneously change their own signs
with a change of the  orientation of $\textbf{v}_{\textbf{k}}$, as indicated by the
change of sign of the respective scalar products
$(\textbf{v}_{\textbf{k}},\vec{\nabla}T)$ or $(\textbf{v}_{\textbf{k}},\vec{\nabla }\mu )$.
Bearing in mind that $\partial f_{\textbf{k}}^{0} / \partial y_{\textbf{k}} <
0$,  we can easily note the correspondence of equations  (\ref{2.5}) and (\ref{2.6})
with Fig. \ref{fig8}.

The requirement of anti-parallelism $\textbf{v}_{\textbf{k}} \downarrow
\uparrow  \textbf{v}_{\textbf{k}}^{\prime}$ bears a more generalized character than the 
singular requirement of anti-parallelism for quasi-momenta $\hbar\textbf{k} 
\uparrow  \downarrow  \hbar\textbf{k}^{\prime}$ of a pair of ES, as indicated under 
sub-point 2 of point 1.5. The requirement $\textbf{k}  \uparrow
\downarrow  \textbf{k}^{\prime}$ is actually satisfied for N and U-processes, near 
the lower and the upper edge of an energy band (cases a and c in Fig.\ref{fig9}).
However it would not be satisfied if an N-process occurred at an 
intermediate location (case b in Fig. \ref{fig9}), whereas the requirement
$\textbf{v}_{\textbf{k}} \downarrow  \uparrow  \textbf{v}_{\textbf{k}^{\prime}}$ would be 
satisfied in any case. Obviously, a consideration of the one-dimensional
spectrum Fig. \ref{fig9} shows that mutually opposed velocity orientations 
correspond to pairs of ES separated by a point corresponding to a
quasi-momentum \textbf{p} (see below) for which the energy distribution 
$\varepsilon (\textbf{k})$ features an extreme value. This means that
for $\textbf{k}= \textbf{p}$ the group velocity 
\begin{displaymath}
\textbf{v}_ \textbf{k}=\frac{1}{\hbar} \frac{\partial
\varepsilon(\textbf{k})}{\partial \textbf{k}} \arrowvert _{\textbf{p}}
\end{displaymath}
vanishes (with $\textbf{p}$ = 0: in case a;
$\textbf{p} = \textbf{k}_{m}$: in case c; and $0 < \textbf{p} < \textbf{k}_{m}$: in case b). 
This way it will be convenient to select pairs of ES in one-dimensional
$\textbf{k}$-space by looking for points $\textbf{p}$ corresponding to 
equilibrium population, which in this case separate the states $\textbf{k}$
and $\textbf{k}^{\prime}$ of an ES-pair.

\section{Definition of boundary surfaces separating inversely occupied states 
for a 3-dimensional electronic spectrum}

We shall now apply the results of point 2.2 for the case of a 3-dimensional 
electronic spectrum. Let $R(\textbf{q}) = \{\arrowvert i\textbf{k}\rangle
,\arrowvert i^{\prime}\textbf{k}^{\prime}\rangle\}$ be sets of equidistant
ES-pairs satisfying the conditions of  (\ref{1.9}), which at the same time are
potentially active during the creation of  phonons with quasi-momenta
$\hbar\textbf{q}$. As already discussed under sub-point 3 of point 1.5, the
contribution of stimulated radiative transitions between ES within a set $R
= \{\arrowvert \textbf{k}\rangle ,\arrowvert \textbf{k}^{\prime}\rangle\}$ can
in some instances become a macroscopic quantity, provided that the number
of ES-pairs in R also is a macroscopic quantity and a population inversion
of the majority of such pairs does exist. For simplicity the radiative 
transitions are supposed to be confined to one electron energy band, and 
for this reason the band indices will be omitted further on. In the case of 
an inhomogeneous spatial electron distribution, a population inversion
would  be possible if ES-pairs with anti-parallel components of velocities
$\textbf{v}  = \hbar^{-1}\vec{\nabla}\varepsilon$ existed along certain 
directions. This requirement could essentially be satisfied if the ES 
belonging to one pair were separated in \textbf{k}-space by certain 
geometrically defined boundary surfaces, including planes, which we shall 
generally denote as P-surfaces. In such case the general condition

\begin{equation}
(\textbf{v}(\textbf{k}),\textbf{n}(\textbf{k})) = 0 
\label{2.8}
\end{equation}
will apply. Here \textbf{n}(\textbf{k}) - unit normal vector in point 
\textbf{k} of surface P. This enables us to allocate to each of the
P-surfaces a totality of ES-pairs  with antiparallel velocity components being
collinear to the axis of  \textbf{n}, being localized in the proximity of a
layer with thickness $\sim q$ (we assume that $q \sim (10^{-3}\div 10^{-1}) 
\pi /a$, $a$ - lattice-parameter). However, not all of the  P-surfaces
actually determine a set of ES-pairs, as the general requirement  of
macroscopicity of the number of inversely occupied ES-pairs, as well as  the
need of compatibility with \eqref{1.9} have to be equally met. Let a given 
orientation of the spatial inhomogeneity be denoted by \textbf{e}. Then, in 
the proximity of those points of the P-surface where \textbf{n} is
collinear to \textbf{e}, an inverted population (IP) of ES-pairs should
actually exist, as in the \textbf{q}-proximity of these points the
existence of states with antiparallel velocity components along the
orientation determined by \textbf{e} would be inevitable. Obviously, the
number and density of these specific points mostly depend on the shape of
the  P-surface, and attain a maximum in the case that the P-surface is a
plane. For this reason, among the totality of P-surfaces, the sub-set of
P-planes (defined by \textbf{n} = \textbf{const}) will be highlighted. 

By definition, the P-plane is a plane in which the velocity component 
$\textbf{v}_{\textbf{n}}$ (along to the normal \textbf{n} of the plane) vanishes for 
all points of the plane. Obviously, planes satisfying this specific 
definition match with the symmetry-planes of the reciprocal lattice. (We 
recall the classical rule of reflection which simply states that all normal 
velocity components conserve their magnitudes and only change their signs, 
before and after reflection at a plane.) Consequently, on the points of a
symmetry plane, the velocity \textbf{v} (as an invariant during a
reflective transformation) does not feature non-vanishing components
collinear to the normal \textbf{n} of that plane.

Of course such symmetry considerations do not exclude in general the 
possible existence of P-planes not coinciding with symmetry planes.
For this reason we need to find certain specific equations which may enable 
us to determine such P- planes not being defined by symmetry 
considerations alone, but still being in accordance with the dispersion law. 
It should be noted that in case that a surface is a P-plane, Eqs. (\ref{2.8}) must 
be satisfied for all points of this plane and the next relationship (\ref{2.9}) 
occurs

\begin{equation}
( \textbf{n}, d\textbf{v}) \arrowvert_{\textbf{p}} = (\textbf{n},(d\textbf{k}, 
\vec{\nabla})\textbf{v})\arrowvert_{\textbf{p}} = 0,
\label{2.9}
\end{equation}
where d\textbf{k} corresponds to the requirement (d\textbf{k},\textbf{n}) = 
0. It is further possible to define two independent vectors $d \textbf{k}= 
d \textbf{k}_{1} , d\textbf{k} = d\textbf{k}_{2}$. Supposing that 
$d\textbf{k}_{1}= [\textbf{n},\textbf{v}]\arrowvert_{\textbf{p}} dl$ and $d\textbf{k}_{2}= 
\textbf{v}\arrowvert_{\textbf{p}} dl$, then we get from (\ref{2.9}):

\begin{equation}
(\textbf{n}, (\lbrack \textbf{n},\textbf{v} \rbrack \vec{\nabla})\textbf{v}) \arrowvert_{\textbf{p}}
= 0,(\textbf{n}, (\textbf{v},\vec{\nabla }) \textbf{v}) \arrowvert_{\textbf{p}} = 0. 
\label{2.10}
\end{equation}
If we write the equation for the P-plane in the form (\textbf{n, k}) = 
C, then it will be possible to find the normal-vector \textbf{n} 
and the constant C by resolving (\ref{2.8}) and (\ref{2.10}) for the components 
of \textbf{n} and C with a given value $\textbf{k} = 
\textbf{k}_{0}$ corresponding to the requirement $\textbf{v}( \textbf{k}_{0 
})\ne  0$. The vector $\textbf{k}_{0}$, which determines a point in the 
P-plane, possesses two independent components, which can be chosen 
arbitrarily. For practical reasons however it is more convenient to choose 
$\textbf{k}_{0}$ in such a way that Eqs. (\ref{2.8}) and (\ref{2.10}) attain their most 
simple form for $\textbf{k} = \textbf{k}_{0}$. 

In the particular case of a curved surface P, both its normal vectors 
\textbf{n} and the velocity vectors \textbf{v} change their orientation from 
point to point, with respect to the orientation of \textbf{e}. Thus the 
number of ES-pairs with anti-parallel velocity components $\textbf{v}_{e}$ 
(collinear to \textbf{e}) located in the \textbf{q}-layer can be small. And 
vice versa, from the point of view of a population inversion, only such 
specific ES-pairs are highlighted whose mutually anti-parallel velocity 
components are oriented collinear to the local spatial inhomogenity 
\textbf{e}. In \textbf{k}-space, such ES-pairs are separated by boundary 
surfaces of a different category (we denote them by S or S-surface) for 
which 

\begin{equation}
(\textbf{v, e}) = 0.
\label{2.11}
\end{equation}

This means that at any point in S, the projection of \textbf{v} onto 
\textbf{e} vanishes, i.e. $\textbf{v}\bot \textbf{e}$. According to
(\ref{2.11}) the surface S comprises the geometric loci of all singularity 
points (i.e. the extremes and flex points) of a function $\varepsilon 
(\textbf{k})$, if \textbf{k} varies parallel with \textbf{e}. Obviously, the 
scalar products $(\textbf{v}(\textbf{k}),\textbf{e}), 
(\textbf{v}(\textbf{k}^{\prime}),\textbf{e})$ will attain different signs for each 
pair of points $\textbf{k}, \textbf{k}^{\prime}$ located at opposed sides of S, but 
still remain in the proximity of S (i.e. where $\arrowvert\textbf{k} - 
\textbf{k}^{\prime}\arrowvert\le 0,1\pi /a$), provided the intersection of the vectors 
$\textbf{k} - \textbf{k}^{\prime}$ with S corresponds to a relative extreme of the 
function $\varepsilon (\textbf{k})$ in such intersections, i.e. for 
$(\textbf{k} - \textbf{k}^{\prime}, \vec{\nabla })\textbf{v}\arrowvert_{s}\ne 
0$. Let $(\textbf{k} - \textbf{k}^{\prime}, \vec{\nabla }) \textbf{v}\arrowvert 
_{s} = 0$ define a flex point of the function $\varepsilon (\textbf{k})$ on 
the surface S. Then, in the proximity of such a flex point, there will not 
occur a change of sign of the scalar product (\textbf{v}(\textbf{k}), 
\textbf{e}). Consequently, ES-pairs $\textbf{k}, \textbf{k}^{\prime}$ corresponding to 
such flex points will not be considered further. 

The equations of surfaces S can also be written in a form not explicitly 
depending on \textbf{e}. It will be reasonable to use such a notion when 
searching for certain S-surfaces with specific characteristics. Doing this 
way, the \textbf{e} vectors will not be given from the outset, but can be 
derived from the already known equations of an S-surface. We note that in 
any point of the S-surface, the velocity vectors \textbf{v}(\textbf{k}) - according 
to (\ref{2.11}) - are oriented parallel to a fixed plane. Consequently, if three 
arbitrary points in the S-surface are defined by vectors \textbf{k}, 
$\textbf{k}_{1}$, $\textbf{k}_{2}$ we get

\begin{equation}
(\textbf{v}(\textbf{k}), [\textbf{v}(\textbf{k}_{1}),\textbf{v}(\textbf{k}_{2})]) = 0. 
\label{2.12}
\end{equation}
Assuming that $\textbf{k}_{1} = \textbf{k} + \Delta \textbf{k}$, 
$\textbf{k}_{2} =\textbf{k} + \Delta \textbf{k}^{\prime}$, we can develop
(\ref{2.12}) by powers of $\Delta \textbf{k}, \Delta \textbf{k}^{\prime}$, 
however confining our decomposition of $\textbf{v}(\textbf{k}_{1})$, 
$\textbf{v}(\textbf{k}_{2})$ to the linear terms.

\begin{equation}
(\textbf{W}(\textbf{k}),[\Delta\textbf{k},\Delta\textbf{k}^{\prime}])\approx 0,
\label{2.13}
\end{equation}
where 
\begin{equation}
\textbf{W} = \frac{1}{2}\sum_{ijn}{\textbf{e}_{i}}\varepsilon_{ijn}
\left(\textbf{v}\,,\left[\frac{\partial\textbf{v}}{\partial
k_{j}},\frac{\partial\textbf{v}}{\partial k_{n}}\right] 
\right)
\label{2.14}
\end{equation}
and $\textbf{e}_{1}$, $\textbf{e}_{2}$, $\textbf{e}_{3}$ are the orthogonal 
unit vectors of a right-hand Cartesian frame of reference, with $\varepsilon 
_{ijn}= (\textbf{e}_{i},[\textbf{e}_{j}, \textbf{e}_{n}])$. 
If $\Delta \textbf{k}, \Delta \textbf{k}^{\prime} \to  0$, then the vector $[ 
\Delta \textbf{k}, \Delta \textbf{k}^{\prime}]$ adopts the orientation of the 
normal \textbf{n} of the surface S in point \textbf{k}. For this reason, the 
vector $[ \Delta \textbf{k}, \Delta \textbf{k}^{\prime}]$ can be replaced by 
\textbf{n} when reaching the limit $\Delta \textbf{k}, \Delta \textbf{k}^{\prime} 
 \to  0$ in (\ref{2.13}). In result, we obtain the general equation for the 
S-surfaces we were searching for:

\begin{equation}
(\textbf{W}(\textbf{k}),\textbf{n}(\textbf{k}) ) = 0. 
\label{2.15}
\end{equation}
In our specific task of search for and identification of the subset of flat 
S-surfaces, we shall essentially use (\ref{2.15}). Formally, our task of finding 
such planes, determined by a vector-field \textbf{W}(\textbf{k}), is 
equivalent to the task of finding P-planes associated with a velocity-field 
\textbf{v}(\textbf{k}), as (\ref{2.8}) and (\ref{2.15}), which define both the P and S 
surfaces, are formally equivalent. Thus their resolution will also lead to a 
resolution of Eqs. \eqref{2.8}, \eqref{2.10}, only by replacing \textbf{v} by \textbf{W}. 
By way of illustration let us analyze a cubic (fcc) lattice, using the 
definition of the basis vectors of the real and of the reciprocal lattice 
given in \cite{Kalluei69}. This will then enable us to use the simplest dispersion law 
in the tight binding approximation for fcc lattice \cite{Slater54}:

\begin{equation}
\varepsilon = \varepsilon_0 - 2\varepsilon_1\sum_{ij}{(1 
- \delta_{ij})\cos{\eta_{i}}\cos{\eta_{j}}},
\label{2.16}
\end{equation}
where $\varepsilon_{0}$, $\varepsilon_{1}$ - constants, $\eta_{i} 
= (ak_{i}/2)$, $a$ - lattice-constant. In this case, the set of P-planes will be confined to planes of symmetry 
with $k_{i }= 2m_{i}(\pi /a)$, $k_{i}\pm k_{j} = 4m_{ij}(\pi/a )$, $i\ne j$, $i,j = 1,2,3$; $m_{i}$, $m_{ij} = 0,\pm  1$,\ldots.
In our further search for S-planes, it will be convenient to distinguish 
between the following 2 variants of planes: 1) planes oriented parallel to 
the coordinate planes (or axes) of the lattice and 2) planes intersecting 
the coordinate planes or axes. The subset of planes oriented parallel to the 
coordinate planes can be defined by the condition $k_{i} = c_{i}$, ($i = 
1,2,3$). The constants $c_{i}$ can be determined from the relation

\begin{equation}
(\textbf{n,W})_{k_{i} = c_{i} ,k_{j} = k_{l} = 0} = 
2B\sin{(\frac{1}{2}{c}_{i}a)}(1 + \cos{\frac{1}{2}c_{i}a})^{2},
\label{2.17}
\end{equation}
where $B = 2a^{5}\hbar^{-3}\varepsilon_{1}^{3}$, being satisfied for 
$c_{i} = 2 m_{i}\pi/a$, $m_{i} = 0,\pm 1,\pm 2$,\ldots. The  specific planes
oriented parallel to a coordinate axis, for example the $k_{3}$ -axis, can
be defined by the condition $n_{1}k_{1} + n_{2}k_{2} = c_{12}$. The unknown
quantities $n_{1}$, $n_{2}$ and $c_{12}$ are determined by 

\begin{equation}
(\textbf{n,W})\arrowvert_{k_{3}=0} = 32 BD^{2}K(n_{1}\sin{\eta
_{1}}+n_{2}\sin{\eta_{2}}) = 0,
\label{2.18a}
\end{equation}

\begin{eqnarray}
(\textbf{n},(\textbf{W},\vec{\nabla})\textbf{W})\arrowvert_{k_{3}=0} =
32aB^{2}D^{4}K\{n_{1}\sin{\eta _{1}}[K(\cos{\eta_{1}}+ K-2 )-
                                           \nonumber            \\
-\sin^{2}{\eta_{1}} - \sin^{2}{\eta_{2}}] + (1\leftrightarrows2)\}=0 
\label{2.18b}
\end{eqnarray}
where\begin{displaymath}
D=\cos{\frac{1}{2}\eta_{1}}\cos{\frac{1}{2}\eta_{2}}, 
K=\frac{1}{2}\cos{\frac{1}{2}(\eta_{1}+\eta_2)} 
\cos{\frac{1}{2}(\eta_{1}-\eta_{2})}.
\end{displaymath}
The factor is K = 0, if $k_{1}\pm  k_{2} = 2(2m+1) \pi /a$, $m = 0, 
\pm 1$,\ldots. We can easily convince ourselves that the term in brackets in 
(\ref{2.18a}) can only vanish if the condition $\arrowvert n_{1}\arrowvert =
\arrowvert n_{2} \arrowvert$ is satisfied, 
i.e. for planes with $k_{1} \pm k_{2} = 4\pi m/a$. The Eqs. $k_{1} \pm k_{2} \pm k_{3}= 4\pi m/a$ defining 
the planes intersecting all coordinate axes can immediately be identified 
taking into account that all sections being cut out from the coordinate axes 
by such planes must be equally sized, as demanded by (\ref{2.17}), 
(\ref{2.18a}), (\ref{2.18b}). 
However, the requirement of (\ref{2.15}) will not be met on these planes, which 
means that they would not be S-planes. 

\begin{figure}[htb]
\centering
\includegraphics[clip=true, width=.6\textwidth]{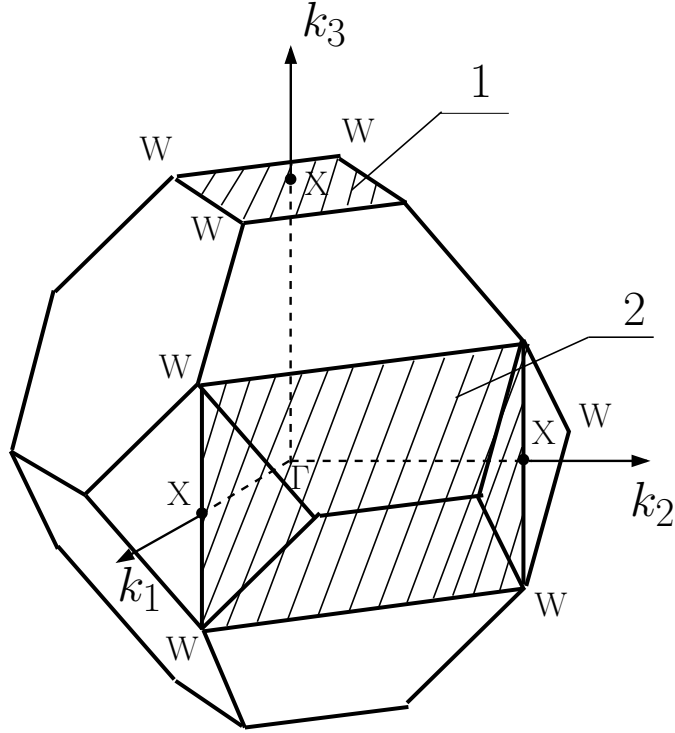}
\renewcommand{\captionlabeldelim}{.}
\caption{Image of the reduced Surface $S_{\lbrack 001 \rbrack}$ for the 
dispersion-law given by (\ref{2.16}); Sheets 1 and 2 are determined by:
$k_{3} =2\pi /a$, $k_{1} + k_{2} = 2\pi/a$, where $a$ - parameter of the 
fcc-lattice.}
\label{fig2.1}
\end{figure}

\begin{figure}[htb]
\centering
\includegraphics[clip=true, width=.6\textwidth]{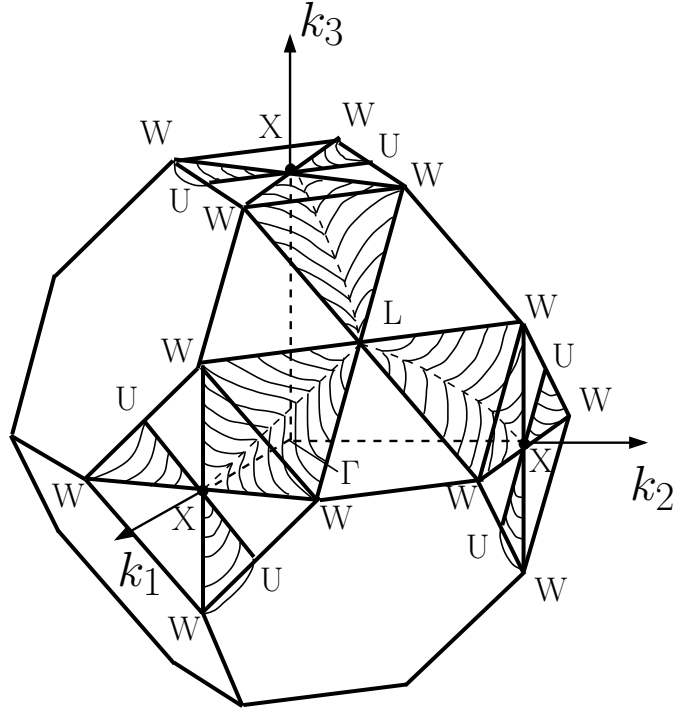}
\renewcommand{\captionlabeldelim}{.}
\caption{ Image of one reduced curved sheet of the $S_{\lbrack 111 \rbrack}$ 
surface for dispersion-law (\ref{2.16}) }
\label{fig2.2}
\end{figure}

\begin{figure}[htb]
\centering
\includegraphics[clip=true, width=.6\textwidth]{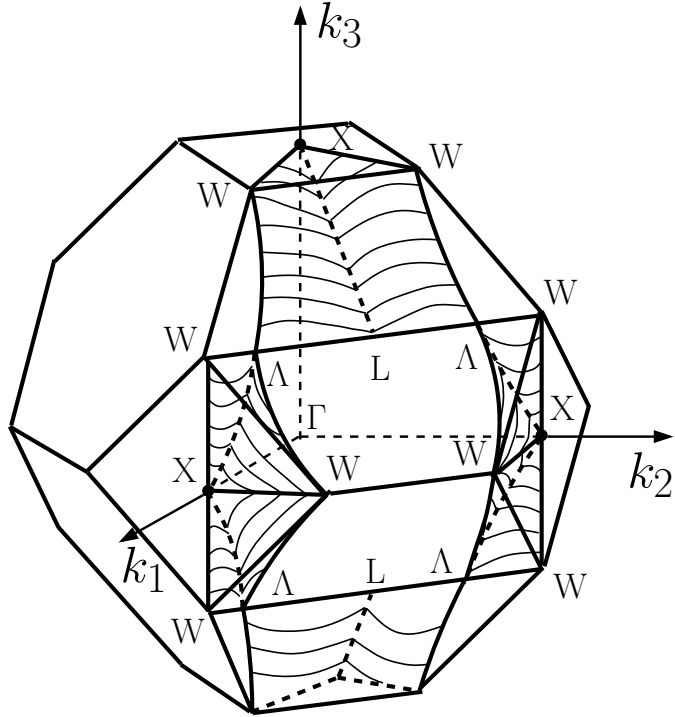}
\renewcommand{\captionlabeldelim}{.}
\caption{ Image of the reduced curved sheet of the $S_{\lbrack 110 \rbrack}$ 
surface for dispersion-law (\ref{2.16})}
\label{fig2.3}
\end{figure}

It should be kept in mind that the requirement for attainment of maximum 
population inversion , which actually selects the flat S-surfaces (planes), 
is not the only one. For this reason, also curved S-surfaces have to be 
included in our considerations. The Eqs. defining such surfaces can be 
derived from (\ref{2.11}) if certain \textbf{e} are given, as each \textbf{e} is 
associated with a multi-sheeted S-surface, whose shape would change with 
varying \textbf{e}. Using the spectrum given by (\ref{2.16}), it is possible to 
obtain the view about a modification of the shape of S-surfaces setting 
\textbf{e} along symmetry axes. For $\textbf{e}   \Vert   \lbrack 001 \rbrack$ (fourfold axis), 
$\textbf{e} \Vert   \lbrack 111 \rbrack$ (threefold axis) and 
$\textbf{e} \Vert \lbrack 110 \rbrack$ (twofold axis) 
the Eqs. describing the S-surfaces are:

\begin{equation}
S_{\lbrack 001 \rbrack} : \sin{(\eta_{3})}\cos{\frac{\eta_{1}+\eta_{2}}{2}}\cos{\frac{(\eta _{1} - \eta
_{2})}{2}} = 0,
\label{2.19} 
\end{equation}

\begin{equation}
S_{\lbrack 111 \rbrack} : \sin{ (\eta_{1}+\eta_{2})} + \sin{ (\eta_{1}+\eta_{3} )} +
\sin{(\eta_{2}+\eta_{3})} = 0, 
\label{2.20} 
\end{equation}

\begin{equation}
S_{\lbrack 110 \rbrack} : \sin{\frac{(\eta_{1}+\eta_{2})}{2}} [\cos{\frac{(\eta_{1}+\eta_{2})}{2}} + 
\cos{\frac{(\eta_{1}-\eta_{2})}{2}}\cos{(\eta_{3})}] = 0. 
\label{2.21} 
\end{equation}
Typical shapes of some sheets of the surfaces (\ref{2.19} -\ref{2.21}), being 
confined within the $1^{st}$ BZ, are shown in Figs. \ref{fig2.1} -\ref{fig2.3}. (In the 
following, we shall denote those specific sheets of S-surfaces being located 
within the 1$^{st}$ BZ as "reduced sheets"). 

For $\textbf{e} \Vert \lbrack 001 \rbrack$, (\ref{2.19}) describes a totality of planes:
\begin{equation}
\eta_{3} = m\pi  \to k_{3} = 2m \pi /a, m = 0, \pm 1,\pm 2,\ldots, 
\label{2.22} 
\end{equation}

\begin{equation}
\eta_{1}\pm \eta_{2} = (2m+1) \pi \to k_{1} + k_{2} = 2(2m+1) \pi/a
\label{2.23}
\end{equation}

In Fig. \ref{fig2.1}, the hatched quadratic sheet of the 1$^{st}$ BZ 
corresponds to the plane defined by (\ref{2.22}) for m = 1 (the opposed quadratic 
face corresponds to m = -1, and the flat region including the point 
$\Gamma$ corresponds to m = 0 ). The vertically hatched rectangle in 
Fig. \ref{fig2.1} corresponds to one of the four planes given by (\ref{2.23}) for 
m = 0: $k_{1}\pm  k_{2} = 2\pi/a$. 

For $\textbf{e} \Vert \lbrack 111 \rbrack$, all sheets of the surface given by 
(\ref{2.20}) are curved. Figure \ref{fig2.2} shows one of the three reduced sheets. 
It is possible to obtain the sheet of similar shape by means of the geometric 
operation of the inversion in relation to point $\Gamma$. (The sheet 
including the point $\Gamma $ is not shown in Fig. \ref{fig2.2}). 

For $\textbf{e} \Vert \lbrack 110 \rbrack$, (\ref{2.21}) is satisfied for S-planes 
with $k_{1} + k_{2} = 4 \pi  m/a$  (the reduced flat sheet corresponds to m = 0)
as well  as for curved S-surfaces (whenever the expression included in
brackets in (\ref{2.21}) vanishes). Figure \ref{fig2.3} shows one of the
reduced curved sheets (the second one is produced by the inversion in
relation to point $\Gamma$. 

\begin{figure}[htb]
\centering
\includegraphics[clip=true, width=.6\textwidth]{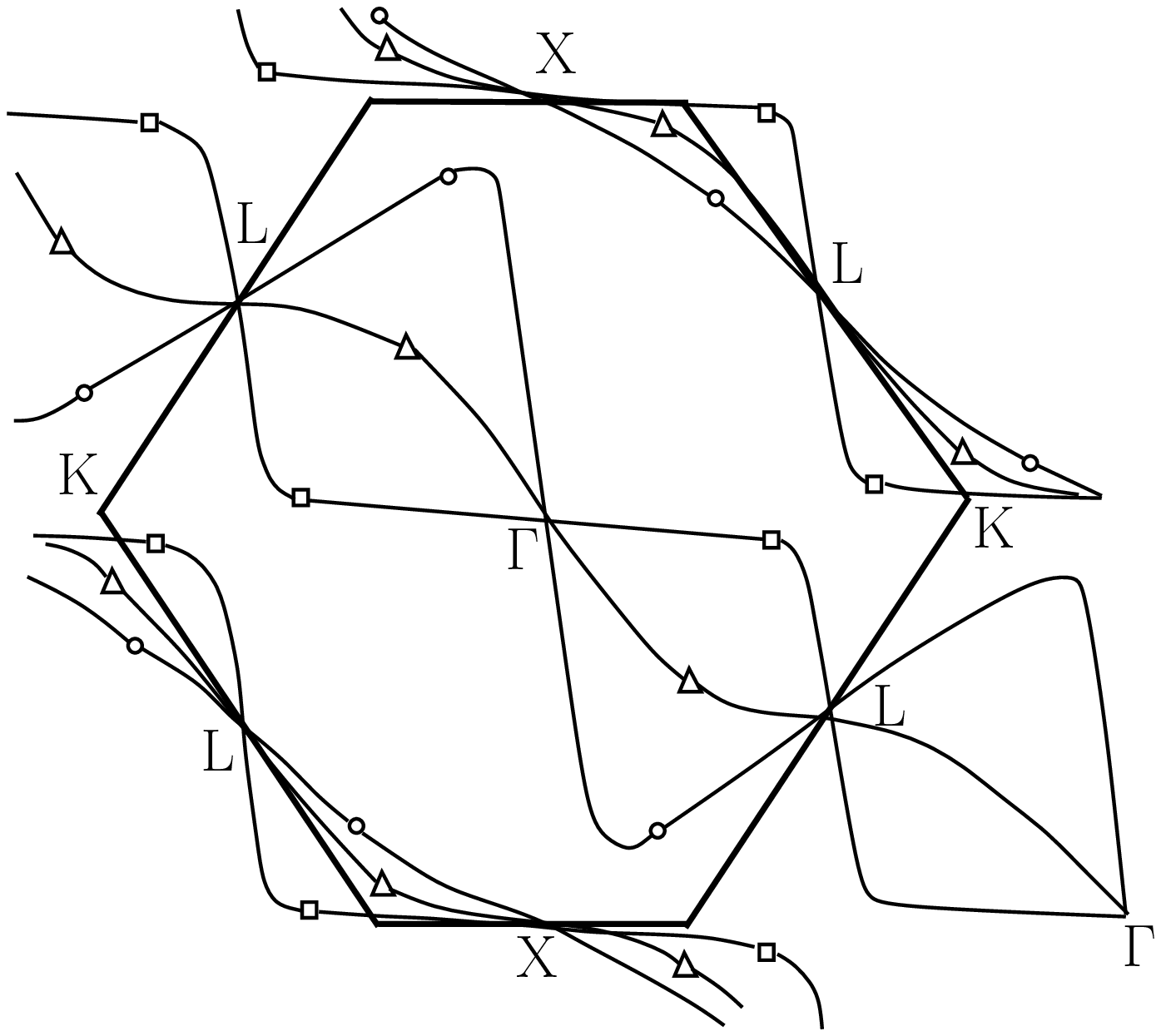}
\renewcommand{\captionlabeldelim}{.}
\caption{Image of the intersection of the $S_{[11\psi]}$
surfaces  with planes $k_{1} = k_{2}$ for dispersion-law (\ref{2.16}):}
\label{fig2.4}
\end{figure}
\begin{figure}[htb]
\includegraphics[clip=true, width=.9\textwidth]{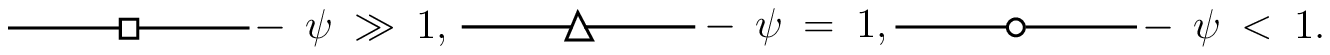}
\end{figure}
In  Fig. \ref{fig2.3}, showing the intersections of the S-surfaces (including 
those not shown in  Fig. \ref{fig2.1} - \ref{fig2.3} with planes $k_{1}= k_{2}$ for 
$\textbf{e} \Vert \lbrack 11\psi \rbrack$, it is possible to track 
their continuous transformation with varying \textbf{e}. For $\psi  >> 1$, the 
traces of the S-surfaces attain a step-like shape with pronounced linear (nearly 
vertical and horizontal) regions, which, for $\psi \to \infty$, will steadily 
transform into groups of perpendicular straight lines, which also represent 
the traces of planes (\ref{2.19}). The trace of that sheet of the S-surface, which 
includes the point $\Gamma$, will be subjected to the most pronounced 
changes. This trace changes over from a nearly horizontal section for $\psi 
 >> 1$, to the almost vertical, for $0 < \psi  << 1$ (for $\psi  = 0$, the 
vertical line can be regarded as trace of the plane $k_{1}= - k_{2}$, 
corresponding to (\ref{2.21})). The particular signs in figure \ref{fig2.4} are used 
to mark and distinguish the different traces (quadrangles, triangles and 
ellipses) and also indicate the proximity of \textbf{e} to the fourfold 
($\psi >> 1$), threefold ($\psi  = 1$) and twofold ($\psi << 1$) axes of 
symmetry.

\section{Limitations imposed by the equidistance relations in (\ref{1.9})}

The first of the implicit conditions in equation (\ref{1.9}), as well as the requirement of 
macroscopicity of the number of ES-pairs with inverted population will now 
enable us to develop an initial conceptual view of the S-surfaces (We should 
however bear in mind that P-planes are a sub-set of the totality of 
S-planes, defined by \textbf{n} = \textbf{e}, and for this reason the 
general symbol S will also include P-planes), justifying the exclusion from 
further consideration of all ES except those for which $R =
\{R^{s}\}=\{\{\arrowvert\textbf{k}\rangle ,\arrowvert\textbf{k}^{\prime}\rangle\}^{s}\}$ 
with wave-vectors $\textbf{k}=\textbf{s} + 1/2\textbf{q} +
\delta(\textbf{q}/q)$ and $\textbf{k}^{\prime} = \textbf{s} - 1/2 \textbf{q} +
\delta (\textbf{q}/q)$, being localized in the proximity of S-surfaces within a layer of thickness  
$\sim q \sim (10^{-3} \div 10^{-1})\pi /a$, with $\delta = \delta(s) =
q^{-1}(\textbf{q} , (\textbf{k}+\textbf{k}^{\prime}) / 2 - \textbf{s})$,
$\arrowvert \delta \arrowvert \le  1/2 q$ (using $\arrowvert \textbf{q} \arrowvert= q$). 
But even within R, not all states are equivalent. Among them, we shall 
highlight those energy states located within an interval $2\Delta$ near 
the Fermi-energy $\mu $, just because they can attain significant population 
differences, as will be shown later on. Let $R_{\mu\Delta}$ denote the set 
of ES-pairs satisfying the conditions

\begin{equation}
\mu - \Delta \le \varepsilon(k), \quad \varepsilon (\textbf{k}^{\prime})  \sim  
\varepsilon (\textbf{s})  \le \mu + \Delta.
\label{2.24}
\end{equation}
where $R_{\mu \Delta }$ represents a unification of the subsets 
$R^{s}_{\mu \Delta }$, in which each set is connected with its particular 
S-surface. For a given $\Delta $, each of these sets is determined by the 
dispersion law $\varepsilon =\varepsilon (\textbf{s})$ on its S-surface.

Obviously, $R^{s}_{\mu \Delta } = R^{s}$, if the following two 
inequalities are satisfied for the S-surface 

\begin{equation}
\varepsilon_{min} \ge \mu - \Delta, \quad \varepsilon_{max}\le \mu + \Delta 
\label{2.25}
\end{equation}
and $R_{\mu \Delta} = R$, if the inequalities (\ref{2.25}) are satisfied for any 
of the S-surfaces. Let us take for example the spectrum described by
(\ref{2.16}): For $\varepsilon_{1}  \approx 0,3$ eV, corresponding to a band width 
of 5 eV, together with $\Delta \approx 0,1$ eV and for $\mu $ values 
close to the upper band boundary, (i.e. $\mu \sim  \varepsilon 
_{max}=\varepsilon_{0} + 4 \varepsilon_{1}$), the whole 
reduced sheet 1 of the surface $S_{\lbrack 001 \rbrack}$ (see Fig. \ref{fig2.1}) is 
confined between the isoenergetic surfaces $\varepsilon =\mu \pm \Delta $, 
thus the inequalities (\ref{2.25}) are satisfied. For ferrous alloys, 
$\Delta \approx 0,24 $ eV may be typical (as shown in Chapter 4) and 
thus large enough for the limitation imposed by (\ref{2.24}) not becoming too 
restrictive.

The set $R_{\mu\Delta }$ includes all ES-pairs whose contribution to 
stimulated radiative transitions may prove to be a potentially macroscopic 
quantity. Thus our next task, being immediately related to the determination 
of the spectrum of generated phonons, will essentially comprise the 
isolation of those sub-sets $R_{\mu \Delta }(\textbf{q})$ from $R_{\mu 
\Delta}$ which contain equidistant ES-pairs being directly involved in the 
emission of phonons with the fixed magnitude and orientation of \textbf{q}.
In resolving this task we shall consistently take into account the 
requirements of : Macroscopicity of the quantity $\arrowvert R_{\mu \Delta
}(\textbf{q})\arrowvert$- i.e. of the number of ES-pairs in $R_{\mu \Delta
}(\textbf{q})$, Maximum of population inversion  and the requirements implied by
(\ref{1.9}), as dominant, while the electronic energy states will be considered as 
stationary quantities. Each of the sub-sets $R^{s}_{\mu\Delta}$ included in $R_{\mu\Delta }$ 
will have to be considered separately. Let us now decompose the energies
associated with the states $\arrowvert \textbf{k} \rangle, 
\arrowvert \textbf{k}^{\prime} \rangle$, being localized close to the S-surface, into powers of the 
deviations $\textbf{k} - \textbf{s}, \textbf{k}^{\prime} - \textbf{s}$, and 
confine our further analysis to the square terms. Using this approximation 
and considering (\ref{2.24}), we get from (\ref{1.9}): 

\begin{equation}
(\vec{\xi},\textbf{v}(\textbf{s})) + \hbar \delta m^{-1}(\textbf{s},\vec{\xi}) = q^{-1}\omega_{j}
(\textbf{q}) = c_{j\vec{\xi}},
\label{2.26}
\end{equation}
where $m^{-1} = \hbar^{-2}(\vec{\xi},\vec{\nabla})^{2}\varepsilon$;
$\vec{\xi}=\textbf{q}/q$, $c_{j\vec {\xi}}$ - velocity of sound.
If we now ignore the second addend in (\ref{2.26}), then (\ref{2.26}) will be satisfied 
anywhere on the S-surface, provided the condition

\begin{equation}
(\vec {\xi}, (\vec{\tau},\vec{\nabla})\textbf{v})\arrowvert_{s} = 0, 
\label{2.27}
\end{equation}
is satisfied. In (\ref{2.27}): $\vec{\tau }$ - arbitrary unit-vector, located in a tangential 
plane of S. According to (\ref{2.27}) and by definition of the S-surface, the 
vector $(\vec{\tau},\vec{\nabla})\textbf{v}$ is oriented 
perpendicular to $\vec{\xi}$ and \textbf{e}. Thus (\ref{2.27}) can be satisfied 
for any $\vec{\xi}$ being non-collinear to \textbf{e}, provided the vector 
($\vec{\tau}$,$\vec{\nabla}$) \textbf{v}, being associated with the 
S-surface, is collinear to fixed vector $\vec{\nu }_0 $ i.e., if the 
velocity field on the S-surface is prescribed by

\begin{equation}
\textbf{v}(\textbf{s}) = \mathrm{v}_{0}(\textbf{s})\vec{\nu}_{0} +\tilde{\textbf{v}}, 
\quad \vec{\nu}_{0} ,\,\tilde{\textbf{v}}\, \bot \,\textbf{e} 
\label{2.28}
\end{equation}
where $\tilde{\textbf{v}}$ is a constant vector. 
Eq. (\ref{2.26}) only applies to the velocity field (\ref{2.28}) if the second term in 
(\ref{2.26}) is omitted, and if $\vec{\xi }$ is perpendicular to $\vec{\nu }_0
$, and if $(\vec{\xi},\tilde{\textbf{v}})\;\approx  c_{j\vec{\xi}} $ . The second term
in (\ref{2.26}) must be considered if $(\vec{\xi}, \textbf{v})\arrowvert_{s} = const \approx 
c_{j\vec{\xi}} $. In this case (\ref{2.26}) is satisfied on the S-surface if, 
for any displacement in the S-surface, the incremental change of the first 
term can be compensated by the incremental change of the second term. The 
dependence of the second term in (\ref{2.26}) on an additional parameter $\delta $ 
enables us to achieve such compensation at any point of the S-surface, only 
by choosing a matching value for $\delta $ out of an interval $[-1/2
q, 1/2 q]$, taking advantage of the less rigid restrictions of the dispersion law 
$\varepsilon  = \varepsilon (\textbf{s})$ in relation to (\ref{2.27}).
This way the parameter $\delta $ can be considered as a function of 
\textbf{s}, not explicitly given by relation (\ref{2.26}). After conversion of 
(\ref{2.26}) with respect to $\delta$, we get

\begin{equation}
\delta = \hbar^{-1}m(\textbf{s},\vec{\xi})[c_{j\vec{\xi}}-(\vec{\xi},\textbf{v}(\textbf{s}))]. 
\label{2.29}
\end{equation}
Obviously the contribution of each term in \eqref{2.26} depends on the relative 
orientation of $\vec{\xi}$ versus \textbf{e}. For $\vec{\xi}  \to  \pm
\textbf{e}$ the first term in \eqref{2.26} will decrease, whereas for $\vec{\xi
} = \pm  \textbf{e}$, only the second term remains unchanged on the S-surface,
provided $\delta = \hbar ^{-1}m (\textbf{s}, \textbf{e}) c_{je}$. We must further note the 
possible existence of certain points on the S-surface, in the proximity of 
which there is a tendency for $m^{-1} \to 0$, corresponding to flex 
points of the function $\varepsilon (\textbf{k})$. Then, the second term in 
(\ref{2.26}) will become small, as the terminating values of $\delta$ cannot 
compensate for the decrease of $m^{-1}$. For the same reason, (\ref{2.26}) 
cannot be satisfied in the proximity of such points, if (\ref{2.26}) was obtained 
by a $2^{nd}$ order approximation of (\ref{1.9}) for \textbf{k} - \textbf{s}, 
$\textbf{k}^{\prime} - \textbf{s}$, so that these points have to be excluded. 

As $\arrowvert \delta \arrowvert \le 1/2 \textbf{q}$, the relation 
(\ref{2.29}) defines certain limitations on the range of variation of the 
functions $m(\textbf{s},\vec{\xi}),\textbf{v}(\textbf{s})$, if certain 
\textbf{q} is given. And vice-versa, if m and \textbf{v} are given, then 
relation \eqref{2.29} determines the lower limit of possible q-values, in any 
point of the S-surface: 

\begin{equation}
q \ge  q_{min}(\textbf{s},\vec {\xi})=2
\arrowvert\delta(\textbf{s},\vec{\xi})\arrowvert. 
\label{2.30}
\end{equation}
In the case of P-planes with $\vec{\xi } \Vert \textbf{n}$ we get $(\vec{\xi 
},\textbf{v}(\textbf{s})) = 0$. Thus from (\ref{2.29}) and (\ref{2.30}):

\begin{equation}
q_{min} = 2 \hbar^{-1}m(\textbf{s},\vec{\xi})c_{j\vec{\xi}}, 
\label{2.31}
\end{equation}
where m corresponds to the $\vec{\xi}$ -component of the effective 
mass-tensor $m = m^{\ast}_{\xi \xi}$. For $m = (1\div 5) m_{0}$ ($m_{0}$ - free electron mass) and assuming 
$c_{j\vec{\xi}} \sim 10^{3}$ m/s we get $q_{min} \sim (10^{-3} \div 10^{-2}) \pi/a$. 
If $\vec{\xi}$ deviates from \textbf{n}, then $q_{min}$ decreases, depending on the 
magnitude of $(\vec {\xi},\textbf{v}(\textbf{s})) \ne 0$. For 
$\arrowvert \textbf{v}(\textbf{s}) \arrowvert < c_{j\vec {\xi}}$, always is
$q_{min} > 0$, but for $\arrowvert \textbf{v}(\textbf{s}) \arrowvert > c_{j\vec {\xi}}$, there exists one 
orientation \textbf{q} for which $q_{min} = 0$. (If $\arrowvert \textbf{v}(\textbf{s}) \arrowvert c_{j\vec 
{\xi}}^{-1} >> 1$, then this orientation is close to \textbf{n} but 
if $\arrowvert\textbf{v}(\textbf{s})\arrowvert c_{j\vec {\xi}}^{-1}\sim 1$, then it 
will deviate significantly). It has to be taken into account that the 
orientation of $\textbf{v}(\textbf{s})$ on the P-plane does not necessarily 
remain constant. In this case the reduction of $q_{min}$, being caused by 
the deviation between the $\vec{\xi}$ and \textbf{n}, will be achieved due 
to the decrease in the number of inversely occupied ES-states, already being 
attached to a smaller region of the P-surface (related to the case $\vec{\xi}  
\,\Vert \, \textbf{n}$). Thus wherever $\vec{\xi}$ is non-collinear to 
\textbf{n}, the P-plane can be partitioned into sections $P_{i}$. In such 
cases each $P_{i}$ will be associated with an individual orientation of 
$\vec{\xi}$. Curved P-surfaces, on the other hand, can only be considered 
if they can be divided into a totality of sufficiently small sections. 
Finally, each of them would approximate flat sections. For example, in the 
proximity of the hexagonal face of the 1$^{st}$ BZ, the shape of the 
P-surface is close to that of the S-surface, as shown in Fig. \ref{fig2.2}. 
On the hexagonal face, the lines LW, which intersect the surface in regions 
with opposite sign of those electron group velocity components being 
oriented normal to the surface, belong to the P-surface. This means that 
those P-regions protruding from the upper side of the hexagonal surface 
alternate with those P-regions protruding from below. \footnote{
We note that such characteristics of a surface are not related to a specific 
kind of dispersion law $\varepsilon$(\textbf{k}), as they rather result 
from pure symmetry considerations \cite{Dzhons68}.}
It is reasonable to infer that, in the proximity of the hexagonal faces, the 
P-surface is divided into roughly six flat regions (with median lines LW), 
being projected towards those faces of the hexagonal face of the 1$^{st}$ BZ 
being limited by the lines LK, LU. The normals \textbf{n}$_{j}$ (j = 1,
\ldots 6) to these regions are grouped in the proximity of one of the
threefold axes $\langle 111 \rangle$, and the degree of deviation from 
\textbf{n}$_{j}$ depends on the magnitude of the velocity component $\textbf{v}_{p}$ 
of the electrons along the lines LW.

Now let us inspect the S-surfaces. The particularity of this case is that 
the first addend in (\ref{2.26}) also exists when S is a plane and $\vec{\xi} = 
\pm  \textbf{n}$ (in the case of P-planes $(\vec{\xi} , \textbf{v}
)\arrowvert_{p} = 0$ for $\vec{\xi}= \pm \textbf{n}$). Induced radiative transitions 
from states of higher energy into states with relatively lower energy 
correspond to such $\vec{\xi}$ pointing towards regions with increasing 
electronic energy. For this reason $\vec{\xi}$ must have such 
characteristics that the product ($\vec{\xi}$, \textbf{v}(\textbf{s})) in 
(\ref{2.26}) results into a positive quantity. According to (\ref{2.30}) $q_{min}$ 
depends on $\vec{\xi}$ and varies from point to point of the S-surface. 
Let us denote by $q_{m}$ ($\vec{\xi})$ the maximal value of $q_{min}$ 
(\textbf{s},$\vec{\xi})$, corresponding to a given $\vec{\xi }$. Among 
the totality of possible $\vec{\xi}$ we shall highlight those specific 
$\vec{\xi}$ satisfying the following conditions:
\begin{enumerate}
\item{ ($\vec{\xi}$, \textbf{v}(\textbf{s})) $ \ge $ 0 for the majority of 
points of the S-surface;}
\item{Maximum population difference with uniform sign at opposed sides of the 
S-surface;}
\item{ Small ($ \le 10^{-1} \pi /a$) magnitudes of $q_{m}$.}
\end{enumerate}
It should however be noted that for a given S-surface it is possible in 
principle that not even a single $\vec{\xi}$ may exist for which compliance 
with all of these conditions would simultaneously be given at all of its 
points. In such cases the possibility of partitioning the S-surface into 
individual sectors $S^{(i)}$ with an individually assigned orientation $\vec{\xi
}  = \vec{\xi}_{i}$ should be considered. The requirement of 
macroscopicity $\arrowvert R_{\mu \Delta }^{s(i)}(\textbf{q}^{(i)})\arrowvert$ 
for each of the desired $\textbf{q}\, = \,\textbf{q}^{(i)}$ must however presuppose the 
correctness of (\ref{2.26}) for the majority of points of the section $S^{(i)}$ of 
an S-surface and can be satisfied at any point of the section for

\begin{equation}
q^{(i)}  \ge q_{m} (\vec{\xi}^{(i)}). 
\label{2.32}
\end{equation}
For a given $\vec{\xi}^{(i)}$ the inequality (\ref{2.32}) determines the 
minimum magnitude of the wave vector of those phonons being emitted during 
stimulated transitions of electrons between ES from the set $R_{\mu \Delta 
}^{s(i)} (\textbf{q}^{(i)})$. 

It should be noted that the requirement of compliance with the conditions
of  equidistance (\ref{1.9}) enables us to compare the S-surfaces being linked
up with  different \textbf{e} and to sort out those ones which, for a given 
\textbf{q}, dispose over the largest total area $\Sigma_{s(\textbf{q})}$ of the 
S(\textbf{q})-regions within the reduced sheets of the S-surface. Obviously 
the number of ES-pairs being active immediately after their generation in 
the proximity of S(\textbf{q}) must be proportional to $\Sigma_{s(\textbf{q})}$. 
Taking the spectrum \eqref{2.16} as an example, there exist for $\textbf{e}\:
\Vert\: \lbrack 001 \rbrack$  and $\textbf{e} \:\Vert \:\lbrack 110 \rbrack$ each a planar (see sheet 2 in
Fig. \ref{fig2.1}) and  a curved (Fig. \ref{fig2.3}) reduced sheet of the
S-surfaces, for which the  vector $\textbf{q}\: \Vert \: \lbrack 110 \rbrack$ is highlighted
from the point of view of maximum  population inversion. Now, let us compare
these particular sheets of the S-surfaces. It is easy to realize that both
of the requirements of maximum population inversion, for  positive amount of
($\vec{\xi}, \textbf{v}$) as well as for conservation  of the sign of the
population difference, are satisfied for $\vec{\xi}=  \pm 1 \,/ \sqrt 2
\,(\textbf{e}_{1} + \textbf{e}_{2})$  in half of  the section $S_{\lbrack 001 \rbrack}$.
However, for small values of $\varepsilon_{1}$ and $q< 0,1\pi/a$, the
majority of points of this section does not satisfy Eq. (\ref{2.26}). This is due
to the fact that in $S_{\lbrack 001 \rbrack }$, the quantity ($\vec{\xi}$, \textbf{v})
varies within the wide range between 0 and $2\sqrt 2 a\varepsilon _{1}$,
whereas the term $m^{-1}= -\hbar^{2}a^{2} \varepsilon_{1}$ remains
constant. Due to the small variations of $\delta $ (not exceeding $0,1 \pi
/a$), (\ref{2.26}) is valid only for a small number of points in sheet 2 of
the surface $S_{\lbrack 001 \rbrack}$. As a result, the associated amount of pairs of
equidistant states is small and would not even  significantly increase with
variations of $\vec{\xi}$. And vice-versa, in  case of a curved surface
(\ref{2.21}) (see also Fig. \ref{fig2.3}) then (\ref{2.26}) could be satisfied
for the same $\vec{\xi}$ in a region being comparable  with that of the
sheet $S_{\lbrack 110 \rbrack}$ of the surface S, located within the  1$^{st}$ BZ. The
dimensions of this region depend on $\varepsilon_{1}$  and q. The first
addend in (\ref{2.26}), being related to the surface (\ref{2.21}), indeed
vanishes for $\vec{\xi}= (1/\sqrt 2 )(\textbf{e}_{1} +  \textbf{e}_{2})$,
while the factor $m^{-1}= - \hbar^{2} a^{2} \varepsilon_{1}
\sin^{2}{1/2(\eta_{1}+\eta_{2})}$ varies. Considering that $m^{-1} \to
0$ for ($\eta_{1} + \eta_{2})\to  0$, we must exclude from the possible
range of  variables $\eta_{1}$, $\eta_{2}$ the specific range

\begin{displaymath}
\arrowvert\eta_{1}+\eta_{2}\arrowvert \le 2 \arcsin{[2\hbar \,c_{j\vec{\xi}}\, 
(q \varepsilon_{1} a^{2})^{-1}]^{1/2}},\quad - \arrowvert\eta_{1} +
\eta_{2}\arrowvert\le \eta_{1} - \eta_{2} \le \arrowvert\eta_{1} +
\eta_{2}\arrowvert, 
\end{displaymath}
in which $\arrowvert\delta \arrowvert\ge\arrowvert\delta\arrowvert _{max} =
q/2$. For example, taking $c_{j\vec {\xi}} = 5 \cdot 10^{3}$ m/s, $a = 3,5
\cdot 10^{-10}$ m, $\varepsilon 
_{1} = 10^{-19}$ J, we obtain $\arrowvert\eta_{1}+\eta_{2}\arrowvert \le  
\pi/4$ for $q = 0,2 \pi /a$, and $\arrowvert\eta_{1}+\eta_{2}\arrowvert\le 
\pi/2$ for $q = 0,02 \pi /a$. The results of this estimate indicate that 
for $S_{\lbrack 110 \rbrack}$, the part of the area to be excluded amounts to about 10 
{\%} in the first case and to about 30{\%} in the second case. As the area 
$\Sigma$ of the sheet $S_{\lbrack 110 \rbrack }$ is large ($\Sigma \sim 10\pi 
^{2}/a^{2}$), the number of equidistant pairs of states in the proximity 
of the sheet $S_{\lbrack 110 \rbrack }$ also becomes large. Thus in the case of the 
dispersion law defined by (\ref{2.16}), and with due consideration of the 
conditions of equidistance, a comparison between flat and curved S-surfaces 
turns out in favor of the curved surfaces.


\section{Potentially active pairs of electronic states in the 
electronic-spectrum of fcc iron}

\begin{figure}[htb]
\centering
\includegraphics[clip=true, width=.5\textwidth]{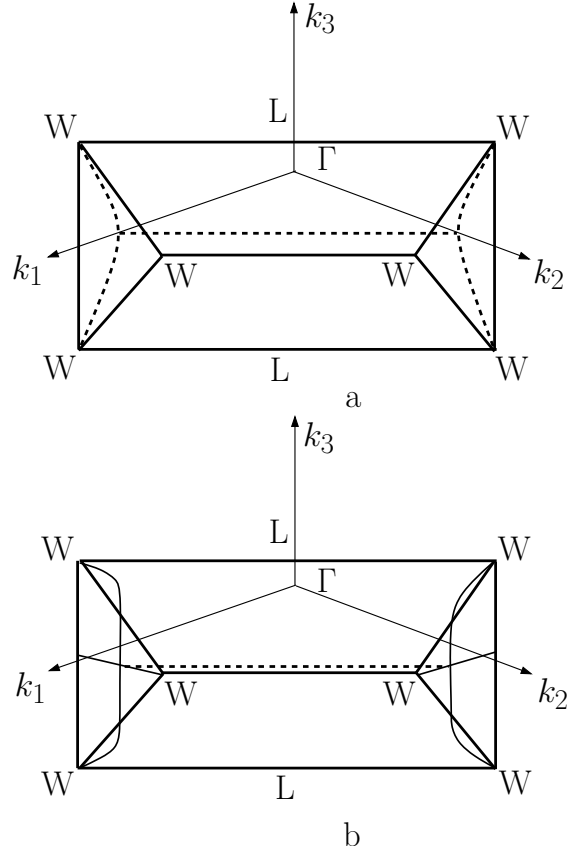}
\renewcommand{\captionlabeldelim}{.}
\caption{ View of reduced curved sheets 2 of the surface 
$S_{\lbrack 001 \rbrack}$ for the spectrum (\ref{2.33}):
a - Parameter $\varepsilon_{2} > 0$; b - Parameter $\varepsilon_{2} < 
0$. The intersection of surface $S_{\lbrack 001 \rbrack}$ with the plane $k_{3} = 0$ is 
shown by the horizontal dashed line.}
\label{fig2.5}
\end{figure}

Within the framework of our final aim to describe the $\gamma  - \alpha 
$- MT, our interest will now focus on an analysis of the S-surfaces of the 
fcc iron lattice. A detailed analysis however appears to be difficult, as 
this would imply substantial knowledge on the spectrum $\varepsilon 
(\mathbf{k})$ as well as on the velocity-fields \textbf{v}(\textbf{k}), 
within a region limited by isoenergetic surfaces $\varepsilon (\textbf{k}) 
= \mu  \pm  \Delta $. The data published in the relevant literature 
however only refer to symmetry lines (predominantly relating the bcc-phase 
of iron), whereas data on the velocity field are totally missing. Thus is 
would be reasonable to investigate to which extent the results obtained for 
the model spectrum (\ref{2.16}) can be used further. Above all we have to remark 
that all symmetry planes of the reciprocal lattice: $k_{i}  \pm  k_{j} 
= 4\pi  m/a$ are S-planes, independent of the underlying spectrum 
$\varepsilon (\textbf{k})$. Yet any of the S-planes corresponds to a vector 
\textbf{e}, oriented in the normal of the plane. This immediately follows from 
symmetry considerations. The velocity \textbf{v}, being an invariant with 
respect to reflective transformations in the symmetry-plane, can't feature a 
non-vanishing component oriented normal to the symmetry-plane and thus must 
completely lie within the symmetry-plane. Based on this consideration it can be 
concluded that the reduced sheet 1 of the surface $S_{\lbrack 001 \rbrack}$ (see
Fig. \ref{fig2.1}) at $\textbf{e}\: \Vert \: \lbrack 001 \rbrack$ also remains conserved for 
the $\varepsilon (\textbf{k})$ -spectrum of iron. However, there exist no 
reasons to surmise that the reduced sheet 2 of this surface maintains its shape. 
Even the slightest modification of the dispersion-law (\ref{2.16}):

\begin{equation}
\label{2.33}
\varepsilon = \varepsilon_{0} - 2\varepsilon_{1}\sum_{i 
j}(1-\delta_{ij})\cos{\eta_{i}}\cos{\eta_{j}} +
2\varepsilon_{2}\sum_{i}\cos{2\eta_{i}}
\end{equation}
taking into account the interaction with the second neighbor from \cite{Slater54}, 
results into a curved sheet 2, associated with an expansion of its area. 
Depending on the sign of $\varepsilon_{2}$, the curvature of sheet 2 
will take shape within the 1$^{st}$ Brillouin-Zone (1$^{st}$ BZ) for 
($\varepsilon_{2} > 0$), or be oriented towards the faces of the 1$^{st}$ 
BZ for ($\varepsilon_{2} < 0$), as shown in Fig. \ref{fig2.5}.
Thus the flattening of sheet 2 follows from particularities of the 
dispersion law. It can further be stated for the iron spectrum that sheet 2 
will intersect the hexagonal surfaces of the 1$^{st}$ BZ in the same 
horizontal WLW-lines as sheet 2 in figs. \ref{fig2.1} and \ref{fig2.5}. 
Symmetry considerations in fact lead to the conclusion that the velocity 
\textbf{v} is oriented along the lines WL (as well as WX), i.e., 
$\textbf{v}\arrowvert_{WLW}  \bot  \textbf{e} \Vert  \lbrack 001 \rbrack$, whereas the 
horizontal lines WLW belong to sheet 2 of surface $S_{\lbrack 001 \rbrack}$. In the case of an 
iron spectrum, the intersecting region of sheet 2, comprising the square 
surfaces of the 1$^{st}$ BZ $k_{1} = 2\pi /a$, $k_{2} = 2\pi /a$ can 
only go along the line WXW if the dispersion $\varepsilon$ vanishes in the 
lines WX, in the same way as for the spectrum (\ref{2.16}), or in other words, if 
the velocity vector \textbf{v} vanishes along these lines. 
Otherwise the intersection would be located either along the curved line 
$\Lambda$, being defined by the requirement $(\textbf{v},\textbf{e} 
)\arrowvert_{\Lambda} = 0$, as shown in Fig. \ref{fig2.5}a, or in singl 
points W, as shown in Fig. \ref{fig2.5}b. We emphasize that from the point of 
view of materialization of a maser-effect the curved sheets 2 imply a promising 
advantage over the flat sheets. The ES-pairs located in the proximity of the flat sheets 2 
cannot contribute to the generation of phonons with $\textbf{q} \:\Vert \:\lbrack 001 \rbrack$, as 
they don't feature inverted populations in this direction, and for 
$\textbf{q}\: \Vert \:\lbrack 110 \rbrack$, their contribution would only be small, as already 
mentioned. This means that for $\textbf{q}\: \Vert \: \textbf{e}\: \Vert \:\lbrack 001 \rbrack$ and 
flat sheets 2 only ES-pairs located in the proximity of the flat sheets 1 would 
have the potential to actively contribute to phonon generation. However, in 
the case of curved faces 2, phonon generation with $\textbf{q}\: \Vert\: \lbrack 001 \rbrack 
$ would not be generally forbidden. Therefore, an appreciable increase of the 
number of ES-pairs being active during phonon generation with $\textbf{q}\: \Vert \: 
\lbrack 001 \rbrack $ can be expected, if compared with the spectrum (\ref{2.16}).

For $\textbf{e}\:\Vert\:\lbrack 111 \rbrack$, the shape of the reduced sheet of surface 
$S_{\lbrack 111 \rbrack}$, as shown in Fig. \ref{fig2.5}, must not differ appreciably from 
the reduced sheet of surface $S_{\lbrack 111 \rbrack}$ of the iron spectrum $\varepsilon 
(\textbf{k})$. This way the lines WLW at the hexagonal face of the 1$^{st}$ 
BZ, and the lines UXU at the square faces of the 1$^{st}$ BZ will also 
belong to the surface $S_{\lbrack 111 \rbrack}$ of iron, as the velocity there is oriented 
perpendicular to $\lbrack 111 \rbrack$, due to the symmetry requirements. It should also be 
noted that the aforementioned remarks on the intersecting lines of face 2 of 
surface $S_{\lbrack 001 \rbrack}$ with the square surface of the 1$^{st}$ BZ also apply to 
the allocation of lines WXW to surface $S_{\lbrack 111 \rbrack}$. 

For spectrum (\ref{2.16}) and for that of iron at $\textbf{e} \:\Vert \:\lbrack 110 \rbrack$, the 
common features of the reduced sheets of the surface $S_{\lbrack 110 \rbrack}$ are the 
vertical lines WXW, being located on the square faces of the 1$^{st}$ BZ,  
$\textbf{k}_{1} = 2\pi /a ,  \textbf{k}_{2} = 2\pi /a$, and the 
horizontal lines WLW located on the hexagonal surfaces of the 1$^{st}$ BZ, 
being an immediate result of the orthogonality relation for the velocity 
$\textbf{v}  \bot  \lbrack 110 \rbrack$ in these lines. It will however be necessary to 
specify the intersection lines of the surface $S_{\lbrack 110 \rbrack}$ with the 
horizontal square surfaces of the 1$^{st}$ BZ (in the same way as in the 
previous cases), as well as the location of the lines WAW (see Fig. \ref{fig2.3}).

\begin{figure}[htb]
\centering
\includegraphics[clip=true, width=.8\textwidth]{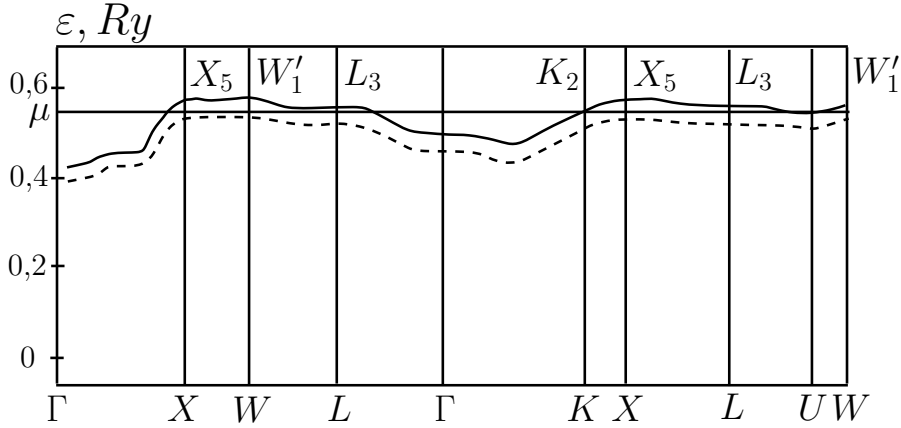}
\renewcommand{\captionlabeldelim}{.}
\caption{ Extract of the calculus in $\lbrack 110 \rbrack$ of the dispersion curves of ferromagnetic 
nickel with fcc lattice. The solid lines indicate spin-down states, dashed lines 
spin-up states, $\mu$ = Fermi-energy.} 
\label{fig2.6}
\end{figure}

Let us now deal with questions related to the possible existence of 
electronic bands in the energy-spectrum of fcc iron, featuring weak 
dispersion at the reduced sheets of the S-surfaces, which at the same time 
dispose of an acceptable energetic interval $\Delta  \sim  0,2 $eV in the 
proximity of the Fermi-energy $\mu$, in accordance with inequality (\ref{2.24}). 
As the bcc lattice modification of iron predominates in the temperature 
range between 0 and 1183 K, most calculus related to structure deal with 
this modification. Nonetheless, the results published in
\cite{Wood62,Bagayoko84,Zornberg70,Connolly67}, dealing 
with the energy spectrum of the fcc lattice modification of iron \cite{Wood62,Bagayoko84}, as well as 
with nickel in \cite{Zornberg70,Connolly67,Jarlborg80}, will enable us to answer the first part of the 
question. The similarity between the energy spectra of iron and nickel, 
becoming evident from a comparison of the results in \cite{Bagayoko84} for spin-up 
sub-bands with the results in \cite{Zornberg70,Connolly67,Jarlborg80}, enables us to use the more 
comprehensive calculus of \cite{Jarlborg80} for Ni, with the aim to reveal those bands 
featuring a weakly pronounced dispersion. Figure~\ref{fig2.6} shows data 
published in \cite{Jarlborg80}, related to those branches of the $\varepsilon_{L_3 } $ 
energy spectrum featuring the weakest dispersion of the lines corresponding 
to the above discussed S-surfaces. As can be seen in Fig. \ref{fig2.6}, the 
energy along the lines LW, LU and LX is close to $\varepsilon_{L_{3}} $, and 
along the line XW is close to $\varepsilon_{X_{5}}$. According to 
\cite{Wood62}, the difference between the energies $\varepsilon _{X_{5}}  - \varepsilon 
_{L_{3}}  \approx 0,02 Ry  \approx  0,27 $ eV. (It should be noted that 
in \cite{Wood62} only the curves $\varepsilon $(\textbf{k}) for the $\gamma $-phase of 
iron along $\Gamma X, \Gamma L, \Gamma K$, as well as the term - energies 
in the points $\Gamma , X, K, L$ are presented).
 
\begin{figure}[htb]
\centering
\includegraphics[clip=true, width=.5\textwidth]{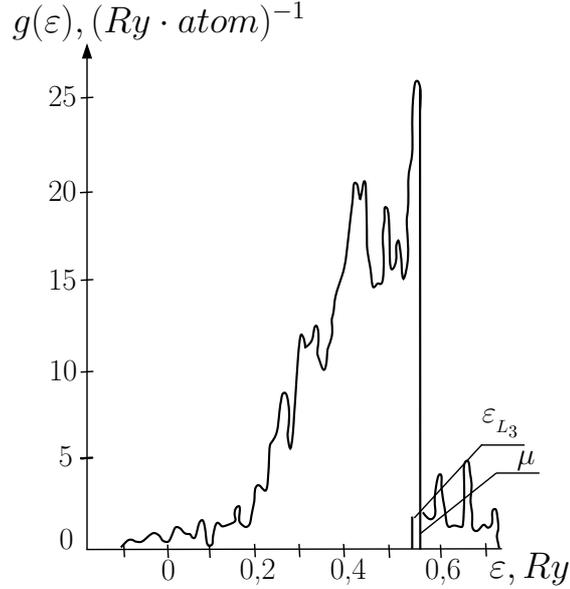}
\renewcommand{\captionlabeldelim}{.}
\caption{ Density of states (DOS) for the non-magnetic state of nickel 
with fcc-lattice $\lbrack 110 \rbrack$. The loci of the Fermi-energy $\mu $ and of the energy 
$\varepsilon _{L_{3}} $ are labeled.} 
\label{fig2.7}
\end{figure}

Thus it can be inferred from the available details of energy spectra that 
there is an appreciable probability for the existence of S-surfaces with 
weak energy dispersion.

\begin{figure}[htb]
\centering
\includegraphics[clip=true, width=.7\textwidth]{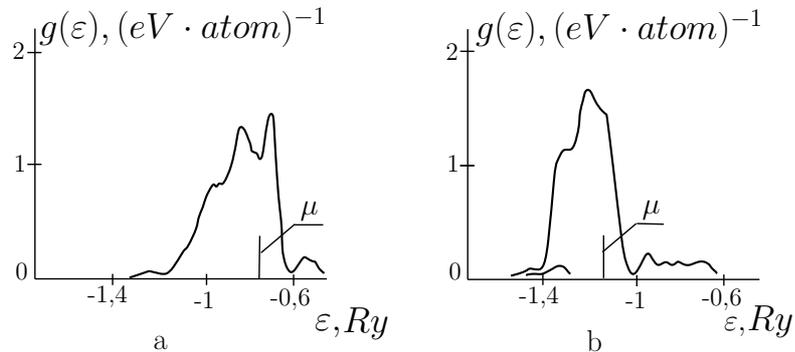}
\renewcommand{\captionlabeldelim}{.}
\caption{ DOS-functions of iron 
with fcc-lattice from \cite{Snow69}, for two different electronic configurations of the 
iron atoms: $a - 3d^{7}4s^{1}$; $b - 3d^{6}4s^{2}$}
\label{fig2.8}
\end{figure}

The second part of our question will be more difficult to answer: Which are 
the mutual arrangements of the energy level $\mu$ and of the energies 
$\varepsilon(\mathbf{s})$ on the S- surfaces or, put in another way: How 
can the size of the interval $\Delta  = \arrowvert \mu  - \varepsilon 
(s)\arrowvert$ be unambiguously determined on the basis of existing calculations 
of electronic spectra, with a required precision of at least $ \le 0,1 eV$. 
If we denote by $\varepsilon_{L_{3}}$ the spatial average energy on the 
S-surface, then we will note from the calculus in \cite{Zornberg70,Connolly67,Jarlborg80}
that $\varepsilon_{L_{3}}$ is localized in the peak area of the DOS $g(\varepsilon)$, whose 
energy is close to the top of the 3d-band, as shown in Fig. \ref{fig2.7}. 
Provided the electronic configuration is known, and after having determined the 
location of $\mu $, it would be reasonable to assume $\Delta = 
\varepsilon_{L_{3}} - \mu $. However, the precise electronic configuration of iron 
atoms in the solid (metallic) state is not precisely known and the quantity 
$\varepsilon _{L_{3}} - \mu $ varies significantly with changes in 
the electronic distribution between the 3d and 4s states. In figure
\ref{fig2.8}, which has been extracted from \cite{Snow69}, the characteristics of 
the curves $g(\varepsilon)$ are compared for the $3d^{7}4s^{1}$, 
$3d^{\,6}4s^{\,2}$ configurations. The change of configurations is associated with a 
reduction of the difference between the peak-energies $\varepsilon  \approx  
 \varepsilon_{L_3} $ (in the proximity of the top of the d-band) and 
$\mu $ from $ \approx $ 0,6 eV down to $ \approx $ 0,2 eV. In the last case, the 
peak degenerates to a step like shape. The intermediate configuration 
$3d^{\,6,5} 4(sp)^{\,1,5}$ of the $\gamma$- and $\alpha$ -phases of iron has 
been used in \cite{Roy77}. (However an explicit form of the $g(\varepsilon)$ is not 
presented in \cite{Roy77}). 

It should be noted here that the results of the calculated 
configurations ($3d^{9,4} 4s^{0,6}$ in \cite{Zornberg70,Connolly67} and $3d^{\,8,6}4(sp)^{\,1,4}$ 
in \cite{Jarlborg80}) differ appreciably for the fcc modification of Ni. If preference was 
given to the self-consistent calculus in \cite{Jarlborg80,Roy77}, then the configurations 
$3d^{6,5} 4(sp)^{1,5}$ for Fe and $3d^{8,6} 4(sp)^{1,4}$ for Ni would be in 
harmony with the lemma of increased probability of population of states with a 
fraction of s-electrons, during passage from the elements of the mean of the 
3d-group towards its end \cite{Samsonov76}. A value of $\Delta
=\varepsilon_{L_{3}} -  
 \mu \approx $ 0,4 eV can be expected, which would be intermediate 
in relation to the $\Delta =\varepsilon_{L_{3}}  -  \mu $, for the 
configurations $3d^{\,7}4(sp)^{\,1} , 3d ^{\,6}4(sp)^{\,2}$. 

One point however, being related to the vague interpretation of paramagnetism of 
the $\gamma$- phase of iron, is important enough to deserve a remark: 
Experimental results published in \cite{Liakutkin81,Liakutkin83} clearly show that the magnetic 
susceptibility of the $\gamma$- phase is less by a factor of about $1.5 \div 2$ 
in comparison with that of the $\alpha$- or $\delta$-phase and, moreover, does 
not obey to the Curie-Weiss law \cite{Dovgopol82}. In \cite{Dovgopol82}, these facts are reasonably 
explained by the Pauli-characteristics (including exchange amplification) of the 
paramagnetism of the $\gamma$- phase. In contrast, the experimental results of 
diffuse magnetic neutron scattering published in \cite{Wilkinson56,Brown83} provide clear 
evidence of a remarkable amount of magnetic moment density in the $\gamma$- 
phase (being of an order of magnitude of Bohr's magneton $\mu_{B}$ per atom). 
As stated in \cite{Dovgopol82}, it is possible to interpret these observations without 
contradiction by the theory of spin-fluctuations, stating that the average 
spin-moment of a given "knot" vanishes, due to permanent statistical 
fluctuations, and thus cannot be observed or derived from magnetic 
susceptibility measurements under static conditions. However, if the 
measurements are performed at a timescale being comparable with the average 
lifetime of these fluctuations (like neutron diffraction), then the spin moment 
fluctuations can clearly be observed. 

In a paramagnetic state, the existence of disordered local magnetic moments 
of finite magnitude implies that the local DOS varies from point to point, 
with the same probability for a spin-up or a spin-down condition of a 
spin-polarized state. Thus, in such cases there must also exist spin-up $ 
\uparrow$ and spin-down $\downarrow$ states located on S-surfaces, with 
identical difference $\Delta = \arrowvert \varepsilon (\textbf{s}) - \mu
\arrowvert$. As described in \cite{Grebennikov81}, the average DOS is broadened and 
characterized by displaced and diffuse peaks of the DOS function, if
compared  with a ferromagnetic (or non-magnetic) state.

We should also recall that in \cite{Tauer61,Kaufman63}, the existence of two atomic states 
($\gamma_{1}$ and $\gamma_{2}$) of iron in the $\gamma$-phase has 
already been hypothesized, essentially stating that a $\gamma_{1}$ state 
coexists with a smaller atomic volume $V_{1}$ and a smaller magnetic moment, 
together with a $\gamma_{2}$ state characterized by a relatively larger 
atomic volume $V_{2} > V_{1}$ and larger magnetic moment. In
\cite{Bagayoko84,Roy77,Poulsen76,Kubler81} (see also \cite{Kulikov82}), 
this hypothesis received some verification 
within the frame of the simple Stoner-model. 
\begin{figure}[htb]
\centering
\includegraphics[clip=true, width=.6\textwidth]{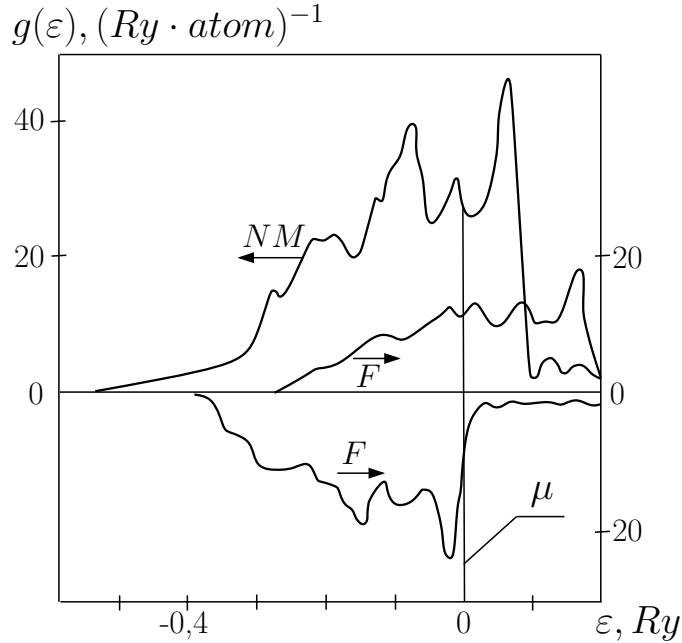}
\renewcommand{\captionlabeldelim}{.}
\caption{DOS-functions of iron with fcc-lattice. NM - non-magnetic 
state, F - ferromagnetic state (lower curve - spin-up, upper curve - spin-down) 
\cite{Kubler81}}
\label{fig2.9}
\end{figure}
As has been shown therein, for 
an existing exchange-splitting of the bands into spin-up $ \uparrow $ and 
spin-down $ \downarrow $ sub-bands, together with the assumption of steady 
increases of the lattice parameters at certain $a = a_{0}$ ($a_{0} \approx  
3,58$ \,\AA \, in \cite{Kubler81} and $3,63 < a_{0} < 3,71 $\,\AA \, in \cite{Bagayoko84}), a stable 
ferromagnetic state with a magnetic moment of $(2,3 \div 3) \mu _{B}$ per 
atom will arise, being separated from states with vanishing magnetic moment by 
an energy gap of 0.1 eV \cite{Roy77} (in \cite{Kaufman63} an energy-gap of 0,037 eV is given). In 
Fig. \ref{fig2.9}, borrowed from \cite{Kubler81}, both the DOS functions $g(\varepsilon)$ 
of the non-magnetic and of the spin-polarized state are shown for $a \approx 
a_{o}$. Obviously, the peak of the nearly filled spin-up $ \uparrow$ sub-band 
is located below the Fermi-level, in the proximity of $\mu $, $\mu  -  
 \varepsilon_{L_{3} \,\uparrow} \approx  0,2$ eV. If we took 
the location of $\mu $ between two large peaks of the DOS function 
$g(\varepsilon)$ of the non-magnetic state as a criterion for selection of the 
electronic configuration, then the electronic configuration of iron would be 
near the $3d^{7}4s^{1}$ configuration (as can be confirmed by comparison with 
Fig. \ref{fig2.8}a). However, due to the existence of exchange-splitting, the 
values of the parameters $\Delta  \sim  0,2$ eV are also possible for 
this configuration. It should be noted that the value $\varepsilon_{L} - \mu  
\approx$ 0,78 eV for the non-magnetic state of the $\gamma$-phase (published 
in \cite{Kubler81}) is larger than the corresponding value of 0,6 eV
published in \cite{Snow69}.

As the martensitic transformation temperature $M_{S}$ of pure iron (or of 
iron with small content of nickel or carbon) is $M_{S} \sim 10^{3}$ K, 
it would be possible for the iron atoms to simultaneously coexist in both states 
$\gamma_{1}$ and $\gamma_{2}$, with the same probability at $T \sim M_{S}$. 
If we assigned to the state $\gamma_{1}$ the configuration $(3d \uparrow)^{3} 
(3d \downarrow)^{3} (4sp)^{2}$, which does not feature a local magnetic 
moment, and to the state $\gamma_{2}$ the configuration $(3d \uparrow)^{5} 
(3d \downarrow )^{2} (4sp)^{1}$, featuring an atomic moment of $3 \mu 
_{B}$, as surmised in the modified Weiss-pattern \cite{Eishinskii84}, then the states located 
above and below the Fermi-level $\mu $, being linked up with the $\gamma_{1}$, 
$\gamma_{2}$ - states of iron, would be interchangeable on the S-surfaces (with 
values $\Delta  \sim  0,2$ eV). 

On the basis of energy spectrum calculus however, substantial difficulties arise 
in unambiguously determining both the magnitude and the sign of the chemical 
potential difference $\mu_{\alpha} - \mu_{\gamma}$ of the electrons in the 
$\alpha$- and $\gamma$ - phase (Let's remind that the specified difference is 
the significant characteristic of a non-equilibrium condition of an electronic 
system). According to \cite{Snow69}, the chemical potentials of non-magnetic phases are 
practically indistinguishable in the configuration $3d^{7}4s^{1}$. For the 
configuration $3d^{6} 4s^{2}$, a value of $\mu_{\alpha} - \mu_{\gamma} 
\approx  - 0,2$ eV is proclaimed. However, the calculus in \cite{Keller74}, which 
was performed by the cluster method (for the configuration $3d^{6} 4s^{2})$, 
delivered $\mu_{\alpha} - \mu_{\gamma} \approx +\,0,2$ eV, the 
sign thus being opposed to the aforementioned sign. However, it should be 
remarked that the limitation to three regions of nearest neighbors, used in 
\cite{Keller74}, presumably is inadequate for a sufficiently precise evaluation of the 
small difference $\mu_{\alpha} - \mu_{\gamma}$. Thus the calculus by the 
augmented plane wave method in \cite{Snow69} should be given preference. 

There still remains one unconsidered aspect: During the $\gamma -\alpha$ - MT 
(with exception of Fe-Ni alloys with a Ni-content of $ > 30$ {\%}), the MT is 
associated with a transition from the paramagnetic $\gamma$- into the 
ferromagnetic $\alpha$ - phase (we recall that the Curie-temperature T$_{c}$ = 
1041 K of the $\alpha$ - phase of iron is higher than the $M_{S}$ temperature at 
the onset of a martensitic transformation). This means that $\mu_{\gamma}$ of 
paramagnetic and $\mu_{\alpha}$ of ferromagnetic iron must be compared among 
themselves. Estimates in \cite{Grebennikov81}, based on a DOS model, indicate that during the 
process of ferromagnetic ordering, the magnitude of $\mu_{\alpha}$ decreases, 
that would be an argument in favor of a negative sign of the difference $\mu 
_{\alpha} - \mu_{\gamma} < 0$. Of course this question could 
most easily be resolved by a proper experimental method, taking advantage of the 
fact that the large temperature hysteresis between martensitic transformations 
of Fe-Ni alloys and their reverse transformation (see Pt. 1.2) enables us to 
measure the difference of the contact potential at one and the same temperature 
between a sample which has undergone practically the total $\gamma - \alpha $ -
MT (as with Fe-Ni alloys containing up to 28 {\%} Ni), and an austenitic sample of 
same composition. In our further evaluations, a value of $\mu_{\alpha} - \mu 
_{\gamma}= - 0,16$ eV will generally be assumed. This latter value was 
determined under adequate consideration of the volume-effect of the MT, as 
described in Pt. 1.5. A more detailed treatment of this question will be 
resumed under Pt. 4.5.3. 

\begin{center}
\textbf{Some additional remarks}
\end{center}

\begin{enumerate}
\item{ In addition to the ES with least dispersion (see Fig. \ref{fig2.6}), 
there may also exist active ES belonging to other energy bands. For example, 
in the proximity of point X, potentially active ES are those 
associated with point $X_{2}$ and energies $\varepsilon_{X_{2}}  \approx 
\varepsilon_{L_{3}}$.}
\item{ From the point of view of compliance with the energetic criterion 
(\ref{2.24}), a curvature of sheet 2 of surface $S_{\lbrack 001 \rbrack}$ oriented towards the 
faces of the 1$^{st}$ BZ, would be more favorable (see Fig. \ref{fig2.5}b ), as 
on the one hand, the dispersion along line $\Gamma K$ is remarkably pronounced, 
and on the other hand, a curvature towards the center of the 1$^{st}$ BZ 
(see Fig. \ref{fig2.5}a) would come into conflict with the conditions implied by 
(\ref{2.24}), for the largest part of sheet 2, and would thus finally lead to a 
contraction of the region adjacent to lines LWL. }
\item{A weak dispersion $\varepsilon$ (\textbf{k}) (comparable with the 
dispersion of the phonon-branches) in the essential sections of lines LW and LV 
of the P-surfaces in the proximity of the hexagonal Brillouin-faces (as well as 
for similarly shaped S$_{\lbrack 111 \rbrack}$ surfaces) opens up the possibility for 
participation of ES-pairs - being mutually separated by these surfaces - on the 
generation of long-wave phonons with quasi-momentum $\hbar \textbf{q}$, whose 
orientation would significantly deviate from the threefold axes of symmetry 
$\lbrack 111 \rbrack$ towards the twofold axes of symmetry $\lbrack 1\bar{1}0
\rbrack$, $\lbrack10\bar {1}\rbrack$, $\lbrack 0\bar {1}1 \rbrack$ 
(also see the remarks following (\ref{2.30}) of Pt. 2.4). We also 
remind that the component of electron group velocity being oriented normal to 
the P-surface would vanish by definition of (\ref{2.8}). This in turn implies that 
the non-equilibrium addends relating to the occupation of states on both sides 
of a P-plane would essentially be proportional to the projection of \textbf{e} 
(\textbf{e} - orientation of spatial inhomogeneity) onto the normal \textbf{n} 
of the P-plane. In contrast to the S-surfaces, which vary with variations of the 
orientation of \textbf{e}, the P-planes are independent of \textbf{e}. Thus only 
the magnitude of occupational inversion would depend on \textbf{e}. Therefore it 
will be necessary to correct our conclusions initially inferred from the 
analysis of ES located in the proximity of S-surfaces, if they should also apply 
to ES-pairs located in the proximity of flat (or nearly flat) sections of 
P-surfaces. To take an example: Instead of the generation of phonons with 
quasi-momenta $ \textbf{q} \:\Vert \:\langle 110 \rangle$ strictly aligned with 
twofold axes of symmetry, which would be most obvious due to the extensiveness 
of the $S_{\langle 110 \rangle}$ surfaces (see Fig. \ref{fig2.3}), there must 
be expected a deviation of \textbf{q} with respect to the axis $\langle 110 
\rangle$. As the area $\Sigma$ of the reduced S-surfaces steadily changes 
with varying \textbf{e}, a deviation of \textbf{e} from the axis $\lbrack 110 \rbrack$ by a 
small angle of let us say $10^{0}, ( \textbf{e}\:\Vert\: [11\eta ], \eta \approx$ 
0,3) could not significantly diminish the area of the reduced surface 
$S_{[11\eta ]}$, if compared with $S_{\lbrack 110 \rbrack}$. The additional contribution to 
the generation of phonons with $\textbf{q}\: \Vert\: [11\eta]$ comprises about one 
sixth of the ES-pairs located in the proximity of the P-planes, in the 
surroundings of the hexagonal faces of the 1$^{st}$ BZ with normal vectors 
\textbf{n} oriented parallel to $[\bar{1}11], [1\bar{1}1]$, corresponding 
to areas of P-planes $ \sim  2\sqrt{3} (\pi /a)^{2}$, thereby not only 
compensating for the losses but leading to a surplus of area. This means that a 
consideration of ES located in the proximity of P-planes ensures a low 
generation threshold for waves whose wave-vectors \textbf{q} lie within 
orientations confined by a cone centered around $\lbrack 110 \rbrack$, with respect to the 
threshold for $\textbf{q} \:\Vert\: \lbrack 110 \rbrack$. The deviation of \textbf{q} from $\lbrack 110 \rbrack$ 
will however be essential for an interpretation of the particularities of the 
morphology of the product of the $\gamma -\alpha$ -MT (see Chap.5), as well as 
for any reasoning about the actually occurring choice of the pathway of a 
martensitic reaction.}

\begin{figure}[htb]
\centering
\includegraphics[clip=true, width=.4\textwidth]{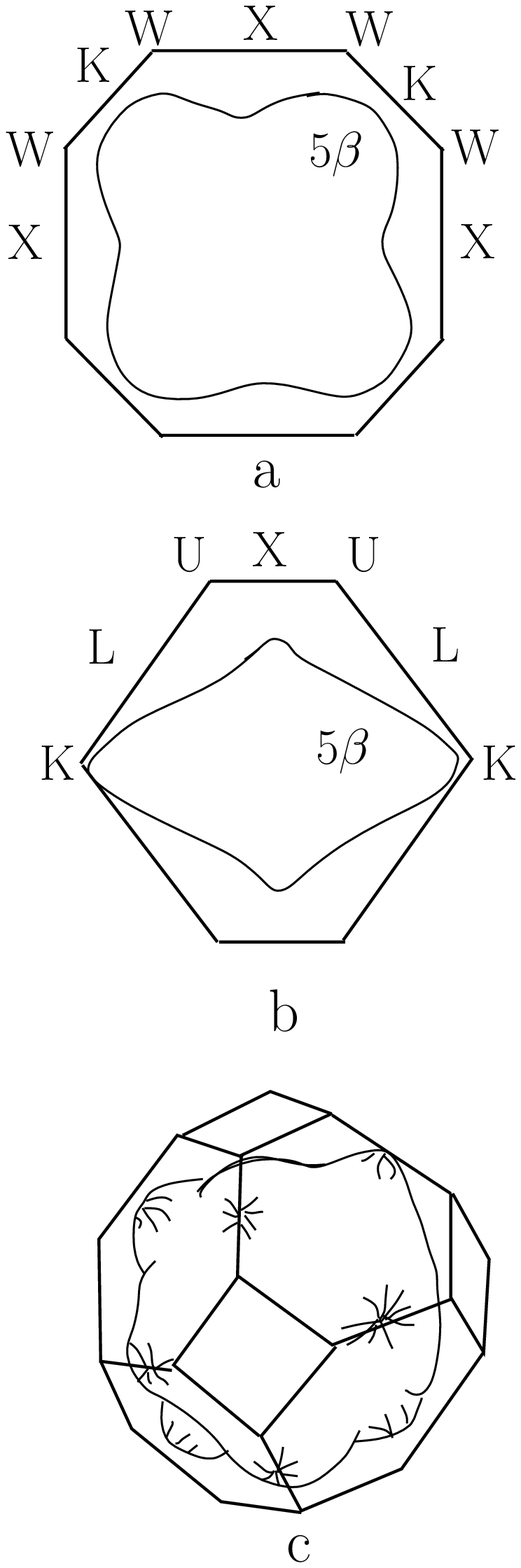}
\renewcommand{\captionlabeldelim}{.}
\caption{ Sheet of the d-like band ($5\,\beta$ with spin-down) for Ni 
with fcc lattice: a - sectional view by the plane (001); b - sectional view by 
the plane (011) \cite{Kondorskii76}; c - general view \cite{Krenkell78}.}
\label{fig2.10}
\end{figure}

\item{ The conclusion on the existence of S-surfaces with weak energy dispersion 
in the proximity of values $\varepsilon =\varepsilon_{L}$ 
means that the shape of the isoenergetic surface $\varepsilon  = const \approx  
 \varepsilon_{L_{3}} $ must be more or less akin to the shape of the 
S-surfaces. As in the case of ferromagnetic Ni discussed in
\cite{Zornberg70}-\cite{Jarlborg80}, the 
location of the Fermi-level with partially filled minority sub-band ( electron 
spin-down $ \downarrow$ oriented against the orientation of the magnetic field) 
is close to the energies $\varepsilon_{L_{3} \downarrow}$ ( $\varepsilon 
_{L_{3}} -  \mu  \approx$ 0,3 eV), the method of Fermi - surface 
calculus can be used. In Fig. \ref{fig2.10}, the intersections extracted from 
\cite{Connolly67} are depicted (see also \cite{Kondorskii76}) and the overview of the sheet corresponding 
to the d-like sheet of the Fermi-surface in \cite{Krenkell78} ($5\beta$ in accordance with 
the designations used in \cite{Connolly67}). Obviously, the curved surface pronouncedly 
protrudes towards the orientation of the WKW-rib of the 1$^{st}$ BZ. Thus 
condition (\ref{2.24}) is satisfied for the similarly curved sheet 2 of surface 
$S_{\lbrack 001 \rbrack}$. In \textbf{k}-space, the sheet of the Fermi-surface is more 
remotely located from sheet 1 of the $S_{\lbrack 001 \rbrack}$ surface than from sheet 2, 
featuring the shape of a double-saddle in the region opposite the square surface 
of the 1$^{st}$ BZ. Obviously with increasing energy, the isoenergetic surface 
will come closer to the faces of the 1$^{st}$ BZ, which in turn will enhance the 
effectivity of those ES-pairs being located in the proximity of the vastly 
extended curved surfaces $S_{\lbrack 110 \rbrack}$, $S_{\lbrack 111 \rbrack}$ 
(see also Figs. \ref{fig2.2}, \ref{fig2.3}).}
\end{enumerate}

\section{Summary of Chapter 2}

A consideration of the crucial conditions for phonon generation to be satisfied 
by a radiative electronic sub-system led us to the following conclusion: 
Potentially active ES in \textbf{k}-space are localized near regions of 
S-surfaces being limited by isoenergetic surfaces $\varepsilon (\textbf{k}) = 
\mu \pm  \Delta $. In \textbf{k}-space such S- surfaces separate ES with the 
opposite signs of group velocity projections $\textbf{v}( \textbf{k}) = 
\hbar^{-1} \vec\nabla_{\textbf{k}}\varepsilon(\textbf{k})$ on the direction of 
the local spatial inhomogeneity \textbf{e}. 

Even if the dispersion law $\varepsilon  = \varepsilon (\textbf{s}) $ on the S- 
surface was arbitrarily chosen, the conditions for inversely occupied and 
equidistant ES-pairs would still demand a slow rate of change of the electron 
energy on the S- surfaces, hence satisfying the inequalities (\ref{2.24}) for the 
macroscopic majority of points on the S- surface. It can thus reasonably be 
expected for systems with high density of electron states in the proximity of 
the Fermi-level to satisfy this requirement. Those S- surfaces (as well as the 
associated orientations of \textbf{e}) for which the greatest part of their area 
is located within the 1$^{st}$ Brillouin-Zone (BZ) (for reduced sheets) are 
highlighted, as they correspond to (\ref{2.26}) and, at the same time, satisfy the 
requirement of sign invariance of the population difference of ES associated 
with a given pair $\textbf{k}, \textbf{k}^{\prime} $, where a phonon-wave-vector 
\textbf{q} connects the pair of points defined by $\textbf{k}, \textbf{k}^{\prime} 
$, at opposed sides from S. 
The quantity \textbf{q} is limited from below, where the minima of \textbf{q} 
are mainly determined by the characteristics of the group-velocities 
$\textbf{v}\arrowvert_{s}$, as well as by the inverse effective masses $m^{- 
1} \arrowvert_{s}$ on the S- surfaces (note that the inverse effective masses 
are essentially independent of the orientation of \textbf{q}, see
eq.(\ref{2.26})). For $\textbf{q} \: \Vert  \:\textbf{e}$, the smallest value of $q = q_{min}$ attains an 
order of magnitude of ($10^{-3}  \div  10^{-2} \pi /a$). 

As the vast majority of points of the reduced sheets of the S-surfaces are 
located within the 1$^{st}$ BZ, and not at its boundaries, the elementary 
processes of phonon-emission with wave-vectors $q  \ge  q_{min}$ can be 
regarded as N-processes, as they are determined by electronic transitions of the 
type $ \textbf{k} \to  \textbf{k}^{\prime} $ being localized in the proximity of 
the S- surfaces. Thus, under the pronounced non-equilibrium conditions locally 
prevailing in a $\gamma - \alpha$ - MT region, there will predominantly be 
emitted or amplified longitudinal (or quasi-longitudinal) phonons, as both the 
scalar products $(\textbf{e}_{q}, \textbf{q})$, and the matrix-elements 
(\ref{2.4}) of electron-phonon-interaction would vanish for pure transversal waves 
related to N-processes. 

The emission of phonons with \textbf{q} vectors collinear with \textbf{e} 
becomes most probable under condition of maximum population inversion. Using
the  conservation law of quasi-momentum (\ref{1.9}) and the non-equilibrium
addenda, to the electronic distribution function expressed in (\ref{2.5}), (\ref{2.6}), 
it is easy to find the relative orientation of
\textbf{q} and \textbf{e} according to the sign of the expression
$m^{-1}\arrowvert_{s}$, $(\varepsilon  -  \mu )\arrowvert_{s}$ on the
S-surface, for two different sources of non-equilibrium. Table~2.1 shows
the relative orientation between \textbf{q} and the  unit vectors defined by
$ \textbf{e}\: \Vert\: \textbf{e}_{T} \:\Vert \: \vec\nabla T$,  $\textbf{e}
\:\Vert \:\textbf{e}_{\mu} \:\Vert\: \vec \nabla \mu $. Positive $m^{- 1}
\arrowvert_{s} \to (+)$ correlate with the points of minimum, while 
negative $m^{-1} \arrowvert_{s}  \to (-)$ correlate with the points of 
maximum of the function $ \varepsilon (\textbf{k})$ on the S-surface.  For
example, for $m^{-1} \arrowvert_{s} < 0 \to  (-)$, $(\varepsilon  -  \mu)
\arrowvert_{\varepsilon} > 0 \to  (+)$, $\textbf{e}\:  \Vert
\:\textbf{e}_{\mu}$ in Table~2.1 we get $\textbf{q} \upuparrows 
\textbf{e}_{\mu}$. The latter statement means that for $\mu _{\alpha} < 
\mu_{\gamma}$, there will be generated (or amplified) the longitudinal
waves  propagating from the growing martensite crystal (into the austenite
area being  free from martensite). Thus it can also be concluded from
Table~2.1 that for $\textbf{e} = \textbf{e}_{\mu}$ the relative
orientation of  \textbf{q} and \textbf{e} is independent of the sign of the
term $(\varepsilon  -  \mu)\arrowvert_{s}$, whereas such a dependence does
exist for $\textbf{e}  = \textbf{e}_{T}$. This latter conclusion immediately
results from the  consideration that for $\nabla \mu \ne 0$, there must
exist an electron flow  oriented from regions with relatively larger
chemical potential $\mu $ towards  regions with lower $\mu$ (thus being
linked up with charge transfer), while for  $\nabla T \ne 0$, there must
exist two mutually opposed electron flows of equal  density, however not
contributing to effective charge transfer in that region  (see also the
legend to Fig. \ref{fig8}).

\begin{table}[htbp]
\begin{center}
\renewcommand{\captionlabeldelim}{.}
\caption{Orientation of the wave-vectors of generated phonons relative to the orientation 
of spatial inhomogeneity}
\vspace{5mm}
\begin{tabular}{|c|c|c|}
\hline
 & \multicolumn{2}{c}{$m^{-1} \arrowvert_{s}$} \vline\\ \cline{2-3}
 \multicolumn{1}{|c|}{$(\varepsilon - \mu )\arrowvert_{s}$}
& \multicolumn{1}{c}{+}\vline & -- \\ \hline
+ &   
$\begin{array}{c}
  \textbf{q}\downarrow\uparrow \textbf{e}_{T}  \\
  \textbf{q}\downarrow\uparrow\textbf{e}_{\mu} \\
  \end{array}$ &
$\begin{array}{c}
  \textbf{q} \upuparrows \textbf{e}_{T}    \\
  \textbf{q} \upuparrows \textbf{e}_{\mu } \\
\end{array}$ \\
\hline 
-- &
$\begin{array}{c}
   \textbf{q} \upuparrows \textbf{e}_{T}  \\
   \textbf{q}\downarrow \uparrow \textbf{e}_{\mu} \\
\end{array}$ &
$\begin{array}{c}
   \textbf{q} \downarrow \uparrow \textbf{e}_{T} \\
   \textbf{q} \upuparrows \textbf{e}_{\mu} \\
\end{array}$ \\
\hline
\end{tabular}
\end{center}
\label{tab2.1}
\end{table}
Considering that $T_{\alpha}  >  T_{\gamma}$ and bearing in mind 
that $\mu_{\alpha}  <  \mu_{\gamma}$, (see also final statement of Point 
(2.5)) the anti-parallelism of $\textbf{e}_{\mu}$ and $\textbf{e}_{T}$ 
becomes evident. Then the effect of the sources of non-equilibrium for the 
states with $(\varepsilon  - \mu )\arrowvert_{s} > 0$ will partially be 
counterbalanced (in that the non-equilibrium addenda of (\ref{2.5}) and (\ref{2.6}) 
are mutually subtracted), whereas they will mutually be amplified for 
$(\varepsilon  -   \mu)\arrowvert < 0$ (i.e. the non-equilibrium addenda of 
(\ref{2.5}) and (\ref{2.6}) add up). Of course, during the process of generation and 
amplification of longitudinal waves propagating ahead of the growing martensitic 
phase, the dominating quantity is the inhomogeneity (gradient) of the chemical 
potential $\mu$ and not the temperature gradient $\vec{\nabla}T$. This 
conclusion is evident as the order of magnitude of the difference $\mu_{\alpha} 
 - \mu_{\gamma}=  - 0,16$eV is about one order of magnitude greater than 
that of the temperature difference $T_{\alpha} - T_{\gamma}$ (after the 
temperature is adequately converted into its energy equivalent). Following an 
evaluation of our electronic band-analysis under Point 2.5 it appears to be most 
likely that the value of the parameter $\Delta$ is approximately some 
tenths of eV for fcc-iron lattice, and that ES-pairs in the proximity of the 
surfaces $S_{\lbrack 001 \rbrack}$, $S_{\lbrack 110 \rbrack}$, $S_{\lbrack 111 \rbrack}$ can be active during the 
generation of longitudinal phonons with orientations \textbf{q} in the proximity 
of symmetry axis. 

The most important results of our analysis of electronic spectra under the 
particular aspect of the general requirements for realization of a phonon maser 
are published in \cite{Petrov78,Kashchenko79,Kashchenko77,
Kashchen84,Kashchenko85,Kashchenko86}.

\chapter{Constitutive equations of the electron-phonon system and the 
threshold wave generation conditions}

The theoretical finding of pairs of electronic states with the inverted 
population (see the previous chapter) is the most essential stage in our aim 
to disclose the phonon-generation mechanism. Next it is necessary to get a 
determination of the physical conditions required for the functioning of a 
phonon maser (see Sub-Pt 4 in Pt.1.5) or, put in another way, to determine 
the threshold value of an inverted population difference. Thanks to the 
already existing comprehensive theory of the maser-effect (for example in 
\cite{Leks75,Khanin75,Khaken74,Khaken80}) this task does not comprise any fundamental 
difficulties and can thus be resolved on the basis of well-proven approaches 
and methods. The simplest approach is semi-phenomenological. Then the 
crucial interaction of a radiating sub-system (in our case the 
non-equilibrium 3-d electrons) with a radiating field (in our case a 
specific class of longitudinal phonons) has to be defined as precisely as 
possible, while the creation of the inverted population (i.e. the pumping), 
as well as dissipation processes, are treated by a phenomenological method.

\section{Threshold conditions for single-mode generation}

Following \cite{Khaken74,Khaken80} we shall start with a brief consideration of the 
simplest derivation of the expression for a threshold-difference of 
population \eqref{eq11}, related to the case of single-mode generation.
As the generation of phonons normally begins with that specific mode for 
which the required population difference is minimal, it would be reasonable 
to start with a consideration of single-mode generation. Thus in this case 
we start with our analysis by sorting out one specific Hamiltonian $H_{1}$ 
from the general set of Hamiltonians \ref{2.1}-\ref{2.3} of an electron-phonon 
system: 

\begin{equation}
H_{1} = \hbar\:\omega_{\textbf{q}}\:b_{\textbf{q}}^{+}\:b_{\textbf{q}}+\sum_{\textbf{p}}
\varepsilon_{\textbf{p}}\:a_{\textbf{p}}^{+}\:a_{\textbf{p}}+\sum_{\textbf{k}}W^{\ast}_{\textbf{q}}\:
b_{\textbf{q}}\:a_{\textbf{k}}^{+}\:a_{\textbf{k}^\prime} + W_{\textbf{q}}\:
b_{\textbf{q}}^{+}\:a_{\textbf{k}^\prime}^{+}\:a_{\textbf{k}}.
\label{3.1}
\end{equation}
Here $b_{\textbf{q}}^{+}$, $b_{\textbf{q}}$ and $a_{\textbf{p}}^{+}$,
$a_{\textbf{p}}$ are the phonon and electron creation and annihilation 
operators, respectively, and $W_{\textbf{q}}$ is the matrix-element 
of electron-phonon interaction. Then, our next step is the decomposition 
of the complete Hamiltonian H of our task into the sum 

\begin{displaymath}
H = H_{1} + H_{2},
\end{displaymath}

The Hamiltonian $H_{2}$ describes the action of other sub-systems like heat 
reservoirs, on the segregated system \eqref{3.1}. In \eqref{3.1}, we omitted the
indices of bands related to electron-energies and -operators (in the case of
a 3d - electron band), as well as the polarization index of phonon-energies and 
-operators (the polarization is supposed to be longitudinal); $\hbar$ - Planck's 
constant.

In the Heisenberg-representation, the following constitutive equations of 
motion can be assigned to any operator x:

\begin{equation}
\label{3.2}
\frac{\partial x}{\partial t} \equiv \dot{x} = \frac{i}{\hbar }\left[ 
{H,x} \right] = \frac{i}{\hbar}\left[{H_{1},x}\right] + \frac{i}{\hbar}
\left[{H_{2},x}\right] \equiv \dot{x}_{1} + \dot{x}_{2}.
\end{equation}

Using the commutation rule

\begin{displaymath}
[b_{\textbf{p}}^{+}, b_{\textbf{q}}] = \delta _{\textbf{q},\textbf{p}},\quad 
[b_{\textbf{q}}^{+}, b_{\textbf{p}}^{+}] = [b_{\textbf{q}}, b_{\textbf{p}}] = 0 
\end{displaymath}

and the anticommutation 

\begin{displaymath}
[a_{\textbf{p}}^{+}, a_{\textbf{q}}]_{+} = \delta_{\textbf{q},\textbf{p}},
\quad [a_{\textbf{p}}^{+},a_{\textbf{q}}^{+}]_{+} = [a_{\textbf{p}}, a_{\textbf{q}}]_{+} = 0,
\end{displaymath}
we find from (\ref{3.1}) and (\ref{3.2}) that the following expressions describe the 
phonon field operators $b_{\textbf{q}}^{+}$, electron-polarization operators 
$d^{+}_{\textbf{k},\textbf{k}^\prime} = a_{\textbf{k}}^{+}a_{\textbf{k}^\prime }$ 
and the operator of the population-difference $\sigma 
_{\textbf{k}\textbf{k}^\prime} \equiv \sigma = a_{\textbf{k}}^{+}
a_{\textbf{k}}^{}
- a_{\textbf{k}^\prime}^{+}a_{\textbf{k}^\prime }^{}$
(where \textbf{k}, $\textbf{k}^\prime$ 
correspond to those used in \eqref{1.9}):

\begin{displaymath}
\dot{b}_{\textbf{q}}^{+} = (i\,\omega_{\textbf{q}} - {\varkappa}_{\textbf{q}})\,b_{\textbf{q}}^{+}
+\frac{i}{\hbar}\,\sum_{\textbf{k}}W^{\ast}_{\textbf{q}}\:d^{+}_{\textbf{k},\textbf{k}^\prime},
\end{displaymath}
\begin{equation}
\dot{d} ^{+}_{\textbf{k},\textbf{k}^\prime} = (i\,\omega_{\textbf{k},\textbf{k}^\prime} - \Gamma )
d ^{+}_{\textbf{k},\textbf{k}^\prime} - \frac{i}{\hbar}\:W_{\textbf{q}}\:b_{\textbf{q}}^{+}\:\sigma,
\label{3.3}
\end{equation}
\begin{displaymath}
\dot{\sigma} = \frac{\sigma_0 - \sigma }{t_\sigma } + 
\frac{2i}{\hbar}\:W_{\textbf{q}} \:d_{\textbf{k},\textbf{k}^{\prime}}\:b_{\textbf{q}}^{+} - \frac{2i}{\hbar}
W^{\ast}_{\textbf{q}}\:d^{+}_{\textbf{k},\textbf{k}^{\prime}}\:b_{\textbf{q}},
\end{displaymath}
\begin{displaymath}
\dot{b}_{\textbf{q}} = (\dot{b}_{\textbf{q}}^{+})^{+}, \quad \hbar \:\omega
_{\textbf{k},\textbf{k}^\prime} = \varepsilon_{\textbf{k}} -
\varepsilon_{\textbf{k}^\prime}.
\end{displaymath}
Here, $\Gamma $ is the electron-attenuation.

Let us now discuss (\ref{3.3}): As our interest is focused on the average values 
of the operators in (\ref{3.3}), we can ignore the fluctuational action of heat 
reservoirs in (\ref{3.3}), without introducing any significant error. Then, the 
effect of dissipation can be considered phenomenologically by introduction 
of relaxation times $t_{\sigma}$ of electron populations, of phonon 
attenuation ${\varkappa}_{\textbf{q}}$, and of electron attenuation $\Gamma$. (It is 
agreed that an attenuation $\Gamma $ for the dipole - moment
$d_{\textbf{k},\textbf{k}^{\prime}}$ of transition between a pair of electronic states (ES-pair) 
\textbf{k}, $\textbf{k}^{\prime}$, emitting phonons with the quasi-momentum \textbf{q}, 
is of the same order of magnitude as the electronic states attenuations, i.e., $\Gamma \sim 
\Gamma _{\textbf{k}}\sim \Gamma_{\textbf{k}^{\prime}}$); $\sigma_{0}$ - inverse 
initial population difference of states \textbf{k}, $\textbf{k}^{\prime}$ located 
on opposite sides of the S-surface, being produced by electronic flows 
(drift-pumping mechanism), as well as by other non-coherent processes in 
absence of maser radiation.

In principle, the constitutive set of equations (\ref{3.3}) represents an analog 
to the Bloch-equations of magnetic resonance theory. This becomes obvious 
from \cite{Makomber79,Khaken78} and by comparing $\sigma $, $d^{+}$, $d$ with the 
projections of the spin-operators $S^{z}$, $S^{+}$, $S^{-}$, and the 
parameters $t_{\sigma}$, $\Gamma^{-1}$ with the longitudinal- and 
transversal relaxation times $T_{1}$, $T_{2}$, respectively. This 
immediately leads to the restriction (\ref{3.4}) already known from \cite{Makomber79}:

\begin{equation}
T_{1} \ge T_{2}: t_{\sigma }\ge \Gamma^{-1}.
\label{3.4}
\end{equation}

In the phonon generation (stimulated emission) regime, the number of quanta 
of the mode to be emitted becomes macroscopic and the operators 
$b_{\textbf{q}}^{+}$, $b_{\textbf{q}}$ represent satisfactory approximations 
of c - number functions of time. From the above presentation it becomes obvious 
that, in the generation regime, the equations of the system (\ref{3.3}), averaged
by a density matrix of the system, are identical with the classical 
expressions, if all operators were replaced by their average values. 
In below equations, this replacement has already been considered. Hence, 
the previously used designations of operators now simply pertain to their 
averages. The oscillatory dependence on time can then simply be eliminated 
by adoption of the quantities

\begin{equation}
\tilde{b}_{\textbf{q}}^{+} = \exp{ [-i \Omega_{\textbf{q}} t]}\,  
  b_{\textbf{q}}^{+}, \quad
\tilde{b}_{\textbf{q}} = \exp{ [i \Omega_{\textbf{q}} t]}\, b_{\textbf{q}},
\label{3.5}
\end{equation}
\begin{displaymath}
\tilde{d}^{+} = \exp{ [-i\Omega_{\textbf{q}} t]}\, d^{+},\quad 
\tilde{d} = \exp{ [i \Omega_{\textbf{q}} t]} \,d,
\end{displaymath}
with consideration of the case of exact resonance:

\begin{equation}
\hbar\Omega_{\textbf{q}} = \hbar \omega_{\textbf{q}}=\varepsilon_{\textbf{k}} - 
\varepsilon_{\textbf{k}^\prime} = \hbar\omega _{\textbf{k},\textbf{k}^\prime}.
\label{3.6}
\end{equation}

Under stationary conditions, we get from (\ref{3.3}) - under consideration of 
(\ref{3.5}) and (\ref{3.6}) - the following system of non-linear algebraic 
equations:

\begin{displaymath}
\varkappa_{\textbf{q}}\tilde{b}_{\textbf{q}}^{+} - \frac{i}{\hbar} W^{\ast}_{\textbf{q}} 
\sum_{\textbf{k}^{\prime}}\tilde{d}^{+}_{\textbf{k},\textbf{k}^{\prime}} = 0,
\end{displaymath}
\begin{equation}
\Gamma \tilde{d}^{+}_{\textbf{k},\textbf{k}^{\prime}} + \frac{i}{\hbar}W_{\textbf{q}}
\tilde{b}_{\textbf{q}}^{+} \sigma = 0,
\label{3.7}
\end{equation}
\begin{displaymath}
\Gamma \tilde{d}_{\textbf{k},\textbf{k}^{\prime}} - \frac{i}{\hbar}W^{\ast}_{\textbf{q}}
\tilde{b}_{\textbf{q}}\sigma  = 0,
\end{displaymath}
\begin{displaymath}
\frac{\sigma _0 - \sigma }{t_\sigma } + \frac{2i}{\hbar}W_{\textbf{q}}
\tilde{d}_{\textbf{k},\textbf{k}^\prime}\tilde{b}_{\textbf{q}}^{+} - \frac{2i}{\hbar}
W^{\ast}_{\textbf{q}}\tilde{d}^{+}_{\textbf{k},\textbf{k}^\prime}\tilde{b}_{\textbf{q}}
= 0.
\end{displaymath}

The stationary value of the population difference $\sigma$, being the 
result of a balance between gains and losses, is the threshold value $\sigma 
_{th}$. Assuming the existence of solutions $\tilde{b}_{\textbf{q}}^{+} \ne  0$, 
we find from the first of two equations of system \eqref{3.7} that 

\begin{equation}
\sum_{\textbf{k}^\prime}{\sigma_{th}} = 
\frac{\hbar^2\Gamma\varkappa_{\textbf{q}}}{\arrowvert{W_\textbf{q}}\arrowvert^2}.
\label{3.8}
\end{equation}
We note that the attenuation $\Gamma$ included in \eqref{3.8} represents certain 
mean attenuation of the dipole-moment \textbf{d} of the pairs of state 
\textbf{k}, $\textbf{k}^\prime$, being localized in the vicinity of the 
S-surface. It is further convenient to introduce an average value $\sigma_{th}$ for those $\sigma _{th}$ 
associated with the S-surface, by substituting in \eqref{3.8} in accordance with the mean-value rule: 

\begin{equation}
\sum_{\textbf{k}^{\prime}}\sigma_{th}=\bar{\sigma}_{th} R_{\textbf{q}}.
\label{3.9}
\end{equation}
Here $R_{\textbf{q}}$ - number of ES-pairs of state \textbf{k}, 
$\textbf{k}^{\prime} $ which emit phonons with quasi-momentum $\hbar\textbf{q}$ 
during $\textbf{k} - \textbf{k}^{\prime} $ transitions. As the energy of the pairs 
of state $\textbf{k},\textbf{k}^{\prime} $ must settle within certain interval $\Delta $ 
(see \eqref{2.24}), the substitution in \eqref{3.9} requires from 
$\bar{\sigma }_{th}$ to be associated with certain constant amount of energy
$\bar{\varepsilon}$ within the interval $\Delta $: $\varepsilon  - \mu \in \Delta $. 
This requirement is equivalent to the replacement of a non-isoenergetic S-surface
by an isoenergetic one. Considering the weakly pronounced dispersion of 
$\varepsilon (\textbf{k})$ near the S-surfaces, such a substitution suggests 
itself as natural. Using \eqref{3.9} and omitting the mean value sign (horizontal 
dash on top of $\sigma_{th})$ we can write \eqref{3.8} in the form

\begin{equation}
\sigma_{th} =
\frac{\hbar^2\Gamma\varkappa_{\textbf{q}}}{\arrowvert{W_\textbf{q}}\arrowvert^2
R_{\textbf{q}}}.
\label{3.10}
\end{equation}
It immediately follows from \eqref{3.10} that $\sigma _{th}$ becomes the less, 
the greater: the lifetimes of the quasi-particles $\Gamma^{-1}$, ${\varkappa}^{-1}$;
electron - phonon interaction; the number of pairs of state with inverted 
population $R_{\textbf{q}}$, being proportional to the area $\Sigma_{S(\textbf{q})}$ 
of the reduced sheet of the S-surface.

If we express (with the aid of the last of the three Eqs. of \eqref{3.7}) the 
value $\tilde{d}^{+}$ by $\tilde{b}_{\textbf{q}}^{+}$ and $\tilde{b}_{\textbf{q}}$ 
and insert the result into the first of Eqs. \eqref{3.7}, then, by consideration of 
\eqref{3.10}, we get

\begin{equation}
\tilde{b}_{\textbf{q}}^{+}\left[{\frac{\sigma _0}{\sigma_{th}}(1 + 
\frac{4t_\sigma\arrowvert{W_\textbf{q}}\arrowvert^2}{\hbar^2\Gamma}
\tilde{b}_{\textbf{q}}^{+}\tilde{b}_{\textbf{q}})^{-1} - 1} \right] = 0,
\label{3.11}
\end{equation}
with the two solutions:

\begin{equation}
\tilde{b}_{\textbf{q}}^{+} = 0 ,\quad \sigma_{0} < \sigma_{th}, 
\label{3.12}
\end{equation}

\begin{equation}
\tilde{b}_{\textbf{q}}^{+} \,\tilde{b}_{\textbf{q}} =
\frac{\hbar^2\Gamma}{\arrowvert{W_{\textbf{q}}}\arrowvert^2 
4t_\sigma }\left[{\frac{\sigma _0}{\sigma_{th}} - 1}\right],\quad
\sigma _{0} > \sigma_{th}, 
\label{3.13}
\end{equation}
which clearly verify that the displacement amplitude

\begin{equation}
u_{\textbf{q}}=\left[\frac{2\hbar^2}{M N \omega_{\textbf{q}}} 
\right]^{\frac{1}{2}}\tilde{b}_{\textbf{q}}  
\label{3.14}
\end{equation}
is vanishing below and non-vanishing above the generation threshold, thus 
indicating essentially classical characteristics of the displacement. In 
fact, it is obvious from \eqref{3.13} and \eqref{2.4} that for 
$\sigma _{0} > \sigma_{th}$ the number of phonons 
$\tilde{b}_{\textbf{q}}^{+} \, \tilde{b}_{\textbf{q}} \sim \arrowvert W_{\textbf{q}}\arrowvert^{-2}
\sim N$ becomes a macroscopic quantity. Consequently, the energy of this 
mode is comparable with the energy of the remaining non-coherent phonons. 

In order to evaluate $\sigma_{th}$, the relaxation-parameters 
$\varkappa_{\textbf{q}}$, $\Gamma$ and the number of pairs of active states
$R_{\textbf{q}}$ must be known. To perform such evaluation, we can use as 
an upper limit of the parameters $\varkappa_{\textbf{q}}$, $\Gamma $ their 
respective values in the high temperature domain, where scattering by 
short - wave phonons predominates ( i.e. where T exceeds the Debye-temperature 
$T_{D})$. 

For an evaluation of $\varkappa_{\textbf{q}}$, let $q << \pi/a$ and use the results
of Woodruff-Ehrenreich (see \cite{Taker75}) for the case $\omega_{\textbf{q}} \tau  << 1$, 
where $\tau $ - average lifetime in the free path of thermal phonons for 
$T > T_{D}$:
 
\begin{equation}
\varkappa_{\textbf{q}} = \frac{\lambda_{l} T\gamma_0^2 \omega_{\textbf{q}}^2 }{\rho c_{\textbf{q}}^4} 
\equiv \frac{\pi \lambda_{l} T \gamma_0^2 }{\rho c_{\textbf{q}}^3 a}\left[{\frac{q
a \omega_{\textbf{q}}}{\pi}}
\right].
\label{3.15}
\end{equation}
In \eqref{3.15}, we used (4.63) of \cite{Taker75}, after multiplication with the speed of 
sound $c_{\textbf{q}}$, in order to get $\varkappa_{\textbf{q}}$ in frequency 
units. In \eqref{3.15}: $\lambda_{l}$ - heat-conductivity of the lattice,
$\gamma _{0}$ - Gruneisen-constant, $\rho $ - density, $a$ - lattice-parameter.
In the case of fcc-iron lattice with 
$\lambda_{l} \approx 0,03\lambda_{t}$ ($\lambda_{t}$ - is the total heat
conductivity), $T \sim 10^{3}$ K, $\lambda_{t} 
\approx  34$ W/(m K) \cite{Spravochnik}, $\gamma_{0} = 2$, $\rho = 7900$ kg/m$^{3}$, 
$c_{\textbf{q}} = 5 \cdot 10^{3}$ m/s, $a = 3,57 \cdot 10^{-10}$ m, 
we get from \eqref{3.15}:

\begin{equation}
\varkappa_{\textbf{q}} \approx 3,5 \cdot 10^{-2}\frac{a q}{\pi}\omega_{\textbf{q}},
\label{3.16} 
\end{equation}
i.e., within our region of interest $q \sim (10^{-3}\div 10^{-2})\pi/a$, 
the extinction would be about $\varkappa_{\textbf{q}}\le (10^{-4}\div 10^{-3})
\omega _{\textbf{q}}$. 

For comparison, let us now find out the extinction ($\varkappa_{\textbf{q}})_{e}$ 
being caused by the interaction of conduction electrons. In the high 
temperature region, with $q << \pi/a$, the condition $q l_{e} << 1$ must be
satisfied, where $l_{e}$ - mean electron free path. Using the Pippard-Formula 
(8.2) in \cite{Taker75} we get:

\begin{equation}
(\varkappa_{\textbf{q}})_{e} \approx \frac{4 n_0 m}{15 \rho \,\tau}(q \,l_e)^{2} = 
\frac{4 n_0 m}{15 \rho \,c_{\textbf{q}}}(q \,l_e) \,\omega _{\textbf{q}}, 
\label{3.17}
\end{equation}
where $n_{0}$ - free electron density, m - mass, v = $l_{e}\tau ^{-1}$ -
electron velocity. Assuming that the quantities $n_{0} = 8 \cdot 10^{28}$ m$^{-3}$,
$m \approx 10^{-30}$ kg, v$ \, \approx \,10^{6}$ m/s, $\rho $, $c_{\textbf{q}}$ are identical
with those used in \eqref{3.15}, we get from \eqref{3.17}:

\begin{displaymath}
(\varkappa_{\textbf{q}})_{e}\approx 5 \cdot 10^{-4}(q l_{e})\omega_{\textbf{q}}. 
\end{displaymath}
In other words, for $q \sim (10^{-3} \div 10^{-2})\pi/a$ and 
$q l_{e} \approx 0,1$, we get a result comparable with that of \eqref{3.16}, 
being equivalent to $l_{e} = (100 \div  10) a$. In the high temperature 
region however, the value $l_{e} \sim 10 a$ may be more realistic, 
therefore our estimate \eqref{3.16} would not significantly be affected by 
adequate consideration of $(\varkappa_{\textbf{q}})_{e}$. The magnitude of 
attenuation of the dipole-moment $\Gamma$ is reciprocal to the average 
lifetime $\tau $ of the states $\textbf{k},\textbf{k}^\prime$, or to the mean 
scattering period $\tau  = \tau_{0}$, as defined in \cite{Gantmakher84}. An evaluation 
of the quantity $\tau_{0}$ however could easily be obtained if the 
following quantities were known: relaxation time $\tilde{\tau}$ of the
electron energy in the states \textbf{k}, $\textbf{k}^{\prime}$, the
non-elastic relaxation coefficient $\delta_{0}$ is the 
reciprocal number of scattering events required for the relaxation 
of an appreciable fraction of energy:
 
\begin{equation}
\tau_{0} \equiv \frac{\tilde{\tau}}{\delta_0^{-1}}.
\label{3.18}
\end{equation}

According to \cite{Gantmakher84} for $T > T_{D}$

\begin{equation}
\delta_{0}^{-1} \approx (2 N_{\textbf{p}} + 1) \frac{\arrowvert \varepsilon -
\mu \arrowvert}{\Delta \varepsilon} \approx \frac{2 k_B T \arrowvert \varepsilon
- \mu \arrowvert}{\hbar^2 \omega_{\textbf{p}}^2},
\label{3.19}
\end{equation}
where $N_{\textbf{p}} \approx k_B \,T / \hbar \omega_{\textbf{p}}$ - 
occupancy of short-wave phonons with frequencies $\omega_{\textbf{p}} \approx 
C\,\pi/a \approx  k_B \,T_D / \hbar$, and $\Delta \varepsilon  
\approx  \hbar\,\omega _{\mathbf{p}}$ - fraction of energy lost during 
non-elastic interaction (for $\varepsilon  - \mu > 0$) or absorbed (for 
$\varepsilon - \mu  < 0$). The presence of a factor $(2 \,N_{\mathbf{p}} + 1)$ 
in \eqref{3.19} becomes obvious by consideration that, for large occupancies 
$N_{\mathbf{p}}$, the processes of induced emission and induced absorption of phonons 
predominate during scattering events, and " \ldots to each $N_{\mathbf{p}} + 1$ events of 
emission there correspond $N_{\mathbf{p}}$ events of absorption, or, put in another 
way, during energy relaxation only one event will be effective out of 
$2\,N_{\mathbf{p}} + 1$ events" \cite{Gantmakher84}. 

The effects of rapid heating of an electron-subsystem by means of short 
pulse of laser radiation, as reported in \cite{Agranat84}, enables us to assess 
$\tilde{\tau} \approx  10^{-10}$ s for nickel, where $\tau$ is a characteristic 
time of electron-lattice relaxation. In this case, the role of a large heat 
reservoir is taken up by the short-wave phonons (which cover the maximum 
spectral density of states in the phonon-spectrum). Under the assumption of 
\cite{Agranat84} that the most part of energy of a laser pulse is absorbed by the 
3d - electrons, it would be reasonable to conclude that the scale of $\tau $ 
in Ni is mainly determined by those electrons with energies near the peak 
area of the density function, i.e. near the upper edge of the 3d - band, as 
the contribution of these electrons makes up for the greatest fraction of 
the heat capacity of an electronic sub-system. We bear in mind that the same 
electronic states are considered to be active during the generation of 
phonons in the case of fcc-iron (see Pt. 2.5). Thus our evaluation of 
$\Gamma \sim \tau _{0}^{-1}$ can be performed with the aid of Eqs. \eqref{3.18} 
and \eqref{3.19}, and under the assumption that in accordance with Pt. 2.5:
$\vert \varepsilon - \mu \vert \sim (0,2 \div  0,3$ eV) or $\vert 
\varepsilon  - \mu \vert \sim (5 \div  8) \hbar \,\omega_{\mathbf{p}}$, 
$k_{B}T \approx  2,5 \hbar \,\omega _{\mathbf{p}}$ and $\tilde{\tau } = 
10^{-10}$ s. Then we get: 

\begin{displaymath}
\delta _{0}^{-1} \approx 30  \div  50,
\end{displaymath}

\begin{equation}
\tau _{0} \approx (3 \div  2) \cdot 10^{-12} s,
\label{3.20}
\end{equation}

\begin{displaymath}
\Gamma \approx (3 \div  5) \cdot 10^{11} s^{-1}.
\end{displaymath}

It should be noted here that the final width of the electronic energy levels 
at $\Gamma  > \omega _{\textbf{q}}$ enables us to use (without getting into 
contradiction to the conservation laws of energy and momentum) as the 
maximum number of electronic states $R_{\textbf{q}}$ the quantity
\begin{displaymath}
(R_{\mathbf{q}})_{max}  \approx  2q \Sigma _{\mathbf{q}} \delta ^{-1},
\end{displaymath}

\begin{equation}
N \delta  \approx  4 \cdot (2\pi /a)^{3} \sim q^{3}_{max}.
\label{3.21}
\end{equation}

In \eqref{3.21}: $\delta $ - volume of the reciprocal space covering one wave 
number, the factor 2 is due to the required consideration of two possible 
spin-orientations, $\Sigma _{\mathbf{q}}$ - area of one sector of the package of 
sectors oriented parallel to the reduced sheet of the S-surface, without 
being more distant from it in \textbf{q} - space than one unit of 
quasi-momentum $\hbar\mathbf{q}$ of the generated phonons, this way ensuring 
the transition of electrons from the volume $\Sigma_{\mathbf{q}} q$ of the states with 
quasi-momenta $\hbar\textbf{k}$ into the same volume with quasi momenta 
$\hbar\textbf{k}^{\prime}$, being located at opposite sides of the S-surface. 
Further assuming that $\Sigma_{\mathbf{q}} \approx 20(\pi /a)^{2}$, we get 

\begin{equation}
\Sigma_{\mathbf{q}} \,\delta ^{-1} = \frac{5\,a}{8\,\pi} \,q\, N ,\quad R_{\mathbf{q}} = 
\frac{5\,a}{4\,\pi } \,q\, N. 
\label{3.22}
\end{equation}
This means that for $q \sim (10^{-3} \div 10^{-2})\pi /a$ the number of pairs 
of state being active during phonon generation is about 
$(10^{-3} \div  10^{-2})$ times from the total number of states 2N. After 
substitution with (\ref{3.22}), (\ref{3.16}), (\ref{2.4}) in (\ref{3.10}) 
and using $ M \approx 10^{-25}$ kg, $G \approx 10^{-19}$ J ($G \approx 0,6$ eV), 
$\Gamma  \sim (10^{11} \div 10^{12})$ s$^{-1}$, we finally get 
$\sigma_{th} \approx  10^{-4} \div 10^{-3}$. 

Using the expressions (\ref{2.5}), (\ref{2.6}), being related to additional 
non-equilibrium terms in the energy distribution of electrons, in 
conjunction with the energy conservation law (\ref{1.9}), we can express the 
inverted initial occupational difference in the pattern

\begin{equation}
\sigma_{0}(\nabla T) = f_{\mathbf{k}} - f_{\mathbf{k}^\prime} \approx 
\frac{y_{\mathbf{k}}}{k_B T}\arrowvert{\frac{\partial f_{\mathbf{k}}^0 }
{\partial y_{\mathbf{k}}}}\arrowvert \left[k_B \tau (\mathbf{v}_{\mathbf{k}^\prime}
 - \textbf{v}_{\mathbf{k}})\vec{\nabla} T - \hbar \omega_{\mathbf{q}} \right],
\label{3.23}
\end{equation}

\begin{equation}
\sigma_{0} (\nabla \mu ) = f_{\mathbf{k}} -  f_{\mathbf{k}^{\prime}} \approx 
\frac{1}{k_{B} T}\arrowvert{\frac{\partial f_{\mathbf{k}}^{0}}{\partial y_{\mathbf{k}}}} 
\arrowvert \left[\tau(\mathbf{v}_{\textbf{k}^{\prime}}
 - \mathbf{v}_{\mathbf{k}})\vec{\nabla} \mu - \hbar \omega_{\mathbf{q}}\right].
\label{3.24}
\end{equation}

Table \ref{table3.1} shows the values of $\sigma_{0}$ obtained for some 
representative $y_{\mathbf{k}} = (\varepsilon_{\mathbf{k}} - \mu )/ k_{B}T$
(after omission of the last term in \eqref{3.23} and \eqref{3.24}), using:
$T = 10^{3}$ K, 
$\nabla T = \Delta T \,(10 $ v$_{\mathbf{k}}\tau )^{-1}$, 
$\nabla \mu =\Delta \mu \, (10$  v$_{\mathbf{k}}\tau )^{-1}$,\, 
$\Delta T = 100$ K, $\Delta \mu  = 0,15 \,eV = 1,74 \, k_{B}T $.

It is obvious from  Table \ref{table3.1} that the crucial condition for excess 
generation of phonons $\sigma_{0} > \sigma_{th}$ is easily satisfied 
for $\sigma_{0}( \nabla T)$, as well as for $\sigma_{0} ( \nabla \mu )$, 
in the latter case even with an excess of the order of magnitude (since 
$2 \cdot 10^{-1} > \sigma_{0}( \nabla \mu ) > 10^{-2} > 10\cdot\sigma_{th}$). We have however 
to remark that for  $\vert \mathbf{q} \vert \sim (10^{-3} \div 10^{-2})\pi /a$, 
the value of the projection $\mathbf{v}_{\textbf{k}}$ towards the orientation 
given by the heterogeneity gradient attains an order of magnitude of about 
$10^{3}$ m/s. At the other hand, if 10 average free path lengths of electrons 
of about $10 \textbf{v}_{\textbf{k}}\tau \sim 10^{-8}$ m were taken as a 
natural scale of the average size of an heterogeneity, and further assuming 
$\tau \sim 10^{-12}$ s, then we would get a length one order of magnitude 
less than half wavelength, as given by the expression 
$\lambda /2 \sim 4\pi /q \sim  10^{-7}$ m. From the point of view that a MT 
can be described as a diffusionless lattice deformation process, then at least 
$\lambda /2 $ would fit into our assumed scale of the heterogeneity 
(see also the treatment of \eqref{eq2} in Pt. 1.3 and Sub-Pt. 7 of Pt. 1.5, 
as well as Chap. 6). In this case however, for an evaluation of $\nabla  T $ 
and $\nabla \mu$, it would be more realistic to use a value of 
$10^{2}$\, v$_{\mathbf{k}} \cdot \tau$, instead of 
10 \, v$_{\mathbf{k}}\cdot \tau $ (thus in this case the values of 
$\sigma_{0}$ in Table \ref{table3.1} would have to be reduced by one order 
of magnitude), and for $\sigma_{th} \sim 10^{-3}$ the condition 
$\sigma_{0} > \sigma_{th}$ would only be satisfied for the 
$\sigma_{0}(\nabla \mu )$. However, assuming $\sigma_{th} \sim  10^{-4}$, 
then the condition $\sigma_{0} > \sigma_{th}$ would easily be satisfied 
for both the thermal and the chemical potential of heterogeneities 
( $\sigma_{0}(\nabla  T)$ and $\sigma_{0}(\nabla \mu )$ ). We shall 
resume our consideration of a realistic spatial scale of an inhomogeneous 
(non-equilibrium) distribution of T and $\mu $ in Pt. 3.3 (from a different 
point of view).

\begin{table}[htbp]
\renewcommand{\captionlabeldelim}{.}
\caption{Inverted initial population difference as a function of the parameter
y}
\begin{center}
\begin{tabular}{|c|c|c|c|c|}               \hline
y  &  $\arrowvert \partial f^0 / \partial y \arrowvert$ & 
$\sigma_{0}(\nabla  T) \cdot 10^{3}$ & $\sigma_{0}(\nabla \mu )\cdot 10^{2}$
& $\sigma_{th }\cdot 10^{3}$  \\ \hline
0 & 0,5 & 0 & 17,4      &    \\ 
1 & 0,197 & 3,94 & 6,86 &     \\  
  &       &      &      & $(0,1 \div  1)$ \\ 
1,54 & 0,156 & 4,78 & 5,43 &     \\ 
2 & 0,105 & 4,2 & 3,65 &    \\ 
3 & 0,045 & 2,7 & 1,57 &  \\ \hline
\end{tabular}
\end{center}
\label{table3.1}
\end{table}

\section{Threshold conditions for two- and three-mode wave generation 
and characteristics of a phase transition of a radiation system}

We shall now deal with multi-phonon processes involved in the generation and 
amplification of waves, using some reasonable and justified approximations 
within the basic objectives of our analysis. In our notion of the $\gamma -\alpha $ -MT, 
the electronic transitions between the states $\mathbf{k}$,$\mathbf{k}^{\prime}$, 
occurring in the vicinity of the $S_{\langle 111 \rangle}$ surface under participation of two 
or three longitudinal phonons, are supposed to be of real importance. 
The quasi-momenta of these phonons are oriented either along the two - and fourfold 
symmetry axes (during a two-phonon process), or along the fourfold symmetry axes (during a 
three-phonon process). This statement is put into relation with the 
assumption that predominantly the longitudinal waves, which propagate in the 
$\langle 001 \rangle$ and $\langle 110 \rangle$ orientations, initiate the process of 
Bain-deformation (see (\ref{fig1})). It would be further interesting to 
find the conditions for multi-mode generation and to compare them with the 
conditions for single-mode generation. Such a comparison would then enable 
us to find a criterion for identification of such cases where multi-phonon 
processes carry out independent or supportive functions for single-mode 
generation, during a $\gamma -\alpha$ - MT. 

To disclose and enlighten the outstanding features of two-mode generation, 
it is convenient from the outset to omit single-phonon processes and define 
a specific model Hamiltonian as the Hamiltonian of the electron-phonon 
system:

\begin{equation}
H = \hbar\omega_{\mathbf{q}}b_{\mathbf{q}}^{+}b_{\mathbf{q}} +
\hbar\omega_{\mathbf{p}}b_{\mathbf{p}}^{+}b_{\mathbf{p}}+
\sum_{\mathbf{k}}\varepsilon_{\mathbf{k}}a_{\mathbf{k}}^{+}a_{\mathbf{k}} 
+ W(\mathbf{q,p})\sum_{\mathbf{k}}b_{\mathbf{p}}^{+}b_{\mathbf{q}}^{+}
a_{\mathbf{k}}^{+}a_{\mathbf{k}^\prime} +
b_{\mathbf{p}}b_{\mathbf{q}}a_{\mathbf{k}^\prime}^{+}a_{\mathbf{k}}.
\label{3.25} 
\end{equation}
Here, $b^{+}$, $b$, $a^{+}$, $a$ are the phonon and electron creation and 
annihilation operators, respectively. During the summation of quasi-momenta 
of the electrons $\hbar\mathbf{k}$ we shall however only consider those states located in 
the vicinity of the surface $S_{\lbrack 111 \rbrack}$ (see (\ref{fig2.2})). In \eqref{3.25}, 
the quasi-momenta $\hbar\mathbf{q}$ and $\hbar\mathbf{p}$ are oriented parallel to 
the axis $\lbrack 001 \rbrack$  and  $\lbrack 110 \rbrack$ ($\mathbf{q} \bot 
\mathbf{p}$), thus the following conservation rules are satisfied:

\begin{equation}
\varepsilon_{\mathbf{k}^{\prime}}-\varepsilon_{\mathbf{k}} -
\hbar (\omega_{\mathbf{q}}+\omega_{\mathbf{p}}) = 0,\quad 
\mathbf{k}^{\prime} - \mathbf{k} - \mathbf{q} - \mathbf{p} = 0,\quad 
\mathbf{Q}.
\label{3.26}
\end{equation}
Here, $\mathbf{Q}$ - vector of the reciprocal lattice, to be considered in the 
case of U-Processes. However, as the quasi-momenta $\hbar\mathbf{q}$,
$\hbar\mathbf{p}$ can be judged as small quantities in relation to the quantity
$\hbar\pi/a$, the importance of U-processes is only appreciable for a small
fraction of the totality of transitions with quasi-momenta $\hbar\mathbf{k}$,
$\hbar \mathbf{k}^{\prime}$, in the immediate proximity of the lines LW, XW.
Thus we will later consider N-processes, selecting 
zero in the right-hand side of the quasi-momentum conservation law \eqref{3.26}.

With quasi-momenta  $\hbar q$, $\hbar p << \hbar \pi / a$, we get the following expression for a matrix-element of 
electron-phonon interaction:

\begin{equation}
W(\mathbf{q,p}) \equiv - \frac{G\;q\;p\;\hbar}{2\;M\;N\;
\sqrt{ \omega_{\mathbf{q}} \,\omega_{\mathbf{p}}}} \equiv W_{2},  
\label{3.27}
\end{equation}
where $M$, $N$ and $G$ have the same meaning as in \ref{2.4}. 

Let us now perform the same sequence of operations as those already 
explained in Pt. 3.1:
\begin{enumerate}
\item {Neglecting the fluctuation effect of heat reservoirs and considering 
their dissipation by the phenomenological parameters $\Gamma $, $\varkappa$ and 
$t_{\sigma}$, we can formulate the Heisenberg equations of motion for the 
operators of the phonon-field $b^{+}$, $b$, the electron-polarization
$d^{+}_{\mathbf{k},\mathbf{k}^\prime} = a^{+}_{\mathbf{k}}a_{\mathbf{k}^\prime}$,
$d_{\mathbf{k},\mathbf{k}^\prime} = a^{+}_{\mathbf{k}^\prime}a_{\mathbf{k}^{}}^{}$, and the 
occupation inversion $\sigma$;} 
\item {By averaging out the equations of motion with the aid of the system 
density-matrix, we can get over to the classical equations for the mean 
values of the relevant operators;}
\item {We shall confine our considerations to the exact resonance case and, 
using equations analog to (\ref{3.5}), we can use the mean values of the operators 
$\tilde{d}^{+}$,\: $\tilde{d}$,\: $\tilde{b}^{+}$,\: $\tilde{b}$, 
which have no oscillation dependence of time. 
In the stationary case, we thus get the following system of non-linear 
algebraic equations:
\begin{displaymath}
\varkappa_{\mathbf{p}}\:\tilde{b}_{\mathbf{p}}^{+} - \frac{i}{\hbar}\:W_{2}\:R\: 
\tilde{b}_{\mathbf{q}}\:\tilde{d}^{+} = 0,
\end{displaymath}
\begin{equation}
\varkappa_{\mathbf{q}}\:\tilde{b}_{\mathbf{q}}^{+} - \frac{i}{\hbar}\:W_{2}\:R\:
\tilde{b}_{\mathbf{p}}\:\tilde{d}^{+} = 0, 
\label{3.28}
\end{equation}
\begin{displaymath}
\Gamma\:\tilde{d}^{+} + \frac{i}{\hbar} \:W_{2}\: \sigma \:
\tilde{b}_{\mathbf{p}}^{+}\:\tilde{b}_{\mathbf{q}}^{+} = 0,
\end{displaymath}
\begin{displaymath}
\frac{\sigma_0 - \sigma}{t_\sigma} +
\frac{2i}{\hbar}\:W_{2}\:(\tilde{b}_{\mathbf{p}}^{+}\:\tilde{b}_{\mathbf{q}}^{+}\:\tilde{d} -
\tilde{b}_{\mathbf{p}}\:\tilde{b}_{\mathbf{q}}\:\tilde{d}^{+}) = 0,
\end{displaymath}
which is expressed in a shorthand notation, after omission of the coupled 
equations obtained by the substitutions $\tilde{b}^{+} \leftrightarrows
\tilde{b}$,\: $\tilde{d}^{+} \leftrightarrows \tilde{d}$,\: $i \to - i$. In
\eqref{3.28}, we denoted by R the number of ES for which an 
electron transition within its own group is associated with the emission of 
a pair of phonons.} 
\end{enumerate}
We can now derive the equation for $\sigma $ from system \eqref{3.28}:

\begin{displaymath}
\sigma^{2} + \sigma_{0}\sigma + \frac{1}{4}\sigma^{2}_{t2} = 0,
\end{displaymath}
from which follows

\begin{equation}
\sigma_{1,2} = \frac{\sigma_0}{2} \pm \frac{\sigma_0}{2}\left[1 - \left( 
\frac{\sigma_{th2}}{\sigma_0}\right)^2 \right]^{\frac{1}{2}},\quad
\sigma _{th2} = \frac{4\:\hbar \:(\Gamma \:t_{\sigma} \:\varkappa_{\mathbf{q}}\:
\varkappa_{\mathbf{p}})^{\frac{1}{2}}}{R\:\arrowvert W_2 \arrowvert}, 
\label{3.29}
\end{equation}
where the aim and object of $\sigma_{th2}$ is the determination of an 
occupational threshold difference for two-mode generation. By \eqref{3.28}, we get 
the following expression for $\tilde{b}_{\mathbf{p}}$:

\begin{equation}
\varkappa_{\mathbf{p}}\,\tilde{b}_{\mathbf{p}}\left[ 1 - \frac{R \,\sigma_0 \,W_2^2 
\,\tilde{b}_{\mathbf{p}}^2}{\hbar^2\,(\varkappa_{\mathbf{q}}\,\Gamma + 4 \,W_2^2
\,\varkappa_{\mathbf{p}} \,t_\sigma \,\tilde{b}_{\mathbf{p}}^4)}\right] = 0,
\label{3.30} 
\end{equation}
which only comprises one singular and valid solution 
$\tilde{b}_{\mathbf{p}}$ = 0, for $\sigma_{0} < \sigma_{th2}$. For $\sigma_{0} > \sigma _{th2}$ , there 
exist two other valid solutions, in addition to $\tilde{b}_{\mathbf{p}} = 0$:

\begin{equation}
(\tilde{b}_{\mathbf{p}})_{1,2} =
\frac{1}{2}\left[\frac{R}{\varkappa_{\mathbf{p}}t_\sigma}(\sigma_0 - \sigma_{1,2})\right]^
{\frac{1}{2}} \ne  0. 
\label{3.31} 
\end{equation}
For $\sigma_{0} = \sigma_{th2}$, we get from \eqref{3.29} and \eqref{3.31}:

\begin{equation}
\sigma_{1} = \sigma_{2} = \frac{\sigma_0}{2} = \frac{\sigma_{th2}}{2};\quad
(\tilde{b}_{\mathbf{p}})_{1} = (\tilde{b}_{\mathbf{p}})_{2} = 
\tilde{b}_{\mathbf{p}th} = \frac{1}{2}\left[\frac{R \:\sigma_{th2}}{2
\:\varkappa_{\mathbf{p}}\:t_{\sigma}}\right]^{\frac{1}{2}}
\label{3.32} 
\end{equation}
The correct expressions for $\tilde{b}_{\mathbf{q}}$ can be obtained by 
replacement of \textbf{p} by \textbf{q} ($\mathbf{p} \leftrightarrows
\mathbf{q}$).

The investigation on the stability of these solutions and their 
interpretations will be performed by considering the transition in the 
generating mode as a particular kind of phase transition of a radiation 
system, as proposed by Haken in \cite{Leks75,Khaken78,Spravochnik}. The role of the ordering 
and temperature parameters will be represented by $\tilde{b}$ and
$-\sigma_{0}$ respectively. The analog expression for free energy B can 
easily be found by considering Eq.\eqref{3.30} as an extreme condition

\begin{displaymath}
\frac{d B}{d \tilde{b}} = 0.
\end{displaymath}
Then, after omission of the irrelevant constant, we get:

\begin{displaymath}
B \equiv B_{2} (\tilde{b}) = \varphi (\tilde{b}_{\mathbf{p}}) 
+ \varphi (\tilde{b}_{\mathbf{q}}),
\end{displaymath}
\begin{equation}
\varphi (\tilde{b}_{\mathbf{p}}) = \frac{1}{2}\,\varkappa_{\mathbf{p}}
\,\tilde{b}_{\mathbf{p}}^{2} - \frac{R\, \sigma_{0}}{16\,t_{\sigma}}
\ln{ \left[1 + \frac{4 \,t_{\sigma}\, W_{2}^{2}
\,\varkappa_{\mathbf{p}}\,\tilde{b}_{\mathbf{p}}^4}{\hbar^{2}\,
\Gamma\, \varkappa_{\mathbf{q}}}\right]}.
\label{3.33} 
\end{equation}
The expression $\varphi (\tilde{b}_{\mathbf{q}})$ can be obtained from $\varphi 
(\tilde{b}_{\mathbf{p}}$ ) by the changing $\mathbf{p} \leftrightarrows \mathbf{q}$.

\begin{figure}[htb]
\centering
\includegraphics[clip=true, width=.6\textwidth]{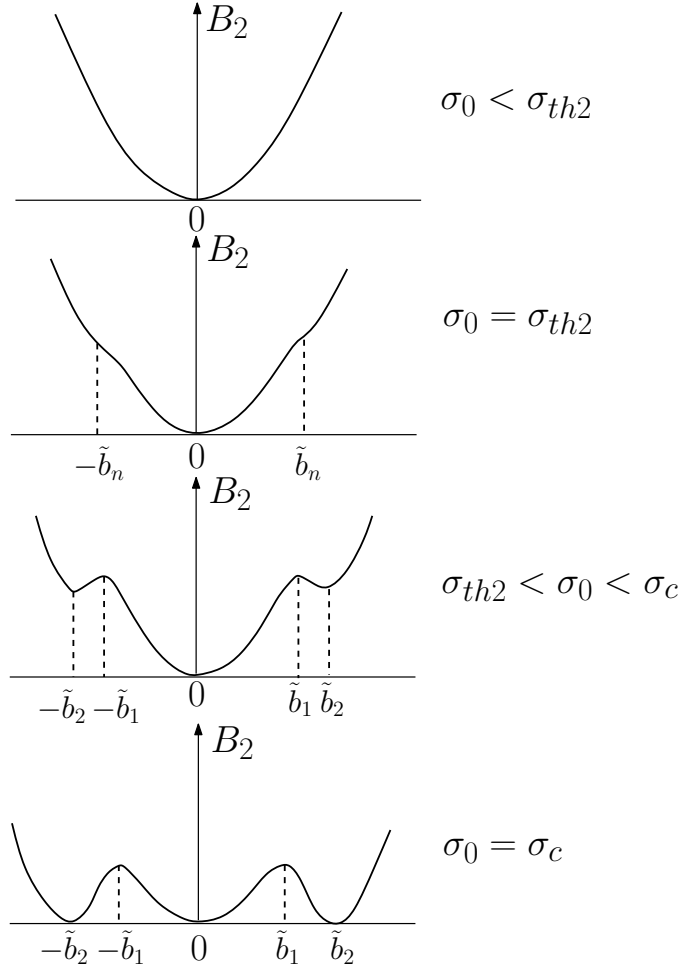}
\renewcommand{\captionlabeldelim}{.}
\caption{Image of the function $B_{2}$ ($\tilde{b})$ at different 
values of occupational difference $\sigma_{0}$.} 
\label{fig3.1}
\end{figure}

Now it is easy to verify that the function $B_{2}(\tilde{b})$ attains a 
single minimum in the region $\sigma_{0} < \sigma_{th2}$, 
corresponding to $\tilde{b} = 0$. For $\sigma_{0} = \sigma_{th2}$, a 
flex point $\tilde{b}_{t}$ arises, being a forerunner of additional 
extremes taking shape in the region $\sigma_{0} > \sigma_{th2}$. The 
additional minimum $B_{2}$ is associated with the value $\tilde{b}_{2}$ 
\eqref{3.31}, while the maximum corresponds to $\tilde{b}_{1}$, which indicates 
an instability of the solution $\tilde{b}_{1}$ and a stability of the 
solution $\tilde{b}_{2}$. This is a distinct indication for the emergence 
of a scenario of a typical 1$^{st}$ order phase-transition (see (\ref{fig3.1})), 
where the value $\sigma_{0} = \sigma_{th2}$ determines the 
limit of absolute loss of stability of the phase with non-vanishing 
magnitude of order parameter. 

The analog of the critical temperature of a 1$^{st}$ - order phase 
transition is the critical value of occupation inversion $\sigma_{0} = 
\sigma_{c}$, at which the minimum values of the function $B_{2}$ match, 
i.e.

\begin{equation}
B_{2} (\tilde{b} = 0) = B_{2} (\tilde{b}_{2} ) = 0.
\label{3.34} 
\end{equation}
We can now determine $\tilde{b}_{2}$ by substituting the expression 
for $\sigma_{2}$ from \eqref{3.29} in \eqref{3.31}. After insertion of $\tilde
{b}_{2}$ in \eqref{3.33}, we are able to express \eqref{3.34} in such a manner that we 
can determine $\sigma_{c}$:
 
\begin{equation}
1 + \left( {1 - \psi^{-2}}\right)^{1/2} - ln \left[ {1 + \psi^2
\left[ {1 + \left( {1 - \psi^{-2}} \right)^{1/2}} 
\right]^{2}} \right] = 0, 
\label{3.35} 
\end{equation}
where $\psi = \sigma_{c}\,\sigma_{th2}^{-1}$ . The numeric solution of
\eqref{3.35} renders:

\begin{displaymath}
\psi \approx 1,25,\quad  
\sigma_{c} \approx 1,25 \,\sigma_{th2}.
\end{displaymath}
For $\sigma_{0} > \sigma_{c}$ we get $B_{2} (\tilde{b}_{2}) < 
B_{2}(0)$, indicating a higher relative stability of the phase 
corresponding to $\tilde{b} \ne 0$. After further analysis, we can 
realize a decrease of the maximum of $B_{2}(\tilde{b}_{1})$, which 
plays the role of a potential-barrier between the states with $\tilde{b} = 
0$ and $\tilde{b}_{2}  \ne 0$, as well as a decrease of the actual 
value of $\tilde{b}_{1}$ with increasing parameter $\sigma _{0}\sigma 
_{c}^{-1} > 1$ $( B_{2}(\tilde{b}_{1}) \to 0$ only for 
$\sigma_{0}\sigma_{c}^{-1} \to \infty $ ). As the value 
of $\sigma_{0}$ is limited ($\sigma_{0} < 1$), the satisfaction of 
Eqs. $\tilde{b}_{1}$ = 0 and $B_{2}(\tilde{b}_{1}) = 0$, indicating 
an absolute loss of stability of the phase with vanishing order-parameter, 
will only be possible at $\sigma_{c}$ = 0. However, if $\sigma_{c} \sim
\sigma_{th} \ne 0$, it is not possible to introduce a value 
$\sigma_{c}$ corresponding to the temperature of absolute loss of 
stability of the high-temperature phase. 

A consideration of three-mode generation with frequencies
$\omega_{\mathbf{k}}$,$\omega_{\mathbf{p}}$ and $\omega_{\mathbf{q}}$
would lead us to qualitatively similar conclusions. For this reason, 
we will only denote the expressions for the threshold- and for the 
critical occupational differences. 

\begin{equation}
\sigma_{th3} = \frac{3}{R}\left[4 \,\Gamma\,
t_{\sigma}^{2}\,\hbar^{2}\,\varkappa_{\mathbf{k}}\,\varkappa_{\mathbf{q}}\,\varkappa_{\textbf{p}}\right]^
\frac{1}{3},\quad \sigma_{c} \approx 1,41\, \sigma_{th3}
\label{3.36}
\end{equation}
where

\begin{displaymath}
W_{3} \equiv W(\mathbf{k},\,\mathbf{p},\,\mathbf{q}) = \frac{i\, G \,k\, p \,q\,
\hbar^{3/2}}{(2 M N)^{3/2} (\omega_{\mathbf{k}} \,\omega_{\mathbf{p}}\,
\omega_{\mathbf{q}})^{1/2}}. 
\end{displaymath}

We can realize the difference between two- and three phase generation on the 
one hand and single-phase generation on the other hand, which can thus be 
defined as a second-order phase transition of a radiation field at threshold 
occupational differences $\sigma_{th1}$ \eqref{3.10}, which we shall write once 
more, for comparison,

\begin{equation}
\sigma_{th1} = \frac{\Gamma \:\varkappa_{\textbf{q}}\:\hbar^{2}}{R \:\arrowvert W_1
\arrowvert^2}, 
W_{1} \equiv W(\mathbf{q}) = \frac{i \hbar^{1/2}\: G \:q}{(2 \:M\: N\:
\omega_{\textbf{q}})^{1/2}} 
\label{3.37}
\end{equation}
and which can be interpreted as an analog to the expression of the 
Curie-temperature. In fact, if we regard \eqref{3.11} in 3.1 as a prerequisite for 
a minimum of the free energy equivalent $B_{1}(\tilde{b})$, then we get:

\begin{equation}
B_{1}(\tilde{b}) = \frac{\varkappa \:\tilde{b}^2}{2} - \frac{\sigma_0 \:R}{8\:
t_{\sigma}}ln \left[{1 + \frac{4\: t_{\sigma}\: \arrowvert W_1 \arrowvert^2\:
\tilde{b}^2}{\hbar^2\: \Gamma}} \right]. 
\label{3.38}
\end{equation} 
It can easily be verified that - in the region $\sigma_{0} < \sigma
_{t1}$ - the function $B_{1}(\tilde{b})$ attains a minimum at $\tilde{b} = 0$,
and in the region $\sigma_{0} > \sigma_{th1}$ another minimum at 
$\tilde{b} \ne  0$. This is a clear indication of a non - stability of the 
ordered state below the generation threshold, and of the disordered state 
above the generation threshold. Moreover, a typical feature 
of 2$^{nd}$ order transitions can be inferred from the smoothly changing 
of the order parameter while approaching to $\sigma_{0} = \sigma_{th1}$. 

The expressions $\sigma_{c} \sim \sigma_{th \,2,3}$ also differ 
significantly from $\sigma_{th1}$. By definition, $\sigma_{c}$ 
implicitly includes a time-variable $t_{\sigma}$, in addition to the 
parameters specifying $\sigma_{th1}$. We thus determine
the relations $\sigma_{th2}\:\sigma_{th1}^{-1}$ and $\sigma_{th3}\:\sigma_{th1}^{-1}$, 
using the assumption that frequencies of a given order of magnitude will be generated 
in a single mode, as well as in two or three modes, respectively. Then we get 
from Eqs. \eqref{3.29}, \eqref{3.36} and \eqref{3.37}, within an accuracy given by
factors of order

\begin{equation}
\label{3.39}
\frac{\sigma_{th2}}{\sigma_{th1} } \sim 
\frac{G}{\hbar}\left(\frac{t_{\sigma}}{\Gamma}\right)^{1/2};\quad
\frac{\sigma_{th3} }{\sigma_{th1}} \sim 
\left[ \frac{G}{\hbar}\left( \frac{t_{\sigma}}{\Gamma} \right)^{1/2} 
\right]^{4 / 3}.
\end{equation}
Thus for $\hbar^{-1}\, G\, t_{\sigma}^{1/2}\, \Gamma^{-1/2} < 1$, multi-mode generation 
could formally start sooner than single-mode generation. It has however to be 
considered that this inequality can only be satisfied by a reduction of G or by 
raising of $\Gamma$ within the region of the threshold condition 
$\sigma_{thj} > 1$, $j = 1, 2, 3$, being unattainable in reality due to the 
limitation $\sigma_{0} < 1$. But also the parameter $t_{\sigma}$ has a natural 
lower limit. According to (\ref{3.4}), $(t_{\sigma})_{\min} = \Gamma^{-1}$, but 
$(\sigma_{th2}\:\sigma_{th1}^{-1})_{\min} \sim G(\hbar\:\Gamma)^{-1}$, and for the 
previously used values $G \sim 10^{-19}$ J, $\Gamma \sim 10^{12}$ 1/s we get 
$\sigma_{th2}\:\sigma_{th1}^{-1} \sim 10^{3}$. Thus the role of processes with 
two and three participating phonons can mainly be attributed to the amplification 
(and possible synchronization) of modes generated by single-phonon processes. 

Our common consideration of single- and two-phonon processes has been 
performed for the most favorable case of generation, being characterized by 
electronic transitions with one or two participating phonons, occurring in 
the proximity of different S-surfaces. To take an example, during the generation of a 
pair of waves along the orientations $\Delta - \lbrack 001 \rbrack$ and $\Sigma -
\lbrack 110 \rbrack$, it is possible to analyze single-phonon processes in the proximity 
of the surfaces $S_{\lbrack 001 \rbrack,\lbrack 110 \rbrack}$, and two-phonon processes between states 
in the proximity of $S_{\lbrack 111 \rbrack}$. Omitting some usual considerations, being 
negligible for our purposes, and considering $\sigma_{0}\:\sigma_{th2}^{-1} \ge 
\sigma_{th1}\:\sigma_{th2}^{-1}$ as a small quantity, we can present a stationary 
solution for the simplest case of equalities: 
$\tilde{b}_{\Delta}^{2} = \tilde{b}_{\Sigma}^{2} = \tilde{b}^{2}$
for the numbers of generated phonons; 
$R_{\Delta} = R_{\Sigma} = R$ for the numbers of pairs of electronic states; 
$\sigma_{0\Delta} = \sigma_{0\Sigma}=\sigma_{0\Lambda}=\sigma_{0}$ and 
$t_{\sigma \Delta} = t_{\sigma \Sigma} = t_{\sigma \Lambda} = t_{\sigma}$ 
(for the longitudinal relaxation times), confined to the states located in 
the proximity of different S-surfaces. Then our solution is:

\begin{equation}
\tilde{b}^{2} =  0,\quad \sigma_{0} < \sigma_{th1},
\label{3.40}
\end{equation}
\begin{displaymath}
\tilde{b}^{2}  \approx  \frac{\hbar^2 \,\Gamma}{4\,\arrowvert W_1\arrowvert^2
\,t_{\sigma}}\left( \frac{\sigma_0}{\sigma_{th1}} - 1 \right)
\left( 1 + \left( \frac{ 2\,\sigma_0 }{\sigma_{th2}}\right)^{2} \right),\quad
\sigma_{0} > \sigma_{th1},
\end{displaymath}
being distinguished from the solution for single-mode generation \eqref{3.13} by a 
factor~$1~+~( 2 \,~\sigma_{0}\, ~\sigma_{th2}^{-1})^{2}$,  
thus clearly indicating the amplification effect by two-phonon processes.

\section{The lattice strain induced by the generated waves }

For realization of martensitic transformation, in accordance with the 
wave-model, it is necessary that the amplitudes of the generated waves have 
exceeded the threshold ($u_{t}(\lambda , T)$), which essentially depends 
on the wave-length $\lambda $ and on the temperature T. The maximum of 
$u_{t}(\lambda)$ can be evaluated by geometrical considerations alone, 
taking into account that - for pure Bain-deformation \cite{Kurdjumov77} - the transition 
to new equilibrium conditions is associated with a relative displacement of 
$0,1 a$ ($a$ - lattice-parameter) of the atoms neighboring in the orientation
$\langle 001 \rangle$. This means that an instability must be expected for 
relative lattice displacements greater than $ \approx 0,05a$. If such 
displacements were associated with a wave, then the amplitude of such waves 
would obviously be $ \sim  0,05 \,\lambda /2 = u_{t}(\lambda)$. The value $u_{t} 
(\lambda)$ being independent of temperature can be put into natural 
correspondence with the threshold $u_{t}(\lambda, T_{0})$, where 
$T_{0}$ - equilibrium temperature between austenite and martensite.

The supercooling below the equilibrium temperature ( for definiteness, the 
straight MT under cooling is considered ) must occur with the decreasing of 
austenite stability and hence:

\begin{displaymath}
u_{t} (T, \lambda) < u_{t} (T_{0}, \lambda).
\end{displaymath}
Let's remind (see Pt. 1.3) that wave-amplitudes creating the 
threshold-deformation $\varepsilon_{th}$ of about $10^{-3}$ 
are anticipated in the proximity of $M_{S}$. 

In evaluating the stationary amplitudes u of the generated waves, we can 
disregard the influence of electron-phonon interaction associated with 
multi-phonon processes. Then, using the link between $u$ and $\tilde{b}$ by 
\eqref{3.14}, as well as the expressions \eqref{3.40} and \eqref{3.37}, we get 
the following relation:

\begin{equation}
u_{\textbf{q}}=\left[ \frac{2 \hbar }{M  N  \omega_{\mathbf{q}}} 
\right] ^{\frac{1}{2}}\tilde{b}_{\textbf{q}} \approx \frac{\hbar 
}{G q} \left[ \frac{\Gamma}{t_{\sigma}}\left( \frac{\sigma_0 
}{\sigma_{th1}} - 1 \right) \right]^{\frac{1}{2}}. 
\label{3.41}
\end{equation}

The deformation $\varepsilon $ in turn is linked up with the amplitude by 
the relation \eqref{eq2} in Pt. 1.3:

\begin{equation}
\varepsilon  = \frac{2}{\pi} u_{\textbf{q}} q \approx 
\frac{2\hbar}{\pi \,G}\left[ \frac{\Gamma}{t_{\sigma}}\left( 
\frac{\sigma_0 }{\sigma_{th1}} - 1 \right) \right]^{\frac{1}{2}}. 
\label{3.42}
\end{equation}
Assuming $t_{\sigma}=\Gamma ^{-1}$, $\sigma_{th1} \sim 10^{-3}$, 
$\left( \sigma_0 \sigma_{th1}^{-1}  - 1 \right) \approx  1,6$ - 
corresponding to the value $\sigma_{0}(\nabla \mu)$,~y~=~2~in 
Table \ref{table3.1} - and using the quantity $10^{2}$ v$_{k}\cdot \tau$ 
as a natural gage of heterogeneity (see end of Pt. 3.1), with 
$G \sim 10 ^{-19}$ J and $\hbar \, \Gamma  \sim  10^{-22}$ J, we get 
from \eqref{3.42}: $\varepsilon  \approx 3,2 / \pi \cdot 10^{-3}$. As 
a result, the existence of waves with amplitudes ensuring a lattice 
deformation near the elastic (yield-) limit, should be reasonably possible. 
We further note that for $\lambda /2  \sim  10 ^{-6}$ m and 
$\varepsilon \sim  10 ^{-3}$ the wave-amplitude 
($u = \varepsilon \,\lambda / 4 \approx  5  \cdot  10 
^{-10}$ m $ \approx  1,4a$) becomes even greater than the 
lattice-parameter. 

Let us now discuss the applicability of our estimates, taking into account 
that they are based on stationary solutions. Above all, the lifetimes of 
spatially inhomogeneous distributions of temperature and chemical potential 
have to be evaluated. Let these times be $t_{\nabla T}$, $t_{\nabla \mu}$ 
and, taking into account that such heterogeneities rapidly fade away, mainly 
as a result of conduction and diffusion processes, we can obtain an estimate 
of these times $t_{\nabla}$, becoming clearly obvious from dimensional 
considerations:

\begin{equation}
t_{\nabla T}  \approx  \frac{l_{T}^{2}}{d_{T}}, \quad t_{\nabla \mu} 
\approx  \frac{l_{\mu}^{2}}{d_{\mu}}, 
\label{3.43}
\end{equation}
where $l_{T,\:\mu}$ - characteristic gages of inhomogeneity, determining 
$\nabla  T  \approx \Delta T \cdot l_{T}^{-1}$, 
$\nabla \mu  \approx \Delta \mu \cdot l_{\mu}^{-1}$; 
$d_{T}$ and $d_{\mu}$ - heat-conduction and diffusion coefficients, 
respectively.

The typical value of $d_{T} \approx  10^{-5}$ m$^{2}$/s for $T > 
T_{D}$ can easily be obtained from below equation \eqref{3.44}, by linking up 
$d_{T}$ with heat-conductivity $\lambda_{t}$, specific heat capacity 
$C_{sp}$ and density $\rho $:

\begin{equation}
d_{T} = \frac{\lambda_{t} }{\rho \,C_{sp} }, 
\label{3.44}
\end{equation}
where $\lambda_{t} = 34$ W/mK, $\rho $ = 7900 kg/m$^{3}$, 
$C_{sp} \approx  4,6  \cdot  10^{2}$ J/kg K (the specific heat capacity 
has been derived from the Dulong-Petit-Law, being substantiated for 
$T > T_{D})$.

Taking into account that the term $d_{\mu}$ is a coefficient of 
proportionality which combines an electric current density j with the 
gradient $\nabla n_{\gamma}$ of conduction electron concentration in the 
$\gamma $ - phase, we are able to express the term $d_{\mu}$ by the specific 
conductivity $\sigma _{\gamma }$ of the $\gamma$ - phase. For this purpose, 
we start with an appropriate conversion of the expression for current-density 
j, presupposing that $\mu \sim  n_{\gamma}^{2/3}$:

\begin{equation}
\label{3.45}
j =  - \frac{\sigma_{\gamma}}{e} \,\nabla \mu  = - \frac{2}{3} \,\frac{\sigma_{\gamma}
\,\mu_{\gamma}}{e\, n_{\gamma}} \,\nabla n_{\gamma}.
\end{equation}
After division of both sides of (\ref{3.45}) by the electron charge e, we
get:
\begin{equation}
\frac{j}{e} =  - \frac{2 \,\sigma_{\gamma}\, \mu_{\gamma}}
{3\,e^2\, n_{\gamma}}\,\nabla n_{\gamma}  \equiv - d_{\mu} \,\nabla n_{\gamma
},\quad d_{\mu} = \frac{2 \,\sigma_{\gamma} \,\mu_{\gamma}}{3\,e^2\, n_{\gamma}}. 
\label{3.46}
\end{equation}
After insertion of $\sigma_{\gamma}  \approx  10^{6}$ m$^{-1}$ Ohm$^{-1}$, 
$n_{\gamma} = 10^{29}$ m$^{-3}$, $\mu_{\gamma} \approx  10 \, eV = 
1,6  \cdot  10^{-18}$ J into \eqref{3.46}, we get 
$d_{\mu}\approx 4 \cdot  10^{-4}$ m$^{2}$/s.\footnote{The order of magnitude 
of the current density j being generated by the chemical potential gradient 
$10^{6}$eV/m in the transformation front is $10^{12}$ A/m$^{2}$ 
for $\sigma_{\gamma}$ about $10^{6}$ m$^{-1}$ Ohm$^{-1}$ and 
substantially greater than any drift current observed in semiconductors.} 

Obviously in our assessment of the quantities $\nabla  T$, $\nabla \mu $ 
the usage of the quantities~$\Delta~T ~ \sim ~ 10^{2}$\,K~$ \sim~( T_{0} 
- M_{S})$ and $\Delta \mu =\arrowvert \mu_{\alpha} - \mu_{\gamma}\arrowvert 
\approx  0,15$ eV can be justified if the nucleation time $t_{N}$ is 
less or of the same order of magnitude than that the time of 
existence of the gradient $t_{\nabla}$. Otherwise, conduction- and 
diffusion processes would smooth down the jumps of temperature and 
concentration at the phase boundary. In the nucleation model of the 
macro-nucleus, we can determine the minimum values $(l_{T})_{min}$ and 
$(l_{\mu})_{min}$ from the requirement $t_{N} = t_{\nabla}$, using 
\eqref{3.43}, further assuming $t_{N} \sim 10^{-11}$ s (see treatise under 
Pt. 1.3), which would correspond to a nuclear radius $r_{N} \sim 10^{-5}$ cm
and to a speed of propagation of the nuclear boundary of $V \sim 10^{3}$ m/s:

\begin{displaymath}
(l_{T})_{min} = ( t_{N}\,d_{T})^{1/2} \sim  10^{-8} m,
\end{displaymath}

\begin{displaymath}
( l_{\mu})_{min} = ( t_{N}\,d_{\mu})^{1/2} \sim  6  \cdot  10^{-8} m.
\end{displaymath}
If $\nabla \mu $ was considered as a stationary quantity, then 
the quantity $l_{\mu}  \sim  10^{-7}$ m would be a natural gage. We 
remind that, at the end of Pt. 3.1, we came to a similar conclusion (which 
however resulted from different considerations). 

The issue related to the applicability of stationary estimates of the 
amplitudes $u = u_{st}$ and deformations $\varepsilon =\varepsilon 
_{st}$ is more complex. In fact, the transition to virtually stationary 
values of $u$, $\varepsilon $, within certain characteristic period $t_{u} = 
t_{\varepsilon }$ , will not depend on the values of $u_{0}$, $\varepsilon 
_{0}$ at the starting moment $t_{0}$, provided $t_{u} << t_{\nabla}$. 
Thus in this particular case, a soft mode of wave excitation with 
$u_{0} << u_{st}$ , $\varepsilon_{0} << \varepsilon _{st}$ would be possible. 
But if $t_{u} >> t_{\nabla}$, the values $u \approx  u_{st}$ and 
$\varepsilon_{0} \approx  \varepsilon_{st}$ cannot be 
materialized by a soft mode of excitation. Consequently, a stationary 
estimate would only be realistic for a hard mode of excitation with 
$u_{0} \le  u_{st}$ , $\varepsilon_{0} \le \varepsilon_{st}$. Of 
course, for $u_{0} > u_{st}$ , $\varepsilon_{0} > \varepsilon_{st}$, 
the stationary estimate then would deliver the lower limit of $u$, 
$\varepsilon $ (for $\sigma_{0} > \sigma_{th1}$).

For an evaluation of $t_{u}$, the non-stationary solution of $\tilde{b}(t)$ 
must be known. However, it cannot be found for the general case, but in the 
vicinity of the threshold for single-mode generation, the equation for 
$\tilde{b}$ leads to the Van-der-Pol equation (see for example \S \,(3,a) in 
the lectures of Haken-Weidlich \cite{Khaken74}): 

\begin{displaymath}
\left( \frac{d}{d t} - D \right)\:\tilde{b}^{+} + 
\beta \,\tilde{b}^{+}\,\tilde{b}\,\,\tilde{b}^{+} = 0,
\end{displaymath}

\begin{equation}
D = \varkappa\left( {\frac{\sigma_0}{\sigma_{th1}} - 1} \right), \quad
\beta  = \frac{4 \,t_{\sigma}\, \varkappa \,\arrowvert W_1 \arrowvert ^{2}
\,\sigma_0}{\Gamma \,\sigma_{th1}\, \hbar^{2}}.
\label{3.47}
\end{equation}

Formally, the Van-der-Pol equation can be obtained if the temporal 
derivatives of system \eqref{3.7} are only re-established for the quantities 
$\tilde{b}_{q}^{+}$, $\tilde{b}_{q}$, by writing - $\dot {\tilde{b}}^{+}_{q}$ 
for the right side of the first Eq. \eqref{3.7}, instead of 
zero, leaving the other equations unchanged (Of course, this is an adiabatic 
approximation, which means that the radiating sub-system will immediately be 
subordinated to the ordering-parameter $\tilde{b}_{\mathbf{q}})$. Then, $\tilde{d}$ 
and $\sigma$ must be excluded, and a decomposition of the resulting 
factor

\begin{equation}
\label{3.48}
\left[ 1 + \frac{4 \:t_{\sigma}\:\arrowvert  W_{1} \arrowvert^{2}}
{\hbar^{2}\:\Gamma}\:\tilde{b}_{\mathbf{q}}^{+}\:\tilde{b}_{\mathbf{q}} 
\right]^{-1} \approx 1 - \frac{4 \:t_{\sigma}\:\arrowvert W_1 \arrowvert^{2}}
{\hbar^{2}\:\Gamma }\tilde{b}_{\mathbf{q}}^{+}\:\tilde{b}_{\mathbf{q}} 
\end{equation}
must be performed with an accuracy up and including to the linear term in 
$\tilde{b}_{\mathbf{q}}^{+}\tilde{b}_{\mathbf{q}}$ . As we already excluded the 
dipole-moment $\tilde{d}$ and the occupation difference $\sigma$ in our 
search for a stationary equation for $\tilde {b}_{\mathbf{q}}^{+}$, we can 
immediately use Eq. \eqref{3.11}. It delivers the non-stationary equation for 
$\tilde{b}_{\mathbf{q}}^{+}$ in an adiabatic approximation, by multiplication 
with $- \varkappa_{\mathbf{q}}$ and addition of $\dot{\tilde{b}}_{\mathbf{q}}^{+}$ to 
the left side:

\begin{equation}
\dot{\tilde{b}}_{\mathbf{q}}^{+} = \varkappa_{\mathbf{q}}\:
\tilde{b}_{\mathbf{q}}^{+} \left[ {\frac{\sigma_0}{\sigma_{th1}}
\left( 1 + \frac{4 \:t_{\sigma}\:\arrowvert W_{1}
\arrowvert^{2}\:\tilde{b}_{\mathbf{q}}^{+}\:\tilde{b}_{\mathbf{q}}}
{\hbar^{2}\:\Gamma} \right)^{-1} - 1 }\right] \equiv - \chi\:
\tilde{b}_{\mathbf{q}}^{+}. 
\label{3.49}
\end{equation}
With consideration of (\ref{3.49}), we immediately get \eqref{3.47} from
\eqref{3.49}. Eq. \eqref{3.49} will be further used in Chapter 6, so that 
we can now deal with to the more simple Eq. \eqref{3.47}.

Assuming that the quantity $\tilde{b}$ attains the value $\tilde{b}_{0}$, 
due to fluctuations at the starting time $t = 0$ , we get from \eqref{3.47}:

\begin{equation}
\label{3.50}
\tilde{b} = \left[ {\left({\frac{1}{b_0^{2}} - \frac{\beta}{D}}\right)
e^{-2\,D\,t} + \frac{\beta}{D}} \right]^{-1/2}.
\end{equation}
It follows from (\ref{3.50}), that for $D > 0$, $t \to \infty $, 
$\tilde{b} \to  (D/\beta)^{1/2} \ne  0$, but $\tilde{b} \to  0$, for $D < 
0$, $t \to \infty $. It also is obvious from (\ref{3.50}) that the only 
relevant characteristic time is:

\begin{equation}
t_{u} = (2\,D) ^{-1}. 
\label{3.51}
\end{equation}
Here, the following observation is striking enough to attract our attention: 
As a general rule, during approach to the starting temperature of a 
second-order phase transition, the tendency of $t_{u}$ to infinity for 
$\sigma_{0}  \to \sigma_{th1}$ would be analog to the critical 
degradation of fluctuations of the ordering parameter. 

Even though the Van-der-Pol equation is not applicable for states located 
significantly beyond the threshold, an evaluation of the order of magnitude 
of  $t_{u}$ can nonetheless be performed on the basis of Eq. \eqref{3.51}. 
Assuming an attenuation $\varkappa \approx  (10^{-4} \div 10^{-3})\omega $, 
in accordance with \eqref{3.16}, we can realize that for $(\sigma_{0}
\,\sigma_{th}^{-1} - 1)$ of order 1 the resulting time $t_{u}$ evaluated on the basis 
of Eq. \eqref{3.51} must be not less than $(10 \varkappa)^{-1}$. Hence only 
the inequality $t_{u} >> t_{\nabla }$ is satisfied in the region of parameters 
interesting for us: $\omega~\sim~10^{10}~\div~10^{11}$~s$^{-1}$, 
$t_{u} \ge (10^{-8} \div 10^{-9})$ s. In accordance with our above treatise, 
it must be concluded from this latter inequality that our stationary 
evaluations of u and $\varepsilon $, having been obtained on the basis of
\eqref{3.41}, \eqref{3.42}, would only be justified for hard modes of excitation, 
i.e. $u_{0} \le  u_{st}$, $\varepsilon  \le  \varepsilon_{st}$.\footnote
{Editorial note: If a transformation energy of ( $0,01 \div  0,1 $) eV 
per atom is assumed, then the maximum possible power-density (power per unit 
surface) of the transformation-wave-front would be of an order of magnitude 
of $ \approx (0,1 \div $1)GW cm $^{-2}$, determining the level of 
external energy flow being necessary for the martensitic transformation 
start.}

\section{Conclusions from Chapter 3}

From the above analysis of the equations for a non-equilibrium 
electron-phonon-system, it is possible to draw the following conclusions:
 
\begin{enumerate}
\item{ Electron-phonon processes with participation of a single phonon play a 
dominating role during the generation of longitudinal lattice-displacement 
waves, while the role of multi-phonon processes is limited to the 
amplification of waves.} 

\item{Wave generation takes place through relatively long-lived ES with an 
average lifetime of about $\tau_{0} \ge 10^{-12}$ s.}

\item{Among two non-equilibrium sources in electronic subsystems, being defined 
by the existence of $\nabla  T$ and $\nabla \mu $, the latter one, being 
caused by the inhomogeneity of the chemical potential $\mu $, is the more 
effective one in generating long-wave phonons ($q \sim  (10^{-3}\div 10^{-2})\pi /a)$.} 

\item{In the case of large-sized reduced sheets of the S-surfaces 
($\Sigma \ge (\pi /a)^{2}$) with weak energy-dispersion near the $\mu$-value, 
as well as for chemical potential gradients $\nabla \mu  \sim 10^{6}$ eV/m 
($10^{10}$ K/m in a temperature-scale) the generation of longitudinal 
elastoplastic waves, being linked up with lattice-deformations close to the 
limits of elasticity of about $\varepsilon \sim 10^{-3}$, becomes possible.}

\item{For some characteristic temporal gauges, the following chain of 
inequality conditions may be suggested:

\begin{equation}
t_{u} >> t_{\nabla} > t_{N} > \tau_{0}  \sim  t_{\sigma}. 
\label{3.52}
\end{equation}
The central term $t_{\nabla} > t_{N}$ of \eqref{3.52} determines a relation 
between $t_{\nabla }$ - the lifetimes of $\nabla T$ and $\nabla \mu$, 
respectively - and the time of macro-nucleation $t_{N} \ge 10^{-11}$ s, 
which, in turn, also implies minimal spatial gauges $l_{T} \sim 10^{-8}$ m, 
$l_{\mu}\sim 10^{-7}$ m for stationary evaluations of 
$\nabla  T = \Delta T \cdot l_{T}^{-1}$, $\nabla \mu = \Delta \mu \cdot l_{\mu }^{-1}$. 
The first term $t_{u} >> t_{\nabla}$ shows that a realization of the 
deformation level obtained from an evaluation of \eqref{3.42} (for stationary 
conditions) only becomes feasible for hard-mode wave-excitation. Within the 
frame of the developed model of the $\gamma -\alpha$ -MT as a deformation-
process, being controlled by the threshold-deformation 
$\varepsilon_{th}  \sim  10^{-3}$ of the carrier-waves, this means that the 
threshold-level $\varepsilon_{th}$ must arise simultaneously with beginning 
growth i.e. for $t_{0} = t_{N}$. This latter conclusion matches very well 
with the interpretation of the observed pronounced supercooling of the 
$\gamma$-phase below the phase-equilibrium temperature $T_{0}$ (including a 
significant free-energy excess for compensation of the energy required for 
built-up of the phase boundary, and of the elastostatic distortion field), 
thus being an essential prerequisite for the excitation of oscillations with 
an amplitude sufficiently large to produce permanent lattice-deformations of 
about $\varepsilon_{0} \sim  10^{-3}$ in the nucleation stage (see Pt. 
1.3 and Pt. 1 of the task defined by Pt. 1.5). Such deformation levels thus 
will be sufficient to overcome the threshold separating the metastable state 
of fcc-lattice from the more stable state of bcc-lattice (at $T = M_{S}$).

The treatise of Chapter 3 is mostly based on the papers \cite{Mints78,Kashchenko80} and, to a 
lesser extent, on \cite{Kashchenko84}, the results of which have been improved in this 
chapter, mainly by the replacement of the P-planes of the simplified model 
by S-surfaces and by the specification of the rather illustrative 
quantitative estimates presented in \cite{Mints78,Kashchenko80}.} 
\end{enumerate}

\begin{center}
\textbf{Some additional remarks}
\end{center}

\begin{enumerate}
\item{ The tasks of development of kinetic equations for a non-equilibrium 
electron-phonon system, as well as the analysis of threshold conditions 
during single-mode generation within the conceptual notion of isoenergetic 
P-surfaces with inhomogeneous distribution of temperature has been treated 
in \cite{Vereshchagin79,Vereshchagin80,Veresh79,Veresh4177,Veresh3455} and summarized in \cite{Veresh80}, 
within the frame of the method of the 
non-equilibrium statistical operator of Zubarev and Kalashnikov
\cite{Zubarev71,Kalashnikov71,Zubarev70}. 
The results presented in these works, as far as they relate to the $\gamma 
-\alpha$ -MT, are alike to the results of Pt. 3.1, at least in a 
qualitative aspect. They can be taken as a justification of the 
semi-phenomenological approach, being characterized by the introduction of 
relaxation constants to consider dissipative processes. Furthermore it has 
been shown in \cite{Veresh3455,Veresh80} that a system of equations with mixed coordinate- 
and momentum-representation, resulting from a reduction of the 
macro-variables f, $\sigma$, d, b by powers of the respective gradients, 
attains a local spatial form matching with (\ref{3.3}), with sufficient precision 
until the first non-vanishing terms. However, in this case the 
macro-variables $\sigma$, d, b must be considered as functions of the 
x-coordinate being oriented parallel to the orientation of highest 
heterogeneity \textbf{e}. This representation highlights that the 
probability for the presence of an electron within in the immediate vicinity 
of a point x, with simultaneous presence of electronic flows, will largely 
depend on the orientation of the group-velocity \textbf{v} of that electron 
in relation to \textbf{e}, and that in one and the same point in space 
states with inverted occupation can exist. The spatially-local nature of 
such describing justifies the application of the findings obtained in the 
quasi-momentum representation to an analysis of the non-equilibrium state in 
the proximity of the propagating phase boundary.}

\item{Generally speaking, the drift mechanism for the creation of a population 
inversion is well known, being used for example at generation and 
amplification of elastic waves in semiconductors (see \cite{Pustovoit69,Bonch-Bruevich77}). 
Usually, the conditions for wave generation by a drift mechanism focus on 
the requirement that the electron drift velocity $\mathbf{v}_{d}$ must 
locally exceed the speed of sound c, i.e. v$_{d} > c$. Thus this condition 
is equivalent to that formulated by Tscherenkov for phonon radiation 
generation by drifting electrons. Within the frame of our designations, 
the condition v$_{d} > c$ is equivalent to the condition $\sigma_{0} > 0$, 
being an essentially weaker requirement than $\sigma_{0} > \sigma_{th1}$. 
Obviously from \eqref{3.24}, the condition $\sigma_{0} > 0$ is equivalent to 
the equation 

\begin{equation}
(\mathbf{v}_{\mathbf{k}^{\prime}} - \mathbf{v}_{\mathbf{k}})\, \tau \;\vec
{\nabla}\,\mu > \hbar\,\omega_{\mathbf{q}} = \hbar\, c \,q. 
\label{3.53}
\end{equation}
Taking into account that the states \textbf{k}, $\mathbf{k}^{\prime}$ are 
located in the vicinity of the S-surface, and that the dispersion-law for 
these states along the direction \textbf{e} - is a quadratic function 
(see Chapter 2), we get:

\setlength\arraycolsep{2pt}
\begin{eqnarray}
\mathbf{v}_{\mathbf{k}^{\prime}} - \mathbf{v}_{\textbf{k}}& = & \hbar \,[
\mathbf{s}- \mathbf{k}^{\prime}
- (\mathbf{s}-\mathbf{k})] \,\arrowvert m \arrowvert_{s}^{-1} = 
                    \nonumber\\
 & & = \hbar \,( \mathbf{k} - \mathbf{k}^{\prime})\, \arrowvert m
 \arrowvert_{s}^{-1} = \hbar \,\mathbf{q} \,\arrowvert m \arrowvert_{s}^{-1},
 \label{3.54}
\end{eqnarray}
where \textbf{s} - vector designating the intersection point \textbf{s} of 
the vector $\mathbf{k} -\mathbf{k}^{\prime} = \mathbf{q}$ with the S-surface,
$\arrowvert m \arrowvert_{s}^{-1}$ - reciprocal effective electron-mass in 
point \textbf{s}. After substitution by \eqref{3.54} in \eqref{3.53}, we get the 
condition for Tscherenkov-radiation:

\begin{equation}
\textrm{v}_{d}=\tau \,\nabla  \mu \, \arrowvert m \arrowvert_{s}^{-1}  >  c.
\label{3.55}
\end{equation}
For $\nabla \mu  \sim  10^{6}$ eV/m =  $1,6 \cdot  10^{-13}$ 
J/m, $\tau \sim 10^{-12}$ s, $\arrowvert m \arrowvert_{s} \approx 3\, m_{0}
\sim  3  \cdot  10^{-30}$ kg, we obtain v$_{d} \approx  5 \cdot 10^{4}$ m/s, 
the magnitude of which exceeds the speed of longitudinal waves by about one 
order of magnitude and thus is sufficient for satisfaction of the threshold 
condition $\sigma_{0} > \sigma_{th1}$, within a temperature region of 
$T \sim 10^{3}$ K. However, the realization of waves carrying a deformation 
amplitude of $\varepsilon_{0} \ge  10^{-3}$ requires the larger values of 
$\nabla  \mu $, corresponding to v$_{d} \sim  10^{5}$ m/s. The region 
of v$_{d} \sim (10^{4} \div 10^{5})$ m/s also is typical of 
semiconductors \cite{Pustovoit69}. To have ended this comparison, we remind that the 
generation of elastic waves in a semiconductor requires the considerable 
drift-velocities as well as the large values of the electron-phonon-
interaction matrix-elements (semiconductors-piezoelectrics). In the case of 
the $\gamma -\alpha$ - MT, this interaction is not large. Instead, the 
number of pairs (\textbf{k}, $\mathbf{k}^{\prime}$) of active electronic 
states is large, whereas in turn this factor is of minor importance for 
semiconductors. Of course, neither the drift mechanism of pumping nor the 
existence of electrical conductivity are exclusive prerequisites for the 
emergence of a phonon-maser-effect. Alternative principles of a maser-effect 
are also known (see for example \cite{Taker75,Zheru76,Kopvillem82}).}

\item{ Finally, we remark that, besides stimulated emission of radiation, in an 
inversely occupied radiation system the spontaneous collective 
super-radiation (Dicke)$^{8}$, with the intensity being proportional to the 
square of the number of pairs of active states, is also possible (see for 
example \cite{Makomber79,Kopvillem82,Andreev80}). As the time for the development of super-radiation 
must be less than the relaxation time $\Gamma^{-1} \sim \tau_{0}$ of the 
dipole-moment, the generation of a super-radiation pulse is not possible 
for the frequencies $\nu$ of phonons below $ \tau_{0}^{-1}$. Then, assuming 
$\tau_{0} \sim 10^{-12}$ s, it is meaningless to discuss this effect for 
phonon-radiation with quasi-momenta $q \sim (10^{-3} \div 10^{-2}) \pi /a$ 
or with frequencies $\nu_{\mathbf{q}} < \tau_{0}^{-1}$.
Thus the maser-effect is the predominant generation mechanism of phonons 
with the relevant region of wave-vectors.}
\end{enumerate}

\chapter{Coordination between of the $\gamma -\alpha$ martensitic~ 
transformation~temperature and the optimum wave generation temperature in 
iron-based alloys }

\section{Definition of the task}

Our previous analysis suggests that, in the growth stage of martensite, the 
existence of lattice-displacement waves with deformation amplitudes of 
$\varepsilon  \sim  10^{-3}$ is a real possibility. Let us remind 
that the evaluation of the fulfilment of the general threshold-conditions 
has been performed for the high-temperature region, where the frequency of 
collisions with short-wave phonons is large. Then there is reason to think 
that this evaluation in order of magnitude is satisfactory not only for pure 
iron but also for the iron based alloys. However, to generalize our 
conclusions and to obtain the optimum conditions of wave generation over the 
full range of concentrations of the alloying elements (for example up to 
10{\%} Mn or up to 34{\%} Ni) and of carbon (up to 1,8 weight{\%}), a 
special investigation is required. 

On the one hand, the existence of a high level of population difference 
$\sigma_{0}$ in accordance with \eqref{3.23}, \eqref{3.24} would require that the 
magnitude of the parameter $y = (\varepsilon  -  \mu )(k_{B}T)^{-1}$ 
must not exceed a few units, as the multiplier 
$\arrowvert \partial f^{0} / \partial y \arrowvert$ rapidly diminishes 
with increasing $y$ (see Tab.\ref{table3.1}), and hence also the value
$\sigma_{0}$. On the other hand, with increasing concentration of Ni the temperature 
($M_{S}$) is decreasing (practically, down to 0 K for 34 {\%} Ni), thus the 
value of $y$ would have to increase. This difficulty cannot be resolved (vide 
infra) by the formal assumption about the necessary reduction of the 
magnitude ($\varepsilon  - \mu $) to ensure practical stability or reduction 
of the parameter $y$.

Moreover, the assumption that the increasing of Ni concentration causes a 
requisite reduction of the difference $\varepsilon  - \mu $, due 
to the filling of the energy band (the model of hard zone) by 
excess-electrons (two electrons per Ni-atom), cannot generally be supported, 
as this model leads to the wrong conclusion about the increasing of $M_{S}$ 
for a Fe-Mn-system, because the difference $\varepsilon - \mu$ must increase 
with increasing Mn concentration. In reality, investigations on binary 
alloys of the Fe-Ni system (up to 28{\%} Ni), as well as of the Fe-Mn system 
( up to 10{\%} Mn) clearly show the decreasing of $M_{S}$ with increasing 
concentration of the alloying element for the common type (the packet-martensite) 
of $\gamma -\alpha$ - transformation, therewith the decrease is more pronounced 
for the Mn-alloys. So, a comparison of data \cite{Kaufman61} and \cite{Bogachev73} 
shows that a reduction of $M_{S}$, down to the level of 500 K, requires 
concentrations of 20{\%} Ni or only 10{\%} Mn. 

Assume that the energies of 3d electronic states, belonging to the 
S-surface, are characterized by a given mean $\varepsilon_{d}$ (see also 
the context to Eq.\eqref{3.9}). Then we can define the following model 
antithetical to the hard-zone model: In alloys of iron with elements of 
close neighbors in the 3d row of the transition metals (substitution 
alloys), the value $\varepsilon_{d} - \mu$ is considered as 
non-depending (or weakly depending) on composition and thus conserving the 
value that is the same as in pure $\gamma$ - iron. This means is that the 
alloying element builds up own 3d sub-bands, which overlap with the 3d - bands 
of iron without a charge exchange flow among the components of alloy. 
(Hypothesis of minimum polarity $\lbrack 100 \rbrack$). Furthermore, we 
will assume that the magnitude of the threshold difference $\sigma_{t}$ 
retains a value not exceeding $10^{-3}$ (as estimated for the region 
of high temperature), in alloys undergoing a martensitic transformation.
Then, fulfilment of the threshold condition $\sigma_{0} > \sigma_{t}$ 
can be justified within of the two-band model, for a wide region of 
concentrations $C_{ae}$ of alloying elements, using the assumption that the 
magnitude of attenuation $\Gamma_{s}$ of mobile s electrons is comparable 
with the value $(\varepsilon_{d} - \mu )\hbar^{-1}$ and 
many times larger than the attenuation $\Gamma_{d}$ of the 3d - electrons, 
which are active during the generation of phonons (the typical lifetime of s - 
electrons $\tau_{s}  \sim  10^{-15}$ s in 3d metals corresponds 
to the value $\hbar\:\Gamma_{s} \sim  0,6 $ eV). The population of states 
with energies $\varepsilon_{d} > \mu $ can be kept up $f^{0}_{d}  \ge 0,1 $ 
by an effective scattering mechanism into d - states of s - electrons with energies 
$\varepsilon_{s}$ that are corresponding to the condition

\begin{equation}
\varepsilon_{d}  -  \mu  -  \frac{\hbar \:\Gamma_{s}}{2}  <  \varepsilon_{s} - \mu
+ \hbar \:\omega  <  \varepsilon_{d} - \mu + \frac{\hbar \:\Gamma_{s}}{2}.
\label{4.1}
\end{equation}
In \eqref{4.1} $\hbar\:\omega$ is the energy of a short-wave phonon, whose 
participation is required for satisfaction of momentum conservation law, if 
the uncertainty of the quasimomentums of s - electrons is small (in \eqref{4.1}, 
the value $ \hbar\:\Gamma_{d}$ is disregarded compared to $ \hbar\Gamma_{s}$).
Thus, for large $\Gamma_{s}$, the magnitude of thermal excitation 
$k_{B} T  \sim (\varepsilon_{s} - \mu )$ can be much smaller 
than $\varepsilon_{d} - \mu$, without reduction of the d-state 
occupation. In a similar way, the mechanism of d - s scattering ensures that 
the states with the energy $\varepsilon_{d}$, located below the 
Fermi-level, despite of an inequality $\mu - \varepsilon_{d}  >  k_{B} T$ 
have the occupations $f^{0}_{d} < 1$ significantly deviating 
from unit at the cost of an additional (nonthermal) broadening of the 
d - electron distribution function.

From the above analysis it is clear that the increase of concentration $C$ of 
the alloying supplement ($C < 1/2$), which causes a rapid 
increase of the attenuation-part $\Gamma_{s}(C)  \sim  C(1 - C)$ 
of total attenuation, linked with the scattering on substitution atoms,  
$\Gamma_{s}$ being combined with a temperature-reduction, by the following 
condition:

\begin{equation}
\hbar\:\omega  + k_{B}\:T + \frac{\hbar \:\Gamma(C,T)}{2} \sim  \arrowvert 
\varepsilon_{d}  -   \mu \arrowvert. 
\label{4.2}
\end{equation}
It has to be noted that the temperature-dependent contribution $\Gamma 
_{s}(T)$ connected with the scattering on thermally activated 
heterogeneities (vacancies, phonons, magnons etc.) is decreasing with 
decreasing T and must smooth down the behavior of $\Gamma_{s}(C,T)$, by 
stabilizing the broadening of the d-electron distribution function, within 
a wide range of T and C. Logically, at first we must obtain a modification 
of the equilibrium distribution function $f^{0} \rightarrow \tilde{f}^{0}$ 
in such a way that $\tilde{f}^{0}$ must simultaneously consider the distribution 
function broadening determined by the factors T and $\Gamma_{S}$. After that it 
is necessary to analyze the behavior of the derivatives $\partial
\tilde{f}^{0} / \partial \mu $, $\partial \tilde{f}^{0} / \partial T$ 
which determine the magnitude of the population difference $\sigma 
_{0}$, together with the gradients $\nabla \mu $ and $\nabla T$. Hence 
we have to find such region of the variables ( T and $\Gamma$) in which 
(for a fixed value of $\arrowvert \varepsilon_{d}  - \mu \arrowvert$) 
the value $\sigma_{0}$ attains a maximum that only slightly varies 
with simultaneous variations of T and $\Gamma$. The temperature belonging 
to this optimum region might then be defined as the optimum temperature
$\tilde{T}$ for the martensitic transformation process. If the electronic 
configurations of the atoms of lattice matrix and of the alloying element are known, 
it will be possible to determine the dependence $\tilde{T}(C)$, by calculating 
$\Gamma_{s}(C)$ and identifying the correlation between $\tilde{T}$ and 
$\Gamma(C,\tilde{T})$. Then, the concentration-dependencies of $\tilde{T}(C)$ 
and $M_{S}(C)$ will be compared. It is also possible to deal with the inverted task: 
\begin{itemize}
\item[i]{ we require a close relation between the dependencies of $\tilde{T}(C)$ 
and $M_{S}(C)$;}
\item[ii]{ we choose the electron configuration of one of the alloying
components;}
\item[iii]{ we find the electron configuration of the atoms of the other component. 
It is the latter definition of task, allowing us to gather the additional 
information which will be used in the fourth chapter, during a comparison of the 
interdependence between $\tilde{T}(C)$ and $M_{S}(C)$, in the substitution 
alloys (Fe-Ni, Fe-Co, Fe-Mn) and in the interstitial alloys (Fe - C).}
\end{itemize}


\section{Modified electronic distributions and their derivatives in the case 
of a rectangular spectral density function}

A statistically disordered lattice of the binary substitutional type can be 
considered as the medium comprising an ideal (periodic) arrangement of 
lattice knots, for which the probability of occupation of a knot (lattice 
point) with atoms of either one of its two components is equal to $C$ and 
$1 - C$, respectively, where $C$ - concentration of the dissolved 
component. The missing of a long-range order in such lattice, being caused by the 
stochastic nature of its single-knot-potentials, causes the states with 
predetermined quasi-momenta to become non-stationary. Therefore - within the 
general notion of some kind of periodic "effective" medium - one of the 
approaches for the description of substitutional alloys is characterized by 
the introduction of an effective Non-Hermitian Hamiltonian with complex 
eigenvalues, the imaginary part of which determines the extinction rate of 
electronic states $\lbrack 100 \rbrack$. The calculus of mean-values of electronic 
operators comprises both thermodynamic and configuration averaging, which 
have to be performed independently from each other for systems without a 
shot-range order. As an example, the modified electronic distribution 
$\tilde{f}^{0}$ of our interest is given by the expression 
\begin{equation}
\tilde{f}_{\textbf{k}}^{0} = \int\limits_{- \infty}^{\infty} [ 1+\exp{( 
\frac{\varepsilon - \mu}{k_{B} T})}]^{-1}
A(\varepsilon, \,\textbf{k}) d\varepsilon,
\label{4.3}
\end{equation}
where the spectral density $ A(\varepsilon, \textbf{k})$  is a 
configurationally averaged probability in a given alloy for the existence of 
an electron of energy $\varepsilon $ in a state with quasi-momentum 
$\hbar\textbf{k}$. Usually, the spectral density obeys the Lorentz (Breit-Wigner) 
formula:
\begin{equation}
\label{4.4}
A(\varepsilon ,\textbf{k}) = \frac{1}{2\,\pi} \,\hbar \,\Gamma
_{\textbf{k}}\left[(\varepsilon - \varepsilon_{\textbf{k}})^{2} +
\left(\frac{1}{2}\,\hbar \,\Gamma_{\textbf{k}} \right) ^2\right]^{-1},
\end{equation}
where $\Gamma_{\textbf{k}}$ - extinction of an electron in a state with energy 
$\varepsilon_{\textbf{k}}$, where the extinction corresponds to the width at half 
of the height of the function $A(\varepsilon, \textbf{k})$. Furthermore, 
the normalization condition is satisfied:
\begin{equation}
\label{4.5}
\int\limits_{-\infty}^{\infty}{A(\varepsilon,\textbf{k})} 
d\varepsilon = 1.
\end{equation}
Obviously from (\ref{4.4}), (\ref{4.5}), the spectral density (SD) function gradually 
transforms into the $\delta$ - function for $\Gamma_{\textbf{k}} \to  0 $
\begin{equation}
\label{4.6}
\lim_{\Gamma_\textbf{k} \to 0} A\left( \varepsilon 
,\textbf{k}\right) = \delta \left( \varepsilon - \varepsilon_{\textbf{k}} 
 \right),
\end{equation}
and the distribution (\ref{4.3}) into the Fermi-Dirac distribution:
\begin{equation}
\label{4.7}
\lim_{\Gamma_{\textbf{k}} \to 0} \tilde{f}^{0} = \tilde{f}_{k}^{0} = 
\left[ 1 + \exp \left( \frac{\varepsilon_{\textbf{k}} - \mu }{k_{B} T} 
\right) \right]^{-1}.
\end{equation}
This is quite reasonable, as the condition $\Gamma _{\textbf{k}} = 0$ would 
implicitly require a re-establishment of long-range ordering, as occurring 
during gradual transition of a binary alloy into a single-component system 
with $C  \to  0$.

A simple analytical expression for $\tilde{f}_{\textbf{k}}^{0}$ can easily be 
found by choosing a rectangularly shaped spectral density function with a 
height of ${\hbar \Gamma_{\textbf{k}}}^{-1}$ and a width of $\Gamma_{\textbf{k}}$ , 
being symmetrically arranged with respect to the energy $\varepsilon 
_{\textbf{k}}$:

\begin{equation}
\label{4.8}
A \left( \varepsilon ,\textbf{k} \right) = \frac{1}{\hbar \Gamma 
_{\textbf{k}}}\left\{ \Theta \left[\varepsilon - \left( \varepsilon 
_{\textbf{k}} - \frac{\hbar \Gamma_{\textbf{k}}}{2} \right) \right] - \Theta 
\left[\varepsilon - \left(\varepsilon_{\textbf{k}} + \frac{\hbar \Gamma 
_{\textbf{k}}}{2} \right) \right] \right\},
\end{equation}
where $\Theta$ - Heaviside unit-function
\begin{equation}
\label{4.9}
\Theta (\varepsilon) = \left\{ \begin{array}{ll}
 1 & \quad  \textrm{$\varepsilon \ge 0$},\\
 0 & \quad   \textrm{$\varepsilon < 0$}.
  \end{array} \right.
\end{equation}

\begin{figure}[htb]
\centering
\includegraphics[clip=true, width=.6\textwidth]{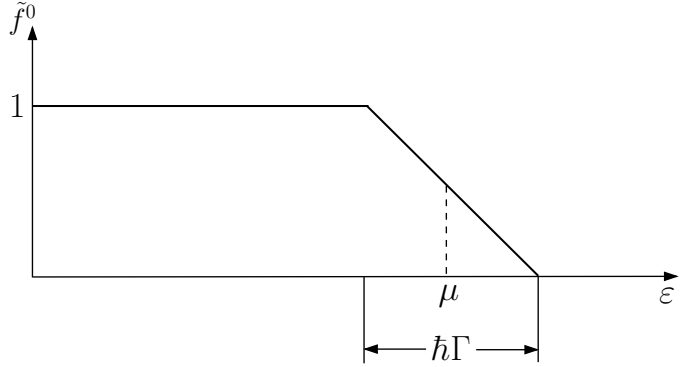}
\renewcommand{\captionlabeldelim}{.}
\caption{Modified SD - function at a temperature of 0 K.}
\label{fig4.1}
\end{figure}

After insertion of (\ref{4.8}) in (\ref{4.3}), we get:
\begin{equation}
\label{4.10}
\tilde{f}_{\textbf{k}}^{0} = 1 + \frac{1}{2\eta_{\textbf{k}}}\ln{\left[\frac{1 
+ \exp{\left( {y_{\textbf{k}} - \eta_{\textbf{k}}} \right)}}{1 + \exp{\left( 
{y_{\textbf{k}} + \eta_{\textbf{k}}} \right)}} \right]},
\end{equation}
where $y_{\textbf{k}} = (\varepsilon_{\textbf{k}} - \mu )(k_{B} T)^{-1}$, 
$\eta_{\textbf{k}} = \hbar \,\Gamma_{\textbf{k}}(2\,k_{B} T)^{-1}$. 
It can easily be verified that the limit condition (\ref{4.7}) is satisfied for 
(\ref{4.10}), so that we get for the low-temperature case $T \to 0$ 
\begin{equation}
\label{4.11}
\lim_{T \to 0} \tilde{f}_{\textbf{k}}^{0} = \left\{ \begin{array}{ll}
 1 &  \quad \varepsilon_{\textbf{k}} < \mu - \, (\hbar\,\Gamma_{\textbf{k}}) / 2
     \\ \\
 1 / 2 + (\mu - \varepsilon_{\textbf{k}})(\hbar\,
 \Gamma_{\textbf{k}})^{-1} &  \quad \mu - (\hbar\,\Gamma_{\textbf{k}}) / 2 \le
 \varepsilon_{\textbf{k}} \le \mu + (\hbar\,\Gamma_{\textbf{k}}) / 2 \\ \\
 0 &  \quad  \mu - (\hbar\,\Gamma_{\textbf{k}}) / 2 < \varepsilon_{\textbf{k}}
  \end{array} \right.
\end{equation}
a distribution with a scattering width $\hbar\,\Gamma_{\textbf{k}}$, as shown in 
\ref{fig4.1}. Moreover, the normalization condition for an electronic 
distribution $\tilde{f}_{\textbf{k}}^{0} = 1 / 2$ at $\varepsilon_{\textbf{k}} = 
\mu$ is satisfied, independent of T. Disregarding the dependency on 
\textbf{k} of the extinction $\Gamma$ within a coherent potential 
approximation $\lbrack 100 \rbrack$, we note that $\tilde{f}_{\textbf{k}}^{0}$ remains constant 
along isoenergetic areas with $\varepsilon_{\textbf{k}} = const$. in 
quasi-momentum space, as $\tilde{f}_{\textbf{k}}^{0}
=\tilde{f}_{\textbf{k}}^{0} 
(\varepsilon_{\textbf{k}})$. Accordingly, we shall design our 
spectral function by $A(\varepsilon, \textbf{k}) = A (\varepsilon, 
\varepsilon_{\textbf{k}})$.

Taking for granted that (\ref{4.3}) represents the energetic distribution function 
$\tilde{f}^{0}\left( \varepsilon_{\textbf{k} \,s}  \right) \equiv
\tilde{f}_{\textbf{k}\,s}^{0}$ of 
the s-electrons in an alloy, we can now show that, for a given 
d-s-scattering mechanism with involvement of short-wave phonons, the 
equilibrium distribution: 
$\tilde{f}_{\textbf{k} d}^{0} = \tilde{f}^{0}\left( \varepsilon_{\textbf{k}\,d} 
 \right)$ of d-electrons of energy $\varepsilon_{d}$ in an alloy of the 
same energy $\varepsilon_{s} = \varepsilon_{d}$ is nearly equal to 
the function $\tilde{f}^{0} \left(\varepsilon_s \right)$. According to 
\cite{Zyrianov76}, the modification of the normalized integral of electron-phonon 
collisions, being required after consideration of scattering at the 
substitutional atoms, can be reduced to a replacement of the 
$\delta$ - function, thus reflecting strict compliance with the conservation of 
energy requirement, by the Lorentz-function:

\begin{eqnarray}
\label{4.12}   
\left({ \frac{\;\;\partial \tilde{f}_{\textbf{k} d}}{\partial\,t}
} \right)_{col}  = \,\frac{2 \,\pi }{\hbar}
 \sum_{\textbf{k}^\prime \textbf{p}}\arrowvert W_{\textbf{p}}\arrowvert^{2}
 \{A(\varepsilon_{\textbf{k}^\prime s},\varepsilon 
_{ \textbf{k} d} + \hbar \,\omega_{\textbf{p}}) \lbrack
N_{\textbf{p}}\, (\tilde{f}_{\textbf{k}^{\prime} s} -
 \tilde{f}_{\textbf{k} d}) + 
                                              \nonumber      \\
 +\tilde{f}_{\textbf{k}^\prime s} (1 - \tilde{f}_{\textbf{k}
 d})\rbrack                                
+ A (\varepsilon_{\textbf{k}^\prime s},\varepsilon_{ \textbf{k} d} -
     \hbar \omega_{\textbf{p}})\lbrack N_{\textbf{p}}\,
     (\tilde{f}_{\textbf{k}^{\prime} s} - \tilde{f}_{\textbf{k} d})                        
 - \tilde{f}_{\textbf{k} d} (1 -
 \tilde{f}_{\textbf{k}^{\prime} s})\rbrack \}
\end{eqnarray}
where $\arrowvert W_{\textbf{p}}\arrowvert$ - modulus of the matrix-element of 
electron-phonon-interaction, $N_{\textbf{p}}$ - phonon-distribution function. 
Proceeding further with the integration of the $\varepsilon_{\textbf{k}^{\,\prime} s}$, under 
consideration of the definitions (\ref{eq1}) and (\ref{eq3}), and in order to find the 
equilibrium-distribution $\tilde{f}_{\textbf{k} d}^{0}$, we require that the 
collision-integral (\ref{4.12}) must vanish. Then we get approximately

\begin{equation}
\label{4.13}
\tilde{f}_{\textbf{k} d}^{0} \, \approx \,\frac{\tilde{f}_s^{0}\,(\varepsilon 
_{ \textbf{k} d} + \hbar \,\omega_{\textbf{p}}) + \lbrack \,\tilde{f}_s^{0} 
\,(\varepsilon_{ \textbf{k} d} + \hbar \,\omega_{\textbf{p}}) +
\tilde{f}_{s}^{0} 
\,(\varepsilon_{ \textbf{k} d} - \hbar\, \omega_{\textbf{p}}) \,\rbrack \,N_{\textbf{p}} 
}{2\, N_{\textbf{p}} + 1 + \lbrack \,\tilde{f}_s^{0} \,( \varepsilon_{ \textbf{k} d} 
+ \hbar \,\omega_{\textbf{p}}) - \tilde{f}_s^{0} \,(\varepsilon_{ \textbf{k} d} - 
\hbar \,\omega_{\textbf{p}})\, \rbrack}.
\end{equation}
If we develop (\ref{4.13}) by powers of $\omega^{\,\prime} = \hbar \,\omega_{\textbf{p}}
(\varepsilon_{ \textbf{k} d} -  \mu)^{-1}$, then we get in a zero-order approximation

\begin{equation}
\label{4.14}
\tilde{f}_{\textbf{k} d}^{0} \approx \tilde{f}_{s}^{0} (\varepsilon_d).
\end{equation}
By consideration of (\ref{4.10}), the linear supplement in the right hand side of 
(\ref{4.14})

\begin{displaymath}
\frac{\partial \tilde{f}_{s}^{0} ( \varepsilon_{d})}{\partial 
(\varepsilon_{d} - \mu )}(1 - 2\,\tilde{f}_{s}^{0}( \varepsilon_d))\hbar\,
\omega_{\textbf{p}} 
\end{displaymath}
can be written in the following form
\begin{equation}
\label{4.15}
\frac{1}{2 \Gamma^{\,\prime}} \frac{\sinh{ \Gamma^{\,\prime}( T^{\,\prime})^{-1}}}
{\cosh{\Gamma^{\,\prime}(T^{\,\prime})^{-1}} + \,\cosh{(T^{\,\prime})^{-1}}}
(1 - 2\,\tilde{f}_{s}^{0})\,\omega^{\prime},
\end{equation}
where
\begin{equation}
\label{4.16}
\Gamma^{\prime} = \frac{\hbar\, \Gamma}{2\,(\varepsilon_{\textbf{k} d} - \mu )},
\quad T^{\prime} = \frac{k_{B} T}{\varepsilon_{\textbf{k} d} - \mu}.
\end{equation}
Let $\varepsilon_{\textbf{k} d} -   \mu   \approx  0,2 $eV, and $\omega 
_{\textbf{p}}$ be of an order of magnitude of half of the Debye-frequency, then we 
get $\omega^{\prime} \le  10^{-1}$. An order of magnitude of about 0,1 also 
applies to the product of the other factors in (\ref{4.15}), so that the 
corrective factor, being linear in $\omega^{\prime} $, becomes at least one order of 
magnitude smaller than in (\ref{4.14}). Therefore, in an analysis of the 
non-equilibrium addenda to the function $\tilde{f}_{d}$

\begin{equation}
\label{4.17}
\Delta\tilde{f}_{\textbf{k} d} (\nabla \mu ) \approx 
\frac{\partial \tilde{f}_{\textbf{k} d}^{0}}{\partial \mu^{\prime} 
}\frac{\tau \,\textbf{v}_{\textbf{k}}\,\vec{\nabla}\mu}{\varepsilon_{d} - \mu 
},\quad \mu^{\prime} = \frac{\mu}{\varepsilon_{d} - \mu},
\end{equation}
\begin{displaymath} 
\Delta \tilde{f}_{\textbf{k} d} ( \nabla T ) \approx 
\frac{\partial \tilde{f}_{\textbf{k} d}^{0}}{\partial T^{\,\prime}}\frac{\tau \,
\textbf{v}_{\textbf{k}}\,\vec{\nabla} T \,k_{B}}{\varepsilon_{d} - \mu }, 
\end{displaymath}
it is possible to replace the function $\tilde{f}_{d}^{0} $ by
$\tilde{f}_{s}^{0}$ 
(\ref{4.3}) for $\varepsilon_{\textbf{k}} = \varepsilon_{\textbf{k} d}$. Under the 
presumption that this replacement has been performed all over, we can omit 
in our subsequent analysis the indices d, zero and tilde in our notation of 
the function $\tilde{f}_{d}^{0}  \to  f$. After having fixed in (\ref{4.17}) the 
values of $\tau  \textbf{v} \vec{\nabla}\mu$, $\tau \textbf{v} 
\vec {\nabla}T$, we only need to analyze the derivatives $\partial 
f / \partial T^{\,\prime},\quad  \partial f / \partial \mu^{\prime}$ for 
our envisaged determination of the optimum values of $\Gamma$ and T (see 
Pt. 4.1), as a function of the dimensionless variables $\Gamma^{\prime} $ and
$T^{\,\prime}$ (\ref{4.16}). It is convenient to use these variables, as the value of 
$\varepsilon_{d} -  \mu$ will remain unchanged, according to our 
assumption in Pt.4.1. In the case of the spectral density (\ref{4.8}) 
we can obtain from (\ref{4.10}) the derivatives $\partial f / \partial
T^{\,\prime},\quad  \partial f / \partial \mu^{\prime}$ by direct differentiation:

\begin{equation}
\frac{\partial f}{\partial
T^{\,\prime}}   =  \frac{1}{2\,\Gamma^{\prime}} \left\{ \ln{ \Big\lbrack \frac{1 + \exp{(y -
\eta)}}{1 + \exp{(y - \eta)}} \Big\rbrack}  +  
\frac{y\sinh{\eta} + \eta(\cosh{\eta} + \exp{y})}{\cosh{y} + \cosh{\eta}}
\right\},
\label{4.18}          
\end{equation}

\begin{equation}
\label{4.19}
\frac{\partial f}{\partial \mu^{\prime}} = \frac{1}{2 \Gamma^{\prime}} 
\frac{\sinh{\eta} }{\cosh{y} + \cosh{\eta}},
\end{equation}
where $\eta = \Gamma^{\prime}\,T^{\,\prime \;-1},\quad y = T^
{\,\prime\, -1}$. 

\begin{figure}[htb]
\centering
\includegraphics[clip=true, width=.6\textwidth]{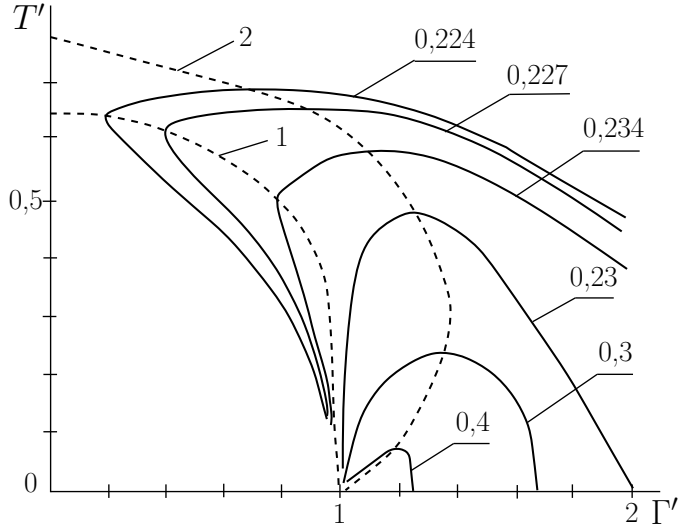}
\renewcommand{\captionlabeldelim}{.}
\caption{Calculated values of the derivative $\partial f / \partial
\mu^{\prime}(\Gamma^{\prime}, T^{\,\prime})$, for the case of a rectangular
SD- function: ------ - iso-lines with constant values of the 
$\partial f / \partial \mu^{\prime}$ - function (Lines 1 and 2 are defined in the
text).}
\label{fig4.2}
\end{figure}

\begin{figure}[htb]
\centering
\includegraphics[clip=true, width=.8\textwidth]{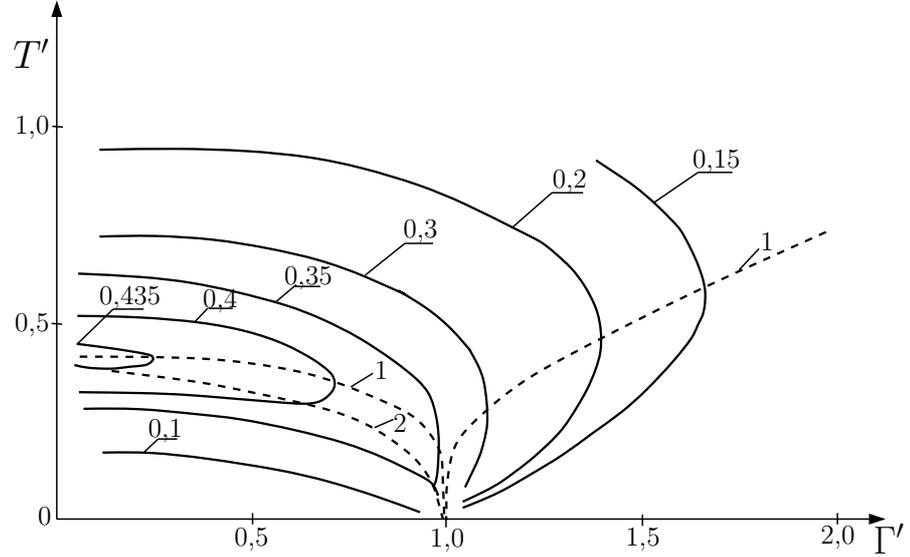}
\renewcommand{\captionlabeldelim}{.}
\caption{Calculated values of the derivative $\partial f
(\Gamma^{\,\prime}, T^{\,\prime}) / \partial T^{\,\prime}$, for the case of a 
rectangular SD-function:-------- - iso-lines.}
\label{fig4.3}
\end{figure}

The results of our calculus of the derivatives $\partial f / \partial \mu^{\prime}$ 
and $\partial f / \partial T^{\,\prime}$ on the basis of 
(\ref{4.18}) and (\ref{4.19}) are presented in Figs. \ref{fig4.2} and  
\ref{fig4.3}. In these figures, the group of thin lines represents the 
constant values of the functions $\partial f / \partial \mu^{\prime}$, 
$\partial f / \partial T^{\,\prime}$ (being labeled at each line), while the dashed 
lines 1 and 2 are defined by the conditions 

\begin{equation}
\label{4.20}
\frac{\partial }{\partial T^{\,\prime}} \left( \frac{\partial 
f}{\partial \mu^{\prime}} \right) \Big\arrowvert_{\Gamma^{\prime }} = 0 , \quad 
\frac{\partial}{\partial \Gamma^{\prime} } \left( \frac{\partial f}{\partial \mu^{
\prime}} \right) 
\Big\arrowvert_{T^{\,\prime}} = 0.
\end{equation}

\begin{figure}[htb]
\centering
\includegraphics[clip=true, width=.8\textwidth]{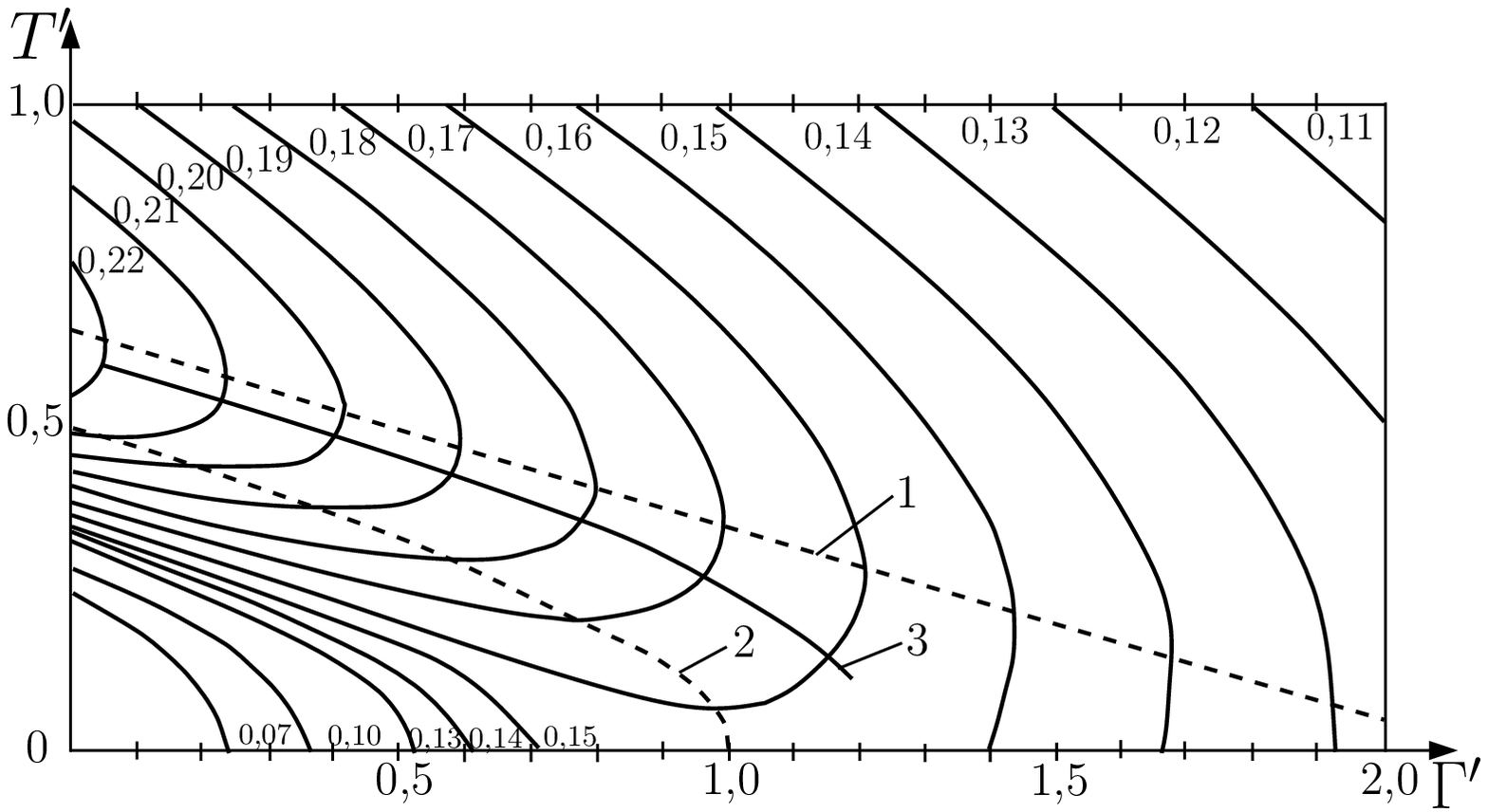}
\renewcommand{\captionlabeldelim}{.}
\caption{Calculated values of the derivative $\partial 
f(\Gamma^{\,\prime},T^{\,\prime}) / \partial \mu^{\prime}$, for the case 
of a Lorentz-shaped SD- function:--------  - iso-lines.}
\label{fig4.4}
\end{figure}

\begin{figure}[htb]
\centering
\includegraphics[clip=true, width=.8\textwidth]{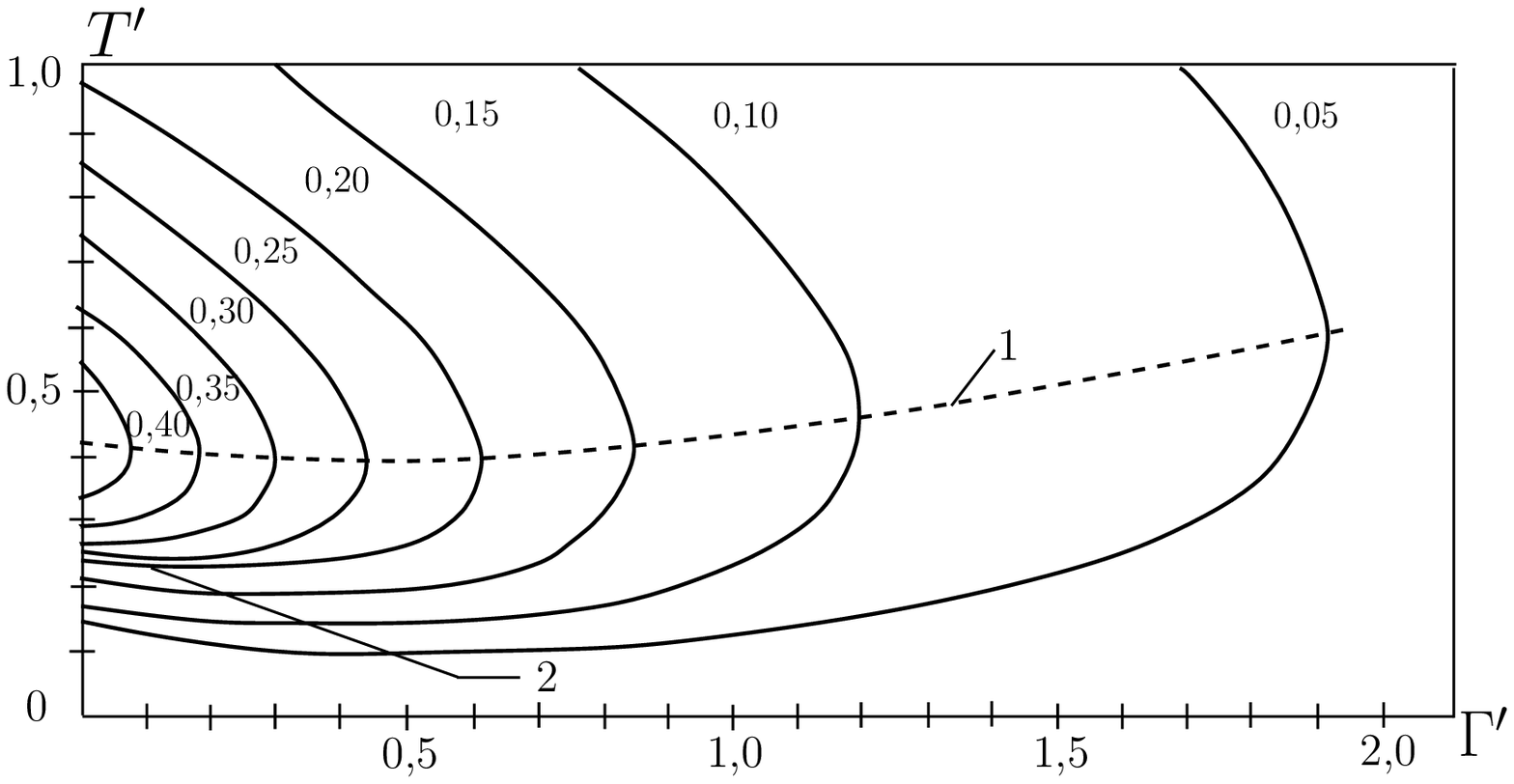}
\renewcommand{\captionlabeldelim}{.}
\caption{Calculated values of the derivative $\partial f(\Gamma^{\,\prime},
T^{\,\prime}) / \partial T^{\,\prime}$ for the case of a Lorentz-shaped 
SD-function : --------  - iso-lines.}
\label{fig4.5}
\end{figure}

These lines correspond to the maximum values of the derivatives  $\partial f
/ \partial \mu^{\prime}$, $ \partial f/ \partial T^{\, \prime}$ of the
variables $T^{\,\prime}$ and $\Gamma^{\,\prime}$ and pass through those
points where the parallels to the vertical and horizontal coordinate axes
touch the  iso-lines. The area enclosed by the lines in Fig. \ref{fig4.2}
represents a  region of values of the parameters T and $\Gamma $ for which
the  occupational inversion $\sigma_{0}$ ($\nabla  \mu $ ) attains maximum 
values, only slowly varying with T and $\Gamma$. Obviously, it is
justified  to define an optimum temperature for phonon-generation, which
decreases with  increasing extinction $\Gamma$, by highlighting within this
region the  point corresponding to pure iron. Thus the behavior of the
derivative  $\partial f / \partial \mu^{\prime}$ suggests the existence of
an optimum temperature  $\tilde{T}_{1} $, the derivative of which obeying to
the condition $\partial \tilde{T}_{1}  / \partial C < 0$, where $C = C_{ae}$
is the concentration of the  alloying element. The jump of the value of
$\partial f / \partial \mu^{\,\prime}$ from 0 up to 0,5, during the
transition from $\Gamma^{\,\prime} < 1$ to  $\Gamma^{\prime} > 1$ (at the
point $\Gamma^{\prime} = 1$ we have $\partial  f / \partial \mu^{\prime} =
0,25$) and for $T^{\,\prime} \to  0$, is obvious from Eq.  (\ref{4.19}),
which attains the following form for $T^{\,\prime} \to  0$:

\begin{displaymath}
\frac{\partial f}{\partial \mu^{\,\prime}} \approx \frac{1}
{2\Gamma^{\,\prime}\lbrack 1 + \exp{(1 - \Gamma^{\,\prime})}(T^{\,\prime})^{-1}\rbrack}.
\end{displaymath}
This jump is due to the extremely rapid decay of the modeled spectral 
density function (\ref{4.8}) within a variation of the energy $\varepsilon
_{\textbf{k}}$ by $\pm \,1/2 \Gamma _{\textbf{k}}$, and there is no reason for 
expecting such a jump for a smoother behavior of $A (\varepsilon , 
\varepsilon _{\textbf{k}})$.

A comparison of Figs. \ref{fig4.3} with \ref{fig4.2} obviously shows 
that the crucial difference among the behavior of  $\partial f / \partial T^{
\,\prime}$ vs. that of  $ \partial f / \partial \mu^{\prime} $ 
is due to the existence of an ascending section of line 1 for $\Gamma ^{\prime} > 
1$, which could serve as the basis for the definition of an optimum phonon 
generation temperature $\tilde{T}_{2}$, which would increase with increasing 
value of $C_{ae}$ : $\partial \tilde{T}_{2}  / \partial C_{ae}  > 
0$. As the abrupt "slump" at $\Gamma^{\,\prime} \to  1$, $T^{\,\prime}  \to  0$, along line 1 
(at point $\Gamma^{\,\prime} = 1$, $T^{\,\prime}  = 0$, $\partial f / \partial
T^{\,\prime} = 1 / 2 \,\ln 2 \approx 0,347$) is caused by the already mentioned 
particularity of the spectral density function (\ref{4.8}). It can be anticipated 
that this "slump" will vanish when getting over to the more smoothly 
changing functions $A(\varepsilon ,\varepsilon_{\textbf{k}})$, and that the 
ascending section of line 1 will broaden out. We also note that the quantity 
$\partial f / \partial T^{\,\prime}$ diminishes more rapidly with 
increasing $\Gamma^{\,\prime} $ than $\partial f / \partial \mu^{\,\prime}$. A 
convincing explanation for this behavior can immediately be concluded from 
(\ref{4.3}), if one considers that derivations on $T$ and $\mu$ lead accordingly to 
factors being odd and even on $\varepsilon$ concerning the point $\mu$ in 
the integrand.

If we summarize the results of our calculus with usage of a rectangularly 
shaped spectral density, it can be concluded that they provide strong 
evidence for the existence of optimum conditions for phonon generation, 
within a wide concentration region of the alloying element, a possibility we 
already tentatively anticipated in Pt. 4.1, on the basis of qualitative 
thoughts. However, a more detailed treatise of the task of Pt. 4.1 could 
reasonably be performed by determination of the derivatives $\partial f / 
\partial T^{\,\prime}$ and  $\partial f / \partial \mu^{\prime}$ on the 
basis of the Lorentz-expression of spectral density, as will be done in the 
next chapter.


\section{Region of $T^{\,\prime}$, $\Gamma^{\,\prime} $ values with optimum conditions for 
phonon-generation in the case of a Lorentz-shaped SD-function}

An explicit determination of $\partial f / \partial \mu^{\,\prime}$, 
$\partial f / \partial T^{\,\prime}$ on the basis of \eqref{4.3} proves to be 
difficult in the case of the Lorentz-form \eqref{4.4} of the function 
$A(\varepsilon, \varepsilon_{\textbf{k}})$. That's why a numeric calculus of 
$T^{\,\prime}$ has been performed, the results of which are presented in 
Figs. \ref{fig4.4},\ref{fig4.5}. Let us now compare in pairs Fig.\ref{4.4} with 
Fig. \ref{4.2} and \ref{fig4.5} with \ref{fig4.3}. A 
comparison of the first pair of results shows that, during a transition to a 
smoothly changing spectral density \eqref{4.4}, the optimum region of parameters 
$T^{\,\prime}$, $\Gamma^{\,\prime} $ is being retained, even though lines 1 and 2 
of the maximum value of $\partial f(T^{\,\prime},\Gamma^{\,\prime})/ \partial 
\mu^{\,\prime}$ exchange their relative positions (in Fig. \ref{fig4.2}, line 2 
is located above line 1 and in Fig. \ref{fig4.4} below line 1). In addition, 
in the region between lines 1 and 2, the values $\partial f / \partial \mu^{\,\prime}$ 
of the Lorentz-form $A(\varepsilon , \varepsilon _{\textbf{k}})$ slowly decrease with 
increasing $\Gamma^{\,\prime} $ and with decreasing $T^{\,\prime}$, while the function 
$\partial f / \partial \mu^{\,\prime}$ in Fig. \ref{fig4.2} behaves 
non-monotonously, especially in the vicinity of the point $T^{\,\prime} \to 0$, 
$\Gamma^{\,\prime} \to 1$. The solid line 3 in Fig. \ref{fig4.4} is the 
projection of the "crest" of the profile of function $\partial f / \partial 
\mu^{ \prime}$ on plane ($\Gamma^{\,\prime}$, $T^{\,\prime}$). The function 
$T^{ \,\prime}$ ( $\Gamma^{\,\prime} $ ) with lines 1, 2, 3 describes the reduction of 
$T^{ \,\prime}$ with increasing  $\Gamma^{\,\prime} $ , in a similar way as relation 
\eqref{4.2}. This means that they can be used as a basis for determination 
of the temperature $\tilde{T}_{1}$. 

The characteristic difference between Figs. \ref{fig4.5} and \ref{fig4.3} consists 
in the missing "slump" of line 1, which - for $0 <  \Gamma^{\,\prime} \le  0,8$
 - slightly changes in the proximity of $T^{\, \prime} \approx  0,4$, 
attaining its minimum at $T^{\, \prime} \approx  0,39$,  $\Gamma^{\, \prime}  \approx 0,42$, 
being approximated by a straight line for  $\Gamma^{\,\prime}  > 0,8$ 

\begin{equation}
T^{ \,\prime}  \approx  0,25 + 0,175  \Gamma^{\,\prime}. 
\label{4.21}
\end{equation}

Of course, also a parabolic approximation might be possible. Obviously, by 
usage of the function $T^{\,\prime}$ ($\Gamma^{\,\prime}$), corresponding to
\eqref{4.21}, it would be feasible to determine the temperature
$\tilde{T}_{2}$, 
which increases with increasing  $\Gamma^{\,\prime}$. Line 2 in
Fig. \ref{fig4.5} is shifted towards the lower $T^{\,\prime}$, 
$\Gamma^{\,\prime}$ region, both in relation to Fig. \ref{fig4.3} and to 
Fig. \ref{fig4.4} and could thus in principle be used for the introduction 
of temperature $\tilde{T}_{1}$, provided the starting temperature of the 
martensitic transformation of the main alloying component is close to it.

Let us surmise that the function $\tilde{T}_{1,2} $ ($\Gamma^{\,\prime}$)
- being the mapping of the function $\tilde{T}_{1,2}(C)$ onto the plane 
($\Gamma^{\,\prime}$, $T^{\,\prime}$) - is known. Then, in order to determine 
the explicit expressions for $\tilde{T}_{1,2}(C)$, we must first discriminate 
those fractions of total electron extinction $\Gamma$ being dependent on the 
concentration of the alloying element:

\begin{equation}
\Gamma(T, C) = \Gamma(T) + \Gamma(C). 
\label{4.22}
\end{equation}

In paramagnetic alloys, at temperatures larger than or close to the 
Debye-temperature $T_{D}$, the extinction $\Gamma(T)$ is mainly caused by 
s-electron scattering at phonons and magnetic moment fluctuations. The first 
of these processes leads to a contribution $\Gamma(T)$ exhibiting linear 
variation with temperature, while the second one only weakly depends on T - 
for temperatures above the Curie-temperature $T_{c}$ (or the 
Neel-temperature $T_{N}$ respectively) - and rapidly diminishes for $T < 
T_{c}$. Measurements of the specific electric resistance $\rho$ \cite{Weiss59} (see 
also \S 3 of Ch. 25 in \cite{Vonsovskii71}) have shown that the contribution to scattering 
by magnetic inhomogeneities for $T > 500$ K exceeds phonon scattering by about 
the same order of magnitude. In the region of temperature $T < 1100$ K, being 
relevant for martensitic transformations, $\rho$ decreases monotonously 
with decreasing T. Even though the function $\rho(T)$ is non-linear, we 
shall use below, for simplification, the linear approximation 
$\hbar \,\Gamma(T) = a_{0}\, k_{B} T$, where $a_{0}$ - dimensionless parameter. 
Obviously, we can already anticipate that $a_{0}$ must be a one-digit number 
greater than 1, as the contribution of pure phonon-scattering already amounts 
$a_{0}  \approx  1$ in the case of transition-metals. In principle, it is 
also possible to include in \eqref{4.22} the part being largely independent 
of T and C, being caused by scattering on static defects (i.e. dislocations, 
alloying atoms of a third element etc.), but we shall assume that the growth 
of martensite crystals will proceed within a region being practically devoid 
of such defects, thus confining ourselves to the contributions of only two 
terms in \eqref{4.22}. The usage of the ratio \eqref{4.22} takes into account 
that phonon scattering leads to a more scattered energy distribution function, 
essentially resulting in the need for replacement of the $\delta$ - function 
by a Lorentz-line of finite width \cite{Auslender75}. In this context we should remind 
that during any consideration of the washout of the Fermi- distribution, the 
equivalence of different scattering mechanisms will always be confirmed, 
which inevitably leads to the question about the justification of a well 
defined Fermi-surface in alloys. Among other arguments, it is being 
emphasized that a case is possible for which "\ldots an alloy in the proximity 
of absolute zero temperature represents a better model of an ideal lattice 
than that of a pure metal lattice at room temperature " \cite{Kheine73}.

The extinction $\Gamma(C)$ is mostly caused by scattering of s-electrons on 
alloying elements. In a modeled alloy with diagonal disorder and weak 
scattering, the following form is true \cite{Erenreikh76}:

\begin{equation}
\hbar\, \Gamma(C) = 2 \,\pi \,g_{s}(\mu)\, \delta^{2}\,C(1 - C) . 
\label{4.23}
\end{equation}
Here $\delta  = (\varepsilon _{s}^{ae}  -  \varepsilon_{s}^{M})$ - 
difference of the energy levels of s-states of alloying components;\\ 
$g_{s}(\mu)$ - density of s-electron-states of the matrix at 
Fermi-level (per spin-orientation). Thus for a numeric assessment of $\Gamma(C)$, 
the quantities $g_{s}(\mu)$ and $\delta$ must be known. They can 
be determined if the width $W_{s}$ of the s-band, the parameter $a_{0}$, and 
the amount of s-electrons donated by the alloying element $Z_{ae}$ (with 
matrix $Z_{M}$) into the common s-band, i.e. if the electronic 
configurations of the alloying elements are known. In fact, we have a 
parabolic s-band:

\begin{equation}
\label{4.24}
g_{s}(\varepsilon)\sim \sqrt \varepsilon ,\quad 
\int\limits_{0\quad}^{W_{s}}{g_{s}(\varepsilon)\;d \varepsilon = 1},
\end{equation}

\begin{displaymath}
g_{s}\left( \varepsilon \right) = \frac{3}{2}\left(\frac{\varepsilon 
}{W_{s}^{3}}\right)^{\frac{1}{2}}.
\end{displaymath}

Considering that the occupied part of the s-band accepts $Z_{M} / 2$ 
electrons, i.e.

\begin{equation}
\label{4.25}
\int\limits_{0\quad}^{\mu} g_{s}(\varepsilon) d \varepsilon = \frac{Z_{M}}{2},
\end{equation}
we get, after insertion of (\ref{4.24}) in (\ref{4.25}):

\begin{equation}
\label{4.26}
\left(\frac{\mu}{W_{s}} \right)^{\frac{1}{2}} = \left(\frac{Z_{M}}{2} 
\right)^{\frac{1}{3}}.
\end{equation}
Finally we can express (\ref{4.26}) and (\ref{4.24}) $g_{s}(\mu)$ as a 
function of $Z_{M}$ :

\begin{equation}
\label{4.27}
g_{s}(\mu) = \frac{3}{2\,W_{s}} \, \left(\frac{\mu}{W_{s}}\right)^{\frac{1}{2}} =  
\frac{3}{2\;W_{s}}\left(\frac{Z_{M}}{2} \right)^{\frac{1}{3}}.
\end{equation}

According to \cite{Erenreikh76}, the parameter $\delta$ is linked up with the difference 
$\Delta Z = Z_{ae} - Z_{M}$ by means of the relation

\begin{equation}
\label{4.28}
\Delta Z = - \frac{2}{\pi} \,\arctan{[ {\pi\,\delta\,g_{s}(\mu)\, 
( 1 - \delta \,I(\mu))^{-1}}]};
\end{equation}

\begin{equation}
\label{4.29}
I(\mu) = \int\limits_{ - \infty \quad \;}^{\infty}{\;\frac{g_{s}( 
 \eta )}{\mu - \eta } \,d \eta } = \frac{3}{2\;W_{s}}\left[x \,\ln{ \Big\arrowvert 
\frac{1 + x}{1 - x} \Big\arrowvert} - 2 \right],
\end{equation}

\begin{displaymath}
x = \left( \frac{Z_{M}}{2} \right)^{\frac{1}{3}}.
\end{displaymath}

In the following, we shall designate by $Z_{M}$ the number of electrons 
donated into the s-band by the iron atoms, using $Z_{M} = Z_{Fe}$.

Obviously, the justification for introduction of the $M_{S}(C)$ - like 
function of $\tilde{T}_{1,2}(C)$, can be given on the basis of the 
analysis performed, by mapping the function $M_{S}(C)$ onto the plane 
$( \Gamma^{\,\prime}, T^{\,\prime})$, with subsequent determination of those 
specific values of the parameters $a_{0}$, $\delta$, 
$\varepsilon_{d}  -  \mu$, for which the obtained representations 
$\tilde{M_{S}}( \Gamma^{\,\prime} )$ are located in the vicinities of the dashed 
lines in Figs. \ref{fig4.4} and \ref{fig4.5}. Doing this way, 
it will be possible to bring in accordance various experimental 
data with the defined above laws.


\section{Mapping of the functions $M_{S}(C)$ into the $T^{\,\prime}$, 
$\Gamma^{\,\prime} $ region of optimum phonon-generation, with analysis of the 
electronic configurations of atoms in binary substitutional alloys}

The variation of $M_{S}(C)$ with concentration mentioned in Pt. 4.1 belongs 
to the kind of results which have been determined with massive specimen. 
Besides these data, there is currently being collected experimental data 
related to super-rapid quenching (up to $5 \cdot 10^{5}$ K/s) of foils 
with a thickness of $\sim  10 ^{-4}$ m (see for example
\cite{Mirzaev73,Shteinberg77,Mirzaev81} for 
binary alloys and \cite{Mirz81,Mirzaev79} for steel). It could be shown that in different 
regimes of cooling-rates of steel and iron-based alloys there can be 
realized four "stages" of martensitic transformations. An extrapolation of 
these data towards pure iron delivers the following temperatures 
$M_{S}^{i}$ of these "stages": $M_{S}^{I} = 820$, $M_{S}^{II} = 720$, 
$M_{S}^{III} = 540$, $M_{S}^{IV} = 430$ \r{ }C. The existence of various 
$M_{S}$ - stages is explained in \cite{Mirzaev81} with the hypothesis of structural 
particularities of the moving phase-boundaries, as well as by different 
diffusion mechanisms in their vicinity. During changes of composition, the 
amount of "stages" will usually decrease, mainly due to the different rates 
of change of $M_{S}^{i}$ during the alloying process, resulting into 
overlapping of the stages.

\begin{figure}[htb]
\centering
\includegraphics[clip=true, width=.5\textwidth]{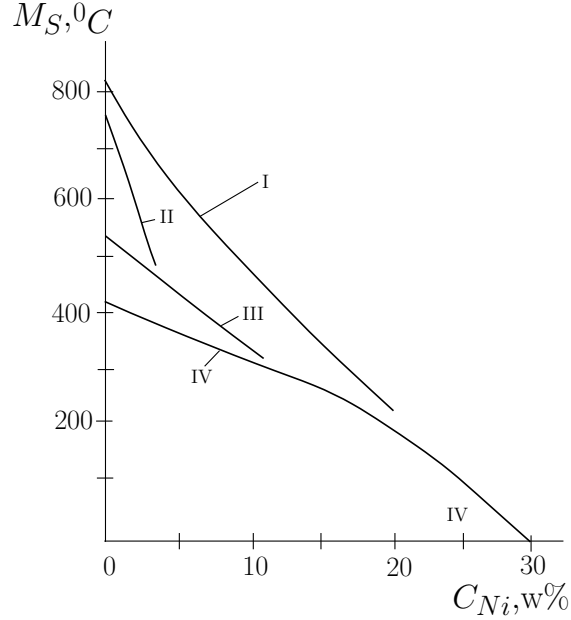}
\renewcommand{\captionlabeldelim}{.}
\caption{ Variation with nickel-concentration (weight {\%}) of the 
temperatures $\tilde{M}_{S}^{i}$, in iron-nickel-alloys \cite{Mirzaev73}.} 
\label{fig4.6}
\end{figure}

\begin{figure}[htb]
\centering
\includegraphics[clip=true, width=.6\textwidth]{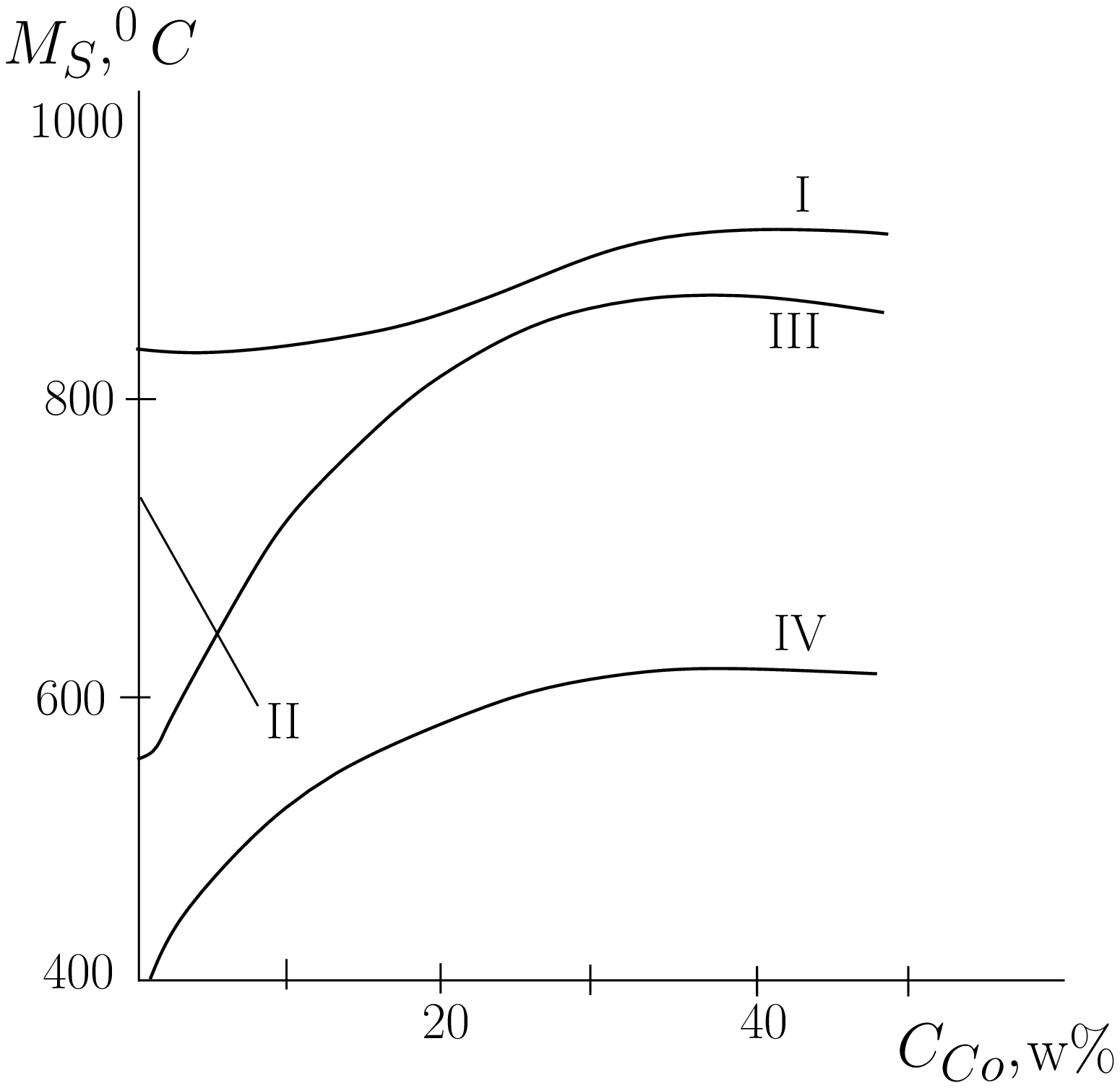}
\renewcommand{\captionlabeldelim}{.}
\caption{Variation with cobalt-concentration (weight {\%}) of the 
temperature $\tilde{M}_{S}^{i}$, in iron-cobalt alloys \cite{Mirzaev81}.}
\label{fig4.7}
\end{figure}

Let us now inspect more in detail the Fe - Ni and Fe - Co systems. For this 
purpose, the corresponding functions $M_{S}^{i}$ are shown in Figs.
\ref{fig4.6} and \ref{fig4.7}. For Fe-Ni-alloys, the following 
inequalities apply for the derivatives (slopes) of the functions 
$M_{S}^{i}(C)$:
\begin{displaymath}
\arrowvert \frac{d\,M_{S}^{II}}{d\,C} \arrowvert > \arrowvert
\frac{d\,M_{S}^{I} 
}{d\,C} \arrowvert > \arrowvert \frac{d\,M_{S}^{III} }{d\,C} \arrowvert > 
\arrowvert \frac{d\,M_{S}^{IV} }{d\,C} \arrowvert ,\quad \frac{d\,M_{S}^{i} 
}{d\,C} < 0.
\end{displaymath}
This means that all $M_{S}^{i}$ decline with increasing concentrations of 
C. By contrast, in Fe-Co-alloys, the temperatures $M_{S}^{I}$, 
$M_{S}^{III}$, $M_{S}^{IV}$ increase with increasing C, while 
$M_{S}^{II}$ decreases:
\begin{displaymath}
\arrowvert \frac{d\,M_{S}^{II}}{d\,C} \arrowvert > \arrowvert 
\frac{d\,M_{S}^{III}}{d\,C} \arrowvert > \arrowvert \frac{d\,M_{S}^{IV} 
}{d\,C} \arrowvert > \arrowvert \frac{d\,M_{S}^{I}}{d\,C} \arrowvert.
\end{displaymath}
If, according to \cite{Mirzaev81}, the second MT-stage was combined with the motion of 
a phase-boundary embodying a Cottrell-atmosphere, then the abnormal 
$M_{S}^{II}$ - behavior could be attributed - at least in the case of 
Fe-Co-alloys - to solute carbon, as the high mobility of carbon atoms can 
contribute to the rapid formation of a Cottrell-atmosphere. We note here 
that for Fe- C-alloys, the slope $d\,M_{S}^{II} / d\,C < 0$ is of an 
order of magnitude of the slope $d\,M_{S}^{II} / d\,C$, thus exceeding 
the corresponding derivatives of Fe-Ni, Fe-Co - alloys. This would be 
evidence for the correctness of our above assumption. As the theory of 
electron-scattering in substitutional alloys also has to take into account 
the effects of lattice-deformation, the function $M_{S}^{II}(C)$ will not 
be analyzed here. 

We also have to point at another particularity of the function $M_{S}^{I} 
(C)$ in Fe-Co-alloys: 
In the region of small concentrations of C, $M_{S}^{I}$ remains almost 
constant, i.e. $d\,M_{S}^{I} / d\,C  \approx  0$. But if $C  \ge  7$ 
{\%}, then a pronounced slope $d\,M_{S}^{I} / d\,C > 0$ appears. Assuming 
an inherent similarity between the functions $\tilde{M}_{S}^{I}(\Gamma^{\,\prime})$ 
and $\tilde{T}_{2}(\Gamma^{\,\prime} )$, this particularity can easily be 
explained: It can be taken for granted that the temperature $M_{S}^{I}(0)$, 
when represented on plane $(\Gamma^{\,\prime} , T^{\,\prime})$, corresponds 
to the coordinates $T^{\,\prime} \approx  0,4$, 
$\Gamma^{\,\prime}  \approx  0,42$, in the vicinity of which (as explained in 
Pt. 4.3) there also is located the minimum of curve $\tilde{T^{\,\prime}}$, 
where the transition of $\tilde{T^{\,\prime}}$ (being essentially independent 
of $\tilde{\Gamma^{\,\prime}}$) towards the function \eqref{4.21} takes place.
Then we choose this point as the starting point and get from \eqref{4.16}:

\begin{equation}
\label{4.30}
T^{\,\prime} = \frac{k_{B} \,M_{S}^{I}(0)}{\varepsilon_{d} - \mu } \approx 
0,4;\quad \Gamma^{\prime}  = \frac{\alpha_{0} \,k_B \,M_{S}^{I}(0)}
{2\;( {\varepsilon_{d} - \mu } )} \approx 0,42,
\end{equation}
from which we can simultaneously determine the values of two parameters: 
$a_{0}  \approx  2,1$ and $(\varepsilon_{d} - \mu) / k_{B}  \approx  
2,5\, M_{S}^{I}  \approx  2750$ K. If these values were considered as fixed 
and characteristic of the fcc phases of iron, then it would be possible 
to perform the mappings of $M_{S}^{i}$ onto plane $(\Gamma^{\,\prime} , 
T^{\,\prime})$, simply by varying the parameter $\Delta Z$, being determined 
by \eqref{4.28}. This way, it has to be considered that in the case of weak 
scattering, the parameter $\delta$, being linked with $\Delta Z$, must be a 
small quantity in relation to the width $W_{s}$ of the s-band. Let 
$W_{s} = 10$ eV, which - for an assumed width $W_{d} = 5$ eV of the d-band - 
would correspond to the 10-fold difference of the average densities of the 
s- and d-electron states. 

For the Fe-Ni - system, we shall use the two nickel-atom configurations 
$3d^{\,9,4} 4s^{0,6}$ and $3d^{\,8,6} 4s^{1,4}$, as configurational 
references, as proposed in \cite{Gudenaf68,Jarlborg80}. During our projection of the graphs 
$M_{S}^{i}(C)$ on plane $(\Gamma^{\,\prime}, T^{\,\prime})$ we choose 5 points 
being borrowed from \cite{Mirzaev73}: For $C_{1} = 0$ (starting points), $C_{2} = 5${\%}, 
$C_{3} = 12${\%} (in the proximity of $C_{3}$, the curves $M_{S}^{III}(C)$ 
and $M_{S}^{IV}(C)$ intersect), $C_{4} = 22${\%} (in the proximity of 
$C_{4}$, the slope of curve $M_{S}^{IV}(C)$ changes), $C_{5} = 30${\%} 
(in the proximity of $C_{5}$ the curves $M_{S}^{I}(C)$ and $M_{S}^{IV}(C)$ 
intersect). We further note that for a Ni-concentration $C_{Ni} > C_{5}$, 
the slope of the graph $M_{S}(C)$ abruptly increases, so that variations 
of Ni-concentrations by about 4{\%} from 30{\%} up to 34{\%} will 
cause a decrease of $M_{S}$ from 250-270 K close to 0 K. Then we have to 
introduce the additional requirement that the point $\tilde{M}_{S}^{I}  = 
\tilde{M}_{S}^{IV}$ of the intersection of the first and second 
stage must fall on the line of extremes 2 (see Fig.\ref{fig4.4}), the end of which, 
close to $\Gamma^{\,\prime}  \approx  1$, exhibits a section of rapid, non-linear 
decrease of $T^{\,\prime}$, with only small variation of $\Gamma^{\,\prime}$. For the a.m. 
parameters $a_{0}$, $W_{s}$, $Z_{Ni} = 0,6$ and $\Delta Z > 0$, it will quite 
easily be possible to show that, on the basis of \eqref{4.16}, \eqref{4.22}, 
\eqref{4.23}, \eqref{4.27} - \eqref{4.29} and by variation of $\Delta Z$, 
the condition for location of the point $\tilde{M}_{S}^{I} = \tilde{M}_{S}^{IV} $ on 
line 2 at $Z_{Fe} = 0,91$ can be met. The magnitude of the slope 
$\partial f / \partial \mu^{\,\prime}  \ge  0,16$ at point $\tilde{M}_{S}^{I}
= \tilde{M}_{S}^{IV}$ remains high. For comparison, we note that at $Z_{Fe} = 1,02$ (point 
$\tilde{M}_{S}^{I}  = \tilde{M}_{S}^{IV} $ on line 1) the slope 
$\partial f / \partial \mu^{\,\prime} \approx  0,13$ is significantly lower. 
Fig. \ref{fig4.8} shows the graphs $\tilde{M}_{S}^{i}$ ($\Gamma^{\,\prime}$) 
for $Z_{Fe} = 0,91$, corresponding to the configuration $3d^{\,7,09} 4s^{0,91}$ 
of iron-atoms. During deviations of the point $\tilde{M}_{S}^{I}
=\tilde{M}_{S}^{IV}$ from line 2, being associated with a variation of Z in the region 
$0,88  \le  Z  \le  0,93$, the magnitude of 
$\partial f / \partial \mu^{\,\prime} \approx 0,16$  remains unchanged. 

For $\Delta Z < 0$, the iron-atoms will attain configurations with a number 
of s-electrons less than 0,6. Even though apparently, such a configuration 
might not be materialized, a consideration of the case $\Delta Z < 0$ may 
still be useful, from a methodological point of view. An analysis of 
\eqref{4.23}, \eqref{4.27} - \eqref{4.29} in fact leads to the conclusion 
that, within our notion, the contribution to total extinction $\Gamma(C)$, 
related to scattering on solute charged atoms, will vanish in the following 
three specific cases:
\begin{figure}[htb]
\centering
\includegraphics[clip=true, width=.8\textwidth]{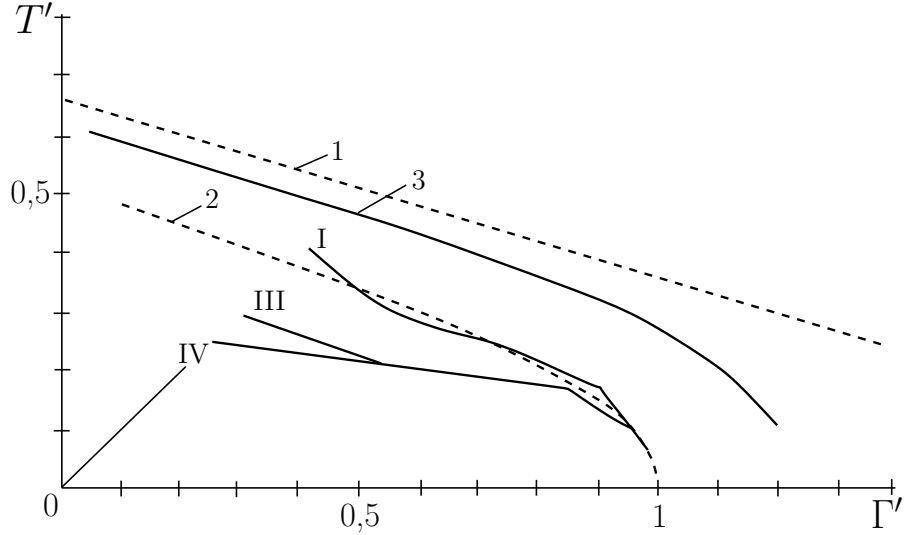}
\renewcommand{\captionlabeldelim}{.}
\caption{ Graphs of $\tilde{M}_{S}^{i} (\Gamma^{\,\prime} )$ for 
Fe-Ni-alloys with $Z_{Ni} = 0,6$, $Z_{Fe} = 0,91$ (or with $Z_{Ni} = 1,4$,
$Z_{Fe} = 1,04$): The graphs $\tilde{M}_{S}^{i}$ were drawn from the starting 
points (I, III, IV), which correspond to pure iron; lines 1, 2 and 3 are 
identical to those in Fig. \ref{fig4.4}.}
\label{fig4.8}
\end{figure}
At $\Delta Z = 0$ (neutral effect), for $Z_{M}  \to 0$, $x  \to  0$, 
$g_{s}(\mu) \to  0$ (in this case, the Fermi-level corresponds to the 
bottom of the s-band coinciding with the baseline of the 
density of states function) and for $Z_{M}  \to  2$, $x  \to  1$ (in this 
case, the Fermi-level corresponds to the upper edge of the fully occupied 
s-band). In these particular cases and in accordance with \eqref{4.22}, 
(\ref{4.30}), total extinction is mainly related to the effects of 
phonon-scattering and magnetic inhomogeneities, so that the relation 
$\Gamma^{\,\prime}  = 1,05 \,T^{\,\prime}$ applies. Point 
$\tilde{M}_{S}^{I} (C_{5}) = \tilde{M}_{S}^{IV}(C_{5})$ has the 
coordinates $T^{\,\prime} \approx  0,1$, $\Gamma^{\,\prime}  \approx  0,1$, 
lying outside of the optimum region, $\partial f / \partial \mu^{\,\prime} \approx
0,04$, thus being one fourth as large as the corresponding values in the 
proximity of line 2 (see Fig. \ref{fig4.8}). For a given value of 
$T^{\,\prime}$, the approach to the optimum region will also depend 
on the increase of $\Gamma^{\,\prime} $, through the scattering effect of the 
impurities. A path being almost equivalent to that of curves
$\tilde{M}_{S}^{i} 
(\Gamma^{\,\prime})$ in Fig. \ref{fig4.8} corresponds to $Z_{Fe} \approx  
0,09$, for a configuration $3d^{\,7,91} 4s^{0,09}$ of the iron atoms, being 
close to the $3d^{\,8} 4s^{0}$ configuration. Full correspondence of the 
paths of these curves however is not possible, as for $\Delta Z < 0$, the 
values of $\Gamma^{\,\prime} $ are limited. The existence of the maximum 
$\Gamma^{\,\prime} $ at fixed $T^{\,\prime}$, for certain $Z_{Fe}$ lying within an 
interval $0 < Z_{Fe} < 0,6$ is obvious, as the extinction $\Gamma^{\,\prime} $ 
takes on minimum equivalent values $\Gamma^{\,\prime}  \approx  T^{\,\prime}$, at 
the borders of the interval ($Z_{Fe} = 0,6 \;(\Delta Z = 0)$, $Z_{Fe} = 0$). 
The condition $\Gamma^{\,\prime}  = \Gamma^{\,\prime}_{max}  \approx  0,936$ 
at $T^{\,\prime} = 0,1$ also matches with the above mentioned value of 
$Z_{Fe}  \approx  0,09$ with the configuration $3d^{\,7,91} 4s^{0,09}$ of 
the iron atoms. It is also obvious from Fig. \ref{fig4.8} that, at 
the point with coordinates $T^{\,\prime} = 0,1$, $\Gamma^{\,\prime} \approx  0,936$, 
the slope is $\partial f / \partial \mu^{\,\prime}  \ge  0,16$, 
but line 2 is out of reach for the point $\tilde{M}_{S}^{I}  =  \tilde{M}_{S}^{IV}$.

For $Z_{Ni} = 1,4$ and $\Delta Z < 0$, the location of point $\tilde{M}_S 
(C_{5})$ on line 2, with coordinates $T^{\,\prime} = 0,1$, 
$\Gamma^{\,\prime}  \approx 0,96$ (the same as for the case with $Z_{Ni} = 0,6$, 
$\Delta Z > 0$), corresponds to the value $Z_{Fe} \approx  1,04$ with 
the iron-atom configuration $3d^{\,6,96} 4s^{1,04}$. After similar
considerations as those done in the case of $Z_{Ni} = 0,6$, $\Delta Z < 0$, 
we can find out for the difference $\Delta Z > 0$, corresponding to the 
inequality $1,4 < Z_{Fe} < 2$, an iron-atom configuration $3d^{\,6,11} 4s^{1,89}$, 
being close to the configuration $3d^{\,6} 4s^{2}$, for which the point 
$\tilde{M}_S(C_{5})$ has the coordinates $T^{\,\prime} \approx  0,1$, 
$\Gamma^{\,\prime}  = \Gamma^{\,\prime}_{max} \approx  0,65$. Even though this 
point does not merely reach line 2, the corresponding slope 
$\partial f / \partial \mu^{\,\prime}  \ge  0,15$ at this point is quite high. 
Thus the assumption on the existence of two iron-configurations in the 
$\gamma $ - phase, being close to the $3d^{\,7} 4s^{1}$ and $3d^{\,6} 4s^{2}$ 
configurations with $Z_{Ni} = 1,4$ (see Pt.2.5), is compatible with the 
requirement of conservation of optimum generation conditions for 
phonon-generation in Fe-Ni-alloys, for both of these configurations.

In \cite{Shteinberg77}, the processing of data for the Fe-Mn - system has been performed in 
a similar way. As reference-configurations, there were used iron-atom 
configurations like those determined above for the Fe-Ni-system. In 
addition, satisfaction of the requirement of conformity for points 
$M_{S}^{IV}$, within the systems Fe-Ni and Fe-Mn, with $C_{Ni} = 22${\%} 
and $C_{Mn} = 11${\%}, respectively, was imperative. The results obtained 
are listed in Table \ref{table4.1}.

\begin{table}[htbp]
\renewcommand{\captionlabeldelim}{.}
\caption{Charge numbers of manganese- and iron-ions}
\begin{center}
\begin{tabular}{|c|c|c|c|c|c|c|}
\hline
$Z_{Fe}$  & 0,09  & 
\multicolumn{2}{c}{0,91} \vline & 
\multicolumn{2}{c}{1,04} \vline & 
1,89                  \\  \hline
$Z_{Mn}$ & 0,84 & 
0,53  & 1,37  & 
1,49  & 0,63  & 
1,19                   \\  \hline
$\Delta Z_{Fe - Mn}$  & 
 - 0,75 & 0,38 & 
 - 0,46 &  - 0,45 & 
0,41  & 0,7            \\  \hline
\end{tabular}
\end{center}
\label{table4.1}
\end{table}

To ease our comparison, Table \ref{table4.2} also shows the specifications 
of the quantities Z and $\Delta Z$, pertaining to the Fe-Ni-system. For 
"average" iron-configurations near $3d^{\,7} 4s^{1}$ (being equivalent to 
$Z_{Fe} = 1$), the following conclusion is obvious: To obtain a slope of the 
relationship $M_{S}(C)$  in Fe-Mn-alloys, doubly exceeding the corresponding 
slope in Fe-Ni-alloys, it will suffice that the manganese-atoms must only 
donate 0,1 more electrons into the s-band than the Ni-atoms. For 
"marginal"-configurations near $3d^{\,8} 4s^{0}$, $3d^{\,6} 4s^{2}$ , this 
difference would amount to 0,2 electrons per atom.

\begin{table}[htbp]
\renewcommand{\captionlabeldelim}{.}
\caption{Charge numbers of nickel- and iron-ions}
\begin{center}
\begin{tabular}{|c|c|c|c|c|}
\hline
$Z_{Ni}$  & 
\multicolumn{2}{c}{0,6} \vline& 
\multicolumn{2}{c}{1,4}  \vline\\  \hline
$Z_{Fe}$  & 0,09  & 
0,91 & 1,04  & 1,89        \\  \hline
$\Delta Z_{Fe - Ni}$  &  - 0,51 & 
0,31  &  - 0,36 & 0,49      \\  \hline
\end{tabular}
\end{center}
\label{table4.2}
\end{table}

\begin{figure}[htb]
\label{fig4.9}
\centering
\includegraphics[clip=true, width=.8\textwidth]{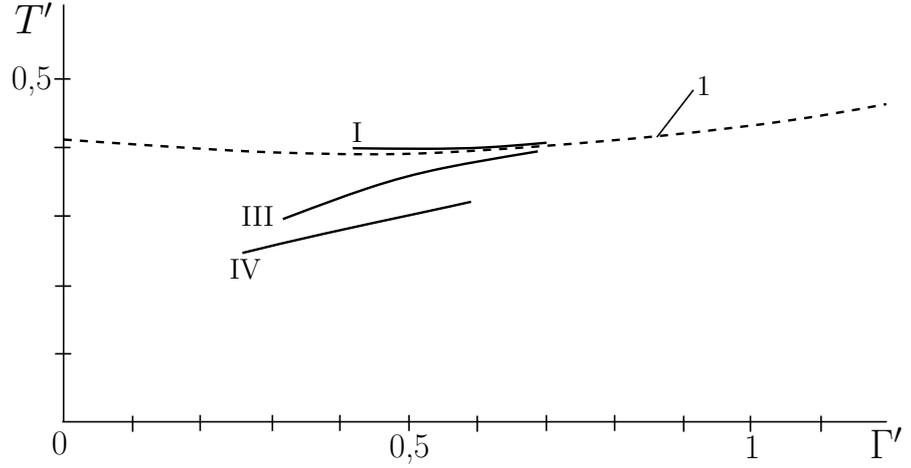}
\renewcommand{\captionlabeldelim}{.}
\caption{ Graphs $\tilde{M}_{S}^{i} (\Gamma^{\,\prime})$ for Fe-Co-alloys 
with $Z_{Fe} = 0,91$, Z$_{Co}$ = 1,12 (or with $Z_{Fe} = 1,04$,
$Z_{Co} = 1,24$); line 1 is identical to that in Fig. \ref{fig4.5}.}
\end{figure}

Now let us use the previously determined iron-atom configurations for our 
foundation of the curves $\tilde{M}_{S}^{i} (\Gamma^{\,\prime})$ of Fe-Co-alloys. 
By varying the parameter $\Delta Z$, it is easy to obtain the lines 
$\tilde{M}_{S}^{i} (\Gamma^{\,\prime})$ for the function
$\tilde{T^{\,\prime}}(\Gamma^{\,\prime})$, which come close to line 1 
in Fig. \ref{fig4.5}. In Fig. \ref{fig4.9}, the curves 
$\tilde{M}_{S}^{i} (\Gamma^{\,\prime} )$ are depicted for an "average" 
iron-atom-configuration $3d^{\,7,09} 4s^{0,91}$ and $\Delta Z = 0,2$, being 
in accordance with the correspondence of the path of curve 
$\tilde{M}_{S}^{I} (\Gamma^{\,\prime} )$ with line 1 (dashed line in Fig.
\ref{fig4.9}). The ends of lines $\tilde{M}_S^i $ correspond to a 
Co-concentration of 40{\%}, and attain maximum values at $M_{S}^{I}$, 
$M_{S}^{III}$ (see Fig. \ref{fig4.7}).

The same general path of curves as those with $\Delta Z = 0,2$ also results 
for $\Delta Z = -0,21$, corresponding to a cobalt-atom-configuration 
$3d^{\,7,88} 4s^{1,12}$. Table \ref{table4.3} shows the specifications of some 
other atom-configurations.

\begin{table}[htbp]
\renewcommand{\captionlabeldelim}{.}
\caption{Charge numbers of cobalt- and iron-ions}
\begin{center}
\begin{tabular}{|c|c|c|c|c|c|c|}     \hline
$Z_{Fe}$  & 0,09 & 
\multicolumn{2}{c}{0,91 } \vline & 
\multicolumn{2}{c}{1,04} \vline & 1,89   \\ \hline
$Z_{Co}$  & 0,29  & 
0,71  & 1,12  & 
0,83 & 1,24  & 1,6                   \\ \hline
$\Delta Z_{Fe - Co}$  & 
 - 0,2 & 0,2  & 
 - 0,21 & 0,21  & 
 - 0,2 & 0,29                        \\  \hline
\end{tabular}
\end{center}
\label{table4.3}
\end{table}

A comparison with Fig. \ref{fig4.4} shows that the coordinates of the ends 
of lines $\tilde{M}_{S}^{I} $, $\tilde{M}_{S}^{III} $, being about $T^{\,\prime} \approx 
0,4$, $\Gamma^{\,\prime}  \approx  0,7$, correspond to a point on line 3, and 
that the coordinates of the end of line $M_{S}^{IV}$ correspond to a point on 
line 2, with a slope $\partial f / \partial \mu^{\,\prime}  \approx  0,18$. In 
this connection, it is also possible to relate the decrease of $M_{S}^{I}$, 
$M_{S}^{III}$ for $C_{Co} > 40${\%} to the onset of shifting of the 
mapped points $\tilde{M}_{S}^{I} $, $\tilde{M}_{S}^{III} $ on line 3, thus 
ensuring a smooth decrease of the slope $\partial f / \partial \mu^{\,\prime}$. 
At the same time, the point $\tilde{M}_{S}^{IV} $ for $C_{Co} > 40${\%} can 
shift along the line of constant slope $\partial f / \partial \mu^{\,\prime} \approx  0,18$, 
with smooth increase of $M_{S}^{IV}$. Obviously, this interpretation requires 
an approximated "equality" of two non-equilibrium sources, during a 
martensitic $\gamma  - \alpha$ - transformation in Fe-Co-alloys, while in 
Fe-Ni- and Fe-Mn-alloys, the chemical potential gradient $\nabla  \mu$ 
serves as the dominating non-equilibrium source.

\section{Discussion of the results on substitutional alloys}

On the basis of reasonable electronic configurations, our consideration 
shows that it is possible, in principle, to obtain satisfactory accordance 
between experimentally $\tilde{M}_{S}^{i} (\Gamma^{\,\prime})$ and the 
theoretically obtained relationship $\tilde{T}_{1,2}^{\,\prime}(\Gamma^{\,\prime})$, 
provided that the spectral density $A(\varepsilon, \varepsilon _{\textbf{k}})$ 
is chosen in the form of a Lorentz-SD-function, as it has to account for the 
complete extinction of s-electrons.

It would also be interesting to discuss the problems related to the amount 
of conformity existing between the parameters used for the $\tilde{M}_{S}^{i}$ - 
mappings and other relevant data, and how sensitive the 
evaluations of the $\Delta Z$ - quantities would behave against variations of 
key parameters, within the frame of an alloying model with diagonal 
disorder.

\subsection{Determination of the parameter ${a}_{0}$ and of
electronic atom-configurations on the basis of electrical and optical 
characteristics of alloying components}

Let us start with the parameter $a_{0}$. To determine the function $\rho(T)$ 
of the electric resistance - within a temperature-interval $0 < T < 1100$ K - 
we use the linear extrapolation for data of \cite{Weiss59}. Then we can determine the average 
slope of the curves $\rho(T)$: $\Delta \rho / \Delta T  \approx 10^{-9}$ Ohm 
$\cdot$ m /K. If we now use the Drude-formula (see for example
\cite{Zaiman74,Akhiezer81})

\begin{equation}
\label{4.31}
\frac{1}{\rho} = \frac{e^{2}\;n}{m^{\ast} \,\tau^{-1}} \equiv \frac{\hbar 
\;e^{2}\;n}{m^{\ast} \,\Gamma } = \frac{\hbar\; e^2\;n}{m^{\ast} \,a_{0}\,
k_{B} T}.
\end{equation}
we get for $a_{0}$:
\begin{equation}
\label{4.32}
a_{0} = \frac{e^{2}n\;\hbar }{m^{\ast}\, k_{B}}\:\frac{\Delta \rho }{\Delta 
T},
\end{equation}
where e, $m^{\ast}$ - charge and effective electron-mass, n - electron 
concentration. If we further hypothesize that each iron-atom donates one 
electron into the s-band ($Z_{Fe} = 1$), we get an s-electron concentration 
of $n_{s} \approx  8 \cdot 10^{28} m^{-3}$. Then we determine 
from (\ref{4.32}) using $m^{\ast}$ = free electron mass = $m_{0} = 9,1 \cdot 10^{-31}$
kg a value of $a_{0}  \approx 17,2$, being 8-times as large as the 
previously mentioned value $a_{0} = 2,1$. For $a_{0}  \approx  17$, the 
extinction $\Gamma $ at point $M_{S}^{I}  \approx  1100$ K would be 
about 1,6 eV (Assumed s-electron lifetime $\tau  \approx  4 \cdot 
10^{-16}$ s). For $\varepsilon_{d}  -  \mu  \approx  0,24$ eV, 
point $\tilde{M}_S^i $ in plane $(\Gamma^{\,\prime} , T^{\,\prime})$ had coordinates 
$T^{\,\prime}  \approx  0,4$, $\Gamma^{\,\prime} \approx  3,3$, and the level of 
the value $\partial f / \partial \mu^{\,\prime}$ would decrease to 0,08. 
We remark that the projection of point $M_{S}^{I}$ into the surroundings of 
the point ($T^{\,\prime} \approx  0,1$, $\Gamma^{\,\prime} \approx  1$), including 
re-establishment of $\partial f / \partial \mu^{\,\prime} \approx  0,15 \div
0,16$ at $\Gamma^{\,\prime} \approx  1,6$ eV, is possible for 
$\varepsilon _{d}  -   \mu  \approx  0,8$ eV, as shown in \cite{Poulsen76} for the case 
of the non-magnetic state of fcc-iron.

Even though in \cite{Beilin74} the value $\tau   \sim  10^{-16}$ s was used for 
an interpretation of the thermoelectric effect in Nickel, the quantity 
$\tau$ most probably had been underestimated, and the quantities $\Gamma$ 
and $a_{0}$ overestimated. This is due to the fact that the Drude-formula 
(\ref{4.31}), being correct for the case of a free-electron gas, does not 
account for the hybridization of s- and d-bands in transition metals. 
As shown in \cite{Snow69} by calculations, the density of the s-states $g_{s}$ is 
only akin to the density of states of free electrons in the proximity of the 
bottom of the s-band. However, in the overlapping region of s- and d-bands, 
the s-state density does not exhibit a monotonous increase with increasing 
energy $\varepsilon$, and thus will only attain values $g_{s}(\mu)$ of an 
order of magnitude of $10^{-2}$ 1/(eV atom), in the proximity of the 
Fermi-level $\mu$. Thus, for correct interpretation of data substantiating 
the function $\rho(T)$, it will be more advisable to use the general formula 
(\ref{4.33}), which links up both quantities $\rho$ and $g_{s}(\mu)$: 

\begin{equation}
\label{4.33}
\frac{1}{\rho} = \frac{2}{3}\,e^{2}\,\textrm{v}^{2}\,\tau  \, g_{s}(\mu) = 
\frac{2}{3}\,\frac{\hbar \;e^{2}\;\textrm{v}^2\;g_{s}(\mu)}{a_{0} \;k_{B}\;T},
\end{equation}
where v$^{2}$ - squared mean value of s-electron velocities at the 
Fermi-surface \cite{Akhiezer81}. The quantity $g_{s}(\mu)$ indicates the number of 
states for a spin-projection within an energy and volume unit interval. For 
an atom concentration of $8 \cdot 10^{28}$ m $^{-3}$, we get the 
following correspondence for different units of energy density:

\begin{equation}
\label{4.34}
\frac{1}{eV \cdot Atom} = 5 \cdot 10^{47}\;\frac{1}{J \cdot m^{3}}.
\end{equation}
Let  v $ \sim  10^{6}$ m/s, $\Delta \rho / \Delta T  \approx  10^{-9}$ Ohm m /K. 
Then we can determine from (\ref{4.33}), (\ref{4.34}) that a value of 
$g_{s}(\mu)  \approx  2,3 \cdot 10^{-2}$ 1/(eV atom) is assigned 
to $a_{0} = 2,1$, thus amounting to only one fifth of the $g_{s}(\mu )$ as 
estimated on the basis of \eqref{4.27}, using $Z_{Fe} = 1$.

Without doubt, the results of optical measurements in the far infrared (at 
wavelengths of about $10,6 \mu$ m) reported in \cite{Vnukovskii82} are interesting, as 
they enable us to evaluate data related to the plasma - $\Omega$ and 
relaxation-frequencies $\tau ^{-1}$ for "pure" s- and hybridized 
s-d-types of the charge-carriers in Fe- and Ni- meltings, and their 
solutions with Cr, at T = 1873 K:

\begin{equation}
\label{4.35}
\begin{array}{l}
\begin{array}{l}
     \Omega_{s}^{2} = 15 \cdot 10^{30}\;s^{-2}, \quad \Omega_{s - d}^{2} 
       \approx 33 \cdot 10^{10}\;s^{-2}, \\ 
     \tau_{s}^{-1} = 2,4 \cdot 10^{14} s^{ - 1},\quad \tau_{s - d}^{-1}
        \approx 4 \cdot 10^{14} s^{-1}, \\ 
    \end{array}   
\Big{\}} \: Fe \\ 
 \\ 
\begin{array}{l}
 \Omega_{s}^{2} \approx 29 \cdot 10^{30}\;s^{-2}\;,\quad \Omega_{s - 
d}^{2} \approx 40 \cdot 10^{10}\;s^{-2},\; \\ 
 \tau_{s}^{-1} \approx 2 \cdot 10^{14}\;s^{-1},\quad \tau_{s - 
d}^{-1} \approx 3 \cdot 10^{14}\;s^{-1}. \\ 
 \end{array}  
\Big{\}} \: Ni \\ 
 \end{array}
\end{equation}
Firstly, we note a remarkable conformity among the above data (\ref{4.35}) and 
those published in [158], related to the specific resistance $\rho$. For 
example we get from [158] $\rho  \approx  1,4 \cdot  10^{-6}$ Ohm m 
for iron at  $T \approx  1900$ K, and an evaluation of the data shown in 
(\ref{4.35}) results in
\begin{displaymath}
\rho_{s - d} = \frac{\tau_{s - d}^{-1} }{\varepsilon_{0} \Omega_{s - 
d}^{2}} \approx 1,35 \cdot 10^{-6}\;Ohm \cdot m, 
\end{displaymath}
\begin{displaymath} 
\rho_{s} = \frac{\tau_{s}^{-1} }{\varepsilon_{0} \Omega_{s}^{2}} \approx 1,8 
\cdot 10^{-6}\,Ohm \cdot m,  
\end{displaymath}
where $\varepsilon_{0} = 8,85 \cdot 10^{-12}$ F/m - dielectric 
vacuum field constant. Secondly, the squared ratios of the plasma-frequencies, 
\begin{displaymath}
\frac{\Omega_{s Ni}^{2} }{\Omega_{s Fe}^{2} } \approx 1,93,\quad 
\frac{\Omega_{s - d Ni}^{2}}{\Omega_{s - d Fe}^{2}} \approx 1,21,
\end{displaymath}
being proportional to the ratios of electron-concentrations in Ni and Fe, 
are greater than 1, a result being in accordance with our initial assumption 
of the proximity of the effective electron-mass $m^{\ast}_{Ni} \sim m^{\ast}_{Fe}$, 
for our chosen configurations $3d^{\,8,6} 4s^{1,4}$ , $3d^{\,7} 4s^{1}$ for Ni and 
Fe, respectively. Thirdly, a calculus of the $n_{s}$ - concentration of 
s-electrons, being included in the formula of the plasma-frequency 

\begin{equation}
\label{4.36}
\Omega_{s}^{2} = \frac{e^{2}\, n_{s}}{m_{s} \,\varepsilon_{0}},
\end{equation}
shows that, using $m_{s}  \approx  m_{0}$ , the value of $n_{s}$, 
amounting to only one sixteenth of $8 \cdot 10^{28}$ m$^{-3}$, corresponds 
to $Z_{Fe} = 1$. This means that the mean density $\bar{g}_{s}$ of s-states 
(being related to the energy-intervals of occupied states in the s-band) is 
smaller by about one order of magnitude than the $\bar{g}_{s}$ value of free 
electrons. This is an immediate indication of the existence of collective 
s-electrons of the iron-atoms, predominantly in hybridized s-d-states. For 
our determination of $m_{s - d}$, we keep in mind the condition $Z_{Fe} = 1$ 
and apply formula (\ref{4.36}) for $\Omega_{s - d}^{2}$. After insertion of 
$n_{s}  \to  n_{s - d}  \approx  7 \cdot 10^{28}$ m$^{-3}$ we 
obtain an effective mass $m_{s - d}  \approx  6,7 m_{0}$. Fourthly, 
considering that the specific resistance of iron $\rho $ decreases 
during a change of temperature T from 1900 K to 1100 K by about $20 \div 
21${\%} \cite{Weiss59}, we get the following relaxation-frequencies of iron at T = 
1100 K, instead of the values shown in (\ref{4.35}):

\begin{equation}
\tau_{s}^{-1}   \approx  1,9 \cdot 10^{14} s^{-1}, \quad \tau 
_{s - d}^{-1}  \approx  3,2 \cdot 10^{14} s^{-1}. 
\label{4.37}
\end{equation}
From this, we can easily evaluate $a_{0}$:

\begin{equation}
\label{4.38}
a_{0\, s} = \frac{\hbar \;\tau_{s}^{-1} }{k_{B} T} \approx 1,3,\quad 
a_{0\;s - d} = \frac{\hbar \,\tau_{s - d}^{-1} }{k_{B} T} \approx 2,2.
\end{equation}

By comparing the results obtained from our interpretation of the 
concentration-dependency $M_{S}(C)$  with measured optical data, it is 
possible to draw the following conclusions: 
\begin{enumerate}
\item{ The above used value of $a_{0}  \approx  2,1$ corresponds to a washed 
out distribution of hybridized s-d-carriers.}

\item{ Out of two reference-configurations of nickel, the configuration 
$3d^{\,8,6} 4s^{1,4}$ should be preferred, and from possible 
iron-configurations, those being close to $3d^{\,7} 4s^{1}$.}

\item{ The value of the parameter $\varepsilon_{d} -  \mu   \approx 0,24$ eV 
indicates a spin-polarized state of iron-atoms, i.e. $\varepsilon 
_{d}  -  \mu = \varepsilon_{d \uparrow }  -  \mu $ (see Pt.2.5).} 

\item{The location of cobalt in the periodical system of elements, i.e. between 
nickel and iron, suggests that out of the two "average" configurations of 
its atoms, with $\Delta Z\, \approx  \, \pm 0,2 $ (see Table 
\ref{table4.3}), the one enabling a rational explanation of the function 
$M_{S}(C)$ in Fe-Co-alloys, being close to $3d^{\,7,8} 4s^{1,2}$, i.e. with 
$\Delta Z \approx   -  0,2$, should be chosen. In accordance with the above 
interpretation of a scenario with plasma-frequencies of s-carriers, it can 
be anticipated, that the inequalities: $\Omega_{s Fe}^{2}  < \Omega_{s Co}^{2} 
  < \Omega_{s Ni}^{2} $, $\Omega_{s - d Fe}^{2}  < \Omega 
_{s - d Co}^{2}   < \Omega_{s - d Ni}^{2}$ will be satisfied. 
However, it has to be taken into account that the partition of the 
frequencies $\Omega_{1,2}$ and $\tau_{1,\;2}^{-1} $ for two 
carrier-groups is no trivial problem, if the frequency of light $\omega  < 
\tau_1^{-1} $, $\tau_2^{ - 1} $ and $\tau_1^{ - 1} $, $\tau_2^{-1}$ 
are of the same order of magnitude \cite{Noskov83}. Thus for example the effective 
values $\Omega_{ef}^{2}  = (32 \pm  1,2) 10^{30}$ s$^{-2}$, and $\tau_{ef}^{-1}  
= (0,8 \pm  0,04) 10^{14}$ s$^{-1}$ , as determined in \cite{Shirokovskii82} 
at T = 295 K for bcc-iron, fit very well with (\ref{4.35}), taken from \cite{Vnukovskii82}, 
for $\Omega_{s-d}^{2}$ and $\tau_{s-d}^{-1}$, under the assumption that 
the absorption of light is of Drude-like-character, and if one considers 
both that the plasma-frequency only weakly varies with temperature, and that 
the relaxation frequency declines with decreasing T (for $a_{0} = 2,1$ and T 
= 295 K we get $\tau^{-1} = a_{0}\, k_{B} T \,\hbar^{-1} = 0,81 \cdot 
10^{14}$ s$^{-1}$ ). The information on frequencies of the s-sub-system 
however remained hidden. Therefore it cannot be excluded that the values 
determined for cobalt $\Omega^{2} = 31 \cdot 10^{15}$ s$^{-2}$, and $\tau 
^{-1} = 0,37 \cdot 10^{14}$ s$^{-1}$  (see \cite{Noskov83,Bolotin73}) 
belong to hybridized s-d-electrons.} 

\item{Assuming that our conclusions drawn from an alloying-model with diagonal 
disorder in the parabolic band of the s-electrons are correct, this would 
suggest that the density of states (DOS) $g_{s}(\mu)  \approx  0,12$ 
1/(eV atom) has been determined realistically, with our model for hybridized 
s-d electrons near the Fermi-level of the $\gamma$ - phase.

Among the papers dealing with the measurement of the specific electric 
resistance of Fe-Ni-alloys, we want to emphasize
\cite{Kaufman56,Sasovskaia69,Bendic78}. In 
\cite{Kaufman56} the "jumps" of the specific electric resistance $\rho$, in the region 
of temperatures $M_{S}$ and $A_{S}$ , are reported, being most probably 
directly linked up with the $\gamma -\alpha$ - MT and the $\alpha -\gamma 
$-reverse MT ($\rho_{\alpha} < \rho_{\gamma}$) occurring during 
the temperature-cycles. During investigations of alloys with 30{\%} Ni 
\cite{Sasovskaia69}, only weak variations of optical data were observed after a $\gamma 
-\alpha $ - MT. The paper \cite{Bendic78} focuses on measurements of the function 
$\rho(T)$ , in the adjacency of magnetic phase transformations.
\cite{Bogachev73}reports on the behavior of $\rho(T)$ in Fe-Mn-alloys, while 
\cite{Kidin69,Andreev69,Kekalo79} is 
related to Fe-C-systems. In our opinion however, a more comprehensive 
consideration of these results would require additional data on $\Omega$ 
and $\tau^{-1}$, which could in principle be obtained from optical 
measurements near the $M_{S}$ - temperatures, for a variety of alloys, in a 
similar way as with the data for $\rho $, being presented in \cite{Kaufman56}. Thus we 
shall confine ourselves to only two qualitative remarks.
  \begin{itemize}
\item[1]{The values of $\rho(M_{S})$ prove to be almost equivalent for various 
alloys with the same $M_{S}$ -temperature, an observation satisfying our 
initial requirement for compliance of the extinction $\Gamma^{\,\prime}$ for Fe-Mn- 
and Fe-Ni-alloys, with identical $M_{S}$ - temperature.}
\item[2]{In Fe-Ni-, Fe-Mn-alloys, the function $\rho(M_{S})$ slowly declines with 
decreasing $M_{S}$, i.e. with increasing Mn and Ni-concentrations, an 
observation which, at the first sight, would contradict with our 
hypothesized increase of s-electron extinction $\Gamma$, with decreasing 
$M_{S}$. However, we have to take into account that, for a given \\ d-s, s-d 
scattering mechanism, the increase of $\Gamma$ delivers the larger 
contribution to conductivity by the d-electrons, leading to a decrease of 
specific impedance. In this case, the increase of $\Gamma$ leads to the 
same class of effect as an increase of photon-frequency, the latter one 
resulting in an increase of the contribution of d-electrons to optically 
induced conductivity \cite{Shirokovskii82}.}
  \end{itemize}}
\end{enumerate}


\subsection{Effect of variations of lattice-parameter and the s-band width  
on the difference of the charge numbers $\Delta Z$ of alloying 
components}

Due to results of energy band analysis (see for example \cite{Wood62}) we will not 
only have to expect significant deviations of $m^{\ast}$ from $m_{0}$, 
but also an anisotropy of the values $m^{\ast}$ and $g_{s}(\mu)$, 
mainly caused by the s-d-hybridization of electronic states. If the value 
$W_{s}$ in formula \eqref{4.27} was varied for the isotropic function 
$g_{s}(\mu)$, is could easily be estimated which electronic configurations 
will lead to the same optimum functions $\tilde{M}_{S}^{i} (\Gamma^{\,\prime})$ 
as previously determined, for new values of $W_{s}$. According to the band 
spectrum of iron \cite{Wood62}, the "widths" of the s-band, related to the orientations 
$\Delta$ and $\Lambda$ of the 1$^{st}$ Brillouin-zone, are close to the 
above used value $W_{s} \approx  10$ eV, whilst the "width" of the s-band in 
$\Sigma$ - orientation is $W_{s} \approx  15$ eV. A calculus shows that, 
for example for $Z_{Ni} = 0,6$, the transition $W_{s}  \approx  10$ eV 
to $W_{s} = 15$ eV is associated with a transition of the configurational 
set $3d^{\,7,1} 4s^{0,9}$, $3d^{\,8,3} 4s^{0,7}$,  $3d^{\,9,4} 4s^{0,6}$ (relating to 
iron, cobalt and nickel, in sequence) into the configurational set $3d 
^{\,7,14} 4s^{0,86}$, $3d^{\,8,32} 4s^{0,68}$,  $3d^{\,9,4} 4s^{0,6}$. 
This means that the variation of the $\Delta Z$ - quantity does not exceed 
0,04. Of course, for known electronic configurations, the deviation of 
$\Delta Z$ could also be larger than 0,04, but we should not overestimate 
its exactness, under a qualitative aspect. We also note that formula
\eqref{4.28}, establishing the link between $\Delta Z$ and the $\delta$ 
parameter (Friedel-summation rule \cite{Erenreikh76}), is only correct in the region of 
small concentrations of an alloying element. Consequently, the description of 
the initial path of the curves $M_{S}(C)$ would be most realistic for small 
C-concentrations. However, if the function $M_{S}(C)$ does not 
significantly deviate from a linear function in a sufficiently large range 
of concentrations, then a more generous application of the results obtained 
by means of \eqref{4.28} would be justified. For alloys with concentrations 
$C = 0,2 \div  0,4$, our results undoubtedly only are of qualitative nature. 
We also note that the parameter $\arrowvert \delta \arrowvert \,W_{s}^{-1}$, 
whose magnitude must be less than 1 (being a prerequisite for the applicability 
of formula \eqref{4.23}), has not exceeded the region of $0,23 \div  0,15$, 
for $W_{s} = 10 \div 15$ eV, with exception of the case $Z_{Fe} = 0,09$, 
$Z_{Mn} = 0,84$, and $\arrowvert \delta \arrowvert \,W_{s}^{-1} = 0,28\div
0,18$ resembling the least probable "marginal" configuration $3d^{\,7,91} 4s^{0,09}$ 
of iron atoms in an Fe-Mn-alloy.

\begin{figure}[htb]
\centering
\includegraphics[clip=true, width=.5\textwidth]{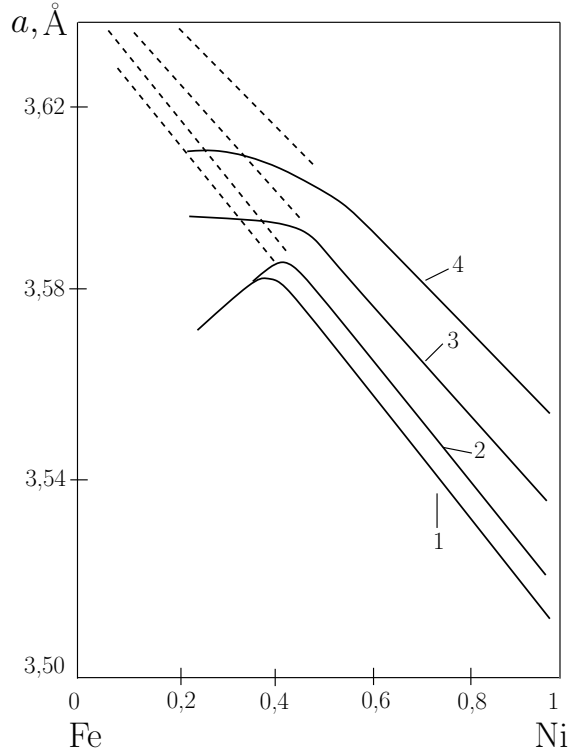}
\renewcommand{\captionlabeldelim}{.}
\caption{ Variation of the lattice-constant of Fe-Ni-alloys with 
Fe-Ni-concentration, at different temperatures \cite{Nakamura76,Nakamura83}: 1 = 0 K, 2 = 288 K, 3 = 575 K, 4 = 
875 K.}
\label{fig4.10}
\end{figure}

Now let us discuss the sensitivity of our calculated results against 
variations $\Delta a$ of the lattice-parameter $a$, essentially depending on 
temperature and composition of an alloy. It is commonly known that a 
homogenous distribution of the impurities \cite{Krivoglaz83} firstly causes a uniform 
change of the lattice parameter of an "average" crystal, as well as of the 
function $a(C)$ of the concentration of a solute, and secondly, the  
inhomogeneous local static displacements of the lattice, in the adjacency of 
an impurity atom. The local variation of volume in the adjacency of the 
impurity atom requires a correction of the quantity $\Delta Z$ \eqref{4.28}, 
which for an isotropic medium, according to \cite{Gantmakher84}, comes to a replacement of 
the function $\Delta Z$ by 

\begin{equation}
\label{4.39}
\Delta Z^{\,\prime} = \Delta Z - Z_{Fe} \frac{\Delta V}{a^{3}} = \Delta 
Z - Z_{Fe} \frac{1 + \sigma_{Fe} }{1 - \sigma_{Fe} } \: \frac{1}{a} \:
\frac{d a }{d C},
\end{equation}
where $\sigma_{Fe} \approx  0,3$ - Poisson's number. The corrected 
expression (\ref{4.39}) reflects the fact that the shielded charge of the impurity 
atom must be corrected by consideration of the excessive volume $\Delta V$ 
occupied by the additional atom (in relation of the matrix-atom). To give an 
example, for Fe-Mn-alloys, the parameter $a(C) = 3,575 - 0,072\, C_{Mn}$ 
(in Angstr\"{o}m), according to \cite{Bogachev73}, where $C_{Mn}$ - Mn-concentration. 
Let $Z_{Fe} = 1$, then we get for the second term in (\ref{4.39}):

\begin{equation}
\label{4.40}
Z_{Fe} \frac{1 + \sigma_{Fe} }{1 - \sigma_{Fe} } \: \frac{1}{a} \: 
\frac{d a}{d C} \approx 0,037.
\end{equation}
According to data in \cite{Nakamura76,Nakamura83}, a function $a(C_{Ni})$ for Fe-Ni-alloys 
is nearly inexistent in the ranges of temperatures above 575 K and 
concentrations $C_{Ni} < 0,3$, being of interest for a comparison of the 
functions $M_{S}(C)$ of Fe-Ni and Fe-Mn-alloys (see Fig. \ref{fig4.10}). 
Thus, if we ignored the corrective factor for $\Delta Z$ in Fe-Ni-alloys, we 
would note that the difference $Z_{Mn}  -  Z_{Ni}  \approx  0,1$, 
which originally had been obtained from a comparison of the values shown in 
Tables \ref{table4.1} and \ref{table4.2} in Pt. 4.4, is correlated by nearly 40 
{\%} with the corrective factor (\ref{4.40}) of Fe-Mn-alloys, being related to the 
case of "average" iron-atom configurations ($Z_{Fe} \approx  1$) and 
$\Delta Z < 0$. In other words, the actual number of electrons donated by 
each Mn-atom into the s-band can be less than values shown in Table 
\ref{table4.1} in Pt. 4.4, thus $Z_{Mn}   -  Z_{Ni}   \approx  0,06$.

\begin{figure}[htb]
\centering
\includegraphics[clip=true, width=.6\textwidth]{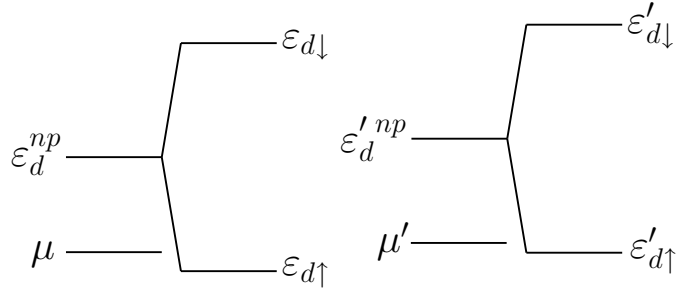}
\renewcommand{\captionlabeldelim}{.}
\caption{ Graph of energy levels showing the opposite signs of the 
increments $\Delta (\varepsilon_{d}^{np} - \mu)  \sim \Delta W_{d} > 0$ 
and $\Delta (\varepsilon_{d \uparrow} - \mu) < 0$, during enlargement 
of the width of the d-band (for $W^{\,\prime}_{d} > W_{d}$) in the case of 
$\varepsilon_{d \uparrow } < \mu $.}
\label{fig4.11} 
\end{figure}

Now let us get over to an estimate of the increment $\Delta(\varepsilon 
_{d} - \mu )$ of the parameter ($\varepsilon_{d} - \mu )$, 
being caused by the homogeneous variation of the parameter $a$, using the 
tight binding approximation. Above all we have to note that the reduction of 
$a$ must be associated with increasing probability of transitions of the 
d-electrons between the knots, with enlargement of the width of $W_{d}$ - 
of the d-band. And vice-versa: The width $W_{d}$ must decrease with  
increasing $a$. If we further assume that the relation 
$ \arrowvert \varepsilon_{d} -  \mu \arrowvert\, W_{d}^{-1}$ will remain 
unchanged during the variation $a   \to   a^{\,\prime} = a + \Delta a$, i.e. 
$\arrowvert \varepsilon_{d}  -  \mu \arrowvert \,W_{d}^{-1} = 
\arrowvert \varepsilon_{d}^{\,\prime} - \mu \arrowvert \,( W^{\,\prime}_{d})^{-1}$, 
then 

\begin{equation}
\Delta (\varepsilon_{d}   -  \mu ) = \arrowvert \varepsilon_{d}   -  
\mu \arrowvert\, W_{d}^{-1}\: \Delta W_{d}. 
\label{4.41}
\end{equation}
It has to be noted that the correspondence of the incremental signs of 
$\Delta W_{d}$ , $\Delta(\varepsilon_{d}  -   \mu )$ in \eqref{4.41} is 
independent of the sign of the difference $\varepsilon_{d} - \mu $, 
as long as the energy $\varepsilon_{d}$ is related to a non-polarized (in 
terms of magnetism) state of electrons $\varepsilon_{d} = \varepsilon 
_{d}^{np}$. However, if polarization exists there are two energy levels 
$\varepsilon_{d \uparrow}$, $\varepsilon_{d \downarrow}$ which emerged after the fission of 
$\varepsilon_{d}^{np}$. The energy level $\varepsilon_{d} = 
\varepsilon_{d \uparrow}$ is characterizing the lower energy level of the pair. 
Then the signs $\Delta(\varepsilon_{d} - \mu )$ and $\Delta 
W_{d}$ will only correspond for $\varepsilon_{d \uparrow} > \mu $. But if 
$\varepsilon_{d \uparrow} < \mu $, then the signs will be opposed (see 
Fig. \ref{fig4.11}). In this case, the ratio \eqref{4.41} must be changed:

\begin{equation}
\Delta (\varepsilon_{d \uparrow}   -  \mu ) = - \arrowvert \varepsilon 
_{d}^{np}   -  \mu \arrowvert\, W_{d}^{-1} \, \Delta W_{d}, \quad
\varepsilon_{d \uparrow} < \mu. 
\label{4.42}
\end{equation}
As already discussed above under Pt. 2.5, it is both possible for the 
$\gamma$ - phase of iron to materialize a non-polarized state with 
$\varepsilon_{d} > \mu $ as well as a spin-polarized state with 
$\varepsilon_{d \uparrow} < \mu $. However, when using \eqref{4.41} in the first case, 
we have to equalize $\varepsilon_{d}  -  \mu $ in both the left and 
right hand side of \eqref{4.41}, but in the second case, the difference 
$\varepsilon_{d \uparrow} - \mu $ can significantly deviate from 
$\arrowvert \varepsilon_{d}^{np} - \mu \arrowvert $. But also 
$\arrowvert \varepsilon_{d}^{np} - \mu \arrowvert$ can deviate from 
$\arrowvert \varepsilon_{d} - \mu \arrowvert$ in \eqref{4.41}. If we further 
took $\varepsilon_{d} - \mu  = \mu - \varepsilon_{d \uparrow}  \approx  2,5\, 
k_{B}\,M_{S}^{I}(0)$, being equivalent to 2750 K (see \eqref{4.30}) at a 
temperature scale or to 0,237 eV at an energy scale, and the quantity 
$\arrowvert \varepsilon_{d}^{np}  -  \mu \arrowvert \approx  0,78$ eV 
(in accordance with \cite{Kubler81}), then we would note that the value 
$\Delta (\varepsilon_{d \uparrow}   -  \mu )$ attains a sign opposed to that of 
$\Delta (\varepsilon_{d} - \mu )$, being about 3,3 times larger in magnitude. 
The quantity $W_{d}$ and its increment $\Delta W_{d}$ can easily be estimated, 
using the relation from \cite{Kharrison83} 

\begin{equation}
\label{4.43}
W_{d} = \frac{6,83\;\hbar^{2}\;r_{d}^{3}}{m_{0} \;r_{0}^{5}},
\end{equation}
where $r_{d}$ - atomic parameter, which, in the case of iron, equates to $0,8 
$\:{\AA}$ \: = \,8 \cdot 10^{-11}$ m, $r_{0}$ - radius of the atomic sphere, 
$m_{0}$ - electron-mass. Taking into account that, for an fcc-lattice,

\begin{displaymath}
\frac{4}{3}\;\pi \;r_{0}^{3} = \frac{a^{3}}{4},
\end{displaymath}
then we are able to express the quantities $W_{d}$, $\Delta W_{d}$ as a 
function of $a$:

\begin{equation}
\label{4.44}
W_{d} = \frac{749,3\;\hbar^{2}\;r_{d}^{3}}{m_0 \;a^{5}},\quad \Delta \,W = - 
5\,W_{d} \frac{\Delta \,a }{a}.
\end{equation}
For $a = 3,6 $\:{\AA}$\: = 3,6 \cdot 10^{-10}$ m we get from (\ref{4.44}) 
$W_{d} \approx  4,85$ eV. Let us now regard the increment $\Delta W_{d}$ 
associated with cooling of the $\gamma$ - phase of iron. According to the 
data in \cite{Kamenetskaia78} we can assume a linear heat expansion coefficient $\beta   
\approx 2,15 \cdot 10^{-5}$ K$^{-1}$. Then, as a result of cooling 
by $\Delta T = -100$ K 

\begin{equation}
\label{4.45}
\frac{\Delta \,a}{a} = \beta \,\Delta \,T = - 2,15 \cdot 10^{- 
3},\quad \Delta \,W_{d} = 1,075 \cdot 10^{-2}\,W_{d}.
\end{equation}
By insertion of (\ref{4.45}) in \eqref{4.42}, \eqref{4.41}, we get

\begin{displaymath}
\Delta (\varepsilon_{d} - \mu ) = 0,237  \cdot  1,075 \cdot 10^{-2} eV  
\approx  2,55 \cdot 10^{-3} eV,
\end{displaymath}
\begin{displaymath}
\Delta (\varepsilon_{d} - \mu ) = -0,78  \cdot  1,075  
\cdot  10^{-2} eV  \approx    - 8,39 \cdot 10^{-3} eV,
\end{displaymath}
or $k_{B}^{-1} \Delta (\varepsilon_{d \uparrow} - \mu )   \approx 30$ K, 
\quad $k_{B}^{-1}\Delta (\varepsilon_{d \uparrow} - \mu ) \approx   - 100$ K at a 
temperature scale. During the cooling process, the starting points 
$M_{S}^{i}(C=0)$, if mapped onto a plane ($T^{\,\prime}$, $\Gamma^{\,\prime}$) 
in accordance with \eqref{4.30}, will settle a straight line 1,05 
$T^{\,\prime} = \Gamma^{\,\prime}$, so that a variation of the parameter 
$\varepsilon _{d} - \mu $ will shift the starting points 
$\tilde{M}_{S}^{III,\;IV} $ along this straight line. In a non-polarized and 
polarized state, respectively, the coordinates of edge-points $\tilde{M}_{S}^{IV}$ 
are ($T^{\,\prime} \approx  0,245$, $\Gamma^{\,\prime}\approx  0,257$) and 
($T^{\,\prime} \approx 0,298$, $\Gamma^{\,\prime} \approx  0,313$), whereas in
Figs. \ref{fig4.8}, \ref{4.9}, the coordinates of the point 
$\tilde{M}_{S}^{IV} $ are: $T^{\,\prime} \approx 0,256$, $\Gamma^{\,\prime} 
\approx  0,269$. In case of a spin-polarized state, the point $\tilde{M}_{S}^{IV} $ 
will shift from the line of constant slope 
$\partial f / \partial \mu^{\,\prime} \approx  0,14$ towards the line of constant 
slope $\partial f / \partial \mu^{\,\prime} \approx  0,17$. Obviously, this 
shifting of the points is favorable for realization of optimum conditions for 
phonon generation. Simultaneously, the shifting of point $\tilde{M}_{S}^{IV} $ 
causes a slight reduction of the gradient $\partial f / \partial \mu^{\,\prime}
\le  0,14$, in the case of a non-polarized state. For this reason, the 
polarized state has certain advantage.

The mappings of points $M_{S}$, corresponding to alloys with a concentration 
$C_{ae}$ greater than a few percent, can be performed by consideration of 
the following: Firstly, we must consider the reduction of the coefficient 
$\beta$, the values of which being about 1,5 to 2-fold smaller (for 
iron-alloys) during approach to room-temperature (and for Fe-Ni-Invar-alloys 
10 times smaller) if compared with $\gamma$ - iron, and secondly, by adequate 
consideration of the dependence of the lattice-parameter on the composition 
of the alloy. Following [181, 182] the value $a  \approx  3,58 $\:{\AA} 
corresponds to a Fe - 30{\%} Ni -alloy. It remains almost invariant with 
varying temperature below 288 K (see Fig. \ref{fig4.10}), whereas at a 
temperature $M_{S}^{I} = 1093$ K, corresponding to the value $\varepsilon 
_{d}  -   \mu $ (see \eqref{4.30}), $a   \approx  3,64 $\:{\AA}. In 
Tables \ref{table4.4}, \ref{table4.5}, there are presented the $Z_{Fe}$ 
for the polarized state of iron-atoms, as determined by the requirement 
of location of point $\tilde{M}_{S}^{I} = \tilde{M}_{S}^{IV} $ on line 2 (at 
$T^{\,\prime} \approx 0,137$, $\Gamma^{\,\prime} \approx  0,92$) and line 3 
(at $T^{\,\prime} \approx  0,137$, 
$\Gamma^{\,\prime} \approx 1,17$ ), taking into account, in each case, the 
increment $\Delta a= - 0,06$\:{\AA}.

\begin{table}[htbp]
\renewcommand{\captionlabeldelim}{.}
\caption{Charge-numbers of nickel- and iron-ions with consideration of the 
lattice parameter variation for the case of location of point 
$\tilde{M}_{S}^{I} = \tilde{M}_{S}^{IV} $ on line 2 (see explanation in the text)}
\begin{center}
\begin{tabular}{|c|c|c|}
\hline
$Z_{Ni}$ & 0,6 & 1,4       \\       \hline
$Z_{Fe}$ & 0,87  & 1,10     \\     \hline
$\Delta Z_{Fe - Ni}$ & 0,27  &  - 0,30   \\     \hline
\end{tabular}
\end{center}
\label{table4.4}
\end{table}

\begin{table}[htbp]
\renewcommand{\captionlabeldelim}{.}
\caption{Charge-numbers of nickel- and iron-ions with 
consideration of the lattice parameter variation for the case of location of point 
$\tilde{M}_{S}^{I} = \tilde{M}_{S}^{IV} $ on line 3 (see explanation in the text)}
\begin{center}
\begin{tabular}{|c|c|c|}
\hline
$Z_{Ni}$ & 0,6  & 1,4        \\   \hline
$Z_{Fe}$ & 0,90  & 1,06       \\   \hline
 
$\Delta Z_{Fe - Ni}$ & 0,30 &  - 0,34 \\  \hline
\end{tabular}
\end{center}
\label{table4.5}
\end{table}

The data in Tables \ref{table4.4}, \ref{table4.5} only relate to "average" 
configurations of iron-atoms, in accordance with the assumption in Pt. 2.5 on 
the polarization of states with a configuration close to $3d^{\,7} 4s^{1}$. A 
comparison of the data are presented in Tables
\ref{table4.2}, \ref{table4.3} also shows that a variation of the 
lattice-parameter $a$ leads to a noticeable effect on the quantity 
$\varepsilon_{d \uparrow} - \mu $, which will diminish by $6,4 \cdot 10^{-2}$ eV 
(or about 27{\%}) for $\Delta a= - 0,06 $\:{\AA}, as well as on the quantity 
$\arrowvert \Delta Z \arrowvert$, the reduction of which will lead to an 
additional deviation of Z from 1, by 0,06. For "marginal"-configurations, if 
put into relation with the non-polarized state of iron, a consideration of 
$\Delta a= - 0,06 $\:{\AA} will result into an increase of 
$ \arrowvert \Delta Z \arrowvert $ by about 1/3 of the corresponding increment 
of $\arrowvert \Delta Z \arrowvert $ in the polarized condition of iron.

This way, in the spin-polarized state of iron-atoms, the $Z_{Fe}$ - values 
for $Z_{Ni} = 0,6$ are lying in a range of $0,87 \div  0,91$, and for 
$Z_{Ni} = 1,4$, in a range of $1,04 \div  1,10$. As to differential-modules 
of electron-numbers, associated with the s-band of atoms of binary 
substitutional alloys, we obtain the following inequalities:

\begin{equation}
\arrowvert \Delta Z_{Fe - Co}\arrowvert  < 
\arrowvert \Delta Z_{Fe - Ni}\arrowvert  < 
\arrowvert \Delta Z_{Fe - Mn}\arrowvert. 
\label{4.46}
\end{equation}

It has to be noted here that the value 
$\arrowvert \Delta Z_{Fe - Co}\arrowvert  \approx 0,2$, being used in Table
\ref{table4.3}, nearly corresponds to an approximated 
conformance of the function $\tilde{M}_S^I $ with the sector of line 1 in 
Fig. \ref{fig4.5} From the point of view of maintenance of a high level 
of the slopes $\partial f / \partial \mu^{\,\prime}$, 
$\partial f / \partial T^{\,\prime}$ during variations of the Co-concentration, 
a conformance of the figures $Z_{Fe}$ und $Z_{Co}$ would also be favorable, 
i.e. $\Delta Z_{Fe - Co} = 0$. For $\Delta Z_{Fe - Co} = 0$, the contribution 
to extinction being caused by scattering at charged impurities, also vanishes, 
i.e., $\Gamma(C) = 0$, while the $\tilde{M}_{S}^{i}$ - points at the $(T^{\,\prime}, 
\Gamma^{\,\prime})$ - plane shift upwards from their initial locations along the 
straight line $1,05 \,T^{\,\prime} = \Gamma^{\,\prime}$. The coordinates of point 
$\tilde{M}_S^I $ change with variations of Co-concentration from C = 0 to 
C = 0,4 from ($T^{\,\prime} \approx  0,4$, $\Gamma^{\,\prime} \approx  0,42$) 
to ($T^{\,\prime} \approx  0,426$, $\Gamma^{\,\prime} \approx  0,447$), and, 
for C = 0,4, the points $\tilde{M}_{S}^{III}$, $\tilde{M}_{S}^{IV} $ shift into 
positions close to that of $\tilde{M}_{S}^{I}$ or $\tilde{M}_{S}^{III} $ for C = 0. 
It can further be seen from Figs. \ref{fig4.4}, \ref{fig4.5} that the 
slopes $\partial f / \partial \mu^{\,\prime}$ and $\partial f / \partial T^{\,\prime}$ 
increase at the same time. For Fe-Ni- and Fe-Mn-alloys, the vanishing of the 
quantities $\Delta Z_{Fe - Ni}$, $\Delta Z_{Fe - Mn}$ is most unfavorable, as 
the downshift of points $\tilde{M}_S^i $ along line 
$\Gamma^{\,\prime} = 1,05 T^{\,\prime}$ will lead to rapid reduction of the 
magnitudes of slopes $\partial f / \partial \mu ^{\,\prime}$, 
$\partial f / \partial T^{\,\prime}$. 

We further note that in the above performed analysis of the scattering by 
impurities, we omitted a consideration of non-diagonal disorder existent in 
binary alloys, being caused by electronic transitions between the knots of 
the components of binary alloys \cite{Erenreikh76}. Obviously, any additional 
contribution to extinction $\Gamma(C)$ will cause a decline of the quantity 
$\Delta Z$. Thus the above determined $\Delta Z$ - values should be regarded 
as estimates lying above the true differences of the number of electrons 
donated into the s-band by the alloying components.


\subsection{Estimation of the chemical potential differences of some 
$\gamma$ - and $\alpha$ - phases, for the sub-system of d-electrons}

Let us use formula \eqref{4.43} in order to calculate the chemical potential 
difference $\Delta \mu = \mu_{\gamma}   -  \mu_{\alpha}$ 
between austenite and martensite, the magnitude of which will determine the 
amount of inverted occupational difference $\sigma _{0}$, in addition to 
the slope $\partial f / \partial \mu^{\,\prime}$. We recall that an elementary 
assessment of $\Delta \mu$, relating to s-electrons, has already been 
performed in Pt. 1.5. With regard of the necessary correspondence of the 
$\Delta \mu$ - assessments for different (s and d) electrons, we now assess 
$\Delta \mu$ for the sub-system of d-electrons. Let us count $\mu$ from 
the bottom of the d-band upwards, assuming that the ratio $\mu \, W_{d}^{-1}$ 
applies for both the $\gamma$- and the $\alpha$ - phases. Then we get:

\begin{equation}
\label{4.47}
\Delta \,\mu = \mu_{\gamma} - \mu_{\alpha} = ( W_{d\,\gamma} - 
W_{d\,\alpha})\;\frac{\mu_{\gamma}}{W_{d\,\gamma}} = \left(1 - 
\frac{W_{d\,\alpha}}{W_{d\,\gamma}}\right)\,\mu_{\gamma}.
\end{equation}
If we regard the already used quantity 
$\varepsilon_{d}^{np}  - \mu \approx  0,78$ eV as the energy-difference 
between the upper edge of the d-band and the Fermi-level of the $\gamma$ - phase, 
we get for $W_{d\,\gamma}  \approx 4,85 $ eV the value $\mu_{\gamma}  \approx  
4,85 - 0,78 = 4,07$ eV. Taking further in consideration that, for a 
bcc-lattice, the volume associated to one atom is

\begin{displaymath}
\frac{4}{3}\;\pi \;r_{0}^{3} = \frac{1}{2}\;a_{\alpha}^{3},
\end{displaymath}
we then can express by formula \eqref{4.43}, in a similar way as previously done 
with \eqref{4.44}, the value $W_{d \alpha}$ by means of the lattice-constant 
$a_{\alpha}$ , and determine the ratio

\begin{equation}
\label{4.48}
\frac{W_{d\,\alpha}}{W_{d\,\gamma}} = 
\Big(\frac{a_{\gamma}}{2^{1 / 3}\,a_{\alpha}}\Big)^{5} \approx 0,961,
\end{equation}
using therein the ratio $a_{\alpha} = 0,8 a_{\gamma}$, being typical of 
the Bain-deformation (see Fig. \ref{fig1} and its legend). If we now 
insert $\mu_{\gamma} = 4,07$ eV in (\ref{4.47}), under consideration of 
(\ref{4.48}), then we get $\Delta \mu \approx  0,16$ eV, being in good 
accordance with the previously performed estimate for the sub-system of 
s-electrons.


\section{The $M_{S}(C)$ function for steel and degree of 
ionization of carbon-atoms}

In principle, one could try to extend the approach for an analysis of the 
concentration-dependence of $M_{S}$, from binary substitutional alloys 
towards interstitial alloys, among which the Fe-C-system would be the most 
interesting one. We can take for granted that carbon-atoms will only occupy 
the octahedral interstitial voids in austenite \cite{Kurdjumov77}. Let us further surmise 
that those iron-ions, in the closest adjacency of which a carbon-atom is 
localized, can be characterized by a charge-number $Z_{Fe C}$, and that these 
iron-ions will adopt the role of the "alloying-element"-ions, substituting 
the "normal" iron-ions by a combined charge number $Z_{Fe C}$, being 
different from $Z_{Fe}$. Then our search for $\Gamma(C)$, which represents 
the contribution to s-electron extintion by scattering at the impurity 
atoms, can be reduced to the task of determination of electron-scattering at 
octahedral clusters of 6 atoms, whose concentration is identical with the 
concentration of carbon. This task however is neither trivial nor has it 
been completely resolved (see for example \cite{Egorushkin82,Zaiman82}). Thus we shall 
confine ourselves on a qualitative reflection and start with an estimate of 
the lower limit by hypothesizing that, during the process of 
cluster-scattering, the s-electrons will only interact with one of the six 
"solute" ions (i.e. that one being nearest to the s-electron). In this case, 
our analysis of $M_{S}(C)$ can be performed in the same way as in the case 
of a substitutional lattice in Pt. 4.4, 4.5. According to \cite{Mirz81}, the rate of 
reduction of $M_{S}$ with increasing carbon concentration is about fourfold 
or twice as large as that for Fe-Ni- or Fe-Mn-alloys. Thus for instance, a 
temperature $M_{S}^{IV}  \approx  473$ K in Fe-C-, Fe-Mn- and 
Fe-Ni-systems corresponds to about 4,6 Atom - {\%} C (1 weight-{\%} C), 10,5 
atom - {\%} Mn and 22 atom - {\%} Ni. In Table \ref{table4.6}, there are 
presented some values of $Z_{Fe C}$, which have been determined for the same parameters 
as those used in Pt. 4.4, i.e. $W_{s} = 10$ eV, 
$\varepsilon _{d}  - \mu  \approx  0,237$ eV, without account on the effect of 
variable lattice parameter.

\begin{table}[htbp]
\renewcommand{\captionlabeldelim}{.}
\caption{Effective charge-numbers of iron-atoms included in octahedral 
clusters with interstitial carbon atoms ( $Z_{Fe C }$), and of iron-ions 
without association to carbon ($Z_{Fe}$ )}
\begin{center}
 \begin{tabular}{|c|c|c|c|c|c|c|}
\hline
$Z_{Fe}$  & 0,09  & 
\multicolumn{2}{c|}{0,91} & 
\multicolumn{2}{c|}{1,04} & 
1,89                             \\ \hline
$Z_{Fe C}$  & 1,08 & 0,42  & 
1,53  & 1,62  & 0,52  & 0,86     \\ \hline
$\Delta Z$ &  - 0,99 & 0,49  & 
 - 0,62 &  - 0,58 & 0,52  & 1,03  \\ \hline
\end{tabular}
\end{center}
\label{table4.6}
\end{table}

By comparison of the data shown in Tables  \ref{table4.6} and
\ref{table4.1}, we note that - for "average" iron-configurations - a twofold 
increase of the rate of reduction of $M_{S}$ would requires an increase of 
$\Delta Z$ - by about $0,1\div 0,15$. We recall that we came to a similar 
conclusion in Pt. 4.4, after having compared the rate of change of the 
$M_{S}$ temperature in Fe-Ni and Fe-Mn-alloys.

The measured magnetic susceptibility of Fe-C-alloys, as reported in \cite{Dovgopol77}, 
points to an increase of electron-concentration in the d-band with 
increasing concentration of carbon. In a local interpretation, d-band 
occupation by carbon-electrons is associated with hybridization of the 2p 
wave-functions of the carbon with the 3d-functions ($e_{g}$ - symmetry) of the 
iron. Such hybridization can associate with a reduction of the charge-number 
of the iron-atoms in an octahedral cluster. Further hypothesizing that the 
difference between $Z_{Fe C}$ and $Z_{Fe}$ is only caused by the additional 
charge transfer from the carbon-atom, it is suggested to keep in mind only 
the following two $Z_{FeC}$ - values of Table \ref{table4.6} : $Z_{Fe C} = 0,42$ 
(at $Z_{Fe} = 0,91$) and $Z_{Fe C} = 0,52$ (at $Z_{Fe} = 1,04$). $Z_{Fe C} = 
0,88$ (at $Z_{Fe} = 1,89$) should be omitted, as the adoption of one electron 
per ion (with $\Delta Z = 1$) would result in complete ionization of the 
carbon-atom, given an equivalence of six iron-ions per cluster. The degree 
of ionization of the carbon atom is 

\begin{equation}
\label{4.49}
6\;\Delta \,Z = 6\,( {Z_{Fe} - Z_{FeC} } ) \approx \left\{ 
\begin{array}{l}
 2,94,\quad Z_{Fe} = 0,91 \\ 
 3,12,\quad Z_{Fe} = 1,04 \\ 
 \end{array} \right.,
\end{equation}
i.e. one carbon-atom may donate about three electrons into the d-band.

Consideration of the homogenous change of the lattice-parameter (see Pt. 
4.5.2) would lead to a reduction of ($\varepsilon_{d \uparrow} - \mu )$ to 
0,215 eV, for Fe-C-alloys (1 weight {\%} C) at $T = M_{S}^{IV}  
\approx  473$\r{ }K. In Table \ref{table4.7}, there are presented the 
corrected $Z_{Fe C}$ values, by adequate consideration of the increments of 
the lattice constant $a$ (the values of $Z_{Fe}$ were extracted from 
Table \ref{table4.4}, as determined from the requirement of location of 
point $\tilde{M}_{S}^{IV} $ in the adjacency of line 2 in Fig.
\ref{fig4.4} (at $T^{\,\prime} \approx  0,19$, $\Gamma^{\,\prime} \approx 
0,82$), relating to an Fe-C-alloy with 1 weight {\%} carbon.

In calculating the corrective factor for inhomogeneous volume change, we 
have to consider that the quantity $\Delta Z$ in Table \ref{table4.7} must 
equate to the quantity $\Delta Z^{\,\prime}$ in \eqref{4.39}.

\begin{table}[htbp]
\renewcommand{\captionlabeldelim}{.}
\caption{Charge-numbers $Z_{Fe}$, $Z_{FeC}$ with consideration of a variable 
lattice-parameter $a$ }
\begin{center}
\begin{tabular}{|c|c|c|}
\hline
$Z_{Fe}$  & 0,87 & 1,1      \\   \hline
$Z_{Fe C}$  & 0,42 & 0,59    \\   \hline
$\Delta Z$ & 0,45 & 0,51      \\   \hline
\end{tabular}
\end{center}
\label{table4.7}
\end{table}

Taking advantage of the concentration-dependency of the lattice-parameter of 
austenite given in \cite{Kurdjumov77}: $a_{\gamma} = (3,578 + 0,645 C) ${\AA} , where C - 
fraction of carbon-atoms, we can determine from \eqref{4.39}

\begin{equation}
\label{4.50}
\Delta \,Z = \Delta \,Z^{\,\prime} + \frac{1 + \sigma _{Fe}}{1 - \sigma _{Fe} 
}\;\frac{Z_{Fe}}{6 a}\;\frac{d a}{d C} \approx \left\{ 
\begin{array}{l}
 0,47,\quad Z_{Fe} \approx 0,87 \\ 
 0,57,\quad Z_{Fe} \approx 1,1  \\ 
 \end{array} \right.
\end{equation}
The factor 1/6 in the second term in (\ref{4.50}) considers that only one sixth of 
the inhomogeneous volume increase caused by the insertion of the carbon-atom 
can be assigned to each iron-ion in the octahedral cluster. The ionization 
degree of the carbon-atoms

\begin{displaymath}
6\;\Delta \,Z \approx 
\Big{\{ }
\begin{array}{l}
2,82,\quad Z_{Fe} \approx 0,87 \\ 
3,42,\quad Z_{Fe} \approx 1,1 \\ 
 \end{array} 
,
\end{displaymath}
which had been determined under consideration of the changes of the 
austenitic lattice parameter, moderately differs from the results of
(\ref{4.49}). The quantity $6 \Delta Z  \approx 3,42$  presumably is the upper margin 
for the degree of ionization of the carbon-atoms, as the hypothesis which 
transforms the effect of s-electron scattering at the complete cluster to 
scattering at only one of its six iron-ions, can only be substantiated for a 
larger number of ions involved in s-electron scattering, which in turn would 
lead to an increase of $Z_{Fe}$, with corresponding decrease of $\Delta Z$. 
However, we note that our conclusion on the maximum value of $6 \Delta Z  
\approx  3,42$ is based on the condition that the point $\tilde{M}_{S}^{IV} $ 
for Fe-C-alloys (1 weight {\%} C) arrives at the adjacency of line 2. If we 
required that the point $\tilde{M}_{S}^{IV} $ arrives at the adjacency of 
line 3 in Fig. \ref{fig4.4} then the degree of ionization would increase to 
$6 \Delta Z  \approx  3,78$. This value is an intermediate in relation to 
the values 3,5 and $3,9 \div  4$, which had been determined in \cite{Dovgopol77} and 
\cite{Kalinovich74}, by analysis of the magnetic susceptibility of Fe-C-systems, within 
the notion of a quasi-rigid d-band with $Z_{Fe}  \approx  1$, and by 
consideration of data of carbon-diffusion in a constant electric field 
(electrotransfer-method).


\section{Summary of Chapter 4}

Our assessments of substitutional (Fe-Ni, Fe-Mn, Fe-Co) and interstitial 
alloys (Fe-C) will now enable us to draw a basic conclusion on the 
possibility of preservation of the conditions for efficient phonon 
generation by non-equilibrium 3d-electrons, within a wide range of 
concentrations of the second alloying component. 

The methodological basis of our investigations was the usage of a 
SD - function, which, besides thermal scattering, also takes into account the 
effect of scattering (washout) caused by the inhomogeneity of the system. We 
note that establishment of a modified equilibrium distribution of 
d-electrons (see Pt. 4.2), whose degree of "washout" is mainly determined by 
the rate of extinction of s-electrons, is limited by the time $\tau_{d - 
s}$ characterizing d-s-electron scattering. As the non-equilibrium addend to 
the electronic distribution cannot come into existence prior to the 
formation of the real equilibrium distribution, the time $\tau_{d - s}$ 
will determine the minimum time $\tau_{\sigma }$ for establishment of an 
inverted occupational difference $\sigma_{0}$ (pumping-time). From this 
perspective, the value of $\tau_{\sigma } \sim  10^{-12}$ s, as 
determined in Chapter 3 in our evaluation of the deformation $\varepsilon $, 
appears quite acceptable. 

Our analysis of the behavior of non-equilibrium addends to the electronic 
distribution functions revealed the existence of certain region of 
temperatures T and extinction $\Gamma$ with optimum conditions for wave 
generation (Electron extinction depends on the concentration C of the second 
alloying component). Within this region, during simultaneous change of the 
key parameters T and C, a high magnitude and slow variation of the 
non-equilibrium addends are typical. 

The next step in our analysis comprised a mapping of the real relationship 
$M_{S}(C)$, between temperature and concentration, into the region of 
optimum values of parameters T and $\Gamma$. As for $\Gamma$, the 
contribution of scattering by impurities has already been emphasized, and 
resulted in our consideration of weak scattering, being intimately related 
to the diagonal disorder in the system matrix. (Characteristic of this 
matrix is that it reflects the effect of solute atoms substituting the 
matrix-atoms in the lattice knots, with conservation of the transition 
integrals among the lattice-knots).

Our choice of the electronic configurations of the Ni-atoms as a reference 
configuration, based on published data, enabled us to determine the 
electronic configurations of iron, cobalt manganese and carbon, in 
accordance with the mapping of the relationship $M_{S}(C)$ into the region 
of optimum values of the parameters $T$ and $\Gamma$. Among the analyzed 
substitutional alloys, the difference $\Delta Z$, indicating the number of 
electrons donated by an alloying atom into a common s-band, proved to be 
least for Fe-Co-alloys: $\arrowvert \Delta Z \arrowvert_{Fe - Co} = (0 \div 
0,2)$ and largest for Fe-Mn-alloys: $\arrowvert \Delta Z \arrowvert_{Fe - Mn} = (0,4 \div  0,5)$, 
for the case of an iron-atom configuration near $3d^{\,7} 4s^{1}$. 
Interestingly, within our model with diagonal disorder, there can be 
realized a twice-fold rate of reduction of $M_{S}(C)$ in Fe-Mn-alloys with 
increasing manganese-concentration, in relation to the rate of decrease of 
$M_{S}(C)$ in Fe-Ni-alloys with $\arrowvert \Delta Z \arrowvert_{Fe - Mn}  -
\arrowvert \Delta Z \arrowvert_{Fe - Ni}  \le  0,1$. This result could also be 
confirmed by comparison of the rates of reduction among the Fe-Mn and 
Fe-C-systems.

Good accordance with the values of $0,2 \div  0,3$ eV, mentioned in Pt. 2.5, 
is also given for the parameters $\varepsilon_{d} - \mu  = (0,17 
\div  0,26)$ eV, which we used for the mapping of the function $M_{S}(C)$. 
We recall that the value $\varepsilon_{d} - \mu \approx  0,237$ 
eV has been chosen as an initial value (at a temperature $M_{S}^{I}(0) = 
1093 $ K), and that the range of the $\varepsilon_{d} - \mu $ values 
depends on the lattice parameter. 

From a quantitative point of view, an evaluation of the derivative $\partial 
f / \partial \mu^{\,\prime} $, which - besides the quantity $\nabla \mu $ - 
determines the initial occupational inversion $\sigma_{0}$, appears to be 
important. It follows from Fig. \ref{fig4.4} that, in the vicinity of 
line 2 of the maximum values of the function $\partial f (\Gamma^{\,\prime}, 
T^{\,\prime}) / \partial \mu^{\,\prime} $, the inequalities 
$0,21 \ge  \partial f / \partial \mu^{\,\prime} \ge  0,15$ are satisfied, 
enabling us to use the value $\partial f / \partial \mu^{\,\prime} \approx  0,15$ 
as a minimum, in our evaluation of $\sigma_{0}$. We further note that the value 
$\arrowvert \partial f / \partial \mu^{\,\prime} \arrowvert \approx  0,1$, 
as used in Pt. 3.3 in our evaluation of the deformation $\varepsilon $ 
(see \eqref{3.42}), was 1,5 -times less than the above value.

\begin{figure}[htb]
\centering
\includegraphics[clip=true, width=.7\textwidth]{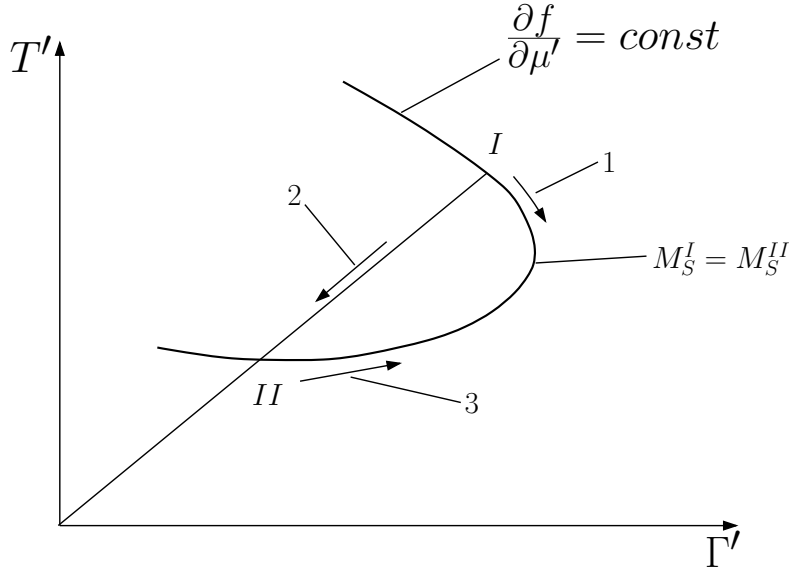}
\renewcommand{\captionlabeldelim}{.}
\caption{One of the variants for description of two stages of the 
concentration-relationship $M_{S}^{I,II}(C)$, mapped on the plane of 
variables $\Gamma^{\,\prime}$, $T^{\,\prime}$: Shifting motion along the isolines 
$\partial f / \partial \mu^{\,\prime}  = const$: Starting at point I and 
proceeding along route 1 corresponds to a reduction of $M_{S}^{I}$, with 
increasing concentration of the alloying elements. The motion along route 3 
from point II is reflected by an increase of $M_{S}^{II}$. Route 2 describes 
a cooling process without variation of composition.}
\label{fig4.12}
\end{figure}

We further note that the character of the isolines in Figs. \ref{fig4.4}, 
\ref{fig4.5} (mirrored C-shaped curves) does not require that the function 
$\tilde{M}_{S}(C)$ strictly follows the path of lines 2 or 3. In the general 
case, which must not necessarily relate to iron-alloys, a variety of 
possibilities is permissible. For instance, the mapping of $\tilde{M}_{S}$ 
on the plane ($T^{\,\prime}$, $\Gamma^{\,\prime}$) (for low $M_{S}$-temperatures and $\varepsilon 
_{d} - \mu $ - values near $\varepsilon_{d}  -  \mu \approx  k_{B}\,M_{S}$) 
can attain the coordinates ($T^{\,\prime} \approx  1$, $\Gamma^{\,\prime} \approx
0$), thus going far beyond of the limits of the region bordered by lines 1 and 2. 
The value $\partial f /\partial \mu^{\prime} \approx  0,2$ however is large in this 
point and remains almost unchanged during a shifting motion along the isolines, 
being caused by increases of $\Gamma^{\,\prime}$ with simultaneous reduction of 
$T^{\,\prime}$. Thus, if the starting-coordinates of $\tilde{M}_{S}$ are located 
in the adjacency of $T^{\,\prime} \approx  0,4$, $\Gamma^{\,\prime} \approx  0$, 
then the value $\partial f /\partial \mu^{\prime} \approx  0,18$ can only be conserved 
with increasing $\Gamma^{\,\prime} $, if the temperature initially decreases and 
later on increases. A situation is also possible in which the temperature 
$M_{S}$, along certain section of the branches (steps) of the functions
$M_{S}(C)$, decreases with increasing C, but increases along other sectors of 
the branches. This situation is illustrated in Fig. \ref{fig4.12}. Obviously, 
the motion away from point I along the isoline leads to a reduction of 
$\tilde{M}_{S}^{I}$ with increasing $\Gamma^{\,\prime}$. Cooling (without change 
of composition) will shift the system away from point I to point II. The shifting 
away from point II, along isoline $\partial f / \partial \mu^{\,\prime} $ = const, 
is associated with an increase of $\tilde{M}_{S}^{II}$, for increasing
$\Gamma^{\,\prime}$, until the intersection of branches $\tilde{M}_{S}^{I}$ and 
$\tilde{M}_{S}^{II}$ is reached. A similar behavior of the branches 
$M_{S}^{i}(C)$ can also be observed in Fe-Mo-alloys \cite{Mirzaev81,Ponomareva77}.

Of course, during an analysis of the functions $M_{S}^{i}$(C) for 
iron-based alloys, the thermodynamic limitation 

\begin{equation}
T_{0}(C) - M_{S}(C)  \ge  \arrowvert \Delta T \arrowvert_{min} \sim  
10^{2} K , 
\label{4.51}
\end{equation}
may not be ignored.

This limitation shows that, during a $\gamma -\alpha$ - MT, the point 
$M_{S}$ is set off by a limited quantity of about $10^{2}$ K, below the 
temperature $T_{0}$ of phase-equilibrium. In the case of a reduction of 
$T_{0}(C)$ with increasing C ($\partial T_{0} / \partial C < 0$), the 
condition $\partial M_{S} / \partial C > 0$ would contradict with the 
limitation \eqref{4.51}, thus all functions $M_{S}^{i}(C)$ must decrease with 
increasing C, as can actually be observed in the cases of Fe-Ni-, Fe-Mn- and 
on Fe-C-alloys. However, if $T_{0}(C)$ increased with increasing C 
($\partial T_{0} / \partial C > 0$), then the $M_{S}^{i}$ could either 
increase with increasing C (Fe-Co-, Fe-Al-alloys), or decrease 
(Fe-Cr-alloys), without contradicting to \eqref{4.51}. At the same time, the 
high level of $\partial f / \partial \mu^{\,\prime} $ remains conserved, while, 
in an alloying-model with diagonal disorder, the choice among these two 
possibilities will be determined by the quantity $\Delta Z$, being the 
difference among the charge-numbers of the alloying - components. For small 
$\Delta Z  \le 0,1$, an increase of $M_{S}$, and for large $\Delta Z > 
0,3$, a decrease of $M_{S}$ would be favorable.

As to the general case, instead of a $\Delta Z$ - comparison, we have to 
compare, for various alloys, the derivatives $\partial \Gamma 
(C) / \partial C$. This way the relations (instead of inequalities \eqref{4.46})

\begin{displaymath}
\frac{\partial \,\Gamma }{\partial \,C} \Big\arrowvert_{Fe - Co} < 
\frac{\partial \,\Gamma }{\partial \,C} \Big\arrowvert_{Fe - Ni} < 
\frac{\partial \,\Gamma }{\partial \,C} \Big\arrowvert_{Fe - Mn} 
\end{displaymath}
are materialized in an alloying model with diagonal disorder.

Thus our presentation of the concentration-relationship $M_{S}(C)$ opens up 
the possibility of sustaining a high level of occupational inversion, in a 
wide range of variations of temperature and concentration of 
alloying-elements, thus delivering a simple explanation for the reasons of 
the different characteristics of the $M_{S}(C)$ functions of iron-based 
alloys, including alloys with elements marking off the limits of the region 
of existence of the $\gamma $ - phase (a more comprehensive list of elements 
than that mentioned in Chapter 4 is given in \cite{Kaufman61}).

The data on which Chapter 4 is based were published in \cite{Kashchenko82,Kash80,Eyshinskiy83}. The 
analysis of the electronic configurations of iron-atoms in Fe - Ni - alloys, 
published in \cite{Eyshinskiy83}, has been used partially \cite{Eishinskii1984}.

\chapter{Interpretation~of~a~variety~of~characteristic morphological 
features of martensite within the notion of a phonon-maser}

In iron-based alloys, the structural product of the $\gamma  - \alpha 
$ - transformation, mainly bcc or bc-tetragonal martensite, exhibits a large 
variety of morphological features (see Pt. 1.2). This morphology undoubtedly 
contains valuable information on the dynamical mechanism of the martensitic 
transformation (MT). Its disclosure will thus be essential in our search for 
potential methods of externally influencing a MT. Thus it would be helpful 
to define certain models which will then enable us to gather such 
information. The notion of a phonon-maser (see Chapter 2.3) focuses on 
non-equilibrium conditions emerging in the adjacency of the boundary of the 
growing phase, as well as on the possibility of generation of 
lattice-displacement waves with amplitudes being sufficiently large to 
initiate the necessary plastic lattice deformation. This enabled us to 
interpret the growth-stage of martensite as a process being controlled by 
certain waves propagating through the premartensitic lattice phase. It 
should be emphasized that our task to explain the mechanism of wave 
generation includes the disclosure of an inherent link between the waves and 
some specific particularities of the electronic spectrum. Within the notion 
of the wave-model, our interpretation of morphological features of 
martensite will define such a link to observable macroscopic features (i.e. 
habit-planes), which will then allow us to analyze the physical causes of 
changes being linked up with variations of composition or of external 
boundary conditions.

\section{Habit-planes in Fe-Ni, Fe-C-alloys}

The qualitative pattern of definition of a habit-plane, which represents a 
fundamental morphological feature of martensite, has already been outlined 
in Sub-Pt.7 of the definition of our task under Pt. 1.5. Within our aim to 
resolve this task, let us start with a qualitative matching of 
characteristic lattice-planes and waves.

\subsection{Matching of a plane with a pair of waves}

It is generally possible, in principle, to link up planes with waves: 
Firstly, by directly associating them with the wave-front of a plane wave, 
and secondly, by associating them with the geometrical pattern of those 
points defined by a frozen-in trace of a moving line of intersection of two 
plane wave-fronts, propagating into different orientations. As the first 
case is rather trivial, let us consider more in detail the second case: Let 
$\textbf{c}_{1}$ and $\textbf{c}_{2}$ be two non-parallel wave-velocity 
vectors, taking into consideration that their vector-product $\lbrack \textbf{c}_{1 
}, \textbf{c}_{2} \rbrack$ is collinear with the line of intersection of the 
wave-fronts, which propagates with velocity $\textbf{c}$. Then we will be able to 
define a vector $\textbf{N}$ , being collinear with the normal vector of the 
plane defined by the vectors $\lbrack \textbf{c}_{1}, \textbf{c}_{2} \rbrack$ and 
$\textbf{c}$, by the following vector-product

\begin{equation}
\textbf{N} = \lbrack  \textbf{c}, \lbrack \textbf{c}_{1}, \textbf{c}_{2} \rbrack \rbrack = 
\textbf{c}_{1}\,\textrm{c}_{2}^{2}  -  \textbf{c}_{2} \,\textrm{c}_{1}^{2}. 
\label{5.1}
\end{equation}

Relation \eqref{5.1} immediately resolves the task of determination of $\textbf{N}$ 
for given $\textbf{c}_{1}$, $\textbf{c}_{2}$. The resolution of the inverted 
task however (i.e. the search for a pair of waves associated with a given 
$\textbf{N}$), is ambiguous from the outset. Nonetheless, with the aid of 
\eqref{5.1}, it is possible to determine conditions limiting the degree of 
ambiguity. As $\textbf{N}$ is a linear combination of the vectors 
$\textbf{c}_{1}$, $\textbf{c}_{2}$, the vectors $\textbf{c}_{1}$, 
$\textbf{c}_{2}$ define a plane. Thus their vector triple product will 
vanish, for any factorial combination of them:

\begin{equation}
\textbf{N} \,\lbrack \textbf{c}_{1}, \textbf{c}_{2} \rbrack = \textbf{c}_{1}\,
\lbrack \textbf{N}, \textbf{c}_{2} \rbrack = \textbf{c}_{2} \,\lbrack \textbf{N}, 
\textbf{c}_{1} \rbrack = 0. 
\label{5.2}
\end{equation}
Inserting the factor $\textbf{c}_{2} \textrm{c}_{1}^{2}$ in \eqref{5.1}, we 
get a representation of \eqref{5.1} in the form:

\begin{equation}
\textbf{N}\: \Vert\: \textbf{n}_{1}\varkappa - \textbf{n}_{2}
\label{5.3}
\end{equation}
where 
\begin{equation}
\textbf{n}_{1} = \textbf{c}_{1} \,\textrm{c}_{1}^{-1}, \quad 
\textbf{n}_{2} = \textbf{c}_{2} \,\textrm{c}_{2}^{-1}, \quad
\varkappa = \textrm{c}_{2} \,\textrm{c}_{1}^{-1}. 
\label{5.4}
\end{equation}

\subsection{Habit (225)}

Getting over to an interpretation of the real habit-planes, we recall (see 
Pt. 1.2) that in Fe-C-, Fe-Ni-systems, the observed habit-planes are: Near 
 $\{5 5 7\}  \div \{1 1 1\}$ (up to 0,6 weight {\%} C, up to 29{\%} Ni), 
$\{2 2 5\}$ (0,6 - 1,4 weight .{\%} C), $\{2 5 9\} \div  \{3 \:\:10 \:\:15\}$ 
(1,4 - 1,8 {\%} C, 29 - 34{\%} Ni). In accordance with Pt. 7, Sub-Pt. 1.5 of the 
definition of our task, as well as with our conclusions from Chapters 2 and 
3, we shall define, in our description of the waves, pairs of longitudinal 
(quasi-longitudinal) waves, propagating close to the perpendicular orientations  
$\langle 001 \rangle$ and $\langle 110 \rangle$ of a fcc-lattice. This way, 
we can hypothesize that the martensitic transformation is initiated by waves 
propagating through the austenitic parent areas, delivering or exceeding the 
required threshold-deformation $\varepsilon  \sim  \varepsilon_{t}$ of 
austenite. In addition, the displacement of lattice atoms, being immediately 
linked up with the waves, feature just the kind of phase relations being required 
to perform a Bain-deformation (the geometric pattern of the Bain-deformation has 
been treated in Pt. 1.4, see also Fig. \ref{fig1}). 

Let us start with a consideration of martensite with habit planes 
$\{225\}$: Our particular interest in this habit group is based, on the 
one hand, on that the crystallographic theory encounters the greatest 
explanatory difficulties \cite{Kurdjumov77} with this group, and on the other hand, on the 
intermediate position of martensite with these habits, in relation to its 
carbon-concentration, which may facilitate a comparison of habit planes of 
martensite belonging to the first and to the third group.

From the outset of our consideration, we note that, if we choose the 
velocity of one of the waves propagating in direction $\textbf{c}_{1}\: \Vert\: 
\lbrack 00\bar {1} \rbrack$ ( Orientation $\Delta $: 4$^{th}$ - order axis of symmetry), 
then, according \eqref{5.2} and \eqref{5.3}, it will be possible to define the type 
(h h \renewcommand{\itdefault}{ui}\textit{l}) habit, where h, 
\renewcommand{\itdefault}{ui}\textit{l}$ > 0$, by choosing the second wave with a velocity 
$\textbf{c}_{2} \:\Vert \:\lbrack 11 \eta \rbrack$, using the condition

\begin{equation}
\varkappa = \frac{c_{2}}{c_{1}} = \frac{\renewcommand{\itdefault}{ui}\textit{l} - \eta
\;h}{h\;\sqrt{\,2 + \eta^{2}}}. 
\label{5.5}
\end{equation}
Moreover, if we choose waves with reasonable long wave-lengths (see the treatise
in Chap. 2), it may be justified to surmise that their role mainly consists in selecting 
those macro-regions (or meso-regions) comprising the most favorable transformation conditions. 
Under such a point of view, it suggests itself to perform our interpretation 
of the particular features of the inner structure of a martensite lamella 
(with a thin structure), using the notion of coordinated action of long- and 
short-waves. It may be reminded that a lamella with habit (225) consists of 
thin twinning lamella (110), among which the main axes $\lbrack 100 \rbrack$ and 
$\lbrack 010 \rbrack$ of Bain-deformation alternate in turn. Thus obviously, 
if we confined ourselves to a description of pure Bain-deformation, it would be 
justified to choose the velocity of one of the longitudinal waves $\textbf{c}_{1} = 
\textbf{c}_{\Delta}$ in direction $\lbrack  0 0 \bar{1} \rbrack$. The coordinated 
action of such wave with the short-length displacement waves will then result in
a tension of the lattice in the orientation $\lbrack 001 \rbrack$, being also required for Bain-deformation 
with compressive-axes $\lbrack 100 \rbrack$ and $\lbrack 010 \rbrack$. 
As to the wave along $\lbrack 11\eta \rbrack$, both type of twins have equal relevance 
to it, like in the previously discussed case. Furthermore, if the wave is of the 
longitudinal type, it will support the Bain-pressure alternating among the 
directions $\lbrack 010 \rbrack$ and $\lbrack 100 \rbrack$, the more, the closer 
$\eta$ is to zero, i.e. the closer $\textbf{c}_{2}$ is oriented towards $\Sigma$. 

\begin{figure}[htb]
\centering
\includegraphics[clip=true, width=.8\textwidth]{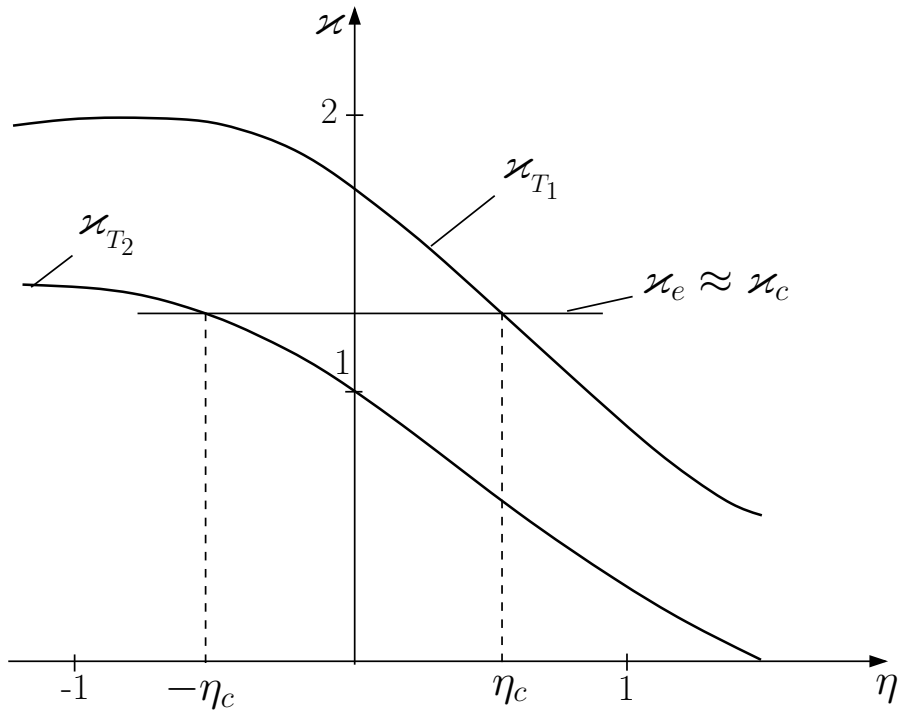}
\renewcommand{\captionlabeldelim}{.}
\caption{Graphic pattern for determination of the $\eta_{0}$, 
which determines the orientation $\lbrack 11\eta_{0}\rbrack$ 
of propagation of a quasi-longitudinal displacement-wave. 
Shown is the "critical" case: The value $\eta_{0}=\eta_{c}$ 
corresponds to the habit plane (225) and the value $\eta_{0}= - 
\eta_{c}$ to habit plane (557). Then functions $\varkappa_{T}(\eta)$ and 
$\varkappa_{e}(\eta)$ are defined in the text.}
\label{Gr pat for det eta}
\end{figure}

For a given alloy, the value $\varkappa_{0} = \varkappa (\eta_{0})$ - 
corresponding to habit (225) - is defined by the intersection between the 
curves determined experimentally $\varkappa_{e}(\eta)$ and theoretically 
$\varkappa_{T1}(\eta)$: $\varkappa(\eta _{0}) = \varkappa_{e}(\eta_{0}) = 
\varkappa_{T1}(\eta _{0})$. According to \eqref{5.5}, the curve 
$\varkappa_{T1}(\eta )$ descents monotonously - within the interval $0 < \eta 
 <  1$ - from $\varkappa(0)  \approx  1,77$ down to $\varkappa(1) \approx 0,87$. 
Obviously, an unequivocal solution $\varkappa_{0}$ exists within the interval $0 
< \eta  < 1$, if the path of curve $\varkappa_{e}(\eta)$ ascends 
monotonously, with an initial value $\varkappa_{e}(0)$ out of the interval: 
$0,87 < \varkappa_{e}(0) < 1,77$. As longitudinal waves normally satisfy the 
inequality c$_{\,\Delta} < \textrm{c}_{\,\Sigma} < \textrm{c}_{\,\Lambda}$, the requirement of 
monotonous ascent of curve $\varkappa_{e}(\eta)$ also is satisfied, in 
addition: $\varkappa_{e}(0) = \textrm{c}_{\,\Sigma} \,\textrm{c}_{\,\Delta}^{-1} > 1$, thus 
the only remaining restriction is $\varkappa_{e}(0) < 1,76$. Under 
consideration of the aforementioned restriction, which most certainly will 
be satisfied for the systems of our interest, there must exist an 
unequivocal solution. The results of our analysis are presented in the first 
part of Fig. \ref{Gr pat for det eta}, where the nearly horizontal curve $\varkappa_{e}$ 
represents the weakly pronounced dependence on orientation of a longitudinal 
wave during a transition from the $\Sigma$ - to the $\Lambda$ - orientation, 
resembling a 3$^{rd}$ order of symmetry axis. As already mentioned, those 
$\textbf{c}_{2}$ -orientations being close to the $\Sigma$ - orientation will 
find our greatest interest. Obviously from Fig. \ref{Gr pat for det eta}, this 
approximation occurs with increasing $\varkappa_{e}(0)$. 

\subsection{Habit (557). Criterion for transition from habitus (557) to 
(225)}

The affiliation of habitus (557) to the type (h h \renewcommand{\itdefault}{ui}\textit{l}), also being common with 
(225), and the possibility of their coexistence in the limit region of 
carbon-concentration near 0,6 {\%} C, suggest to describe these lamellae by 
a pair of waves $\textbf{c}_{1} \:\Vert\: \lbrack 00\bar {1}\rbrack$,
$\textbf{c}_{2} \:\Vert \:\lbrack 11 \eta \rbrack$, in a similar way as habitus (225). 
Even though such a description is possible, the solution of the equation 
$\varkappa(\eta_{0}) = \varkappa_{T2}(\eta_{0}) = \varkappa_{e}(\eta _{0})$ 
is defined in the negative region $-1 < \eta_{0} < 0$. In the left side of 
Fig. \ref{Gr pat for det eta}, the curve $\varkappa_{T2}(\eta)$ has been drawn in accordance with 
\eqref{5.5}, for the habit (557), which ascends monotonously from $\varkappa_{T2}(0) 
\approx  1$ up to $\varkappa_{T2}(-1) \approx  1,386$. The larger 
$\varkappa_{e}(0) > 1$ becomes, the more the direction of velocity 
$\textbf{c}_{2}$ will deviate from that of \,$\Sigma$. In other words, it 
behaves just the opposite way as in the case of habit (225). Obviously, 
there must exist a value $\eta_{c}$ (let us call it the critical value), 
for which $\varkappa_{T1}(\eta_{c}) = \varkappa_{c} = \varkappa_{T2}( - 
\eta_{c}) \approx \varkappa_{e}(0)$. This equation corresponds to 
$\eta_{c} \approx  0,55$, with $\varkappa_{c} \approx 1,285$. The 
quantitative criterion for separation of these cases, for which apparently 
the dynamical conditions for development of only one of the habits must be 
more favorable, has a simple structure:

\begin{equation}
\label{5.6}
\begin{array}{l}
(225) -- \varkappa_{e}(0) > \varkappa_{c},\\ 
(557) -- \varkappa_{e}(0) < \varkappa_{c}.\\ 
\end{array}
\end{equation}
Obviously, both habits can coexist in the region $\varkappa_{e}(0) \approx  
\varkappa_{c}$. Further assuming that an increase of carbon-concentration leads 
to a slow increase of $\varkappa_{e}(0)$, and that the value $\varkappa_{c}$ is 
attained near 0,6 {\%} C, then any further increase of $\varkappa_{e}(0) > 
\varkappa_{c}$ must be associated with a transition from habit (557) to 
(225). 

This specific effect of carbon can be explained in a qualitative manner: 
Those carbon atoms accumulating in voids \cite{Kurdjumov77} hybridize their 2p - functions 
with the 3d-functions of the iron atoms with $e_{g}$ - symmetry (as 
shown under Pt. 4.6), and thus reduce the effective positive charges Z of 
the ions, in the field of which the free 4s-electrons are moving. This must 
cause a local reduction of longitudinal velocity of sound, as - in 
accordance with the "jelly"-model \cite{Shriffer70} - the velocity of sound is \,$\textrm{c} \sim  
Z^{1 / 2}$. Presumably due to the anisotropy of the 2p-functions in 
Fe-C-systems, the increase of the parameter $\varkappa_{e}(0) = \textrm{c}_{\,\Sigma} 
\textrm{c}_{\,\Delta}^{-1}$ is mainly associated with the decrease of 
$\textrm{c}_{\,\Delta}$ with increasing carbon-concentration (note that any 
comparison of the $\textrm{c}_{\,\Lambda}$ - values of different alloys must be 
performed at the same temperature). In the more general case, a situation is 
also conceivable in which an increase of $\varkappa_{e}(0)$ is related to the 
more rapid increase of the velocity $\textrm{c}_{\,\Sigma}$ in relation to 
$\textrm{c}_{\,\Delta}$. 

Remarkably, the lack of a noticeable region of Ni-concentration for habit 
(225) in Fe-Ni-systems is in accordance with our criterion \eqref{5.6}. 
Experimental evidence reported in \cite{Haush73} shows that $\varkappa_{e}(0)$ can only 
attain the value $\varkappa_{c}$ for 33 {\%} Ni in ferromagnetic austenite. In 
this case, $\varkappa_{e}(0)$ attains the values $1,12 \div  1,15$ until 
reaching magnetic ordering, clearly being less than $\varkappa_{c}$.

Having applied this approach for the case of perpendicular waves 
$\textbf{c}_{1} \:\Vert\: \left[{\eta / 2 \;\: \eta / 2 \;\: \bar {1}} 
\right]$, $\textbf{c}_{2} \:\Vert \:\lbrack 1\, 1 \,\eta \rbrack$ - where $\textbf{c}_{1}$ already is 
non-parallel to $\lbrack 00\bar {1} \rbrack$ - and with consideration of the
variation of habit from (557) to (111), we get the results listed in Table \ref{table5.1}.

\begin{table}[htbp]
\renewcommand{\captionlabeldelim}{.}
\caption{Values $\varkappa_{c }$, $\eta _{c}$ , $\theta _{c}$ , for three pairs of 
habits}
\begin{center}
\begin{tabular}{|c|c|c|c|}
\hline
habit-pairs & $\varkappa_{c}$ &
$\eta_{c}$  & $\theta_{c}$ , Degree   \\   \hline
(225), (557)& 1,3083 & 0,1962 & 7,8966 \\   \hline
(225), (667)& 1,1923 & 0,2619 & 10,4913 \\   \hline
(225), (111)& 1,1061 & 0,3166 & 12,6197  \\   \hline
\end{tabular}
\end{center}
\label{table5.1}
\end{table}
Here, $\theta_{c}$ - angle between $\lbrack 11\eta_{c} \rbrack$ and $\lbrack 110 \rbrack$; 
$\eta_{c}$ being determined on the basis of equivalence of the velocity-modules
ratios $\varkappa = \textrm{c}_{2} \,\textrm{c}_{1}^{-1}$, for different habits of the types 
$(h_{1} \,h_{1}\, l_{1})$ and $(h_{2} \,h_{2}\, l_{2})$, as given by

\begin{equation}
\label{5.7}
\eta_c = \sqrt {\,2 + b^{2}} - b,\quad b = 
\frac{2\,h_{1} \,h_{2} + l_{1} \,l_{2}}{h_{1} \,l_{2} - h_{2} \,l_{1}},
\end{equation}
while $\varkappa_{c}$ can be determined by substituting $\eta = \eta_{c}$ 
in the expression

\begin{equation}
\varkappa = \frac{c_{2}}{c_{1}} = \frac{(l - h\,\eta)\sqrt{2}}{2\,h + l \,\eta}. 
\label{5.8}
\end{equation}

In (\ref{5.7}) and \eqref{5.8} the indices $l = l_{2} = 5$, $h = h_{2} = 2$ correspond to 
positive $\eta$. As can be seen from the data in Table \ref{table5.1} for 
perpendicular wave-pairs $\textbf{c}_{1}$ and $\textbf{c}_{2}$, an increase 
of $\varkappa$ supports the transition from the habit (557) to (225), in 
conjunction with a reduction of the angle of deviation $\theta$ among the 
axes of symmetry of second and fourth order. We also note that the deviation 
of the first wave from the axis $\lbrack 00\bar {1} \rbrack$ opens up the possibility to 
significantly reduce the deviation of the second wave from the axis 
$\lbrack 110 \rbrack$, in comparison with the case 
$\textbf{c}_{2}\: \Vert \:\lbrack 0 0 \bar {1} \rbrack$. 
Of course, for $\varkappa \approx \varkappa_{c}$, a coexistence of crystals with 
habits (557) and (225) is possible. This way, e.g. in \cite{Mirz81} a similar kind of 
coexistence has been observed during rapid quenching of steels with $< 0,6$ {\%} C

\subsection{Habits $\{15 \:3 \:10\} \div  \{ 9 \:2\: 5 \}$}

The characteristic feature of martensite with these habit-planes is 
associated with internal twinning. For the habits (15 3 10) and (9 2 5), the 
Bain-pressure axes in the twins correspond to $\lbrack 010 \rbrack$ and 
$\lbrack 001 \rbrack$. Thus, in analogy with habit (225), we have to choose 
as the first wave contributing to lattice expansion of both twins during Bain-deformation, 
the longitudinal wave with velocity $\textbf{c}_{1}\: \Vert\: \lbrack \bar {1}00 \rbrack$. 
Then we determine from \eqref{5.2} the velocity of the second wave: 
$\textbf{c}_{2} \:\Vert\: \lbrack \eta \:\: 0,3\:\:1 \rbrack$ and 
$\textbf{c}_{2}^{\prime} \:\Vert\: \lbrack \eta \:\: 0,4\:\:1 \rbrack$ (in the following, we mark with 
$\prime$ the results relating to the description of habit (9 2 5)). If we construct, 
on the basis of \eqref{5.4}, the curves $\varkappa_{T}(\eta)$ within a region $0  
\le \eta \le  0,3$, and $\varkappa_{T}^{\prime}(\eta)$ within a region $0 
\le \eta \le  0,4$, then we can easily convince ourselves that they 
descend monotonously, attaining the following values at the limits of those 
regions: $\varkappa_{T}(0) \approx  1,44, \:\:\varkappa_{T}(0,3) \approx  
1,1$; \,$\varkappa_{T}^{\prime}(0) \approx  1,67$,\, $\varkappa_{T}^{\prime}(0,4) \approx  1,22$. 
Obviously, there exist unequivocal solutions of the equations $\varkappa_{e}(\eta ) 
= \varkappa_{T}(\eta)$, $\varkappa_{e}(\eta) = \varkappa_{T}^{\prime}(\eta)$, 
corresponding to the experimentally determined curves $\varkappa_{e}(\eta)$,
$\varkappa_{e}^{\prime}(\eta)$, which ascend monotonously with increasing 
$\eta $, in the regions: $1,1 < \varkappa_{e}(0) < 1,44$;\, $1,22 < \varkappa_{e}^{\prime}(0) 
< 1,67$. Let us now take the data of \cite{Haush73} for Fe - 33{\%} Ni and use them to 
calculate the longitudinal velocities of sound, related to the axes of 
symmetry of cubic crystals \cite{Fedorov65}, in order to determine $\varkappa_{e}(0) 
\approx  1,1$, $\varkappa_{e}(0,3) \approx  1,22$ for the second wave: 
$\textbf{c}_{2} \:\Vert\: \lbrack 0,22 \quad 0,3 \quad 1 \rbrack$, 
$\textbf{c}_{2}^{\prime} \:\Vert\: \lbrack 0,34 \quad 0,4 \quad 1 \rbrack$. It 
should be noted here that the twins no longer can be treated equally in 
relation to the second wave, if compared with the case of habit (225). 
Obviously, the second wave delivers more pressure towards direction 
$\lbrack 001 \rbrack$. The experimental result published in \cite{Kurdjumov77} gives 
evidence for our conclusion that the thickness of the respective twins 
effectively is more in comparison to the width of the 
twins with Bain-pressure directed in axis $\lbrack 010 \rbrack$.

A very convenient form of notation of axes and planes, being particularly 
useful for the description of morphologic characteristics of twinned martensitic
crystals, can 
easily be obtained for any given habit, out of a totality 
\{h k \renewcommand{\itdefault}{ui}\textit{l}\} under the assumption 
$\arrowvert $h$ \arrowvert < \arrowvert $k$ \arrowvert < \arrowvert 
\renewcommand{\itdefault}{ui}\textit{l} \arrowvert$.

Then the direction $\langle 0 \:\:0\:\: \,\renewcommand{\itdefault}{ui}\textit{l} / \arrowvert 
\renewcommand{\itdefault}{ui}\textit{l} \arrowvert \, \rangle$ for 
 \{h k \renewcommand{\itdefault}{ui}\textit{l}\}-habits  $\{3\, 10 \,15\} \div   \{2 \,5\, 9\}$ (the position of the 1 
corresponds with the position of the largest modulus of the index 
\renewcommand{\itdefault}{ui}\textit{l}\,) will be collinear with the direction of wave-propagation 
with velocity $\textbf{c}_{1}$, which initiates the $\gamma -\alpha$ - transformation 
in the phase of tension. The axis $\langle 0 \:\:$k$ / \arrowvert $k$ \arrowvert \:\:0  \rangle$ 
(position of the 1 corresponds to the position of the intermediate module of 
index k) determines the direction of the main Bain-pressure axis and the 
axis $\langle  $h$ / \arrowvert $h$ \arrowvert\:\: 0\:\: 0 \rangle$ (the position of the 1 
corresponds to the position of the smallest module of index $h$) the direction 
of the Bain-pressure axis for the volumetrically smaller twin components. 
The twinning-plane can be noted in the form \{ h$/\arrowvert$h$\arrowvert \:\:
$k$ / \arrowvert $k$ \arrowvert \:\: 0 $\}. 

The same symbols can also be used for definition of the direction of the 
normals $\langle$ h k \renewcommand{\itdefault}{ui}\textit{l}~$\rangle$ to habits 
\{ h k \renewcommand{\itdefault}{ui}\textit{l} \}, in a stereographic projection, being 
commonly used in crystallographic analysis. Namely: the pole 
$\langle$ h k \renewcommand{\itdefault}{ui}\textit{l} $\rangle$ is located within a stereographic 
triangle, its corners 

\begin{equation}
\langle 0 \:\:0 \:\:\frac{\renewcommand{\itdefault}{ui}\textit{l}}
{\arrowvert \renewcommand{\itdefault}{ui}\textit{l} \arrowvert} \rangle,
\quad \langle 0 \:\:\frac{\renewcommand{\itdefault}{ui}\textit{h}}
{\arrowvert \renewcommand{\itdefault}{ui}\textit{k} \arrowvert} \:\:
\frac{\renewcommand{\itdefault}{ui}\textit{l}}{\arrowvert 
\renewcommand{\itdefault}{ui}\textit{l} \arrowvert} \rangle, \quad 
\langle \frac{\renewcommand{\itdefault}{ui}\textit{h}}{\arrowvert 
\renewcommand{\itdefault}{ui}\textit{h} \arrowvert}\:\: 
\frac{\renewcommand{\itdefault}{ui}\textit{k}}{\arrowvert 
\renewcommand{\itdefault}{ui}\textit{k} \arrowvert} \:\:
\frac{\renewcommand{\itdefault}{ui}\textit{l}}{\arrowvert 
\renewcommand{\itdefault}{ui}\textit{l} \arrowvert} \rangle 
\label{5.9}
\end{equation}
correspond with the projections of the axes of symmetry of the fourth, 
second and third order, respectively. Following \eqref{5.9}, e.g. the normal $\lbrack 15\:\: 
\bar {3} \:\:10\rbrack$ corresponds with the stereographic triangle $\lbrack 100 \rbrack$ - 
$\lbrack 1 0 1 \rbrack$ - $\lbrack 1 \bar{1} 1 \rbrack$. We can easily convince ourselves 
that there exists a totality of twenty-four orientational variants, all of them 
corresponding with the normals $\langle$ h k \renewcommand{\itdefault}{ui}\textit{l}
$\rangle$. By consideration of the notes in the previous section we see that 
the corner $\langle 0 \quad 0 \quad \renewcommand{\itdefault}{ui}\textit{l} / 
\arrowvert \renewcommand{\itdefault}{ui}\textit{l} \arrowvert \rangle$
of the triangle, being confronted with the \{ h k \renewcommand{\itdefault}{ui}\textit{l}\} - 
habit, indicates the direction of the austenitic expansion during formation of 
a martensite crystal. We also note that the corner $\langle \,$h$ / \arrowvert
$h$\arrowvert \quad $k$ / \arrowvert $k$ \arrowvert \quad 
\renewcommand{\itdefault}{ui}\textit{l} / \arrowvert 
\renewcommand{\itdefault}{ui}\textit{l} \arrowvert  \rangle$  
of the stereographic triangle points into the direction of the 
normal of the plane of densely packed austenite, out of a totality of four 
planes $\{1 1 1\}$, belonging to the orientational relationship of the 
lattices of the $\gamma$ - and $\alpha$ - phases (see Pt. 1.2), with respect 
to a materialized habit of the \{ h k \renewcommand{\itdefault}{ui}\textit{l}\} 
family, which also feature the smallest angle with the habit-plane in relation 
to the remaining $\{ 1 1 1 \}$ planes. The latter circumstance most likely 
is linked up with the minimization of free energy of the phase-boundary, 
during its gradually staged microscopic coupling, as commonly known for 
grain-boundaries \cite{Novikov75}.


\section{Grouping laws for packet-martensite crystals}

Research on lath- or packet-martensite is a highly demanding scientific
task. The papers \cite{Izotov1972,Marder69,Schastlivtsev72,Schastlivtsev74,
Schastlivtsev76,Izot72,Izotov75,Eterashvili78,Eterashvili79,Andreev77,Andreev83,Sandvic82} 
provide us with a relatively comprehensive and contemporary image on the 
development of concepts and the structure of packet-martensite. The habits of 
packet-martensite near to $\{557\}$ were firstly mentioned in \cite{Marder69}, 
for Fe - 0,2{\%} C-, Fe - 0,6{\%} C-systems (and confirmed in \cite{Schastlivtsev74,Sandvic82}), 
where the conformance of a pair of Miller-indices
only is approximated. Thus in reality, the normal of the habit-plane is
located within the stereographic triangle (see end of Pt. 5.1) corresponding
to its habit, and not at the border of two triangles. In our wave-pattern
used in Pt. 5.1 for a description of habits \{ h h \renewcommand{\itdefault}{ui}\textit{l} \} - $\{557\} \div 
\{ 2 2 5 \}$, an unequivocal orientation of the habits near $\{557\}$,
$\{ 2 2 5\}$ will arise if the directions defined by $\textbf{c}_{2} \:\Vert\:
\langle 1 1 \eta \rangle$ are "split" into pairs $\langle 1 \pm  \eta_{1} ,\:  1 \mp 
\eta_{2} ,\: \eta \rangle$ , $\arrowvert \eta_{1,2} \arrowvert < \eta$. Then we get
instead of \{h h \renewcommand{\itdefault}{ui}\textit{l}\}  the corresponding expression \{h $\pm  \delta
_{1}$ , h $\mp  \delta_{2}$, \renewcommand{\itdefault}{ui}\textit{l}\}, where 
$\arrowvert \delta _{1,2} \arrowvert < 1$. Using \eqref{5.3} and the relation 
$\textbf{c}_{1}\: \Vert\: \lbrack 0 0 \bar{1} \rbrack$,
$\textbf{c}_{2} \:\Vert\: \lbrack 1 + \eta_{1},\: 1 - \eta_{2}, \:\eta \rbrack$ we
can determine the vector $\textbf{N}$, being collinear to the normal of the 
habit-plane

\begin{equation}
N \:\Vert \:\lbrack \,1 + \eta_{1} ,\: 1 - \eta_{2} ,\: \varkappa\sqrt{( 1
+ \eta_{1})^{2} + ( 1 - \eta_{2})^{2} + \eta^{2}} + \eta \, \rbrack, 
\label{5.10}
\end{equation}
where $\varkappa = c_{2}\,c_{1}^{- 1}$. Then we can determine from \eqref{5.10}, 
e.g. for $\eta_{1} = \eta_{2} = 0,03$, $\eta = -\, 0,21$, $\varkappa =
1,2$, the habit (4,79 \,4,51\, 7), whose normal is located within the 
stereographic triangle $\lbrack 0 0 1 \rbrack$ - $\lbrack 1 0 1 \rbrack$ - 
$\lbrack 1 1 1 \rbrack$.

Let us further assume that, for initiation of a $\gamma -\alpha
$ - transformation resulting in packet-martensite, there are required certain 
phase-relationships among the relative displacements of the lattice atoms
associated with the waves, in a similar way as for lamellar martensite (see 
Pt. 5.1.4). Then we compare in the crystals with habits of the type $\{5 +
\delta_{1} , \: 7,\: 5 - \delta_{2}\}$, for $\delta_{1,2 } > 0$, 
the direction of expansion - $\langle 0 1 0 \rangle$ and the direction of 
compression $\langle 1 0 0 \rangle$, where positions 1 correspond with those 
habit-plane indices where the modules attain the largest or the mean value, 
respectively.

Further assuming that the strain-amplitudes of the waves are sufficiently
large to initiate plastic deformation, (see Chap.3), then it will be 
possible to determine the direction of macro-shear \textbf{S} (lacking the
correct sign), as the pair of waves propagating into directions $\langle 001 \rangle$, 
$\langle  1 + \eta_{1}, \:1 - \eta_{2},\: \eta  \rangle$, will prefer one of the
twelve slip-systems \{111\} $\langle 1 \bar{1} 0 \rangle$ of the fcc-lattice. 
The real selection will then be performed in accordance with the Schmid-relationships
for shear-stress $\delta_{c}$ (see e.g. \cite{Urusovskaia81}): 

\begin{equation}
\delta_{c} \sim \cos{ \varphi_{1}} \cos{\varphi_{2}}, 
\label{5.11}
\end{equation}
resulting from the requirement of maximum angular factor modulus \eqref{5.11} 
for the mechanical stress associated with the waves. $\varphi_{1}$, $\varphi 
_{2}$ in \eqref{5.11} are the angles between the direction of applied normal
stress (being close to the direction of propagation of the longitudinal 
wave) and the directions of the normal $\langle 1 1 1 \rangle$ to the slip plane
- \{111\} or shear $\langle 1 \bar{1} 0 \rangle$, respectively. As a result of this 
analysis, the habits  $\{5 \pm \delta ,\: 5 \mp \delta ,\: 7\}$ will naturally 
form groups of sextets \{ h k \renewcommand{\itdefault}{ui}\textit{l}\} close to the
common slip planes \{111\}. An example of such grouping is presented in Table
\ref{table5.2}, for $\eta < 0$. 
\begin{table}[htbp]
\renewcommand{\captionlabeldelim}{.}
\caption{Directions of wave propagation and some characteristics of 
packet-martensite (slip-plane (111))}
\begin{center}
\renewcommand{\arraystretch}{1.5}
\begin{tabular}
{|c|c|c|p{0.6in}|p{0.9in}|p{0.6in}|}
\hline
Nr. &  Habit  & 
Direction of $\textbf{c}_{2}$  &
Direction $\textbf{c}_{1}$  &
Compressive-axis  &
Direction \textbf{S}                      \\    \hline 
1  & $(5+\delta_{1} , 5-\delta_{2} , 7)$ & 
$[1+\eta_{1}, 1-\eta_{2}, \eta ]$ & 
\quad$[0 0 \bar{1}]$ & \quad\quad$[100]$ & 
\quad$[\bar{1} 0 1]$        \\   \hline
2  & $(5-\delta_{1} , 5+\delta_{2} , 7)$  &
$[1-\eta_{1}, 1+\eta_{2}, \eta ]$  &
\quad$[0 0 \bar{1}]$ & \quad\quad$[010]$  &
\quad$[0 \bar{1} 1]$         \\   \hline
3  &  $(7, 5+\delta_{1} , 5-\delta_{2})$  & 
$[\eta , 1+\eta_{1}, 1-\eta_{2}]$ & 
\quad$[\bar{1} 0 0] $ & \quad\quad$[010]$  & 
\quad$[1 \bar{1} 0]$            \\   \hline
4  & $(7, 5-\delta_{1} , 5+\delta_{2})$  &
$[\eta ,1-\eta_{1}, 1+\eta_{2}]$  &
\quad$[\bar{1} 0 0]$  & \quad\quad$[001]$  &
\quad$[1 0 \bar{1}]$             \\   \hline
5  & $(5-\delta_{1} , 7, 5+\delta_{2})$  & 
$[1-\eta_{1}, \eta , 1+\eta_{2}]$  & 
\quad$[0 \bar{1} 0]$ & \quad\quad[001]  & 
\quad$[0 1 \bar{1}]$              \\   \hline
6  & $(5+\delta_{1} , 7, 5-\delta_{2})$  &
$[1+\eta_{1}, \eta , 1-\eta_{2}] $ &
\quad$[0 \bar{1} 0]$  & \quad\quad[100]  &
\quad$[\bar{1} 1 0]$             \\   \hline
\end{tabular}
\end{center}
\label{table5.2}
\end{table}
This grouping corresponds to a package of 
crystals \cite{Schastlivtsev72,Schastlivtsev74,Schastlivtsev76} with six orientations 
close to the common plane (111), forming part of the orientational relationship. 
Obviously there exist three similar groups 
of crystals with six orientations in each group, comprising other common planes 
out of the totality of \{111\}. It is also obvious from Table
\ref{table5.2} that, if the crystals with six orientations are equally 
partitioned in such a package, then, during a martensitic transformation, the 
region occupied by such a package must be subjected to a quasi-isotropic volumetric 
change, as the directions of maximum compressive and tensile stress correspond 
to all three variants of Bain-deformation, while the alternating 
orientations of lattice-displacement \textbf{S} make possible a macroscopic
neutralization of totalized local displacements within a packet. In 
\cite{Schastlivtsev72,Schastlivtsev74,Schastlivtsev76}, just these rules were pointed out.

Due to a missing more stringent definition of the term "Packet-martensite
crystal" it came about that, apart from methodological problems, different 
results were obtained in the determination of the number of orientations
associated with martensite packet crystals. Thus in \cite{Izot72}, one or two, in 
\cite{Sandvic82} only one, and in \cite{Schastlivtsev72,Schastlivtsev74,Schastlivtsev76}, 
six orientations within a common plane
\{111\} were considered. If one describes the term "packet" as a  totality
of martensite-crystals (with habits near \{557\}) and with six orientations
close to a common plane out of \{111\}, as practically done in  
\cite{Schastlivtsev72,Schastlivtsev74,Schastlivtsev76}, then the
determined reduced number of orientations must be associated with a certain
local zone of the packet, which would comprise a  reduced number of
orientations. In \cite{Eterashvili79}, the existence of four zonal types was determined in
one packet. Type 1 represents triplets of orientations, i.e.  habit-triplets
of numbers 1, 3, 5, or 2, 4, 6 as shown in Table \ref{table5.2}. Type 2
represents pairs of orientations, i.e. habit-pairs 1, 4; 2, 5; 3, 6 as  shown
in Table \ref{table5.2}. Type 3 represents pairs of twin-orientations,
corresponding to pairs of habits 1, 2; 3, 4; 5, 6 as shown in Table
\ref{table5.2}. Type 4, with only one orientation, corresponds to any of the
habits shown in Table \ref{table5.2}. For equal volumetric fractions of
crystals with  different orientations in zone of type 1, a quasi-isotropic
change of volume in conjunction with neutralization of shear-deformation may
be possible. In  the zone of type 2 (as well as in the zones of type 3 and
4), the volumetric changes become anisotropic, but a neutralization of
displacement-deformation  still is possible. This however is no longer
possible for zones of type 3 and 4, that is becoming evident by comparison of the
\textbf{S}-directions in Table \ref{table5.2}. Obviously, the used definition of
the packet implicitly takes for granted that the transforming austenite
comprises a sufficiently large  volume. In other words, this definition is
critical with respect to the austenitic volume size. The results obtained
with pseudo-monocrystals in  \cite{Andreev77,Andreev83} fit well with the findings in
\cite{Schastlivtsev72,Schastlivtsev74,Schastlivtsev76}. In \cite{Izotov75}, 
being related to the case of coexistence of lamellae
with habits \{225\} and laths, a package is  defined as a totality of "blocks" composed
of lamellae and laths (each lamellae being connected with laths embodying
two twinning-orientations), in  such a way that within this definition, the
requirement of six orientations lives on for packet-martensite. A comparison
with the a.m. data \cite{Eterashvili79} shows  that, due to the presence of lamellae with
habits \{225\}, to which a dominant role during formation of packets is
assigned in \cite{Izotov75}, only the  zone of type 3 is segregated out of four types
of local packing zones in lath-shaped crystals. We also note that the
association of plane (111) in  Table \ref{table5.2} with the plane of
least slip resistance correlates well with the concepts in \cite{Kurdjumov77} on
stress-relaxation, during the isothermal  martensitic transformation by
shear-processes within the austenite-matrix surrounding the martensite
crystal. The orientation \textbf{S}, being precisely determined by the
Schmid-rule (with exception of the sign), correlates with the macroscopic
direction of displacement of a martensite crystal of this orientation
\cite{Izotov75,Eterashvili78}.
\vspace{\stretch{1}}

\section{Effect of the magnetic state of austenite and of an externally 
applied magnetic field on the martensitic $\gamma -\alpha$ - transformation }

\subsection{Possible causes for the transition of habit-planes}

After our matching of the habits with pairs of waves, it will be easy to 
estimate which parts of the electronic spectrum are actively involved in a 
martensitic transformation. Following such general characteristics, it is 
possible to associate crystals with habit-planes near \{557\}, 
\{225\}, as the pairs of waves, propagating near the axes of symmetry of 
the second and fourth order, can be put in relation with the electronic 
states being localized in the vicinity of the S-surfaces: $S_{\langle 001 \rangle}$, 
$S_{\langle 110 \rangle}$ (see Chap. 2). In the case of the habit (15 3 10), already 
mentioned under Pt. 5.1.4, the direction of $\textbf{c}_{2}$ defines an 
angle $\theta_{1} = 20,3$\r{ }, $\theta_{2} = 69,8$\r{ }, $\theta 
_{3} = 34,7$\r{ } in relation to the axes $[001]$, $[110]$, $[111]$. It is 
therefore justified to assume that the habits \{15 3 10\} need only to 
be matched with the waves propagating near the fourth-order symmetry-axes, 
being linked up with the $S_{\langle 0 0 1 \rangle}$ surfaces. Thus the causes for the first 
and second change of habits, i.e. $\{557\} \to \{225\}$ in Fe-C-systems, and 
for the second change, i.e. $\{225\} \to \{259\}$ in Fe-C-systems and  
$\{557\}  \to  \{3 \:10\: 15\}$ in Fe-Ni-alloys, apparently must be 
different. 

In the first case, no change of the class of waves governing the growth of a 
martensitic lamellae occurs; while the velocity-vector $\textbf{c}_{2}$ of 
the second wave remains within the directional cone near $\langle 110
\rangle$, and only the sign of the projection $\textbf{c}_{2}$ on the 
direction $\langle 0 0 \bar {1} \rangle$ (close to direction of propagation 
of the first wave (see Pt. 5.1.3)) changes. The causes for this behavior may be:
\begin{enumerate}
\item{Increase of the velocity-ratio $\varkappa = \textrm{c}_{2} 
\textrm{c}_{1}^{-1}$ with increasing carbon-concentration, leading to 
improved conditions for Bain-deformation for $\varkappa > \varkappa_{c}$ (see 
Table \ref{table5.1}) (i.e. to smaller threshold-deformation $\varepsilon 
_{t})$, in case of changing sign of the projection of $\textbf{c}_{2}$ on 
$\langle 0 0 \bar{1} \rangle$.}

\item{The appearance of a short-wave displacement component, being responsible 
for the thin (twinned) structure of macro-lamellae in the case of crystals with habits 
\{225\}. From a dynamical point of view, the appearance of short-wave 
displacement (of about half the wave-length with an order of magnitude of 
about the thickness of twins) at lower temperatures (related to the 
$M_{S}$ - temperature of lath-martensite) should be possible in principle, as 
a reduction of T will inevitably lead to a reduction of phonon-extinction 
and, consequently, also of the threshold-value $\sigma_{t}$ of an 
inverted occupational difference \\ (see \eqref{3.10} in Pt. 3.1). The latter argument fits 
well with the commonly used interpretation in \cite{Kurdjumov77}, stating that during a 
martensitic transformation, the relaxation of internal stress at low 
temperatures is mainly linked up with twinning-processes.}
\end{enumerate}
The second change of habits apparently is caused by a reduction of the 
activity of electronic states localized near the $S_{\langle 110 \rangle}$ surfaces, 
in relation to the states located near $S_{\langle 001 \rangle}$. The relative 
passiveness near $S_{\langle 110 \rangle}$ may be related to the magnitudes of the 
parameters $\Gamma$ and $\varepsilon_{d} - \mu$, at the surfaces 
$S_{\langle 110 \rangle}$ and $S_{\langle 001 \rangle}$.
We recall that $\varepsilon_{d}$  represents certain mean-value 
$\bar{\varepsilon}$ of the energies of the d-electrons on the S-surface and $\Gamma
$ is an extinction of s-electrons of energy $\varepsilon_{s} \approx  
\varepsilon_{d}$. As a common cause of the different values of the 
$\Gamma$ parameter at different S-surfaces, it is possible to define the 
anisotropy of s-d-scattering, the degree of which varying with concentration 
of the alloying elements. The variation of $\varepsilon_{d} - \mu$ 
for non spin-polarized states with $\varepsilon_{d} - \mu  > 0$ 
(see Pt. 4.5) is mainly caused by a the variation of the lattice-parameter 
$a$, whereas the rate of variation of $\varepsilon_{d} - \mu$ may 
differ for different S-surfaces, due to the non-correspondence of the rates 
of change of energies $\varepsilon_{d}$. If, for example, the 
tight binding approximation in \cite{Slater54} was used with consideration of the 
first and second neighbors, as well as the expressions of the 
matrix-elements $V_{ij} \sim  a^{-5}$ for the electronic transitions 
among the knots given in \cite{Kharrison83}, then it would easily be recognized that the 
characteristic energies (see Pt. 2.5) $\varepsilon_{X_{5}} ,\;\varepsilon
_{X_{2}} \,,\;\varepsilon_{L_{3}}$ decrease with increasing $a$, while the rate 
of decrease is largest for $\varepsilon_{X_{5}} $ and least for $\varepsilon 
_{L_{3}}$. For spin-polarized states, the value of $\varepsilon_{d \uparrow} - 
 \mu  < 0$ will also depend on the magnitude of exchange-splitting, which 
probably increases at temperatures below the magnetic ordering temperature, 
and in the case of anisotropic splitting, may be an additional cause of 
passivation of electronic states near $S_{\langle 110 \rangle}$ for the Fe-Ni-system 
(with $C_{Ni} > 29${\%}), in which magnetic ordering of the $\gamma$ - phase 
precedes the martensitic transformation.

\begin{figure}[htb]
\centering
\includegraphics[clip=true, width=.6\textwidth]{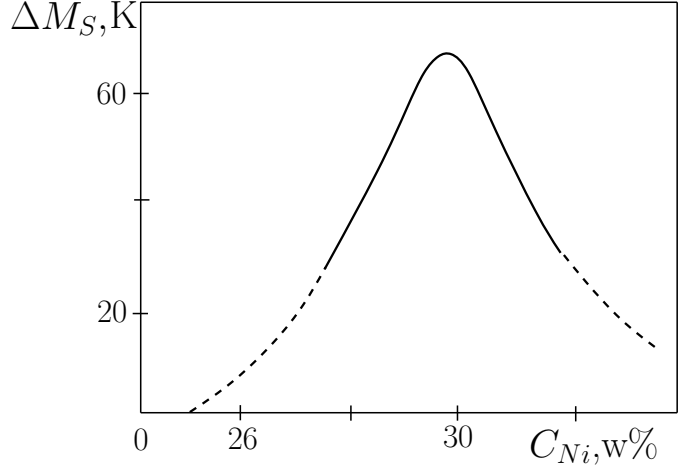}
\renewcommand{\captionlabeldelim}{.}
\caption{Relationship between the shifting of $M_{S}$ in a magnetic field of 
$2,6 \cdot 10^{7}$ A/m and Ni-concentration (weight {\%}) \cite{Krivoglaz77}}
\label{figure5.2}
\end{figure}

The influence of the magnetic state of austenite on reaction kinetics has 
been highlighted in \cite{Georgieva72} and \cite{Men'shikov75}, 
in connection with a qualitative 
consideration of the causes of a selective influence of a magnetic field on 
the kinetics of the $\gamma -\alpha$ - transformation. The most important 
law published in \cite{Krivoglaz77} is related to the effect of magnetic field bursts with 
field-strengths H up to $2,8 \cdot 10^{7}$ A/m ($\approx 350$ ke) on 
Fe-Ni-alloys. This law states that a magnetic field stimulates the athermal 
transformation into lamellar-martensite of Fe-Ni-alloys with 
Ni-concentration of more than 25 {\%}. In addition, such magnetic field 
burts cause a non-monotonous increase of the point $M_{S}$ by $\Delta 
M_{S}$, as depicted in Fig. \ref{figure5.2}. 

\begin{figure}[htb]
\centering
\includegraphics[clip=true, width=.6\textwidth]{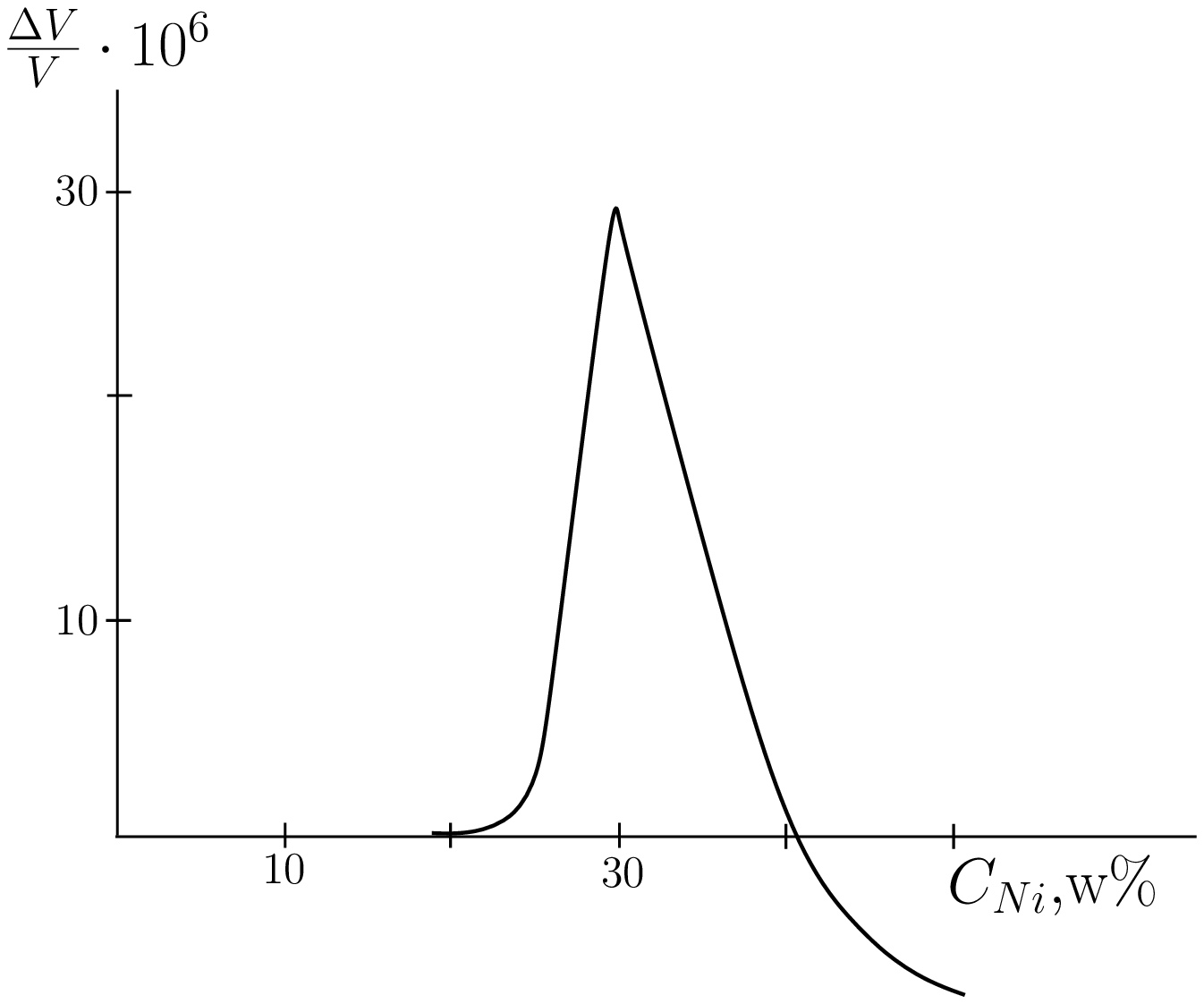}
\renewcommand{\captionlabeldelim}{.}
\caption{Relative change of volume due to magnetostriction in a magnetic 
field of $8,4 \cdot 10^{4}$ A/m [211] (Ni-weight {\%})}
\label{figure5.3}
\end{figure}
\begin{figure}[htb]
\centering
\includegraphics[clip=true, width=.6\textwidth]{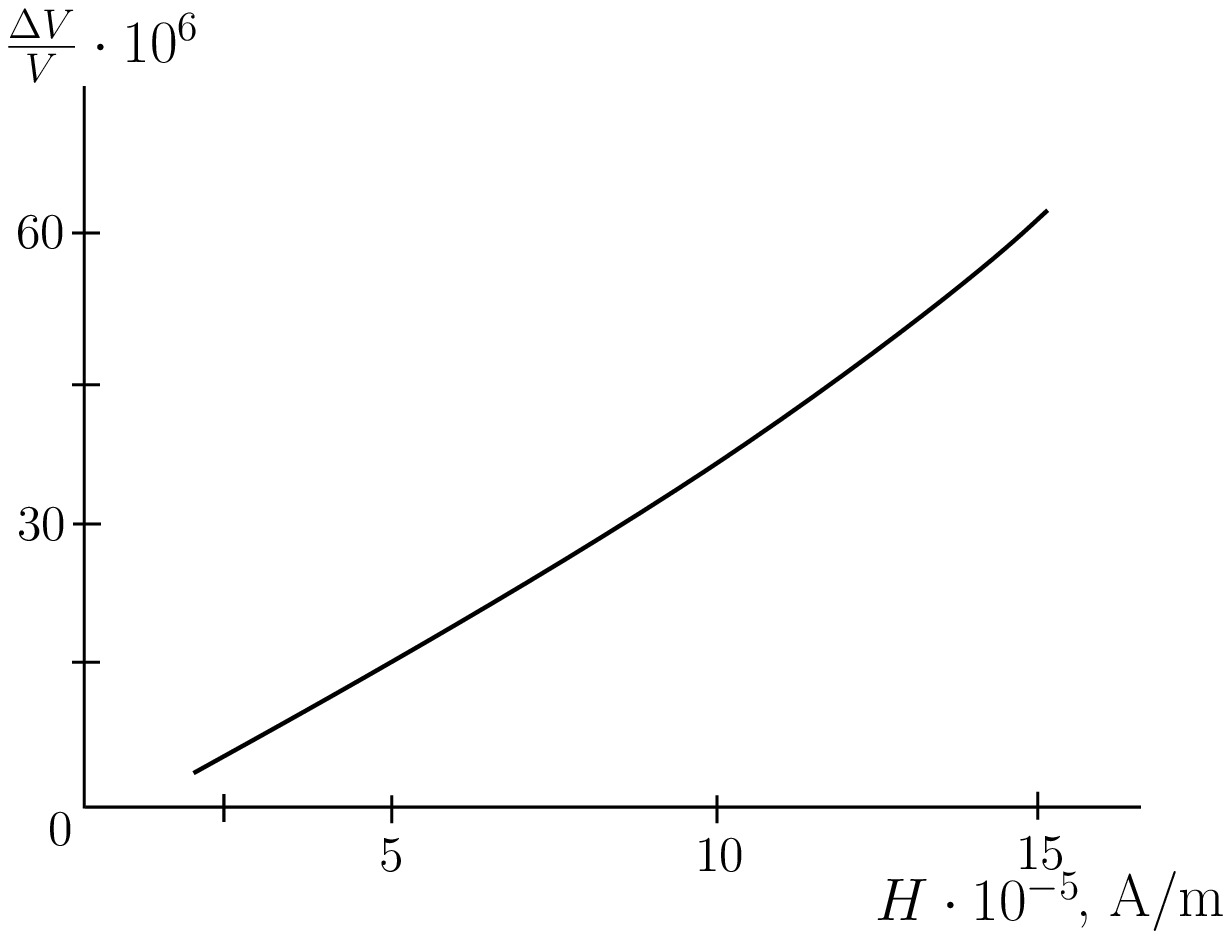}
\renewcommand{\captionlabeldelim}{.}
\caption{Relative change of austenite volume in the steel 40X2H20 
in a magnetic field up to $1,5 \cdot 10^{6}$ A/m , at temperature of 77 K 
\cite{Zolotarevskii79}}
\label{figure5.4}
\end{figure}

According to \cite{Krivoglaz77}, the positive shifting $\Delta M_{S} > 0$ is deducible from the larger inherent 
magnetization of the $\alpha$ - phase, and, consequently, from the larger 
incremental free-energy reduction of the $\alpha $ - phase in a magnetic 
field, in relation to the $\gamma$ - phase. Non-monotonous character of 
the shifting $\Delta M_{S}$ is attributable to the exceptionally large volumetric 
magnetostriction of the para-process defined in \cite{Zolotarevskii79,Kosenko79}, with 
consideration that the lower concentration limit of the non-monotonous 
magnetic field relationship corresponds with the lower limit of the $\Delta 
M_{S}$ - relationship (see Fig. \ref{figure5.3}, borrowed from \cite{Belov51}). A 
similar correlation also exists for steel with a super-paramagnetic $\gamma 
$ - phase \cite{Zolotarevskii79,Romashev82} as well as for Fe-Ni-Mn-alloys with $Fe_{ 70 + x} Ni_{ 
30 - 2x} Mn_{x}$ ($0 < x < 4$), which have been investigated in
\cite{Zolotarevskii81,Zolotarevskii83}. 
A linear extrapolation of the relationship between the volumetric increment 
$\Delta V/V$ and the quantity H, as shown in Fig. \ref{figure5.4} (borrowed from [209]), 
results in a value of $\Delta V/V \sim 10^{-3}$ for $H = 2,8 \cdot 
10^{7}$ A/m. The growth of specific volume in a magnetic field also 
correlates well with an increase of magnetization \cite{Zolotarevskii81}. This suggests that 
"quasi-ferromagnetic" ordering exists in a strong magnetic field, being also 
associated with the transition towards athermal transformation kinetics 
\cite{Men'shikov75}. The emergence of this state presumably is caused by magnetic-field 
induced exchange-splitting, which in turn results from the dependence of 
the exchange integral on the atomic distances (lattice parameter) or 
specific volume, respectively. As a reasonable basis for analysis and 
description of this effect, the exchange-striction model may be suggested 
(see e.g. \cite{Zavadskii80}).

\subsection{Single peak density of states (DOS) model, exchange-splitting and electronic 
redistribution }

\begin{figure}[htb]
\centering
\includegraphics[clip=true, width=.5\textwidth]{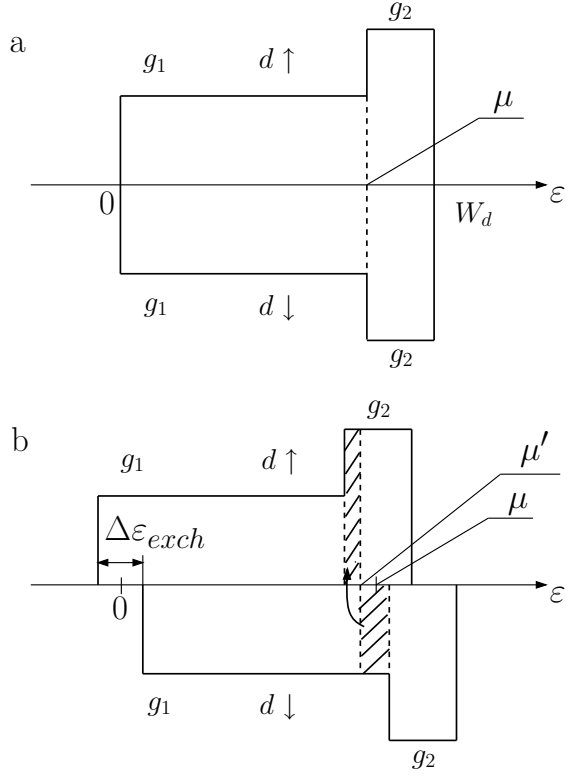}
\renewcommand{\captionlabeldelim}{.}
\caption{ Modeled (single-peak) DOS-function of d-electrons, in 
the paramagnetic (a) and ferromagnetic (b) phase}
\label{fig5.5}
\end{figure}

If the Fermi-level $\mu$ of the non-magnetic state of the $\gamma$ - phase 
of a Fe-Ni-alloy is located at the boundary of a step-like change of the 
DOS-function (right in front of the last peak of the density of states 
function of the d-band in Fig. \ref{fig2.9}), then the process of exchange 
splitting must be associated with an overflow of a part of the s-electrons 
into d-states \cite{Eyshinskiy81,Kash81}. By comparison of a modeled DOS with two 
characteristic values $g_{1}$ and $g_{2}$, as shown in Figs. 
\ref{fig5.5}a and \ref{fig5.5}b, it becomes evident that the 
exchange-splitting $\Delta \varepsilon_{exch}$ must lower the 
Fermi-level by forcing the spin-up states into states with greater 
occupational density $g_{2}$. The new Fermi-level $\mu^{\prime} $ of the sub-system 
of d-electrons (dashed line in Fig. \ref{fig5.5}b) can be determined using 
the condition of equality of the hatched areas for the sub-bands $d \uparrow 
$, $d \downarrow$ as shown in Fig. \ref{fig5.5}b. With consideration that 
the difference of the areas of the $d \uparrow$, $d \downarrow$ sub-bands 
- below the level $\mu^{\prime}$ - determines the atomic magnetic momentum 
$\bar{\mu}\,\mu_{B}^{-1}$ (expressed in units of Bohr's magneton $\mu 
_{B})$, then we get for the ferromagnetic phase

\begin{equation}
\label{5.12}
\Delta \,\varepsilon_{exch} = \frac{g_{1} + \,g_{2}}{g_{1} \,g_{2}} \, \frac{\bar{\mu} 
}{2\,\mu_{B}},\quad \quad \mu - \mu^{\prime} = \frac{g_{2} - g_{1}}{g_{1}
\,g_{2}}\,\frac{\bar{\mu}}{4\,\mu_{B}}.
\end{equation}
In the parabolic s-band, being characterized by a monotonously changing 
density of states $g_{s}$ (see Pt. 4.3), the variation of the 
Fermi-level $\mu = \mu_{s}$ of the sub-system, being related to 
exchange-splitting of the s-electrons, can be ignored, i.e., a separate 
consideration of the s- and d-band will deliver $\mu^{\prime}  = \mu_{d} \,<\, 
\mu_{s} = \mu$. If an equality of the Fermi-levels of d- and 
s-electrons was required, then it would be possible to estimate the number 
of electrons flowing over from the s- into the d-band (related to a singe 
atom):

\begin{equation}
\Delta n_{s}  \approx  2\,g_{s}(\mu)\, (\mu - \mu^{\prime}), 
\label{5.13}
\end{equation}
where the factor 2 accounts for both projections of the electron-spin (note 
that in Pt. 4.3 we have $g_{s}(\mu)$  with only one spin-projection). 
For a width of the s-band $W_{s} = 10$ eV, and $Z_{M} 
\approx 1$, we get from \eqref{4.27} $g_{s} \approx  0,12 eV^{-1}$ atom$^{ 
-1}$. Further assuming that the width of the d-band $W_{d} = 4,5$ eV, and 
the width of the rectangular peak 0,5 eV, $g_{1} \approx  1$ eV$^{-
1}$ atom$^{-1}$, $g_{2}  \approx  2$ eV$^{-1}$ atom$^{-1}$, $\bar 
{\mu}  \mu_{B}^{-1} = 1,7$, which, in accordance with \cite{Crangle63}, almost 
represents the alloy $Fe_{0,65} Ni_{0,35}$, then

\begin{equation}
\Delta \varepsilon_{exch} \approx 1,28 \: eV, \quad \quad \mu - \mu^{\prime} 
\approx  0,21 \: eV,\quad \Delta n_{s} \approx 0,05. 
\label{5.14}
\end{equation}
Of course this pattern features an illustrative character, in that it only 
gives us an estimated order of magnitude $\mu - \mu^{\prime}  \sim  0,1$ 
eV, $\Delta n_{s} \sim  10^{-2}$. Presumably, the value $\Delta 
n_{s} \approx 0,05$ is too large, as firstly, in the non-magnetic case 
shown in Fig. \ref{fig5.5}a, the shape of the DOS-peak, as well as the 
location of the Fermi-level, correspond to the maximum value of $\Delta 
n_{s}$, and, secondly, the relative shift of the d- and s-bands has not 
been properly taken into consideration. According to \cite{Snow69}, the transition 
from configuration $3d^{6} 4s^{2}$ to $3d^{7} 4s^{1}$ of the $\gamma 
$ - phase of iron is associated with an upward shift of the d-band by 5,5 eV - 
at an energetic scale - due to an increase of the repulsive energy-potential 
of the d-electrons. Then, due to an s-electron overflow of $\Delta n_{s} 
=  10^{-2}$ into the d-band, we get the following estimate for the upward 
shift of the d-band: $\Delta \mu^{\prime} = 5,5 \cdot 10^{-2}$ eV. If we 
further consider that for $g_{s} \approx  0,12$ eV$^{-1}$ atom$^{-1}$, 
an s-electron overflow of $\Delta n_{s} = 10^{-2}$ would reduce 
the value of $\mu = \mu_{s}$ by $4,2 \cdot 10^{-2}$ eV (using the 
same $g_{1}$ and $g_{2}$), we got a value of only $\Delta n_{s} \approx 
0,022$, being less than half of our initial estimate \eqref{5.14}. Thus, 
the reduction of the Fermi-level $\mu - \mu^{\prime}$ proves not to be 
greater than 0,11 eV. 

\begin{figure}[htb]
\centering
\includegraphics[clip=true, width=.6\textwidth]{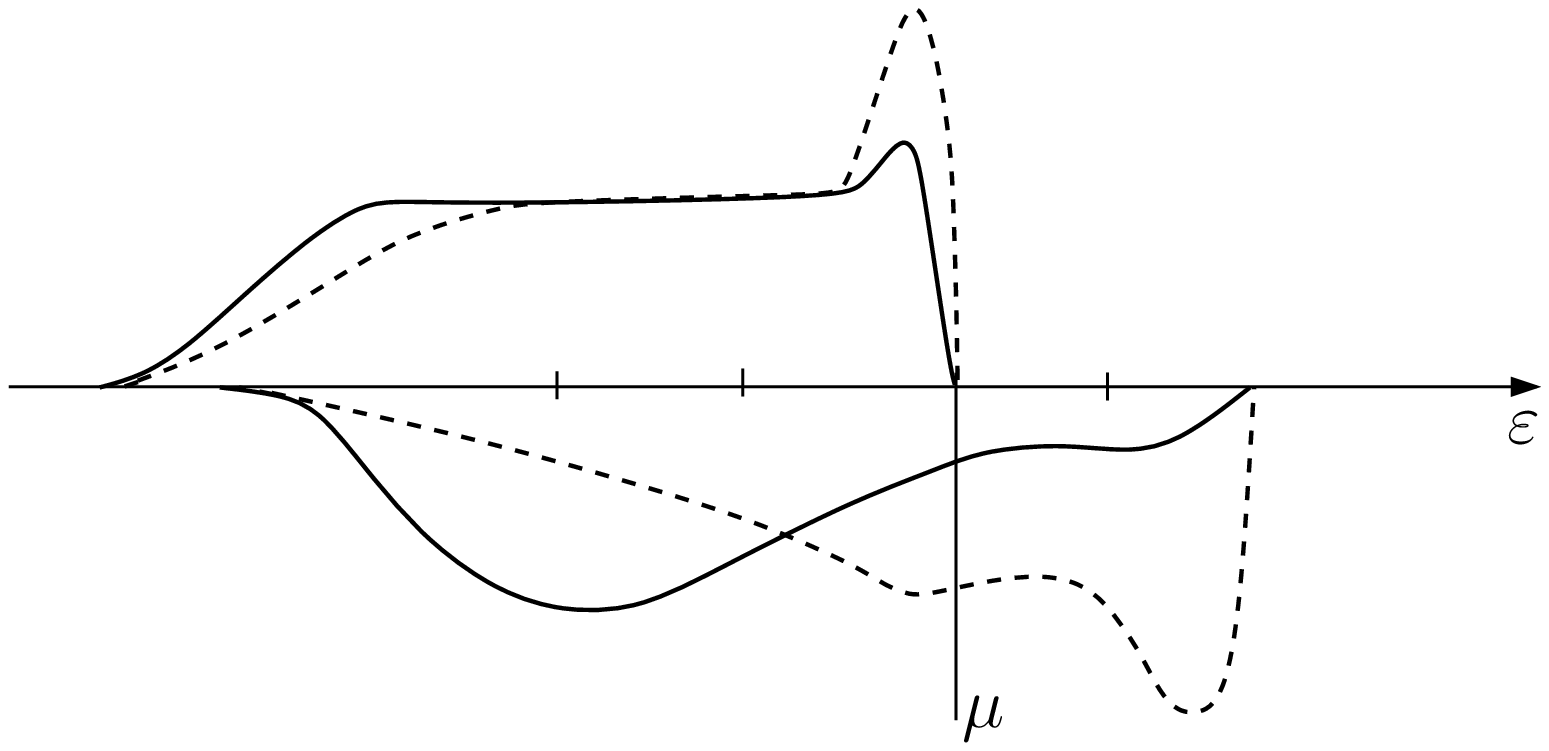}
\renewcommand{\captionlabeldelim}{.}
\caption{Contribution of iron - (dashed line) and 
nickel-atoms (solid line) to the single-peak DOS- function of the $Fe 
_{0,6} Ni_{0,4}$ - alloy \cite{Tikadzumi83}}
\label{fig5.6}
\end{figure}

Of larger interest however is the shift of the Fermi-level being related to 
the bottom of the d-band: In the ferromagnetic state, the bottom of the 
d-band is equal to the bottom of the $d \uparrow$ - sub-band. Let us 
designate by $\Delta \mu_{d \uparrow}$ the shift (towards the upper edge of the 
band). Then the quantity $\Delta \mu_{d \uparrow}$ can be determined, using the 
condition of equality of the hatched areas in Fig. \ref{fig5.5}b:

\begin{equation}
\Delta \mu_{d \uparrow} = \frac{g_{1} \,\Delta \,\varepsilon_{exch} }{g_{1} +
g_{2}} \approx \frac{1}{3}\,\Delta \,\varepsilon_{exch} \approx 0,43  eV, 
\label{5.15}
\end{equation}
where the value $\Delta \varepsilon_{exch}$ has been taken from \eqref{5.14}. 
A substantial difference among the values $\Delta \mu_{d \uparrow}$ and $\mu  
- \mu^{\prime}$, namely the weak dependence on the relative shift of the s- 
and d-bands, (note that the shift $\Delta \mu_{d \uparrow}$ is mainly caused by 
re-distribution of electrons within the $d \downarrow  \to d \uparrow 
$ sub-bands) as well as disappearing difficulties related to the selection 
of the origin of energy-reference, are arguments in support of $\Delta \mu  
_{d \uparrow}$, when compared with the measurement conditions of the parameters 
$\Delta \mu_{d \uparrow}$ and $\mu - \mu^{\prime}$. Obviously in our model, the 
heating process, being associated with the destruction of ferromagnetic 
ordering, will result in a relocation of the Fermi-level relative to the 
d-band, by about the same magnitude as \eqref{5.15}, linked with a shift 
corresponding to the inverted electron overflow $d \uparrow \to  d 
\downarrow (d \to s)$. These findings are in accordance with data 
presented in \cite{Zakharov76,Zakharov82}, in which a relationship between the Fermi-level and 
the temperature of invar-alloys has been observed, during photo-electronic X-ray 
spectroscopy. During heat-up of two Fe-Ni-alloys with about 34,87 atom-{\%} 
Ni and 35,8 atom-{\%} Ni, from 100 K to 600 K, the shift $\Delta 
\mu_{d \uparrow}$ towards the bottom of the d-band was about 0,51 eV and 0,36 
eV, respectively. Considering of the simplified character of our model, this 
correspondence of the measured values with our estimate by \eqref{5.15} appears 
fairly well.

We should however note here that, during magnetic ordering, the associated 
$s \to d$ redistribution of electrons must result in a reduction $\Delta 
C_{l}$ of the elastic modulus $C_{l}$, which, according to the 
Bohm-Staver-formula \cite{Shriffer70}, is proportional to the ion-charge-number Z, while 
a variation of Z must correspond with the quantity - $\Delta n_{s}$. 
Thus obviously

\begin{displaymath}
\Delta \,C_{l}  \approx  \frac{\Delta \,Z}{Z}\;C_{l},
\end{displaymath}
and with $\Delta Z  \approx  - \,0,02$, $Z \approx 1$, we get $\Delta 
C_{l} \approx - \,0,02 C_{l}$, being about 10{\%} of the of the 
elastic modulus reduction\footnote{The basic contribution to a softening of
elastic modulus gives the redistribution of d-electrons. So e.g. at
$\bar{\mu}\,=\,1,7\mu_B$ from states with small density $g_1$ into states with
major density $g_{2}$ \quad 0,85 electron/atom overflows. Then $\Delta Z/ Z \approx \Delta C
/C \approx -0,85/7$.} associated with the magnetic ordering process, as 
reported in \cite{Haush73}.

Despite their low amount, the number $\Delta n_{s}$ of electrons 
participating on the $s \to d$ - transition can be regarded as an 
informative parameter, due to the high sensitivity of the isomer-shift 
$\delta$ of the $\gamma$ - nuclear resonance absorption-line, in conjunction 
with to the $s \to d$ - transition. In particular, the $\gamma$ - resonance 
increases during $s \to d$  - transitions and decreases during $d \to s$ - transitions 
(see e.g. \cite{Vertkheim66}). If we now use the relationship between the isomer-shift of $Fe^{57}$ 
and the charge-density of the 3d- and 4s-electrons (see Fig.16 in \cite{Vertkheim66}), 
in the configuration $3d^{8 - z} 4s^{z}$, for  $Z \approx 1$, $\Delta Z 
 \approx - \,\Delta n_{s} \approx - \,0,022$, then we can 
determine that the isomer-shift increases by $\Delta \delta  \approx  
6,4 \cdot 10^{-5}$ m/s. This estimate matches fairly well within the 
interval of $\Delta \delta = (0,6 \div  1) \cdot 10^{-4}$ m/s, as 
reported in \cite{Iurchikov1971} for the isomer-shift in $Fe - (28 \div 75)${\%} 
Ni-alloys during magnetic transitions (however a somewhat less 
$\Delta \delta  = (4 \div  5)  \cdot 10^{-5}$ m/s has been reported 
in \cite{Hesse80}, for Fe-33{\%} Ni-alloys). 

Single-peak DOS-models have successfully been developed for the fcc-lattice 
structure of Fe-Ni-alloys (see e.g. \cite{Nakamura1976,Hasegawa71,Tikadzumi83}). Using a realistic (nearly 
triangular) shape of the peak, the following characteristics of such 
DOS-functions were obtained, on the basis of a coherent-potential 
approximation: 
\begin{itemize}
\item An appreciable difference among the DOS-functions for 
sub-bands $d \uparrow$ and $d \downarrow$, being evident in Fig.
\ref{fig2.9};
\item Changes of the shape of the sub-band DOS-functions 
(in most cases, the shape of the $d \downarrow$ - band) will vary 
with the composition of the alloy;
\item Splitting of the DOS into partial contributions of Fe - and Ni - atoms. 
An example for this splitting, pertaining to the $Fe_{0,6} Ni_{0,4}$ - alloy, 
borrowed from \cite{Tikadzumi83}, is shown in Fig. \ref{fig5.6}.
\end{itemize}
The relative success of the hard-zone model used for these estimates can mainly 
be deduced to the conservation of the shape of the $d \uparrow$ band for 
partial components of the iron- and nickel-atoms, as well as to the weakly 
pronounced variation of the DOS of the $d \downarrow$ band at the Fermi-level, 
with varying alloying composition. It has to be noted however that the considered 
mechanism of variation of the isomer-shift through $s \to d$ - overflow is
limited to such Ni-concentrations starting from which the process of $d \downarrow 
 \to  d \uparrow$ - electron redistribution doesn't cause an appreciable 
downshift of the Fermi-level of the d-electrons subsystem. According to 
the data in \cite{Tikadzumi83}, this particular concentration is about 40{\%} Ni, 
thus the process of $s \to d$ - overflow can contribute the most part to 
$\Delta \delta$, within the region of concentration of our interest of 
$29 \div 34${\%} Ni.

Our qualitative notion of these processes however deserves at least another 
remark, namely: Due to significant inhomogeneities of concentration and 
magnetic order (see e.g. \cite{Men'shikov77,Goman'kov79}), the transition into a magnetic state 
of order is remarkably blurred. Thus, with decreasing temperature, the 
structure of an inhomogeneity can vary from a quasi-paramagnetic up to a 
chaotically-frozen spin distribution, akin to a spin-glass (\cite{Kuz'min80}). 
According to \cite{Sidorov65,Iziumov70}, the cause of the emergence of a non-collinear 
magnetic structure is the blended character of exchange-interaction, being a 
common feature of many binary transition-metal alloys. For example, the 
exchange-integrals of pairs of Fe-Fe atoms in Fe-Ni - alloys are of negative 
sign, i.e. $J_{Fe - Fe} < 0$, and of Fe - Ni -, Ni - Ni- pairs, of positive sign, 
i.e. $J_{Fe - Ni} > 0$, $J_{Ni - Ni} > 0$. The notion of an 
anti-ferromagnetic character of the exchange interaction of Fe-Fe in iron 
and its alloys with fcc-lattice has for the first time been postulated in 
\cite{Kondorskii59}. The values of exchange-interactions $J_{Fe - Fe}= - \,9$ meV, 
$J_{Fe - Ni} = 39$ meV, $J_{Ni - Ni} = 52$ meV, as determined in \cite{Hatherly64}, are 
evidence in support of these concepts, i.e., below the Curie temperature 
$T_{c}$, strictly (collinear) magnetic structure not exists in the 
pre-martensitic austenite, thus the assumption of a ferromagnetic state of 
the fcc-austenite should only be regarded as a reasonable approximation. 

During various neutron-spectroscopic investigations on austenitic 
Fe-Ni - invar alloys (see e.g. \cite{Dubinin83,Dubinin85,Teploukhov85} 
and further references therein), 
there have been discovered super-structures of the $\gamma^{\prime}$,
$\gamma^{\prime \prime}$, $\gamma^{\prime \prime \prime}$ type. 
The first of them already appearing in the paramagnetic region $T > T_{c} > M_{S}$, 
the second one at $T \le T_{c}\quad (T > M_{S})$, being linked up with the onset 
of ferromagnetic ordering, while the third one appears at $T \le T_{N}$ 
($T_{N}$ - Neel-temperature, $T_{N} < M_{S}$) and reflects the onset of 
antiferromagnetic long-range ordering. We note that the $\gamma^{\prime
\prime}$ -phase has been considered as an intermediate phase in
\cite{Dubinin81,Dub81}, 
thus ensuring the necessary symmetry-link between the $\gamma$ - and the 
$\alpha$ - phases, which will finally lead to the Nishiyama orientational 
relationship of the $\gamma$ - and $\alpha$ - phases.
 
Evidence of a rather complex nature of the magnetic transition is also given 
by the results of \cite{Bukhalenkov83} (more detailed see \cite{Bukhal83}), 
in which two different stages of the magnetic transition of austenite have 
been determined, in a temperatures range of $0,8 \le T \cdot T_{c}^{-1} \le 1$, on 
the basis of a very detailed analysis of the isomer-shift of Fe-Ni-alloys 
with ($28,5 \div 31,3$) wt.-{\%} Ni. The first stage occurs in a temperature 
range of about $25 \div 30$ K below $T_{c}$ and is characterized by an 
increase of the isomer-shift of $Fe^{57}$ nuclei in the ferromagnetic areas of 
the fcc-phase, by about $(\Delta \delta )_{fm} \approx 5 \cdot 10^{-5}$ m/s, 
while the $\delta$-values of the paramagnetic areas remain unchanged, i.e. 
$(\Delta \delta)_{pm} = 0$. In the second stage, the isomer-shift decreases, 
exhibiting slightly larger incremental decreases of 
$(\Delta \delta)_{pm} \approx  - 6 \cdot 10^{-5}$ m/s in the paramagnetic 
areas than in the ferromagnetic areas, with 
$(\Delta \delta)_{fm} \approx - 4,5 \cdot 10^{-5}$ m/s. The positive shift 
$(\Delta \delta)_{fm}$ in the first stage matches fairly well with the 
estimates of the shift $\Delta \delta$ being associated with $s \to d$ - electron 
overflow. However, experimental results reported in \cite{Bukhalenkov83,Bukhal83}, 
which aimed at an explanation of the particularities of the second stage of 
the magnetic transition, by relating them to the increases of hydrostatic 
pressure during formation of an "infinite" ferromagnetic cluster within the 
system, provided $\Delta \delta$-values which would only correspond to an 
isomer-shift of about 15 to 20 {\%} of the observed values. Thus it cannot be 
excluded that the missing (main) part of the observed effect is related to 
the processes of $d \to d$ - electron redistribution between iron- and 
nickel-atoms, being mainly caused by the significant relative differences 
among their affected spectral density regions, i.e. where the partial 
contributions of the iron- and nickel-atoms are largest. Thus for example 
it is obvious from Fig. \ref{fig5.6} that the increase of exchange-splitting 
must lead to a reduction of the number of $d \downarrow$ - electrons of the 
iron-atoms, and to a corresponding increase for the nickel-atoms. This kind 
of $d \downarrow \to d \downarrow $ - electron redistribution among alloying 
components must result in a decrease of isomer-shift, as the decrease of the 
number of d-electrons of iron-atoms must elevate the charge-density of 
s-electrons at the nucleus \cite{Vertkheim66}. Such minor effects however, 
as well as those of a possible polarization of quasi-paramagnetic areas, 
in conjunction with the formation of internal or external antiferromagnetic 
order (being put in relation with the ferromagnetic areas), will not deserve 
further consideration in our further analysis.

\subsection{Oriented lattice-growth of athermal martensite in an externally
applied magnetic field}

The orientational effect of an externally applied magnetic field on the 
growth of martensite crystals mainly consists in a preferred orientation of 
the long axes of martensite crystals parallel to the magnetic field, as 
highlighted in \cite{Krivoglaz77,Bernshtein68}. The conclusions in \cite{Ermolenko68} 
on the orientational effect of a magnetic field had been drawn from 
investigations of the magnetic anisotropy of a specimen exposed to a magnetic 
field during and after a $\gamma -\alpha$ - MT. A thermodynamical analysis 
in \cite{Krivoglaz77} explains the predominance of small-angle orientations 
among martensite-lamellae in cubic crystals, mainly by attributing them to 
the minimization effect on the magnetic field energy contribution to the 
thermodynamic potential, being closely related to the proper magnetic field 
produced by the magnetic spin-moments of electrons in the ferromagnetic 
$\alpha$ - phase. As commonly known for the specific case of an ellipsoidal 
magnetic specimen, this energy resembles a quadratic function of all 
magnetization-components, the coefficients of which being proportional to 
the components of the demagnetization-tensor. For small-angle orientations, 
the proper magnetic energy is proportional to the square of magnetization 
and to the least of the three eigenvalues of the demagnetization-tensor, 
being many times less than 1 for small slices (lamellae). And vice-versa, 
for lamellae with large-angle orientation, the demagnetization factor is 
of an order of magnitude of 1, thus its contribution by proper magnetic 
energy becomes largest. It has however to be considered that the orientational 
influence of the demagnetization-fields, as well as that of magnetic anisotropy, 
on the growth of martensite crystals not exceeding $8 \cdot 10^{5}$ A/m 
$\sim 10^{4}$ Oe, dwindles with increasing externally applied magnetic field 
strength, finally disappearing for magnetic fields of $H  \ge  10^{7}$ A/m 
$ \sim  10^{5}$ Oe \cite{Zolotarevskii79}. Within this context, in case of 
strong magnetic fields, those factors which increase with increasing external 
magnetic field can come to the fore. Among them, there is the non-equilibrium 
addend to the electronic distribution function.

Let us assume that the athermal martensitic transformation would only be 
related with those electronic states being localized in the proximity of the 
$S_{\langle 001 \rangle}$ surfaces (see Pt. 5.3.1). Thus we shall consider the effect 
of a magnetic field \textbf{H} on the non-equilibrium addend $f - f^{0}$ 
of the electronic distribution function $f^{0}$. In accordance with 
\cite{Blatt71}, we can use the following approximation, being linear in H: 

\begin{equation}
\label{5.16}
f - f^{0} = - \,\Phi^{\prime}\:\frac{\partial f^{0}}{\partial \,\varepsilon},
\end{equation}

\begin{displaymath}
\Phi^{\prime} = (\tau \,\textbf{P} \,,\,\textbf{v}) \,+ \, \frac{e\,\tau\,}
{c_{0}\,\hbar}\: (\,\textbf{H},\,[ \textbf{v}\,,\vec{\nabla}]\,)\: ( \tau 
\,\textbf{P}\,,\,\textbf{v}),
\end{displaymath}
where e - electron charge; $\tau$ - average free path time of existence of 
an electron; $c_{0}$ - speed of light; $\vec{\nabla}$ - gradient-operator 
in \textbf{k} - space, while the explicit form of the vector \textbf{P},
including the temperature-, chemical and electrical potential-gradients, not 
being worth to be considered in our further analysis.

We shall further confine our analysis to the calculus of the contribution of 
non-equilibrium addends to the electronic distribution functions, 
related to states in the proximity of the reduced 1$^{st}$ face of the 
$S_{[001]}$ surface, which in this case corresponds with the square-shaped 
border of the 1$^{st}$ BZ. In our calculus of the group-velocity \textbf{v} 
and its derivatives, it will be justified to use a dispersion law in the 
tight binding approximation, which considers fairly satisfactory the 
particularities of the electronic energy spectrum within a weakly dispersed 
band, being enumerated by $X_{5}$ on sheet Nr.1. For example, we can obtain 
from \eqref{2.16} $\varepsilon = 4 \,\varepsilon_{1} = const$ along the line 
$X_{5} W$ (weak dispersion will only appear after consideration of the 
resonance-transition integrals with the second neighbor). Let
$\textbf{H} \:\Vert\: [001]$ and assume that the predominant 
contribution to the rate of change with \textbf{k} under the numeral of 
$\vec{\nabla}_{\textbf{k}} $ brings in \textbf{v}, then we get from \eqref{2.16} and 
(\ref{5.16}):

\begin{equation}
\label{5.17}
\Phi^{\prime} - (\tau\, \textbf{P}\,,\,\textbf{v}) = \frac{2\, e \,\tau^{2}\, 
\varepsilon_{1}^{2} \,a^{3}}{c_{0} \;\hbar^{3}}\:H\,( \,J_{x} + J_{y} + J_{z}\,),
\end{equation}
\begin{displaymath}
J_{x} = \sin{x}\, (\cos{x} + \cos{z}) \, (1 + \cos{z} \cos{y}) \,P_{y},
\end{displaymath}
\begin{equation}
J_{y}= - \sin{y} \, (\cos{y} + \cos{z}) \, (1 + \cos{z} \cos{x}) \,P_{x}, 
\label{5.18}
\end{equation}
\begin{displaymath}
J_{z} = \sin{z} \,\sin{x}\, \sin{y} \, (\cos{x} - \cos{y}) \,P_{z},
\end{displaymath}
\begin{displaymath}
\textrm{v}_{x} = 2 \,\varepsilon_{1} \,a\, \hbar^{-1} \,\sin{x} \, (\cos{y} + \cos{z}),
\end{displaymath}
\begin{equation}
\textrm{v}_{y} = 2 \,\varepsilon_{1} \,a \,\hbar^{-1} \,\sin{y}\, (\cos{x} + \cos{z}), 
\label{5.19}
\end{equation}
\begin{displaymath}
\textrm{v}_{z} = 2 \,\varepsilon_{1} \,a \,\hbar^{-1}\, \sin{z}\, (\cos{x} + \cos{y}).
\end{displaymath}

In \eqref{5.18}, \eqref{5.19} we used instead of the designations $\eta_{1}$, $\eta
_{2}$, $\eta_{3}$ in \eqref{2.16} the designations x, y, z, without index. 
During U-processes among states in the proximity of the sheets Nr. 1, the 
velocity component oriented normal to sheets Nr. 1 inverts its sign ( the 
$V_{x,y,z}$ in \eqref{5.19} include odd sine-functions), and the differences of 
the non-equilibrium terms $\sim (\tau\, \textbf{P}\,,\,\textbf{v})$ will sum up, which, in 
effect, leads to maximum inverted occupational difference (IOD) for 
ES-pairs. As a result of \eqref{5.18}, the field-terms containing the $J_{x}$, 
$J_{y}$ and also deliver an additional contribution to the IOD, being 
related to ES - pairs located in the proximity of the sheets Nr. 1 with normals 
$[100]$ and $[010]$, have the same characteristics. The term $\sim J_{z}$ 
however is odd with respect to the inversion $x \leftrightarrows y$, 
Therefore, ES - pairs near sheet Nr. 1 with normal $[001]$ do not contribute to IOD
as $J_{z}$ is integrated in the region which is symmetric with relation to 
the $x \leftrightarrows y$ inversion. Within usual definitions, this behavior of the 
field-supplements reflects transversal thermo-magnetic and galvano-magnetic  
effects. As a result, a field $\textbf{H} \:\Vert\: [001]$ will support 
the generation of waves with \textbf{q} near $[100]$ and $[010]$. Thus, 
in accordance with the two-wave pattern, there can be expected habits 
\{\renewcommand{\itdefault}{ui}\textit{l} k \underline{h}\}, whose smallest angle with 
\textbf{H} will be ($\arrowvert$ \renewcommand{\itdefault}{ui}\textit{l} 
$\arrowvert > \arrowvert$ k $\arrowvert > \arrowvert$ h $\arrowvert$), where the 
underlined indices mark their position being fixed, i.e.; \{\renewcommand{\itdefault}{ui}\textit{l} 
k~\underline{h}\} - (\renewcommand{\itdefault}{ui}\textit{l} k h), 
(k \renewcommand{\itdefault}{ui}\textit{l} h), (k \renewcommand{\itdefault}{ui}\textit{l}
$\bar{\textrm{h}}$) \ldots Accordingly, \{\renewcommand{\itdefault}{ui}\textit{l} h \underline{k}\} and 
\{ h k \underline{\renewcommand{\itdefault}{ui}\textit{l}} \} represent the 
totalities of the medium- and large angle habits. For $\textbf{H} \:\Vert\:
\langle 111 \rangle$, this orientational effect must vanish. This latter 
conclusion is important in order to clarify the causes of the observed 
orientations, as there also exist thermodynamical reasoning \cite{Krivoglaz77} in favor 
of the small-angled orientation of the first martensite lamellae, being 
independent of the orientation of \textbf{H}. The lack of an appreciable 
orientation at $\textbf{H} \:\Vert\: \langle 111 \rangle$, in relation to 
$\textbf{H} \:\Vert\: [001]$, would imply that the dynamic mechanism plays 
a decisive role. Of course, the effect of H on the generation of 
lattice-displacement waves becomes appreciable as soon as the 
magnitude of field-supplement becomes comparable with the initial supplement 
$(\tau\,\textbf{P\,,\,v})$. It is however more convenient to average the 
supplemental terms when performing such comparison, by integrating the 
expressions (\ref{5.17}) and \eqref{5.18} over those areas being parallel to faces Nr. 
1. Then their ratio will be

\begin{equation}
\label{5.20}
\psi \,\approx \,\frac{e\,\tau \,a^{2}\,\varepsilon_{1}}{5\,c_{0} \;\hbar 
^{2}}\:\frac{P_{y}}{P_{x}}\:H.
\end{equation}
For $P_{y}  \approx  P_{x}$ , $\tau  \sim  10^{-12}$ s, $a = 3,5 
\cdot 10^{-10}$ m, $\varepsilon_{1} = 5 \cdot 10^{-20}$ J 
(corresponding to the width of the band $\Delta \varepsilon  = 16 
\varepsilon_{1} = 5$ eV), $H = 8 \cdot 10^{6}$ A/m $ \sim  10^{5}$ Oe
(being typical experimental results published in \cite{Krivoglaz77}), we get from
(\ref{5.20}) $\psi \approx  0,2$. Thus the field-supplement is a non-negligible 
quantity. 

It has however to be emphasized here that the conclusion related to the 
transversal character of the thermo- and galvano-magnetic effects at 
$\textbf{H} \:\Vert\: [001]$, having been drawn during our previous consideration of 
the electronic states in the proximity of the $S_{\langle 001 \rangle}$ surfaces, will 
remain valid during an analysis of the non-equilibrium terms for states in 
the proximity of the curved sheets Nr. 2 of surfaces $S_{\langle 001 \rangle}$. To 
convince oneself, it will suffice to recall the definition \eqref{2.11} of a 
S-surface, taking into consideration that the vector $[\textbf{v}, \vec{\nabla }]$ 
appearing in the field-supplement (\ref{5.16}) is perpendicular to
$\textbf{v}$. Consequently, the scalar product $(\,\textbf{H}, [\textbf{v},
\vec{\nabla }]\,)$ vanishes for $\textbf{v}\: \Vert\: \textbf{H}$ attaining a maximum for
$\textbf{v}\:\bot  \:\textbf{H}$. This means that our conclusion of the preference of 
small-angled orientations of martensite lamellae for $\textbf{H}\: \Vert\: [001]$ also 
applies for the field supplements to the electronic distribution function f in 
the proximity of sheets Nr. 2 of the $S_{\langle 001 \rangle}$ surfaces, thus 
being a general conclusion.

In the case of magnetically ordered austenite, the local field-induction 
vector has to be used instead of vector \textbf{H} and also the effect of 
magnetostriction has to be taken into account as the emerging lattice deformations 
can modify the orientational effect. At $\textbf{H} \:\Vert\: [001]$ and for a 
magnetostrictive constant of $\lambda_{100} < 0$ there will be produced compressive 
stress in the orientation $[001]$ and, in the orientations $[100]$, $[010]$ equal 
magnitudes of tensile strains (see \S 2, Chap. 23 in \cite{Vonsovskii71}). Considering the
rule of matching between the compressive and tensile axes and the habit planes
indices, as discussed in Pt. 5.1.4, then there can also be expected medium-angled 
\{\renewcommand{\itdefault}{ui}\textit{l} h \underline{k}\} habits, in addition to the 
small-angled \{\renewcommand{\itdefault}{ui}\textit{l} k \underline{h}\} habits. It is 
worth to note that if \textbf{H} is oriented near $[001]$, so that a larger 
dilatation along one of the transversal axes is ensured, e.g. along 
$[010]$, then the emergence of \{k \underline{\renewcommand{\itdefault}{ui}\textit{l}} \underline{h}\} 
and \{h \underline{\renewcommand{\itdefault}{ui}\textit{l}} \underline{k}\}, with two 
fixed indices positions will be most likely for the small- and medium-angled habits . 
For $\lambda_{100} > 0$ (i.e. dilatation along $[001]$), also large-angled 
habits \{h k \underline{\renewcommand{\itdefault}{ui}\textit{l}}\} must arise, apart 
from the small-angled habits.

We emphasize that the above orientational effect of an externally applied 
magnetic field can be discriminated with satisfactory reliability from 
orientational effects related to uni-axial elastic stress (like that 
produced by magnetostriction or by externally applied stress), which would 
result in the appearance of medium-angled (for compressive stress) and large 
angled (for tensile stress) orientations of the lamellae in athermal 
martensite. 

In sufficiently strong magnetic fields, the contribution of a 
non-equilibrium electronic field-supplement to the orientational effect, 
which would have to become more pronounced with increasing \textbf{H}, can 
be discriminated from the background of statical orientational effects are caused
by the existence of demagnetization- and anisotropy magnetic fields, the general impact of 
which will decrease with increasing \textbf{H}.

In \cite{Leont'ev84}, the predominance of small-angled orientations of martensitic 
lamellae has been reported, based on reproducible experiments with samples 
of 50X2H22, 77X2H22 steel, being characterized by low autocatalysis of the 
$\gamma -\alpha$ - transformation, both in magnetic field bursts and in 
constant field conditions. In these cases, the orientational relationship is 
most pronounced if \textbf{H} is oriented parallel to the $\langle 001 \rangle$-axes. This 
matches fairly well with our above prediction drawn from our analysis of the 
effect of the non-equilibrium field supplement. Characteristic of this 
effect is that an increase of \textbf{H} over a critical magnitude of 
$\textbf{H}_{c} = 1,1 \cdot 10^{7}$ A/m (related to the steel 50X2H22, 
where $\textbf{H}_{c}$ is the magnetic field for which, at a given 
temperature $T_{exp} = 203$ K, the first martensite crystals emerge, i.e. 
$T_{exp} - M_{S} \approx 100$ K) results in an increase of the 
orientational effect, if $\textbf{H}\: \Vert\: [001]$. Similar results were obtained 
for the steel 77X2H22 (at $T_{exp} = 77$ K, with a critical field $H_{c}
\approx  8,76 \cdot 10^{6}$ A/m). Unfortunately, the results reported in \cite{Leont'ev84}, 
pertaining to $\textbf{H}\: \Vert\: \langle 111 \rangle$, only relate to the specific case 
$\textbf{H} = \textbf{H}_{c}$, thus the dependence of the orientational 
relationship on the magnetic field strength $\textbf{H}$ has not been 
investigated. In accordance with the above considerations, it can also be 
concluded that in the special case of $\textbf{H} \:\Vert\: \langle 111 \rangle$ 
an increase of \textbf{H} must lead to a reduction of the orientational effect. 
In general, the results presented in \cite{Leont'ev84} give evidence in 
support of our findings based on the notion of the control of the growth 
process of athermal martensite by longitudinal waves propagating near the 
$\langle 001 \rangle$ directions and, accordingly, also on the dominating role 
of the electronic states being localized near the $S_{\langle 001 \rangle}$ 
surfaces.

\section{Summary of chapter 5}

Following our interpretation of the morphological characteristics of 
martensite within the notion of a two-wave pattern, the main conclusions 
that we can draw here are:
 
\begin{enumerate}
\item{The habits of the Fe-Ni- and Fe-C-systems can be associated with a two-wave 
pattern, where one of the waves, belonging to the longitudinal type, 
propagates in the $\langle 001 \rangle$ direction, while the second wave is of a 
quasi-longitudinal type, which can be determined on the basis of 
experimental data of elastic-moduli at a given temperature $M_{S}$.} 

\item{The habits near \{557\} and \{225\} are characterized by the pairs 
of waves propagating near the perpendicular axes of symmetry of the fourth 
and second order $\langle 110 \rangle$, $\langle 001 \rangle$. 
By means of the criterion discriminating the most favorable conditions 
for materialization of habits \{557\}: $\varkappa = 
c_{\,\Sigma} c_{\,\Delta}^{-1} < \varkappa_{c}$ and \{225\}: $\varkappa = 
c_{\,\Sigma} c_{\,\Delta}^{-1} > \varkappa_{c}$ enables us, in principle, 
to explain the default of habit \{225\} in Fe-Ni-systems.}

\item{The habits near \{259\}, \{3 10 15\} are only associated with those 
waves propagating near the fourth-order axes of symmetry. Possible causes 
for the passivation of electronic states (located near the surfaces 
$S_{\langle 110 \rangle}$), being associated with wave-generation in the  
$\langle 110 \rangle$ - direction, can be the anisotropy of s-d-scattering 
processes and differences of the parameter $\varepsilon_{d} - \mu$, 
depending on the size of lattice-parameters as well as on the magnetic 
state of austenite.} 

\item{The large-scale growth process of martensite is being controlled by long 
displacement waves, acting in coordination with short displacement waves 
which form the fine structure. It is possible e.g. to describe the 
formation of a twinning-plane \{110\} analogous to that of a habit 
plane, in such a way that a twinning-plane resembles the imprinted path of 
the moving line of intersection of a pair of perpendicular short-wave 
fronts, including the pairs of waves propagating perpendicular to $\langle 001 \rangle$. 
This of course does not impede the description of the transformation of 
small areas within the concept of stationary waves, as short-wave 
displacements can perform a multitude of oscillations within half a period 
of the long-wave oscillations, whenever favorable transformation conditions 
materialize, thereby enabling the lattice to perform the transformation 
under optimized conditions. The oscillation period can easily be estimated 
if one considers that the widths of the macro- and micro-lamellae of the 
twins is within the order of magnitude of half of a wave-length each of the 
long and the short displacements.

Of course, the general possibility of twinning being driven by the need for 
periodic discharge of accumulated elastic strain energy, through lattice 
re-arrangement, is no contradiction to our two-wave lattice control pattern, 
which only emphasizes the main component.

On the other hand, the long-wave displacements, featuring a slower rate of 
decay than short-wave displacements, are effective carriers of information 
controlling the transformation of various regions of the specimen. Long-wave 
displacements thus can attain a leading role in the formation of crystal 
ensembles.}

\item{The knowledge of phase-relationships characterizing atomic displacements 
being most favorable for the initiation of the martensitic 
transformation by waves enables us to define rules related to the comparison of 
habits with the directions of compressive and tensile stresses of the 
Bain-deformation (based on a stereographic projection) (see end of Pt. 5.1.4). 
Using these rules it is easily possible, among other things, to determine the 
predominant orientation of martensite crystals, based on an orientational relationship 
with the unidirectional elastic stress field arising at temperatures above 
the $M_{S}$ - temperature.}

\item{The rules derived from observations on packet-martensite are in accordance 
with the conceptual view of martensite growth in which growth is controlled 
by pairs of elasto-plastic waves. Within this notion, the deviations of the 
habit planes of lath-crystals from the relevant pairs of Miller-indices, 
resulting in unequivocal orientation of martensite lamellae, can be put in 
relation with the orientational deviations from the $\{1\bar{1}0\}$ 
planes of the waves propagating near the $\langle 110 \rangle$ axes.}

\item{The new mechanism of controlled lattice growth of athermal martensite within 
an externally applied magnetic field, having been predicted after an 
analysis of the non-equilibrium field-supplement to the electronic distribution 
function, is confirmed by experimental analysis on the morphology of ensembles of 
martensitic mono-crystals, emerging within the austenitic single-crystals when 
being subjected to the orientational action of an external magnetic field.}
\end{enumerate}

\begin{center}
\textbf{Some additional remarks}
\end{center}

\begin{enumerate}
\item{The simple formalism establishing a link between the habit-plane and a given 
pair of waves can be used for a phenomenological analysis of the involved 
longitudinal and transversal waves, independently of causes and 
conditions of their emergence, like generation under non-equilibrium conditions, 
local lattice deformation associated with the nucleation of phases, or even 
excitation by external sources. To take an example reported in \cite{Kashchenko82}, 
experiments have been performed which aimed at an interpretation of habits 
of the type \{h h \renewcommand{\itdefault}{ui}\textit{l}\}, being observable during the $\beta   \to
\gamma^{\prime}$, $\beta^{\prime \prime}$ - martensitic transformation 
of copper and gold based alloys \cite{Varlimont80}, using a combined transversal and 
longitudinal two-wave pattern.}

\item{An experimental verification of the relationship between a two-wave pattern 
and the habits must, in the first step, comprise the determination of the 
$\varkappa$ - parameters for a given pairs of waves at $M_{S}$, as well as the 
selection of the preferable parameters. Of course it suggests itself to 
confine our measurements to those velocities of sound (for cubic crystals, 
at the orientations $\Delta$, $\Sigma$, $\Lambda$) being sufficient for 
determination of the elastic moduli. This will then enable us to calculate 
$\varkappa$ for any given pair of waves. In the second step, it will be possible, 
by means of a pair of external hypersound generators (\textit{$\nu$}$ \sim  10^{9} 
\div 10^{10}$ s$^{-1}$), to excite the selected pair of waves in the 
single crystal. This would raise, during a crystallographic analysis, the 
probability for discovery of a habit-plane (already being known at the 
outset). It should be noted here that the knowledge on the orientations of 
propagation and the types of the waves which materialize a thermodynamically 
relatively stable habit, should make it possible to perform such experiments 
even with low-power generators. Moreover, using high-power generators, it 
should be possible to achieve habits not occurring under natural 
conditions, in a given type of crystal. For example, a habit (225) in the 
Fe-Ni-system. Most promising would be an experiment for the case of a 
parameter $\varkappa$ being close to the critical value $\varkappa_{c}$. This way, 
using a predetermined orientation of $\textbf{c}_{1}$, e.g. 
$\textbf{c}_{1} \:\Vert\: [0 0 \bar {1}]$, it should be possible to materialize the 
cases shown in Table \ref{table5.3}, simply by varying the orientation of 
$\textbf{c}_{2}$.

\begin{table}[htbp]
\renewcommand{\captionlabeldelim}{.}
\caption{Miller-Indices of some habit-planes formed by a pair of waves, 
for a variety of orientations of the second wave velocity 
$\textbf{c}_{2}$ ($\textbf{c}_{1} \:\Vert\: [0 0 \bar {1}]$)}
\begin{center}
\begin{tabular}
{|c|c|c|}
\hline
Orientation of $\textbf{c}_{2}$ & 
$\eta$ & 
Habit                                    \\   \hline
$[1+\eta_{1},\, 1-\eta_{2},\, \eta]$  &  & 
$(5+\delta_{1} ,\, 5-\delta_{2} ,\, 7)$     \\  
$[1-\eta_{1},\, 1+\eta_{2},\, \eta]$  & 
$ - \arrowvert \eta_{c}\arrowvert  \le  \eta < 0$ & 
$(5-\delta_{1} ,\, 5+\delta_{2},\, 7)$      \\  \hline
$[1+\eta_{1},\, 1-\eta_{2}, \,\eta]$  &   & 
$(2+\delta_{1}\, , 2-\delta_{2}\, , 5)$       \\ 
$[1-\eta_{1},\, 1+\eta_{2},\, \eta]$  & 
$0 < \eta \le  \arrowvert \eta_{c}\arrowvert$  & 
$(2-\delta_{1}\,, 2+\delta_{2}\,, 5)$       \\   \hline
\end{tabular}
\end{center}
\label{table5.3}
\end{table}
A large variety of possibilities for exertion of direct influence on 
directed generation and growth of martensitic structures in austenitic 
single crystals is opened up by the combined application of magnetic, 
hypersound and conventional treatment. We also want to point at future 
potential for development of methods with two or three sources of ultrasound 
aiming at direct control of the direction of crystallization immediately out 
of the melting (upon cooling) and in the amorphous condition (upon heating), 
while these methods can be of interest both for stationary as well as for 
moving waves. In the latter case, it should be possible e.g. to produce of a 
chessboard-pattern structure with alternating amorphous and 
crystalline regions, by usage of two external sources of perpendicular 
waves.}

\item{The attractiveness of the two-wave pattern is mainly due to its potential to 
materially demonstrate the possibility of flexible behavior of a dynamic 
system, still being able to preserve the conditions for realization of 
thermodynamically favorable phase-coupling, in spite of variations of key 
parameters (in our case $\varkappa_{e}(0)$). The materialization e.g. of habit 
(557) could thus be attained due to the increasing deviation of the 
orientation of the velocity vector $\textbf{c}_{2}$ from the direction 
$\langle 110 \rangle$, occurring coincidentally with an increase of the parameter 
$\varkappa_{e}(0)$. As the energy being released during the transformation 
serves as the source of a non-equilibrium, those waves producing the most 
favorable conditions of generation (as needed for preservation of 
temperature- and chemical potential gradient) will develop and predominate. 
Due to the continuous feedback effect, the system will enter into a 
condition enabling it to flexibly follow and adapt itself to the most 
favorable thermodynamical path, on the one hand, by its ability to tune the 
spectrum of generated waves (i.e. mode selection), and on the other hand, to 
select various velocities $\textbf{c}_{2}$ out from an orientational cone 
around the $\langle 110 \rangle$ axis. Obviously, this point of view is free of 
contradictions with general thermodynamical reasoning and broadens them by an 
image of transformation dynamics.} 

\item{Within the approach combining martensite growth with wave generation by 
non-equilibrium electrons, it is obvious that the conceptual view of nuclear 
growth can be complemented by means of a soliton-like description of the 
motion of the phase-boundary, as the conditions required for their 
generation (i.e. temperature- and chemical potential gradients) exist in the 
proximity of the boundary.

The most important results relating to an interpretation of morphological 
characteristics of martensite, within the frame of a wave-pattern, have been 
published in \cite{Mints77},\cite{Kashchenko84},\cite{Kashchenko1982,Kash82,Kashchen82}. 
The re-distribution of electrons during the development of ferromagnetic 
ordering within a single-peak DOS environment is discussed in \cite{Eyshinskiy81,Kash81}, 
the results of which were used in \cite{Eishinskii84}.} 
\end{enumerate}

\vspace{\stretch{1}}

\chapter{Wave-model of motion for a martensitic lattice boundary}

The wave model under development for reconstructive martensitic 
transformations is essentially based on the conceptual view of the stage of 
growth of the new phase as a lattice deformation propagating in a wave-mode. 
As already discussed in Chap. 1, there exists sufficient experimental 
evidence for a metastable condition in the lattice of the pre-martensitic 
$\gamma$ - phase at $M_{S}$ temperature. This means the transition into the 
new lattice structure is inevitably associated with the need to overcome the 
interfacial energy barrier, for any degree of supercooling, only becoming 
possible if the magnitude of lattice deformation $\varepsilon$ exceeds a 
given threshold value $\varepsilon_{th}(M_{S})$. In our wave-model, we 
got $\varepsilon  = 4\, u \lambda^{-1}$, where $u$ - wave-amplitude and 
$\lambda$ - wavelength, further assuming $\varepsilon_{th}(M_{S}) \sim  10^{-3}$. 

Under a condition of strongly pronounced supercooling, the transformation of 
the austenitic lattice is associated with release of transformation heat and 
with a sudden (jump-like) increase of volume. From the outset, the emerging 
martensite crystal exerts a thermal, electrical and mechanical influence on 
the interfacial region $B_{\gamma - \alpha}$ between the phases, thereby 
causing a deviation of the local thermodynamic state from equilibrium 
conditions. Under such non-equilibrium conditions, amplification of atomic 
oscillations is generally possible, through the generation and selective 
amplification of displacement-waves by means of non-equilibrium electrons 
(i.e. instability of lattice oscillations by phonon-maser action). The 
displacement waves in turn ensure a lattice-deformation close to 
threshold-deformation. With regard to the aforementioned, it is obvious to 
consider the non-equilibrium region $B_{\gamma - \alpha}$ as an active and 
highly excited medium, being capable to generate, preserve and amplify 
cooperative waves, which in turn induce a transition at $\varepsilon   
\sim \varepsilon_{th}(M_{S})$. This way the phonon-maser effect can 
be defined as an effect based on positive feedback, by conversion of a 
fraction of the released transformation energy into the energy of 
displacement waves. Further surmising that these waves in turn ensure the 
overcoming of the energetic barrier, then it is conclusive to interpret 
martensitic lattice growth as a self-sustaining wave-like process, during 
which the transforming $B_{\gamma - \alpha}$ region propagates in 
conjunction with lattice displacement waves.

\vspace{\stretch{1}} 

\section{Coordinated propagation of a displacement-wave with a switching-wave 
of chemical potential or temperature}

We designate in general by a vector \textbf{u} the displacement of atoms 
participating on an arbitrary wave motion in a non-transformed lattice area. 
Let us further theorize that the wave is longitudinal with wave-length 
$\lambda  = 2\,\pi / q >> a$, where $a$ - lattice-parameter; q - modulus of the 
wave-vector $\textbf{q}$. Then, in a continuous lattice model, we can express 
the equations of motion for displacements \textbf{u} in the following way, 
using a frame of reference where the x-axis is collinear to \textbf{u}:

\begin{equation}
\label{6.1}
\ddot{u} - c^{2}\,u^{\prime \prime}_{xx} = \chi \,c\,u^{\prime}_{x},
\end{equation}

\begin{equation}
\chi = - \varkappa_{\textbf{q}} \,\left\{1 - \frac{\sigma_{0}}{\sigma_{th} 
} \left[1 + \frac{G_{e}^{2} \;t_{\sigma}}{\hbar^{2}\;\Gamma_{e}} 
( u^{\prime\,2}_{x} + q^{2}\,u^{2}) \right]^{-1} \right\}. 
\label{6.2}
\end{equation}
Here, c - velocity of sound; $\varkappa_{\textbf{q}}$ - wave-extinction without 
phonon-generation; $t_{\sigma}$ - period for emergence of an occupational 
inversion $\sigma_{0}$; $\sigma_{th}$ - threshold-inversion; $\Gamma 
_{e}$ - extinction of electrons active during generation; $G_{e}$ - 
resonance overlap integral, which determines the width of an electronic band 
in tight binding approximation. Eq. (\ref{6.1}) can most easily be derived by 
confining our consideration of phonon generation to single-mode (related to 
the case of perfect resonance) and to adiabatic approximation, leading to 
Eq. \eqref{3.49} for slowly varying phonon-field amplitudes $\tilde{b}_{\textbf{q}}^ 
{+}$, $\tilde{b}_{\textbf{q}}$. The expression of effective extinction
\eqref{6.2} is the same as in \eqref{3.49}. The transition to expression
\eqref{6.2} can easily be performed by insertion of the explicit form
\eqref{2.4} of matrix-element $W_{1}$ in \eqref{3.49}, using the relations

\begin{displaymath}
\tilde{b}_{\textbf{q}}^{+}  \approx  \left[\frac{M\;N\;\omega_{\textbf{q}}
}{2\;\hbar} \right]^{\frac{1}{2}} \exp{\left[ - i (\omega 
_{\textbf{q}}\,t - q\;x) \right]\;\left(\frac{\dot{u}}{i\;\omega_{\textbf{q}}} + u \right)},
\end{displaymath}

\begin{displaymath}
\tilde{b}_{\textbf{q}}  \approx \left[\frac{M\;N\;\omega_{\textbf{q}} 
}{2\;\hbar} \right]^{\frac{1}{2}}\exp{ \left[ i\;(\omega 
_{\textbf{q}}\,t - q x ) \right] \left(u - \frac{\dot{u}}
{i\;\omega_{\textbf{q}}} \right)},
\end{displaymath}

\begin{displaymath}
\frac{\partial \,u}{\partial \,t} \equiv \dot{u} \approx - c\,u^{\prime}_{x} 
\end{displaymath}
for the transformation of $\tilde{b}_{\textbf{q}}^{+}$, $\tilde{b}_{\textbf{q}}$ 
in \eqref{3.49}. 

Eq. (\ref{6.1}) can be resolved, provided the class of relationship among the 
parameters appearing therein and the parameters x and t can be 
substantiated. Most important is the relationship $\sigma_{0} = \sigma 
_{0}(x,t)$, whereas the relationships of the remaining parameters $(c, 
\varkappa_{\textbf{q}}, \sigma_{th}, t_{\sigma}, \Gamma_{e}, G_{e})$ with 
x and t is more or less irrelevant and can thus be ignored, without 
introducing any significant error. According to \eqref{3.23}, \eqref{3.24}, the quantity 
$\sigma_{0}(x,t)$ implicitly also depends on local temperature T, 
chemical potential $\mu$ and on their local gradients. Assuming steady 
motion of the phase-boundary with a velocity \textbf{V} $\sim \textbf{c}$, 
then it suggests itself to describe temperature and chemical potential by 
means of "solitary fronts" which propagate with the velocity of sound: i.e. 
switching-waves (see e.g.\cite{Rabinovich84,Marri83}). Thus the frontal width 
$l$ of such a switching wave practically determines the thickness of the 
transforming region $B_{\gamma - \alpha}$, being characterized by highly 
pronounced gradients $\nabla T$, $\nabla \mu$, which in turn satisfy the 
threshold-conditions $D_{0} \equiv \sigma_{0}\sigma_{thM}^{-1} - 1 > 0$ 
for phonon generation. Taking this into proper 
consideration, we can assume $\sigma_{0} = \sigma_{0}(x - $V$ 
t)$. As the amplification of a displacement-wave only is possible within the 
region of $B_{\gamma - \alpha}$, propagating at velocity V, a 
coordinated propagation of stationary switching waves and displacement waves 
can only be realized if V = c. Taking this for granted, 
then we can search the solution of Eq. (\ref{6.1}) in form of expression

\begin{equation}
u = u(x - c\,t)  \equiv  u(\xi). 
\label{6.3}
\end{equation}
After insertion of \eqref{6.3} in (\ref{6.1}) we get a set of equations

\begin{equation}
\label{6.4}
\frac{\partial \,u}{\partial \,\xi} = 0,\qquad \left(\frac{\partial 
\,u}{\partial \,\xi} \right)^{2} + q^{2}\;u^{2} - \frac{D_{0}(\xi)}{\beta_{0}} = 0,
\end{equation}
where $\beta_{0} = \hbar^{-2}\,G_{e}^{2}\, t_{\sigma}\, \Gamma
_{e}^{-1}$. The first equation in (\ref{6.4}) delivers a trivial solution, 
not matching with our expected spatially inhomogeneous distribution of local 
displacement at the starting moment. Solutions of the second equation of 
(\ref{6.4}) only exist for $D_{0} \ge 0$, i.e. in the $B_{\gamma - \alpha}$ 
region. If we characterize the region of $B_{\gamma - \alpha}$ by means of 
inequalities $-l  \le  \xi  \le  0$, further assuming that T and $\mu
$ linearly depend on $\xi$ - within the $B_{\gamma - \alpha}$ region - 
then we can use the following expression for $D_{0}$ 

\begin{equation}
\label{6.5}
D_{0}(\xi) = \left\{\begin{array}{l}
const_{1} \ge 0, \qquad \xi \in [-l, \,0]; \\
\\
const_{2} < 0, \qquad  \xi \notin [-l, \,0].
  \end{array} \right.
\end{equation}
Thus the solution of the second equation in (\ref{6.4}), being confined to the 
region $B_{\gamma - \alpha}$, will look as follows:

\begin{equation}
\label{6.6}
u\,(\xi) = \frac{1}{q}\,\left[ \frac{D_{0}(\xi)}{\beta_{0}} 
\right]^{\frac{1}{2}}\,\sin{( q\;\xi + 
\varphi)}.
\end{equation}
The value of the phase-term $\varphi$ in (\ref{6.6}) can be determined from the 
additional conditions imposed on u at the starting moment $t = 0$, at the 
boundaries of the segment $[-l,\, 0]$.

Solution (\ref{6.6}) is based on the requirement of equality of the velocities of 
the displacement-wave and of the switching-waves for T and $\mu$. Within 
this context it is reasonable to discuss here the limitations imposed on our 
model-parameters by the above requirement. According to \cite{Rabinovich84}, a 
switching-wave can be defined by a non-linear equation of the 
diffusion-type:

\begin{equation}
\label{6.7}
\dot{\psi} = F(\psi) + d\, \psi^{\prime \prime}_{xx},
\end{equation}
where d - diffusion coefficient, $\psi$ - dimensionless variable of the 
type $\psi = \psi_{T} = (T - T_{\gamma}) / (T_{\alpha}  - 
 T_{\gamma})$, or $\psi = \psi_{\mu} = (\mu  -  \mu_{\gamma})
  / (\mu_{\alpha}  -   \mu_{\gamma})$ and $F (\psi)$ - 
non-linear function. Assuming that $F(\psi) \ge  0$ and Eq. $F(\psi) = 0$ 
only has two solutions, namely $\psi  = 0$ and $\psi  = 1$, 
corresponding to the values of $\psi$ in the $\gamma$ - and 
$\alpha$ - phases, then we get for velocity V \cite{Rabinovich84}:

\begin{equation}
\label{6.8}
2\,(d\,F_{\psi}^{\prime}(0))^{\frac{1}{2}} = 2 \,\left(d\, \frac{d F}{d
\psi}\Bigg{\arrowvert}_{\psi = 0} \right)^{\frac{1}{2}} \equiv V_{\min} \le
\textrm{V}.
\end{equation}
In accordance with (\ref{6.8}) the equality V = c is possible for V$_{min} \le 
c$. We estimate V$_{min}$ by assuming that V$_{min}  \ne  0$. For it, 
$F(\psi)$ must be materialized. A function $F(\psi)$, being compatible with 
the conditions $F(0) = F(1) = 0$, $F^{\prime}_{\psi}(0)  \ne  0$, can always be 
expressed in the form

\begin{equation}
F = \tau^{-1}\, \psi \, (1 - \upsilon (\psi)), 
\label{6.9}
\end{equation}
where $\upsilon (\psi) = 1$ only for $\psi  = 1$ and $\upsilon(0) = 0$, 
$\tau$ - constant with temporal dimension. It follows from \eqref{6.9}, that 
$F^{\prime}_{\psi}(0) = \tau^{-1}$ and V$_{min} = 2(d / \tau)^{1 / 2}$:

\begin{equation}
\label{6.10}
V_{\min} = 2\,\left(\frac{d_{T,\,\mu}}{\tau} \right)^{\frac{1}{2}}.
\end{equation}
In the case of a temperature-wave (T-wave) the coefficient $d  \equiv  
d_{T}$ is the temperature- conductivity ($d_{T}  \sim  10^{-5}$ 
m$^{2}$s$^{-1}$ for iron-alloys). In the case of chemical potential ($\mu
$ - wave), the coefficient $d  \equiv  d_{\mu}$ can be determined in 
accordance with \eqref{3.46} by means of the specific conductivity $\sigma
_{\gamma}$ in the form of

\begin{displaymath}
d_{\mu} = \frac{2}{3}\frac{\sigma_{\gamma} \;\mu_{\gamma}}{e^2\;n_{\gamma}}\quad 
and \quad d_{\mu} \sim 4 \cdot 10^{-4}\;m^2 / s.
\end{displaymath}

Let us come back to the condition V = V$_{min} = c$ and estimate $\tau$ 
from (\ref{6.10}) for $c = 5 \cdot 10^{3}$ m/s. This delivers $\tau  \equiv  
\tau_{T}  \sim  2 \cdot 10^{-12}$ s, $\tau \equiv  \tau_{\mu}  \sim  6 \cdot 10^{-11}$ s. 
Using a linear approximation for the profile of the frontal region of the switching wave, 
we get for $l$ the following relationship with $\tau$, using the condition V$_{min} = c$: 
$l = A\, \tau\,  c$, where A - a figure not larger than a few units, depending 
on the form of expression $\upsilon(\psi)$ in \eqref{6.9}. For $\upsilon 
(\psi ) = \psi $, e.g., we have A = 4, according to \cite{Marri83}, thus $l = 
l_{T}  \sim  4 \cdot 10^{-8}$ m or $l = l_{\mu} \sim 10^{-6}$ m, respectively. 
Taking into account that the optimum size of the region being deformed by the 
displacement-wave corresponds to $\lambda / 2$ in the wave model, then we 
can assess the minimum value q in (\ref{6.6}) using 
the relation $l \sim \lambda / 2 = \pi /q$: $q = q_{T}  \sim 
 10^{-2} \,\pi / a$ for $l = l_{T}$ and $q = q_{\mu}  \sim  (10^
 {-4} \div  10^{-3}) \pi / a$ at $l = l_{\mu}$.

Strictly speaking, expression (\ref{6.5}), when used for resolving (\ref{6.4}), would 
require the following form of the function $\psi$:

\begin{equation}
\psi  = \theta( -\eta - 1) - \eta\: [\theta(\eta  
+ 1) - \theta(\eta)], 
\label{6.11}
\end{equation}

\begin{displaymath}
\eta = \frac{( x - c\,t )}{l} \equiv \frac{\xi}{l},
\end{displaymath}
which will only be satisfied after replacement of expression \eqref{6.9} for 
$F\,(\psi)$ by a rectangularly shaped pulse-function:

\begin{equation}
\label{6.12}
F\,(\psi) = \frac{1}{\tau}\:\theta \,(\psi)\,\theta 
\,(1 - \psi).
\end{equation}
In \eqref{6.11}, (\ref{6.12}) the function $\theta$ - Heaviside unit-function:

\begin{displaymath}
\theta \,(\psi) = \left\{\begin{array}{l}
 \,1\;,\quad \psi > 0, \\ 
 \,0\;,\quad \psi < 0. \\ 
 \end{array} \right.
\end{displaymath}
Such an approximation of the wave profile and of $F(
\psi)$ should be fully satisfactory for a qualitative consideration, 
after having determined the speed of propagation of the switching-wave.

\section{Coordinated propagation of a pair of displacement-waves with the 
switching-wave of $T(\mu)$}

\begin{figure}[htb]
\centering
\includegraphics[clip=true,width=.8\textwidth]{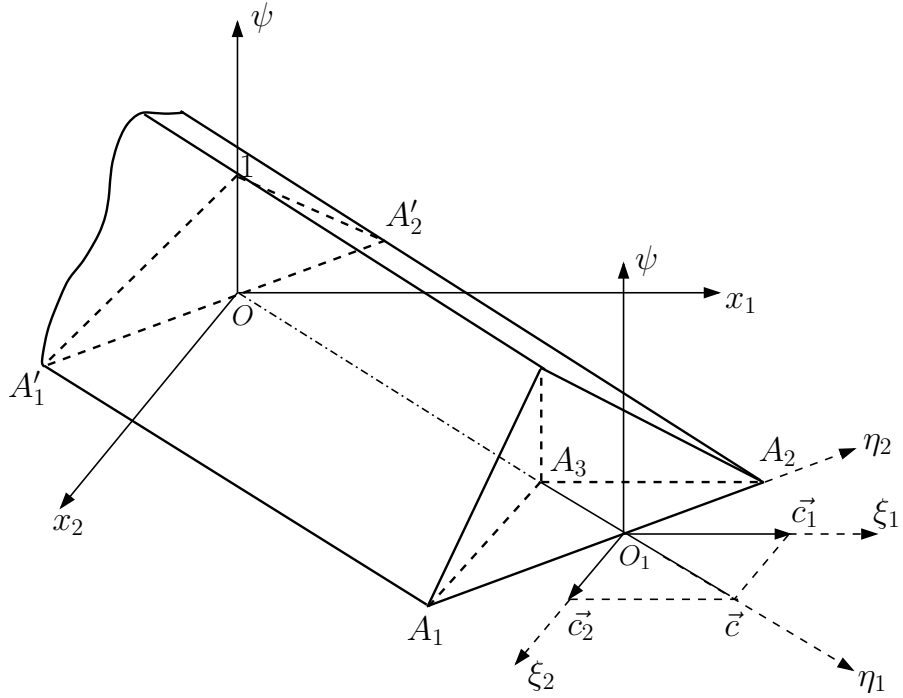}
\caption{Model of a switching-wave propagating in conjunction with 
two wave-beams. The coordinates of the points are: $A_{1}$: 
($\eta_{1} = 0$, $\eta_{2} = -\,c_{1} / c_{2}$ ), $A_{2}$: ($\eta_{1} = 0$, 
$\eta_{2} = c_{2} / c_{1}$ ), $A_{3}$: ($\eta_{1} = - 1$, $\eta_{2} = 0$); 
further explanations are given in the text.}
\label{fig6.1}
\end{figure}

Let us extend the results of Pt. 6.1 to the case of a pair of longitudinal 
waves propagating with the velocities $\textbf{c}_{1}$ and $\textbf{c}_{2}$ 
in non-collinear directions of axes x$_{1}$ and x$_{2}$, in that we surmise, 
as done in Chapter 5, that the formation of a martensite lamellae of 
thickness $ \sim \lambda_{1}$, $ \sim \lambda_{2}$ ($\lambda
_{1,2}$ - wave lengths) is performed by the association of the areas 
subjected to synchronized tensile and compressive stress, the main axes of 
which being oriented parallel to x$_{1}$ and x$_{2}$, respectively. 
If we start our conceptual view with a pair of waves of type \eqref{6.6}, 
featuring infinite flat fronts, it will be obvious that the specific region 
in which the combination of tensile and compressive stress being required 
for a martensitic transformation (also combined with the release of 
transformation energy) is granted, is identical with the region of 
wave-imposition, propagating in the same direction and with the same 
velocity $\textbf{c} = \textbf{c}_{1} + \textbf{c}_{2}$ (if 
$\textbf{c}_{1} \:\bot\: \textbf{c}_{2}$), that is the line of 
intersection of the (perpendicular) wave fronts. Consequently, the 
stationary wave pattern must correspond with the region $B_{\gamma - \alpha
}$ bound in both x$_{1}$ and x$_{2}$ directions and moving with velocity 
$\textbf{c}$. In other words, a stationary propagation of the pair of 
displacement-waves in cooperation with the switching wave is feasible, if 
the latter one propagates with a velocity \textbf{V} = $\textbf{c} = 
\textbf{c}_{1} +\,\textbf{c}_{2}$. This also means that, in the case of a 
displacement-wave, the requirement \textbf{V} = $\textbf{c}$ is to be replaced 
by the requirement V$_{x1} = c_{1}$, V$_{x2} = c_{2}$. Within our 
following considerations, the actual orientation of the axes x$_{1}$ and 
x$_{2}$ will be of no fundamental importance. Thus we shall further on 
consider x$_{1}\:\bot\: $ x$_{2}$. With consideration of \eqref{6.6} it is 
possible to note the expression for the displacements $u_{i}$, with $i$ = 
x$_{1}$, x$_{2}$ in the form

\begin{equation}
\label{6.13}
u_{i} = \frac{1}{q_{i}}\;\left[\frac{D_{0\,i} }{\beta_{0\,i} } 
\right]^{\frac{1}{2}}\;\sin{(q_{i} \,\xi_{i} + \varphi_{i})}.
\end{equation}
However, the term $D_{0i}$ already depends on two variables $\xi_{1}$ = 
x$_{1} - c_{1}t$ and $\xi_{2}$ = x$_{2}  -  c_{2}t$, as the 
region of $B_{\gamma - \alpha}$, in which $D_{0\,i} > 0$, is in simultaneous 
motion both in the x$_{1}$ - and x$_{2}$ - directions, while featuring 
limited dimensions $ \sim  l_{1}$, $l_{2}$ in both directions. This means 
that each of the expressions (\ref{6.13}) describes a stationary wave bundle 
moving together with the switching wave. The most simple switching-wave 
model (see Fig. \ref{fig6.1}), corresponding to the wave beams (\ref{6.13}), can be 
prescribed by the expression 
\begin{eqnarray}
\label{6.14}
\psi & = &\left[1 + \frac{c_{2}}{c_{1}}\,\eta_{2}  \right]\;\left[ \theta 
\left( \eta_{2} + \frac{c_{1}}{c_{2}} \right) - \theta(\eta_{2}) \right]\,
\theta \left(-\,\eta_{1} - \frac{c_{2}}{c_{1}}\,\eta_{2} - 1 \right) + {}                                                                    
									 \nonumber\\
									\nonumber \\						 
 & & {}+ \left[1 -
 \frac{c_{2}}{c_{1}}\,\eta_{2}\right]\;\left[\theta(\eta_{2}) 
 - \theta \left(\eta_{2} - \frac{c_{2}}{c_{1}} \right) \right]\;
 \theta \left({ -\,\eta_{1} + \frac{c_{1}}{c_{2}}\;\eta_{2} - 1} \right) - {} 
                                                                              \nonumber  \\
                                                                              \nonumber  \\									 
  & & {} - \eta_{1} \;[ \theta(\eta_{1} + 1 ) - \theta 
\,(\eta_{1})\,]\;\left[\theta \left(\eta_{1} + 
\frac{c_{2}}{c_{1}}\;\eta_{2} + 1 \right)                         
 - \theta \left(-\,\eta_{1} +    
\frac{c_{1}}{c_{2}}\;\eta_{2} - 1 \right) \right], \nonumber  \\ 
                                                   \nonumber  \\                                                                                                                            	                       
\end{eqnarray}
\vspace{-7mm}
\begin{displaymath}
\eta_{1} = ( c_{1} \,\xi_{1} + c_{2} \,\xi_{2})\;\tilde{d}^{-
1}\;,\quad \eta_{2} = ( c_{2} \,\xi_{1} - c_{1} \,\xi_{2})\;\tilde{d}^{-1},
\end{displaymath}
where $\tilde{d} = c_{1} l_{1} + c_{2} l_{2}$; the coordinates $\eta_{1}$, 
$\eta_{2}$ are mutually perpendicular and located in the plane $\xi
_{1}$ and $\xi_{2 }$, where the axis $\eta_{1}$ is oriented parallel 
to \textbf{V}. If the definition of the Heaviside $\theta$ - function is 
taken into consideration, it will be easy to understand that the first row 
of Eq. (\ref{6.14}) describes the distribution of the reduced temperatures (chem. 
potentials) $\psi $ in the region $OA_{3}A_{1}A_{1}^{\prime}$, the second row 
in the region $OA_{3}A_{2}A_{2}^{\prime}$, and the third row in the region 
$A_{1}O_{1}A_{2}A_{3}$ as illustrated in Fig. \ref{fig6.1}. Put in 
another way, (\ref{6.14}) is sub-divided into terms corresponding to the flat 
sectors of the triclinic surface $\psi$.

The peripheral points of the plane (x$_{1}$, x$_{2}$), supposed to be 
reached by the switching-wave at an instant t, are the points of segment 
$A_{1}A_{2}$ of the straight line $\eta_{1} = 0$ (see Fig. \ref{fig6.1}) 
on which $\psi  = const = 0$. Consequently, this segment belongs to 
the switching-wave front (note that in an orientation perpendicular to 
x$_{1}$ and x$_{2}$ , the switching-wave front is unlimited). The velocity 
of motion of the points belonging to the segment $A_{1}A_{2}$, related to 
the frame of reference x$_{1}$, x$_{2}$, is easy to find, after having 
noted that the conditions $\dot{\eta}_{1}  = 0$, $\dot{\eta}_{2}  = 0$ are 
satisfied for any point of segment $A_{1}A_{2}$, from which it can be 
concluded that - in the frame of reference x$_{1}$, x$_{2}$ - the velocity 
$\mathbf{\dot{r}}$ of the wave-front is equal to \textbf{V} = 
$\textbf{c}_{1} + \textbf{c}_{2}$. Obviously, for $\dot{\eta}_{2}  \to 
 \pm  0$ the expression (\ref{6.14}) transforms into \eqref{6.10}. Then it will be 
possible to determine the expressions for the time $\tau$ and the 
coefficient of diffusion d, assuming that the quantity V = $(c_{1}^{2} + 
c_{2}^{2})^{1 / 2}$ corresponds to the minimum velocity of the 
switching-wave 2 (d$\tau^{-1})^{1 / 2}$, and by consideration of the 
relationship $\dot{\psi} = \psi_{\eta_{1}}^{\prime}\, \dot{\eta}_{1} = - 
\dot{\eta}_{1} = \tau^{-1}$, resulting from \eqref{6.7}, \eqref{6.10} and 
\eqref{6.11}, where $\dot{\eta }_{1} = -\,c^{2} / \tilde{d}$:

\begin{equation}
\label{6.15}
\tau = \frac{\tilde{d}}{V^{2}},\qquad d = \frac{1}{4}\,V^{2}\,\tau = 
\frac{1}{4}\,\tilde{d}.
\end{equation}
For a given d, relationship (\ref{6.15}) delivers the connection between $l_{1}$ 
and $l_{2}$, and thus also among the lengths of the displacement-waves, 
propagating along the x$_{1}$ and x$_{2}$ -directions, because  $l_{1}   
\sim   \lambda_{1} \sim  q_{1}^{-1}$ and $l_{2}  \sim \lambda_{2}  \sim  
q_{2}^{-1}$.

Finally we have to note that, within the notion of a solitary wave 
front separating regions with different stationary values of $T$ and $\mu$, 
the description of $T$ and $\mu$ only is an approximation, as the 
diffusion-processes proceeding perpendicular to \textbf{V} will modify the 
profile of a two-dimensional wave with increasing time. Thus the real 
profile of the $T$ - and $\mu$ - waves would only feature a shape near their 
stationary value if they were located close to the $B_{\gamma - \alpha}$ 
region.

\section{Stationary wave of relative spatial deformation $\tilde{\varepsilon}$ 
during the $\gamma -\alpha$ - transformation}

\subsection{$\tilde{\varepsilon}$ - trigger-switching - wave in absence of 
displacement-waves}

The displacement-waves generated in the $B_{\gamma - \alpha}$ region ensure 
the lattice deformation required for a cooperative overcoming of the 
energy-threshold separating the structural $\alpha$ and 
$\gamma$ states. These structural $\alpha$ - and $\gamma$ - states are 
characterized by different values of static (permanent) displacements. 
However, the states of the $\alpha$ - and $\gamma$ - phases feature different 
$T$ -, $\mu$ - values, a fact not having been explicitly considered in our 
previous treatise. We shall now focus our analysis on the static 
deformations and only consider the relative change of volume during the 
transformation or, put in other words, characterize the $\gamma$ - and 
$\alpha$ - phases by their relative volume deformation $\tilde{\varepsilon
}$, where $\tilde{\varepsilon}_{\gamma}  = 0$ and $\tilde{\varepsilon
}_{\alpha}  = 2,4 \cdot 10^{-2}$, being a value typical of the volume 
effect of the Bain-deformation (see also Pt. 1.4). The tilde-sign ($^\sim$) 
is introduced to distinguish the relative volume deformation 
$\tilde{\varepsilon}$ from the associated linear deformation $\varepsilon$ 
being produced by a displacement-wave. In a system comprising two stable states 
($\tilde{\varepsilon} = \tilde{\varepsilon}_{\alpha}$, $\tilde{\varepsilon} 
= \tilde{\varepsilon }_{\gamma}$~), being separated by certain barrier, 
the propagation of a trigger-switching wave (TSW) is possible in principle 
\cite{Rabinovich84,Iakhno81}. However, for its generation, an initial 
level of excitation ($\tilde{\varepsilon} = \tilde{\varepsilon}_{0})$ 
is required, which must at least exceed the threshold-value ($\tilde
{\varepsilon} = \tilde{\varepsilon}_{th}$). It is also known that an 
$\tilde{\varepsilon}$ - wave of the triggering type, propagating along 
the x-orientation, satisfies the relationship

\begin{equation}
\label{6.16}
\dot{\tilde{\varepsilon}} = d_{\tilde{\varepsilon}} \,\tilde{\varepsilon
}^{\prime \prime}_{xx} + F\,( \tilde{\varepsilon}),
\end{equation}
for which, in contrast to \eqref{6.7}, the non-linear function $F(\tilde{\varepsilon 
})$ for $\tilde{\varepsilon} = \tilde{\varepsilon}_{\gamma} = 0$, 
$\tilde{\varepsilon} = \tilde{\varepsilon}_{th}  > 0$, 
$\tilde{\varepsilon} = \tilde{\varepsilon}_{\alpha}  > 
\tilde{\varepsilon}_{th}$ must vanish. The most simple expression for 
$F(\tilde{\varepsilon})$ thus may look as follows:

\begin{equation}
\label{6.17}
F\,(\tilde{\varepsilon }) = - \frac{1}{\tau_{\tilde{\varepsilon}}}\,
\tilde{\varepsilon}\,(\tilde{\varepsilon} - 
\tilde{\varepsilon}_{th})\,( \tilde{\varepsilon} - 
\tilde{\varepsilon}_{\alpha}),
\end{equation}
where $\tau_{\tilde{\varepsilon}}$ - constant with the dimension of time. 
The stationary solution $\tilde{\varepsilon} = \tilde{\varepsilon
}\;( x - V_{\tilde{\varepsilon}} \,t ) \equiv \tilde{\varepsilon}\;
(\xi)$ of wave-equation (\ref{6.16}), comprising a 
non-linearity (\ref{6.17}), will look as follows:

\begin{equation}
\label{6.18}
\tilde{\varepsilon} = \frac{1}{2}\;\tilde{\varepsilon}_{\alpha} \;[
1 - \tanh{( 2\;\xi \;l_{\tilde{\varepsilon}}^{-1}})],
\end{equation}

\begin{equation}
\label{6.19}
l_{\tilde{\varepsilon}} = \frac{4}{\tilde{\varepsilon}_{\alpha}}\,( 
2\;d_{\tilde{\varepsilon}} \;\tau_{\tilde{\varepsilon}})^{\frac{1}{2}}.
\end{equation}
In contrast to the $T$ - and $\mu$ - waves - resembling so called 
phase - switching waves (PSW) \cite{Iakhno81}), which dispose of a velocity-continuum V 
$ \ge $V$_{min}$ - a TSW only features a single (discrete) velocity:

\begin{equation}
\label{6.20}
\textrm{V}_{\tilde{\varepsilon}} = \left[ \,\frac{d_{\tilde{\varepsilon}}}
{2\;\tau_{\tilde{\varepsilon}}} \right]^{\frac{1}{2}}\;
(\tilde{\varepsilon}_{\alpha} - 2\;\tilde{\varepsilon}_{th}).
\end{equation}

In order to interpret these results, we recall that Eq. (\ref{6.16}) can be 
expressed in the notation of the Landau-Halatnikov-Eqaution (see e.g. 
\cite{Patashinskii82}):

\begin{displaymath}
\dot{\tilde{\varepsilon}} = -\,\Gamma\;\frac{\delta \,\Phi}{\delta 
\,\tilde{\varepsilon}},
\end{displaymath}
where $\Gamma$ - kinetic coefficient (not to be confounded with 
electron-extinction); $\Phi$ - fraction of free energy depending on 
$\tilde{\varepsilon}$; $\delta \,\Phi / \delta \,\tilde{\varepsilon 
}$ - functional derivative of $\Phi $ by $\tilde{\varepsilon}$,

\begin{equation}
\label{6.21}
\Phi = \int{\,d\,V\;\left\{{\,\frac{A}{2}\,\tilde{\varepsilon}^{\,2} + 
\frac{1}{3}\,B\;\tilde{\varepsilon}^{\,3} + \frac{1}{4}\,G\;\tilde
{\varepsilon}^{\,4} + \frac{1}{2}\,P\;\left( {\frac{\partial \,\tilde 
{\varepsilon}}{\partial\,x}} \right)^2} \right\}},
\end{equation}
where $A / G =  \varepsilon_{th}\;\tilde{\varepsilon}_{\alpha },
\quad B / G = - \left(\tilde{\varepsilon} _{\alpha} + 
\tilde{\varepsilon}_{th} \right),\quad \Gamma G = \tau^{-1}
_{\tilde{\varepsilon}}$,\quad $\Gamma \:P = d_{\tilde{\varepsilon}}$.
Our model (\ref{6.21}) for $\Phi$ represents a 1$^{st}$ order 
phase - transformation, because $A > 0$, $B < 0$, $G > 0$. If, in addition, 
$dB / dT < dA/dT < 0$, then the elastic moduli will feature a normal temperature 
dependence during cooling, $\tilde{\varepsilon}_{th}$ and the level of 
the energy-threshold $\Phi_{0}(\tilde{\varepsilon}_{th})$, 
which separates the states $\tilde{\varepsilon} = \tilde{\varepsilon 
}_{\gamma}$, $\tilde{\varepsilon} = \tilde{\varepsilon}_{\alpha}$, will 
subside and remain a finite quantity. If we use the notation $\Phi_{0}$ 
for $\Phi$ at $\partial \,\tilde{\varepsilon } / \partial \,x = 0$, 
and consider that

\begin{equation}
\label{6.22}
\Phi_{0}(\tilde{\varepsilon}_{\gamma}) = 0, \quad \Phi_{0}
(\tilde{\varepsilon}_{\alpha}) = \frac{V\;A\;\tilde{\varepsilon
}_{\alpha} ^{\,2} \;( 2\,\tilde{\varepsilon}_{th} - \tilde{\varepsilon}_{\alpha} )}
{12\;\tilde{\varepsilon}_{th}},
\end{equation}

\begin{displaymath}
\Phi_{0}( \tilde{\varepsilon}_{th}) = \frac{\textrm{V}\;A\;\tilde{\varepsilon
}_{th}^{\,2} \;
( 2\,\tilde{\varepsilon}_{\alpha} - \tilde{\varepsilon}_{th})}
{12\;\tilde{\varepsilon}_{\alpha} },
\end{displaymath}
where V - volume of the system, it becomes obvious from (\ref{6.20}) and
(\ref{6.22}), that V$_{\tilde{\varepsilon}} \sim ( \Phi_{0}(
\tilde{\varepsilon}_{\gamma}) - \Phi_{0}( \tilde{\varepsilon}_{\alpha}))$, 
being proportional to the driving-force of the transformation, and 
V$_{\tilde{\varepsilon}} = 0$ at $T = T_{0}$, being 
corresponded by the value of $\tilde{\varepsilon}_{th} = 
1 / 2\,\tilde{\varepsilon}_{\alpha} $. We should however 
remark that, in order to excite a TSW, being in motion with non-vanishing 
velocity V$_{\tilde{\varepsilon}}$ at a temperature $T$ near $T_{0}$, there 
would be required large initial values of $\tilde{\varepsilon}_{0}$ of 
about $\sim 1 / 2\,\tilde{\varepsilon}_{\alpha}$.

The frontal width $l_{\tilde{\varepsilon}}$ of a TSW, as defined by
(\ref{6.19}) (with linear approximation of its profile), is determined by 
parameters A and P in the function (\ref{6.21}), in the form

\begin{equation}
\label{6.23}
l_{\tilde{\varepsilon}} = 4\;\left[ \frac{2\;\tilde{\varepsilon}_{th} \;P}
{A\;\tilde{\varepsilon}_{\alpha}}\right]^{\frac{1}{2}}.
\end{equation}
To perform an evaluation of $l_{\tilde{\varepsilon}}$, let us first express 
P by means of the specific surface-energy $E_{s}$ and A by the additional 
assumption

\begin{equation}
\label{6.24}
E_{s} = \int\limits_{- \infty}^{\infty}{d\,x\;P\;\left( {\frac{\partial 
\,\tilde{\varepsilon}}{\partial \,x}} \right)^{2}}.
\end{equation}
After substitution by (\ref{6.18}) in (\ref{6.24}), and with consideration 
of (\ref{6.23}) we get:

\begin{equation}
\label{6.25}
E_s = \frac{1}{6}\;\left[\frac{\tilde{\varepsilon}_{\alpha} 
\;P\;A}{2\;\tilde{\varepsilon}_{th}}\right]^{\frac{1}{2}}\;
\tilde{\varepsilon}_{\alpha}^{\,2},
\end{equation}
i.e.

\begin{displaymath}
P = 72\;E_{s}^{2} \;\frac{\tilde{\varepsilon}_{th}}{A\;\tilde{\varepsilon
}_{\alpha}^{\,5} }.
\end{displaymath}
Using (\ref{6.22}), we can express $\tilde{\varepsilon}_{th}$ 
by the following notation: 
\begin{equation}
\tilde{\varepsilon}_{th}  = A \tilde{\varepsilon}_{\alpha}^{\,3} 
 \Bigg[2\,A \tilde{\varepsilon}_{\alpha}^{\,2}  -  12 \Bigg(\frac{\Delta \Phi_{0}}{
 V}\Bigg)\Bigg]^{-1}. 
\label{6.26}
\end{equation}
According to the calculus in \cite{Krizement61}, it is justified to assume for the $\gamma 
-\alpha$ - transformation of iron-based alloys: $\Delta \Phi_{0} 
(M_{S}) /V  \approx  (4 / 3)\,Q$, where $Q$ - specific reactive heat (it will 
be surmised that one fourth part of $\Delta \Phi_{0} / V$ is needed for 
the formation of the phase-boundary, thus providing an explanation for the 
factor (4 / 3) associated with $Q$). E.g. for the alloy H30: $Q \approx 3 
\cdot 10^{8}$ J m$^{-3}$ \cite{Vinnikov74}. Thus we get for $A \approx 2 \cdot 
10^{11}$ J m$^{-3}$ - being a typical value of the compressive modulus - 
from \eqref{6.26}: $\tilde{\varepsilon}_{th}  \approx  5,4 \cdot 
10^{-4}$. Taking a value of 0,2 J m$^{-2}$ for $E_{s}$ \cite{Missol78}, we can 
determine from (\ref{6.25}) and (\ref{6.23}): $P \approx 10^{-6}$ J m$^{-1}$, 
$l_{\tilde{\varepsilon}} \approx  1,9 \cdot 10^{-9}$ m $ \approx 5\,a$, 
i.e., the width of the front $l_{\tilde{\varepsilon}} $ is about 5 times 
as large as the lattice-parameter of the $\gamma$ - phase.

Some additional remarks:
\begin{enumerate}
\item{The data presented in \cite{Missol78} mainly relate to the surface energy 
of a static phase-boundary, whereas, in our above compilation of $E_{s}$, 
we used the solution (\ref{6.18}) pertaining to a moving phase-boundary, which, 
strictly speaking, would only be justified for the case of a temperature 
close to $T_{0}$, and with V$_{\tilde{\varepsilon}} \approx 0$. In case 
of an arbitrary temperature, the value $E_{s}$ of a static phase-boundary 
can however be calculated using the general approach related to the 
determination of the boundary energy of a ferromagnetic domain in \cite{Landau82}:
\begin{eqnarray}
\label{6.27}
E_{s} & = &\Bigg(\frac{PA}{2} \Bigg)^{\frac{1}{2}}\;\frac{\varepsilon 
_{\alpha}^{2}}{27\;n^{2}}\;\Bigg\{6\;\sqrt{\;2\;(n - 1)}\;( n^{2} - n + 1)
-{}                                                                \nonumber\\                
& & {} - \sqrt{\;3\;(n - 2)} \;(n - 2)\;(n - 1) + \frac{2}{\sqrt{\,n} }\;( 2\,n - 1 
)\;(n - 2 )\;(n + 1)\;\times {}                                   \nonumber\\                
& & {} \times \;\ln{\left| \frac{2\;( 2\,n - 1) + 3\,
\sqrt{2\,n\;(n - 1)} }{n - 2 + \sqrt{3\,n\;(n - 2 )}} \right|} \Bigg\} 
 \end{eqnarray}
where $n = \tilde{\varepsilon}_{\alpha} \ {\tilde{\varepsilon}_{th}}$. 
A comparison between (\ref{6.27}) and (\ref{6.25}) 
shows that for $T = T_{0}$ , ($\tilde{\varepsilon}_{\alpha} = 2\;\tilde
{\varepsilon}_{th})$, (n = 2) the expressions for $E_{s}$ match, 
as required. For $T = M_{S}$, if $n >> 1$, we get from (\ref{6.27}): 

\begin{eqnarray}
\label{6.28}
 E_{s} & \approx & \frac{(PA\,n)^{\frac{1}{2}}}{27\,\sqrt{\,2}}\;
\tilde{\varepsilon}_{\alpha}^{\,2} \;\Bigg\{6\,\sqrt{\,2} - \sqrt{\,3} +
{} \nonumber\\ 
& & {}+ 4\;\ln{\left| {\,\frac{4 + 3\,\sqrt{\,2}}{\sqrt{\,3} + 
1}} \right|} \Bigg\} \approx \;0,3\;( P\,A\,n)^{\frac{1}{2}}\;
\tilde{\varepsilon }_{\alpha}^{\,2}, 
\end{eqnarray}
being about 2,5 - fold larger than the result obtained from (\ref{6.25}). Thus we 
obtain with $E_{s} \approx  0,2$ J m$^{-2}$ the values $P  \approx  4 
\cdot 10^{-7}$ J m$^{-1}$ and $l_{\tilde{\varepsilon }} \approx 2\,a$.}

\item{In the papers \cite{Roitburd72,Roitburd78}, dealing with $E_{s}$, a ratio of

\begin{displaymath}
E_{s} \approx \frac{\Phi_{0}(\tilde{\varepsilon}_{th})}{V}\;
l_{\tilde{\varepsilon}},
\end{displaymath}
was used, which can also be expressed by 

\begin{equation}
\label{6.29}
E_{s} \approx \frac{1}{3}\;\left({\frac{P\,A\,\tilde{\varepsilon}
_{th}}{\tilde{\varepsilon}_{\alpha}}} \right)^{\frac{1}{2}}\;\left({2 - 
\frac{\tilde{\varepsilon }_{th} }{\tilde{\varepsilon }_{\alpha} 
}} \right)\;\tilde{\varepsilon}_{th}^{\,2},
\end{equation}
after consideration of (\ref{6.22}), (\ref{6.23}). At $T = T_{0}$, $\tilde{\varepsilon 
}_{\alpha} = 2\;\tilde{\varepsilon}_{th}$, Eq. (\ref{6.29}) 
realistically delivers the order of magnitude of the quantity $E_{s}$, 
however at $T = M_{S}$ and $\tilde{\varepsilon}_{\alpha} \ge 10\;
\tilde{\varepsilon}_{th}$, the ratio between $E_{s}$, as determined from
(\ref{6.25}) or (\ref{6.27}), and the value of $E_{s}$ derived from
(\ref{6.29}), is of an order of magnitude of $( \tilde{\varepsilon
}_{\alpha} / 2\;\tilde{\varepsilon}_{th})^{3} \ge 10^{4}$, thus we can 
conclude that the assessment from (\ref{6.29}) is not realistic in
principle.}

\item{If the motion of the phase-boundary resembles a thermally 
activated process, being controlled e.g. by a diffusion-process, then the 
quantities V$_{\tilde{\varepsilon}}$ and thus also $\Gamma$, 
$d_{\tilde{\varepsilon}} $ and $\tau_{\tilde{ \varepsilon}} $ should exhibit a 
strongly pronounced (exponential) dependence on temperature.}

\item{In order to determine $\Gamma$, $d_{\tilde{\varepsilon}} $ and 
$\tau_{\tilde{\varepsilon}} $ by means of our model (\ref{6.21}) of the potential 
$\Phi $, it should suffice to know the experimentally determined value 
V$_{\tilde{\varepsilon}}$ , provided the quantities A, $\tilde{\varepsilon 
}_{\alpha}\;,\;\tilde{\varepsilon}_{th}$, P are previously given. 
We shall use the expressions suffixing Eq. (\ref{6.21}), and write 
V$_{\tilde{\varepsilon}} $ in the form

\begin{equation}
\label{6.30}
\textrm{V}_{\tilde{\varepsilon}} = \Gamma \;\,\left[ {\frac{P\,A}
{2\,\tilde{\varepsilon}_{th} \,\tilde{\varepsilon}_{\alpha}}} 
\right]^{\frac{1}{2}}\;\left({\tilde{\varepsilon}_{\alpha} - 2\,
\tilde{\varepsilon}_{th}} \right)
\end{equation}
thus highlighting the proportionality between V$_{\tilde{\varepsilon}} $ and 
$\Gamma$.} 

\item{Besides the potential (\ref{6.21}) there is frequently used a 
potential which only depends on even powers of $\varepsilon$, i.e. 
$\tilde{\varepsilon}^{\,2}\;,\;\tilde{\varepsilon
}^{\,4}\;,\;\tilde{\varepsilon}^{\,6}$. 
In this particular case, being qualitatively analogous to that relating to 
(\ref{6.21}), it is easily possible to perform a consideration of the 
triggering-switching wave (TSW), using the results of \cite{Iakhno76}.}
\end{enumerate}

\subsection{Effect of the displacement-waves on the velocity of the 
$\tilde{\varepsilon}$ - switching-wave}

\begin{figure}[htb]
\centering
\includegraphics[clip=true,width=.8\textwidth]{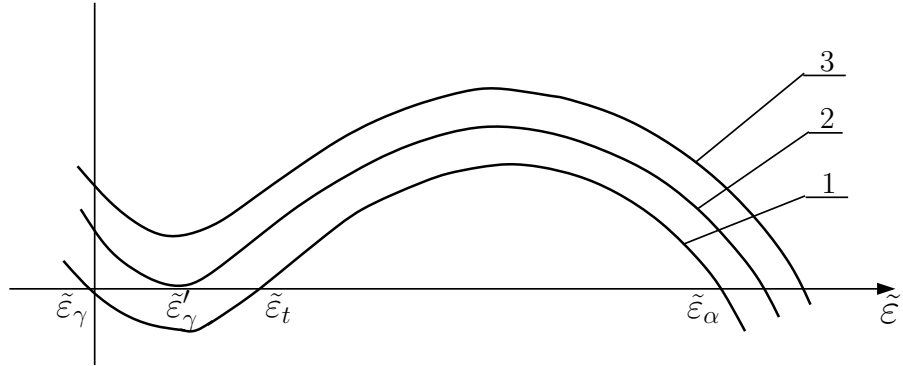}
\caption{ Development of a "Catastrophe" of the folding-type: Curve 
1 - path of the function \eqref{6.17} in absence of displacement waves; Curves 2 
and 3 are associated with the confluence and disappearance of discrete 
points $\tilde{\varepsilon}_{\gamma} \;,\;\tilde{\varepsilon 
}_{th}$ , respectively.}
\label{fig6.2}
\end{figure}

The relationship between the velocity V$_{\tilde{\varepsilon}} $ and the 
temperature, in conjunction with the parameters included in \eqref{6.20}, is not 
typical of the $\gamma -\alpha$ - MT. That is no surprise, as the mechanism 
being responsible for the cooperative character of the transformation is not 
expressed explicitly neither in Eq. \eqref{6.16} nor in Eq. \eqref{6.21}. In the 
two-wave model however, the cooperative nature of the lattice-transformation 
is linked up with displacement-waves, which, during their propagation, 
ensure the attainment of the threshold-deformation-level in the $B_{\gamma - 
\alpha}$ region. As soon as this threshold-deformation is surpassed, 
irreversible deformations up to the value of $\tilde{\varepsilon} = 
\tilde{\varepsilon}_{\alpha} $ will be produced, resulting in the development of an 
$\tilde{\varepsilon}$ - wave. Obviously, this conceptual view of martensite 
crystal growth must take for granted the motion of the $\tilde{\varepsilon
}$-wave into x-direction with a velocity $\textbf{c}$, being equal to the 
geometric sum of the velocities of each individual displacement wave. That 
becomes a possibility when, as a result of the interaction of the 
$\tilde{\varepsilon}$ - TSW and the displacement waves, the 
$\tilde{\varepsilon}$ - TSW is converted into a wave featuring more than one 
discrete velocity V$_{\tilde{\varepsilon}}$, i.e. a velocity continuum, 
like a PSW. In mathematical terms, the effect of the displacement waves  at the
points of the $B_{\gamma - \alpha}$ region is expressed in such a way that 
there occur in sequence the events of approaching, confluence and finally, 
of disappearance (so-called "Folding-Catastrophe" \cite{Gilmor84}) of two singular points 
$\tilde{\varepsilon}_{\gamma}$ and $\tilde{\varepsilon}_{th}$. As a result, 
only a singular stable point $\tilde{\varepsilon}_{\alpha}$ will remain, to 
which the system tends (see Fig. \ref{fig6.2}).

In the latter case, the $\tilde{\varepsilon}$ - wave can generally propagate 
with any arbitrary velocity, since the region deformed by the displacement 
waves becomes unstable against any positive fluctuation. As the 
aforementioned vanishing process of particular points, propagating with 
velocity $\textbf{c}$ in space, is caused by the action of displacement waves, 
just this velocity proves to be the only one highlighted in the velocity 
continuum of the modified $\tilde{\varepsilon}$ - wave. The vanishing of the 
discrete points $\tilde{\varepsilon}_{\gamma}$ and $\tilde{\varepsilon
}_{th}$ of the function $F(\tilde{\varepsilon})$ can be taken 
into consideration by adoption of an addend of the "field-type" in
\eqref{6.21}: $ - \tilde{\sigma}\,\left({\tilde{\varepsilon}_{u}} 
\right)\;\tilde{\varepsilon}$, where $\tilde{\sigma}\,\left( 
{\tilde{\varepsilon}_{u}} \right)$ - pressure and $\tilde{\varepsilon 
}_{u}$ - volume deformation associated with the displacement-waves. 
If we further infer that $\tilde{\sigma}\,\left({\tilde{\varepsilon}_{u}} 
\right) = K \tilde{\varepsilon}_{u}$, with $K > 0$, 
and equalize the conditions for confluence $\tilde{\varepsilon}_{\gamma} \;,\;
\tilde{\varepsilon}_{th} \to \tilde{\varepsilon}^{\prime}_{\gamma} =
 \tilde{\varepsilon}^{\prime}_{th}  \approx  \tilde{\varepsilon}_{th} / 2$ of 
the discrete points $\tilde{\varepsilon }_{\gamma}$ and 
$\tilde{\varepsilon}_{th}$ with the condition $\tilde{\varepsilon}_{u} = 
\tilde{\varepsilon}^{\prime}_{\gamma}$, then we can determine $K = A/2$. 
Thus, within the framework of our considered model, $\tilde{\varepsilon}
_{th} \left({M_{S}}\right)\; = \;\tilde{\varepsilon}^{\prime}_{\gamma} \approx 
\tilde{\varepsilon}_{th} / 2$.

This way our consideration of displacement waves, which ensure the 
cooperative lattice transformation in the growth-stage, leads to a 
conceptual view of the motion of the martensite lattice-boundary with a 
velocity $\textbf{c}$, only weakly depending on $T$, its magnitude corresponding 
to that of the velocity of sound. More than that, if $\textbf{c}$ is the 
geometric sum of $\textbf{c}_{1}$, $\textbf{c}_{2}$, than $\textbf{c}$ can 
even exceed the velocity of sound in the direction given by $\textbf{c}$. We 
recall that supersonic speed of growth has been observed during the passage 
of a detonation wave through steel \cite{Lokshin1968,Lokshin1967}.

\subsection{The $\tilde{\varepsilon}$ - Modified-Switching-Wave}

Let us now consider in more detail the $\tilde{\varepsilon
}$ - Modified-Switching-Wave ($\tilde{\varepsilon}$ - MSW), related to the case of the 
confluence of the discrete points $\tilde{\varepsilon}_{\gamma}$, $\tilde
{\varepsilon}_{th}$ (Curve 2 in Fig. \ref{fig6.2}): $\tilde{\varepsilon
}^{\,\prime}_{\gamma} = \tilde{\varepsilon}^{\,\prime}_{t} \approx 
\tilde{\varepsilon}_{th} / 2$. Such a wave can generally be described by an 
equation of the form

\begin{equation}
\label{6.31}
\dot{\tilde{\varepsilon}} = d_{\tilde{\varepsilon}} \;\tilde{\varepsilon
}^{\,\prime \prime}_{xx} + \frac{1}{\tau_{\tilde{\varepsilon}} }\;\left( 
 \tilde{\varepsilon} - \tilde{\varepsilon}^{\,\prime}_{th} \right)^{2}\;
 \left(\tilde{\varepsilon}_{\alpha} - \tilde{\varepsilon} \right),
\end{equation}
in which, by way of contrast to \eqref{6.16}, the non-linear function 
$F(\tilde{\varepsilon})$ only features two values $\tilde{\varepsilon}_{th}^{\,\prime}$, 
$\tilde{\varepsilon}_{\alpha}$, corresponding to the requirement $F(\tilde
{\varepsilon}) = 0$\footnote{Notice that equation \eqref{6.31} has the 
solution \eqref{6.18} for the solitary wave front propagating with
velocity \eqref{6.20}}. Let us now introduce the designations

\begin{equation}
\label{6.32}
Z = \frac{\tilde{\varepsilon} - \tilde{\varepsilon}^{\,\prime}_{th} 
}{\tilde{\varepsilon}_{\alpha} - \tilde{\varepsilon}^{\,\prime}_{th} 
},\quad \tilde{\tau} = \frac{\tau_{\tilde{\varepsilon}}}{\left(
\tilde{\varepsilon}_{\alpha} - \tilde{\varepsilon}^{\,\prime}_{th}  
\right)^{2}},
\end{equation}
and a dimensionless time $\tilde{t}$ and spatial coordinate $\tilde{x}$:

\begin{equation}
\label{6.33}
\tilde{t} = \frac{t}{\tilde{\tau}},\quad  \tilde{x} = x\;\left( 
\frac{1}{d_{\tilde {\varepsilon}}\,\tilde{\tau}} 
\right)^{\frac{1}{2}}.
\end{equation}
Then we can write Eq. (\ref{6.31}) in the form

\begin{equation}
\label{6.34}
\dot{Z}_{\tilde{t}} = Z^{\prime \prime}_{\tilde{x} \tilde{x}} + Z^2\;\left( 1 - Z 
\right).
\end{equation}
The solution Z is of the travelling-wave type $Z = Z(\xi)$

\begin{equation}
\label{6.35}
Z\,(\xi) \equiv Z\,\left(\tilde{x} + \tilde{c}\,\tilde{t} 
\right),
\end{equation}
where $\tilde{c}$ - dimensionless wave-velocity

\begin{equation}
\label{6.36}
\tilde{c} = c\;\left( \frac{\tilde{\tau}}{d_{\tilde{\varepsilon}} } 
\right)^{\frac{1}{2}},
\end{equation}
corresponds to the equation

\begin{equation}
\label{6.37}
Z^{\prime \prime}_{\xi \,\xi } - \tilde{c}\,Z^{\prime}_{\xi} + Z^{2}\;
\left( 1 - Z \right) = 0.
\end{equation}

In case of large values $\tilde{c}^{\,2} >> 1$ it is possible to use a 
simplified procedure for resolving Eq. (\ref{6.37}), by introduction of a row of 
additive terms with increasing powers of the small parameter $\tilde{c}^{-
2}$. This procedure is described in detail in \cite{Marri83}. Taking the velocity $c = 
\arrowvert \textbf{c}_{1} + \textbf{c}_{2} \arrowvert$ as given 
(see Pt.6.3.2) and the condition $\tilde{c}^{2} >> 1$ as satisfied, then we 
will be able to assess the frontal width of the $\tilde{\varepsilon}$ - SW. 
Following \cite{Marri83}, we designate

\begin{equation}
\label{6.38}
\varphi = - \tilde{c}\,Z^{\prime}_{\xi} 
\end{equation}
and transform (\ref{6.37}) into the coordinate frame of a phase plane:

\begin{equation}
\label{6.39}
\frac{1}{\tilde{c}^{\,2}}\,\frac{d\,\varphi}{d\,Z} = \frac{\varphi - 
Z^2\,\left( 1 - Z \right)}{\varphi}.
\end{equation}
Let us now search for a solution of (\ref{6.39}) in the form of

\begin{equation}
\label{6.40}
\varphi \,\left(Z,\,\tilde{c}^{-2} \right) = g_{0} \,(Z)
 + \tilde{c}^{-2}\;g_{1}\,(Z) + \left( \tilde{c}^{-2} \right)^{2}\;
g_{2} \,(Z) + \cdots.
\end{equation}
After insertion of (\ref{6.40}) in (\ref{6.39}) and equalizing the terms 
with identical powers ($\tilde{c}^{-2})$ we get:

\begin{equation}
g_{0}(Z) = Z^{2}(1 - Z), 
\label{6.41}
\end{equation}

\begin{equation}
\label{6.42}
g_{1}\,(Z) = g_{0} \,\frac{d\,g_{0}}{d\,Z} = Z^{3}\;(1 - Z)\;\left( 2 - 3\,Z \right),
\end{equation}

\begin{equation}
\label{6.43}
g_{2}\,(Z) = \;\frac{d\,}{d\,Z}\left(  
 g_{0} \;g_{1}\right) = 2\,Z^{4}\;(1 - Z)\;\left(1 
- 2\,Z \right)\;\left( 5 - 6\,Z \right).
\end{equation}

We determine the dimensionless frontal width $\tilde{l}$ of the 
$\tilde{\varepsilon}$ - PSW from \cite{Marri83} by the ratio

\begin{equation}
\label{6.44}
\tilde{l} \sim \frac{1}{Z^{\prime}_{\xi}(Z_{0})},
\end{equation}
in which $Z_{0}$ is a solution of the equation

\begin{equation}
\label{6.45}
\frac{d\,\varphi}{d\,Z} = 0
\end{equation}
corresponding to the flex point of the function $Z(\xi)$, i.e. $Z^{\prime
\prime}_{\xi\,\xi}  = 0$ for $Z = Z_{0}$. From (\ref{6.45}) and (\ref{6.40})
- (\ref{6.43}) we can immediately find

\begin{equation}
\label{6.46}
Z_0 = \frac{2}{3} + \frac{14}{27}\,\tilde{c}^{-2} + O\,\left( 
\tilde{c}^{-4}\right).
\end{equation}
After insertion of (\ref{6.46}) in (\ref{6.40}), and with consideration of 
(\ref{6.38}), we get

\begin{equation}
\label{6.47}
Z^{\prime}_{\xi}(Z_{0}) = \frac{1}{\tilde{c}}\,\frac{4}{27}\, + 
O\,\left(\tilde{c}^{-5} \right).
\end{equation}
Making use of (\ref{6.44}), (\ref{6.47}), we get

\begin{equation}
\label{6.48}
\tilde{l} \sim \frac{27}{4}\,\tilde{c}.
\end{equation}
If we turn over to dimensional quantities with consideration of (\ref{6.36}) 
and (\ref{6.33}), we get from (\ref{6.48})

\begin{equation}
\label{6.49}
l \sim  7\,c\,\tilde{\tau } = \frac{7\,c\,\tau_{\tilde{\varepsilon 
}} }{\left(\tilde{\varepsilon}_{\alpha} - \tilde
{\varepsilon}^{\,\prime}_{th} \right)^{2}} \approx \frac{7\,c\,\tau 
_{\tilde{\varepsilon}}}{\tilde{\varepsilon}_{\alpha}^{\,2}},
\end{equation}
in which the condition $\tilde{\varepsilon}^{\,\prime}_{th} < < \tilde{\varepsilon 
}_{\alpha}$ is appropriately considered.

Up to now, the constant $\tau_{\tilde{\varepsilon}}$ with the dimension of 
time played the role of a free phenomenological parameter, the determination 
of which would have to consider certain additional condition. This condition 
results from the general requirement of coordinated propagation of atomic 
displacement waves (the waves ensure the attainment of a threshold-deformation 
$\tilde{\varepsilon}^{\,\prime}_{th}$) with coexistent temperature $(T)$ chemical 
potential $(\mu)$ and $\tilde{\varepsilon}$ - switching waves, as this 
requirement takes for granted that the a.m. conditions of equalized 
wave-propagation velocity as well as that of equalized spatial scale of the 
frontal widths of the $T$ -, $\mu$ - and $\tilde{\varepsilon}$ - SW's. In 
fact, an unambiguous relationship between electrochemical potential $\mu$ 
and electron-concentration (and thus also with specific volume) inhibits the 
materialization of a PSW of the $\mu$ - type with frontal width 
$l_{\mu}$, which would substantially differ from the frontal width of 
the $\tilde{\varepsilon }$-wave. It is nonetheless justified to assume that 
also the process of heat-emission, being represented by the non-linear 
function $F(\psi)$ introduced in Eq. \eqref{6.7}, will occur in a region with 
$\tilde{\varepsilon} > \tilde{\varepsilon}^{\,\prime}_{th}$, i.e. in the frontal 
area of the $\tilde{\varepsilon}$ - MSW. Thus it is justified to introduce 
the following additional condition: 

\begin{equation}
l \sim l_{\mu} \sim l_{T}  \sim   \lambda / 2, 
\label{6.50}
\end{equation}
where $\lambda$ - length of a displacement-wave with magnitude of about 
($10^{-7}  \div  10^{-6}$) m. For the aforementioned region of 
$\lambda $ we can determine from \eqref{6.50} and (\ref{6.49}) with $c  \sim  5 \cdot 
10^{3}$ m/s $\tau_{\tilde{\varepsilon}}   \sim  2\,(10^{-15}\div 
 10^{-14})$ s. The time $\tilde{\tau} = \tau_{\tilde{\varepsilon}} / 
( \tilde{\varepsilon}_{\alpha} - \tilde{\varepsilon}^{\,\prime}_{th})
 ^2$, which, according to (\ref{6.49}), determines the frontal width $l$, is of 
 an order of magnitude of $10^{-11}$s for $\tilde{\varepsilon}_{\alpha} 
 \sim  2,4 \cdot 10^{-2}$.

Now it is easy to convince ourselves that, for $\tau_{\tilde {\varepsilon}
} \sim  (10^{-15}  \div  10^{-14})$ s, our initial assumption 
$\tilde{c}^{2} >> 1$, used for our above analysis, is satisfied. Thus we can 
also use the relationships linking up the quantities $d_{\tilde{\varepsilon 
}}$, $\tau_{\tilde{\varepsilon}}$ with $\Gamma$, P, A, 
$\tilde{\varepsilon}_{\alpha}$, $\tilde{\varepsilon}_{th}$, 
appearing in Eq. \eqref{6.21}, to determine

\begin{equation}
\label{6.51}
\frac{d_{\tilde{\varepsilon}}}{\tilde{\tau}} =
\frac{P\,\tilde{\varepsilon}_{th} 
\,\tilde{\varepsilon}_{\alpha}}{\tau_{\tilde{\varepsilon}} \,\tilde{\tau} \,A} 
\approx \frac{P\,\tilde{\varepsilon}_{th} \,\tilde{\varepsilon
}_{\alpha}^{\,3}}{A\,\tau_{\tilde{\varepsilon }}^{2}}.
\end{equation}

From (\ref{6.51}), we can determine for $P  \sim  4 \cdot 10^{-7}$ J/m, $A = 
2 \cdot 10^{11}$ J/m$^{3}$, $\tilde{\varepsilon }_{th}  \sim  10^{-3}$, 
$\tilde{\varepsilon}_{\alpha}  \sim 2,4 \cdot 10^{-2}$, $\tau
_{\tilde{\varepsilon}} \sim  (10^{-15} \div  10^{-14})$ s the value of 
$d_{\tilde{\varepsilon}} / \tilde{\tau}   \sim  (10^{2}  \div  10^{4})$ m$^{2}$/s$^{2}$ . 
Then we obtain from (\ref{6.36}) the quantity $\tilde{c}^{\,2}$ for 
$c \sim  5 \cdot 10^{3}$ m/s, delivering $\tilde{c}^{\,2}  \sim  (10^{5}  \div 
 10^{3}) >> 1$. 

Remarkably, for the same values of P, A, $\tilde{\varepsilon} 
_th $, $\tilde{\varepsilon}_{\alpha}$, and $\tau_{\tilde{\varepsilon}}  
 \sim  10^{-14}$ s, we obtain a kinetic coefficient  
$\Gamma  \sim  10^{-2}$ m$^{3}$ /(J s), and then, in accordance with 
\eqref{6.30}, the velocity of the TSW V$_{\tilde{\varepsilon}}   \sim  10$ 
m/s is almost three orders of magnitude smaller than the speed of growth of a 
martensite crystal. 

\begin{figure}[htb]
\centering
\includegraphics[clip=true,width=.8\textwidth]{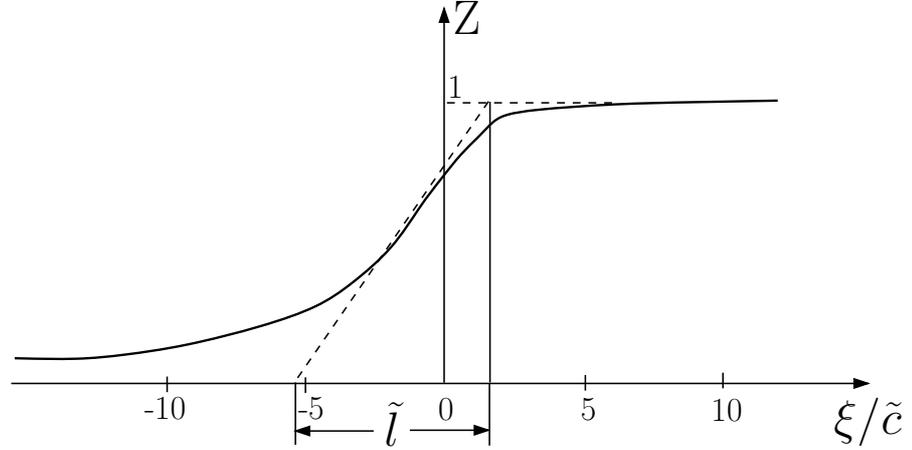}
\caption{Phase-Switching-Wave (PSW) profile of relative volume 
deformation.}
\label{fig6.3}
\end{figure}

Large values of $\tilde{c}^{\,2}$ justify the omission of two addends of an 
order of magnitude of $\tilde{c}^{-2}$ and, all the more, those of an 
order of magnitude of $\tilde{c}^{-4}$, $\tilde{c}^{-5}$ appearing in 
Eqs. (\ref{6.47}), (\ref{6.46}) without introducing any significant error. This also 
means that, when breaking down Eq. (\ref{6.40}), it is justified to confine our 
further analysis to the first addend. This approximation corresponds to an 
omission of the second derivative $Z^{\prime \prime}_{\xi \,\xi} $ in Eq. 
(\ref{6.37}). After elementary integration of Eq. (\ref{6.37}) we are able 
to find the expression defining the shape of the $\tilde{\varepsilon}$ - 
modified-wave:

\begin{equation}
\label{6.52}
\frac{Z}{(1 - Z)\;\exp{(Z^{-1})}} = \exp{ 
\left( {\frac{\xi}{\tilde{c}}} \right)}.
\end{equation}

As a result of (\ref{6.52}), we obtain $Z  \to  0$ for $\xi \to  -  
\infty$, and $Z \to 1$ for $\xi \to +\infty$. Fig. \ref{fig6.3} 
shows the profile of the $\tilde{\varepsilon}$ - wave, as obtained from Eq. 
(\ref{6.52}), as well as the width $\tilde{l}$ as obtained from Eqs.
(\ref{6.44}) and (\ref{6.48}). It is obvious from Fig. \ref{fig6.3} that the wave-profile is asymmetric in 
relation to the point with coordinates $Z = 1 / 2$, \quad $\xi / \tilde{c}= -  2$, 
which does not correspond to the flex point ($Z =  2 / 3$, $\xi / \tilde{c}  \approx   -  0,807$) 
of the function (\ref{6.52}), and is characterized by a more moderate variation at 
$Z \to 0$ in relation to a variation at $Z \to 1$. This in turn 
justifies to some degree the usage of the band-structure of the fcc-phase of 
iron for an analysis of the conditions for displacement-wave generation.

\section{Summary of Chapter 6}

Let us summarize as follows:
\begin{enumerate}
\item{Within the framework of the wave-model, our qualitative approach as 
proposed in Chap. 6 is mainly based on the presumption that several of the 
mutually - connected processes, which are propagated in a wave regime, 
determine the growth mechanism of martensite crystals.}

\item{In the particular case of missing atomic displacement-waves, the 
width of the moving boundary between the adjacent phases is mainly 
determined by the frontal width $l_{\tilde{\varepsilon}}$ of the 
$\tilde{\varepsilon}$ - TSW, corresponding to some lattice-constants. 

In the case of non-vanishing atomic displacement-waves, which ensure the 
required threshold-deformation, the $\tilde{\varepsilon}$ - wave transforms 
into a modified-switching wave of frontal width $l$, being characterized by a 
displacement-wavelength $\lambda$. For $\lambda /2  \sim  (10^{-7} 
\div  10^{-6})$ m, the quantity $l$ is about two to three orders of 
magnitude larger than $l_{\tilde{\varepsilon}}$. The region $B_{\gamma - 
\alpha}$, comprising highly pronounced inhomogeneities of temperature and 
chemical potential, basically corresponds with the frontal structure of the 
modified $\tilde{\varepsilon}$ - wave.

The complicated structure of the wave-process justifies the introduction of 
the special term "transformation-wave", closely related to the growth stage 
of martensite, also having been used in \cite{Meyers90}, but without specification of 
the wave-structure.}

\item{The cooperative character of the structural transformation, 
being the most typical phenomenological feature of martensitic 
transformations, is ensured by a phonon-maser effect occurring in the 
$B_{\gamma - \alpha}$ - region, which transforms a fraction of total 
reaction energy into energy of the atomic displacement waves.}

\item{The specification of the pattern of growth control of a 
martensite crystal by a pair of waves, as developed in Chap. 5, is reduced 
with the replacement of the lattice displacement waves with infinite frontal 
dimensions by wave-beams. The notion of wave-beams, i.e. of 
displacement-waves with finite frontal dimensions, suggests itself for the 
description of the growth process of a cylindrical macronucleus (see Pt. 
1.3), whose lateral dimension will just determine the minimum dimension of 
the wave front.}

\item{In order to provide further evidence for our 
interpretation by the notion of wave- beams, it may be of crucial interest 
to investigate how efficiently crystal-growth can be stimulated by means of 
two external sources of hypersound, and its relationship with the dimensions 
of the wave-front sources, with the aim of an improved specification of the 
transformation pattern treated in the final part of Chap. 5. In particular, 
getting over to closely located and properly oriented linear emitters with 
lateral dimensions of about $\le  10^{-6}$ m might make possible the 
simulation of a generation mechanism with long-wave components of the 
displacement waves, during the development of a single nucleus. It cannot be 
excluded that a single linear emitter (with the special orientation) may be a successful tool 
for simulation of the development of a single cylindrical macronucleus. 

It is obvious from the description of the habits near \{557\}, 
\{225\} in the two-wave pattern that, in order to excite waves 
propagating near the axes $\langle 0 0 \bar{1} \rangle$, $\langle 110 \rangle$, 
the emitter must be oriented in the direction of $\langle 1\bar{1}0 \rangle$, 
lying in the habit-plane. The success of such an experiment might be 
evidenced by the emergence of crystals with a limited choice of habits 
(i.e. from 1 to 4) of the type $(h \pm  \delta , h  \mp   \delta, \pm  l)$ with 
$\delta  < 1$. Such result would testify the benefit of nucleation of the 
martensitic crystals in elastic fields of dislocations with typical 
Burgers-vectors of fcc-lattice structures. It would also be promising to try to 
simulate the process of energy irradiation during the nucleation process, by 
means of narrow (with lateral dimensions of about $\sim  10^{-6}$ m) directed 
electron-, ion- or photon- beams, with a profile ensuring a linear trace of 
interaction on a single crystal-surface, with a period of interaction of 
$t_{i} \sim 10^{-10} \div 10^{-11}$ s.} 

\item{The mutual interaction of wave beams among themselves and with 
PSW of the $T$- or $\mu$- type is expressed by the requirement of their 
coordinated propagation, i.e. ~\textbf{V}~=~$\textbf{c}~=~\textbf{c}_{1}~+
~\textbf{c}_{2}$.}

\item{In the case of the $\gamma -\alpha$ - MT, the volumetric effect 
as well as the macroscopic shear with a typical magnitude of 0,2, serve as 
macroscopic characteristics. It can however easily be shown that the 
aforementioned model of thermodynamic potential \eqref{6.21}, being related to the 
macroshear, will deliver a threshold-level of macroscopic shear exceeding 
the threshold of volume deformation by one power, even for an A-value 
equivalent to the minimum shear-modulus $C^{\:\prime} = (C_{11} - C_{12})/2$. 
Thus it can be expected that macroscopic shear-distortion will only play a 
subordinate role, in that it would only appear in regions which already lost 
their stability by volume deformation, due to the action of longitudinal 
displacement waves. In other words, the relative volumetric change (actually 
playing the role of an ordering-parameter) might be the better choice as a 
macro-parameter. This point of view is close to the position of the author 
of \cite{Meyers90}, highlighting the leading role of longitudinal waves during the 
development of the central zone (the "midrib") of lenticularly shaped 
martensite crystals.}

\item{In the case of lenticularly shaped martensite crystals, the 
quantity c determines the greatest (frontal) midrib-growth velocity and thus 
features an important characteristic of the process. This circumstance 
implies the requirement for a precise measurement - by more than only of its 
order of magnitude - of the velocity c by an independent measuring 
principle. For example, the method described in \cite{Bunshah1953} could be improved by 
triggering the martensitic transformation of certain specimen exhibiting a 
pronounced two-stage transformation, by means of a strong magnetic field 
\cite{Sadovskii78,Schastlivtsev83}. On the one hand, this would allow a distinguished identification 
of the first, rapidly processing transformation stage and, on the other 
hand, a reduction of possible orientations of martensite, by proper 
consideration of the orientation of the pre-martensitic austenitic single 
crystals (see Pt. 5.3.3), in addition to allowing an improved determination 
of the dimensions of the crystals.}
\end{enumerate}
\begin{center}
Some additional remarks:
\end{center}
\begin{enumerate}
\item{In \cite{Liubov60}, the temperature-field at the boundary of the growing
martensitic crystal with the shape of an ellipsoid 
cylinder with small ratio of the cross-section axes has been treated. The 
task treated in \cite{Liubov60} is like the freezing-problem of humid ground \cite{Lykov67}: 
The equation of heat-conduction is devoid of sources, and the speed of 
boundary motion is assumed to be proportional to the temperature-gradient 
within the boundary. However, this approach differs remarkably from the one 
we used. Nonetheless, it is interesting to note that the estimate of the 
radius of curvature of the martensite-boundary $\rho  \sim  10^{-6}$ 
m (as determined in \cite{Liubov60}), at which adiabatic crystal growth mode becomes 
effective, confirms the possibility of adiabatic growth of the cylindrical 
macro-nucleus with a radius of an order of magnitude of about $10^{-6}$ m.}

\item{The description of martensitic growth as a lattice-deformation 
process suggests itself as most natural. In our above consideration, we 
distinguished between the deformation controlling the growth process of 
martensite, by means of displacement-waves, and the observable macroscopic 
deformation. Obviously, from a phenomenological point of view, this 
interpretation represents one of some possible variants of description, 
within the framework of a pattern of cooperating micro- and macroscopic 
ordering parameters \cite{Iziumov1984}. We also drew the attention to 
\cite{Likhachev83}, focusing on 
the selection of an ordering-parameter not being deducible to the observed 
macro-deformation, and on the applicability of a striction-model for an 
interpretation of thermodynamic laws, being typical of thermoelastic 
martensite.} 

\item{For large propagation velocities of the phase boundary, it is 
generally required to take into consideration the systematic motion of the 
electrons with the velocity of local lattice motion analogously to the 
systematic motion of the electrons with the local velocity of ions in an 
elastic wave (see e.g. Chap. 3 in \cite{kniga74}). The results published in
\cite{Robin1982,Bobrov83} give evidence for the dragging-effect of electrons 
by a moving boundary. However it is easy to prove that the contribution of 
such dragging effect on the occupational inversion would only be of an order 
of magnitude of $\Delta \sigma_{0} \sim 10^{-4}$ and thus have no 
appreciable effect on the rate of phonon-generation, as becoming obvious 
after comparison with the $\sigma_{0}$ - values in Table \ref{table3.1}.}

\item{In addition to the description of the phase boundary by means of 
a functional of type \eqref{6.21} we also would like to draw attention on the work 
of Lihachev \cite{Likhachev82}, which, within the framework of continuum-mechanics, 
suggests to consider boundary layers in the solid state (without specifying 
their structure) as independent planar-defects.

The results related to the qualitative analysis of coordinated propagation 
of switching- and displacement-waves, which simulate the motion of the 
phase-boundary during the $\gamma -\alpha$ - MT, are published in papers 
\cite{Kashchenko1986,Kashchenko1985,Vereshchagin85,Veresh85}. The substantiation of Chap. 6 mainly follows the work published 
in \cite{Kashchenko1985}.}
\end{enumerate}

\chapter*{Synopsis and outlook}
\addcontentsline{toc}{chapter}{Synopsis and outlook}

Based on analysis of experimental data and theoretical notions, a new 
conceptual approach for the description of rapid (frontal) crystal growth 
during fcc - bcc ($\gamma - \alpha$) martensitic 
transformations in ferrous alloys has been developed. The particularities in 
the course of this transformation (e.g. supersonic velocity of growth, 
anomalous magnitude of supercooling below the point of phase-equilibrium, 
missing phonon mode softening in the temperature region before the onset of 
a martensitic transformation) are considered as strong evidence for setting 
off the spontaneous martensitic transformation as a pecular kind of 
"limiting case" among diffusionless transformations, contrasting most 
pronouncedly with the opposed "limiting case" of structural transformations, 
described most conveniently under the notion of soft-phonons. 

Basically, the martensite growth stage is considered as a process of 
propagation of lattice deformation, being controlled by wavelike atomic 
displacements of relatively large wavelengths $\lambda  \sim  
(10^{-7} \div  10^{-6})$ m. However, the cooperative character of atomic 
displacement can only become effective if significant deviations from 
equilibrium conditions in the sub-system of 3d-electrons are present, 
thus implying the requirement of the existence of 
3-d-electron sub-systems embodying a macroscopic number of pairs of 
equidistant, inversely occupied states. This way, electronic transitions 
among inversely occupied electronic states lead to the generation of 
cooperative macroscopic lattice-displacement waves by stimulated emission of
phonons (phonon-maser effect). 
Because, from a quantum-mechanical point of view, the propagating atomic 
displacement wave represents a macroscopic amount of phonons of 
non-vanishing frequency, the previously defined notion of martensitic 
transformations, being essentially based on the propagation of displacement 
waves, is equivalent to a description within the conceptual framework of 
"hard" phonon modes, conforming to the new scientific tendency of 
theoretical research on $\gamma - \alpha$ - martensitic transformations. 

The survey of results presented in the final sections of Chaps. 2 to 6 of 
this monograph is showing that the tasks determined in Pt. 1.5 have been 
resolved, at least at a qualitative level. Nonetheless, let us now review 
once more the most important findings considered essential for the new 
description of the $\gamma - \alpha$ - martensitic transformation in ferrous 
alloys, as well as some remarkable quantitative key-evaluations obtained for 
the stage of nucleation of martensite, as well as some predictions derived 
from the theory and their future developments, respectively. 

\begin{center}
Most important findings
\end{center}
\begin{enumerate}
\item{ In the stage of rapid martensite crystal growth there exists a boundary 
area between the phases being characterized by intensive electron currents 
within coexisting strongly pronounced temperature and  even more important - 
chemical potential gradients.}

\item{ An electronic drift current leads to an inverted occupation of those 
pairs of electronic states being localized in the proximity of the 
S - surfaces in quasi-momentum space. S - surfaces are defined by the condition 
that the projection of electronic group velocity towards the orientations of 
$\vec{\nabla} T$ or $\vec{\nabla} \mu$ must vanish at all points of the 
S-surfaces.}

\item{The number of pairs of inversely occupied electronic states of the 
3d - bands of iron is a macroscopic quantity.} 

\item{The process of generation of atomic displacement waves is energized by 
stimulated emission of phonons during transitions of the non-equilibrium 
3d-electrons between the inversely occupied states. This process is similar 
to the radiation of photons in a maser.}

\item{The displacement waves controlling the process of martensitic crystal 
growth are of the longitudinal type (or quasi-longitudinal) with frequencies 
of $\nu \sim 10^{10}$ s$^{-1}$ (region of hypersound) 
and amplitudes ensuring the required level of lattice deformation of 
$\varepsilon \sim 10^{-3}$ needed for initiation of the 
$\gamma - \alpha$ - martensitic transformation. The mode of initial 
excitation of waves during the nucleation stage of the $\alpha$ - phase 
is a hard mode.}

\item{Certain combinations of displacement waves are important but not 
separate waves. Thus for instance, the stage of rapid growth of a 
martensitic lamellae is correlated with the propagation of a pair of 
perpendicularly oriented waves, stimulating the process of flat lattice 
deformation of a combined tensile-compressive type.}

\item{The displacement waves exist in the shape of the wave bundles propagating 
in coordination with the spatially limited front of a wave of relative 
volume deformation and making the function of "pilot-waves", paving the 
way for the martensitic reaction in their wake.}
\end{enumerate}

\begin{center}
Key quantitative evaluations
\end{center}

\begin{enumerate}
\item{Obviously, the electronic (chemical) potential gradient 
$\vec{\nabla} \mu$ is the main non-equilibrium source within 
an electron-subsystem. The assessment of the quantity $\vec{\nabla} \mu$
 requires that the difference of chemical phase potentials 
 $\Delta \mu = \mu_{\gamma} - \mu_{\alpha}$ as well as the 
width $l$ of the region among the phases are known. The value $\Delta 
\mu \approx  0,16$ eV has been determined for the s - and 
d - electron sub - systems under Pts. 1.5 and 4.5.3. The quantity $l$ can be 
assessed by means of Eq. \eqref{6.49}. Taking into account that the constant 
$\tau_{\tilde{\varepsilon}} $ included in \eqref{6.49} plays the role of 
the minimum time-constant characterizing the Bain-deformation process, 
it suggests itself to take the physically consistent value $\tau _{\tilde
{\varepsilon} } \sim  10^{-14} s$, corresponding to the period in which 
a lattice - atom will be displaced - at near-sound velocity - by a distance 
of about $0,1\,a$ ($a$ - lattice-parameter). Then we get from \eqref{6.49} 
the value $l  \sim  10^{-6}$ m and $\nabla  \mu    \sim  \Delta  \mu  / 
l   \sim  10^{5}$ eV/m.}

\item{Within the frame of the proposed wave-model of martensite-growth the 
requirement for satisfaction of the condition $\sigma_{0} > 
g_{th}$ for displacement-wave generation, as outlined 
under Pt. 3.1, attains a key role. An assessment of the inverted population 
difference of a pair of electronic states, associated with an electron drift 
in a field $\vec{\nabla} \mu$, delivers $\sigma_{0}(\nabla \mu) \ge 10^{-3}$. 
The threshold-value $\sigma_{th}  \sim  10^{-4} \div  10^{-3}$ had 
been determined under Pt. 3.1 from Eq. \eqref{3.10}. After further evaluation 
of $\sigma_{th}$, it becomes obvious that the number of pairs of inversely 
occupied states $R_{\textbf{q}}$, being proportional to the area
$\Sigma_{\textbf{q}}$ of the reduced sheet of the S-surface, is of paramount 
importance. For $\Sigma_{\textbf{q}} \approx  20( \pi / a)^{2}$ , the condition 
$\sigma_{0} > \sigma_{th}$ is satisfied for ordinary values of the 
matrix-elements of electron-phonon-interection $W_{\textbf{q}}$ ($W_{\textbf{q}}$ 
is taken in the tight binding approximation ). We emphasize that the 
instability of a metal lattice, being linked up with the amplification of an 
acoustic wave by the mechanism of stimulated emission of phonons (see e.g. 
\cite{Pains65}) has generally been supposed as not realizable in metals.}

\item{ The magnitude of maximum lattice deformation $\varepsilon_{m}$ 
maintainable in a displacement wave also is of fundamental 
importance. An evaluation of $\varepsilon_{m}$ has been 
performed under Pt. 3.3 by means of Eq. \eqref{3.42}, being correct for stationary 
conditions. Using this methodology, a value of $\varepsilon 
_{m} \sim  10^{-3}$ was obtained for $\sigma_{0}  \approx  2,5 \,\sigma_{th}$. 

The inequality $t_{u} >> t_{\nabla}$, being included in Pt. 3.3, 
refers to the settling time - $t_{u}$ - of the stationary amplitude (or of 
the deformation $\varepsilon_{m})$ within the displacement 
wave. The lifetime of the chemical potential gradient $t_{\nabla}$ 
shows that a usage of the stationary $\varepsilon_{m}$ - 
estimate is only justified for a hard mode of wave excitation. With this 
approach, it is possible to link up the growth of martensite with 
displacement waves, while we believe that near $M_{S}$ the austenitic 
lattice remains essentially stable against deformations being less than $10 
^{-3}$, thus suggesting the following conclusion: 
\begin{center}
the nucleation process of martensite crystals is associated with the hard mode excitation of 
displacement waves of an estimated strain amplidude 
$ \varepsilon_{m} \sim 10^{-3}$.
\end{center}}

\item{An evaluation of the threshold value $\tilde{\varepsilon }_{th} 
(M_{S})$ of relative volume deformation $\tilde{\varepsilon}$ at temperature 
$M_{S}$ also is of crucial importance. The point here is that coordinated 
propagation of displacement waves with an interphase boundary would only 
be possible if the relative volume deformation $\tilde{\varepsilon}_{u}$ 
within the displacement waves exceeded the value 
$\tilde{\varepsilon}_{th}(M_{S})$. Otherwise, ($\tilde{\varepsilon }_u < \tilde
{\varepsilon }_{th}(M_{S})$), the motion of the phase boundary would be 
described by a triggering-switching wave with the sole velocity being distinctly
smaller than the magnitude of the sound velocity. In Pts. 6.3.2, 6.3.1, there 
has been shown that $\tilde{\varepsilon}_{th}(M_{S}) \approx  2,5  \cdot 10^{-4}$, 
so that the condition $\tilde{\varepsilon}_{u}  >
\tilde{\varepsilon}_{th}(M_{S})$ 
is easily satisfied for $\varepsilon_{m} \sim 10^{-3}$.}
\end{enumerate}
\begin{center}
Proposed experiments for further verification of the growth model
\end{center}
\begin{enumerate}
\item{Measurement of the contact-potential difference $\Delta 
\varphi$ among the $\gamma$ - and $\alpha 
$ - phases (see Pt. 2.5). The value and sign of $\Delta \varphi$ 
are needed to determine the inverted occupational difference 
$ \sigma_{0} \sim  \Delta \varphi$, as well as the direction of 
propagation of the displacement waves in relation to $\vec{\nabla} \mu$ 
(see legend of table 2.1).}

\item{Measurement of the ratio of the velocity moduli of the longitudinal waves 
$\varkappa_{e} = c_{\langle 110 \rangle}c_{\langle 001 \rangle}^{-1}$ 
in austenite at the temperature $M_{S}$, as a function of the concentration 
of the alloying element $C_{ae}$, in order to check the criterion for the 
first change of the habit-planes ($\{557\} \to \{225\}$). A positive result 
of the experiments would be an increase of $\varkappa_{e}$ with increasing 
$C_{ae}$ while approaching to the concentration at which the habit 
\{225\} obviously becomes stabilized (see Pt. 5.1.3).}

\item{Precise measurement of the rapid frontal growth velocities of martensite 
crystals and comparison with the values $\arrowvert \textbf{c} \arrowvert 
\approx \arrowvert \textbf{c}_{1} + \textbf{c}_{2} \arrowvert$ predicted by 
theory, while the velocities $\textbf{c}_{1}$ and $\textbf{c}_{2}$ could be 
calculated on the basis of measured elastic moduli of austenite at $M_{S}$.}

\item{Observation of oriented growth of the martensite crystals during 
simultaneous influence of hypersound and strong magnetic fields, as well as 
of the effects of laser radiation pulses impinging on the surface of a 
specimen, directed along a trace near to a linear form.

We note here that the discovery of a predicted orientational effect in a 
magnetic field of $H  \sim  10^{7}$ A/m \cite{Leont'ev84}, exceeding in 
order of a magnitude both demagnetization- and magnetic anisotropy fields, 
delivers strong evidence in favor of the maser-effect during 
phonon-generation, and, moreover, confirms our conclusion on the connection 
of the second change of habit $\{557\} \to  \{3 \:10\: 15\}$ in 
the Fe-Ni-system, and $\{225\}  \to  \{259\}$ in the 
Fe-C-system, with exclusion of the quasi-longitudinal waves with velocity 
directions near the second-order axes of symmetry of the fcc-lattice from 
the total spectrum of generated waves.

If we mention further research, we want to note, above all, that a 
comprehensive understanding of the processes occurring in the stage of 
growth of martensite crystals should also contribute to the definition of 
physical models of the preceding stage of nucleation, as well as of the 
subsequent stage of accommodation of the coexisting phases. This way, for 
instance, the requirement for coordination of the nucleation stage with the 
growth stage, caused by the non-equilibrium conditions at the interphase 
boundary and being controlled by displacement waves, will lead to new 
findings on the rapid emergence of the macro-nucleus, as treated under Pt. 
1.3. 
\begin{center}
Synthesis of concepts of the heterogeneous nucleation and of the wave growth
of martensitic crystals
\end{center}
The presently achieved success in improved understanding of the nucleation 
stage can mainly be attributed to the consolidation of the notions of 
heterogeneous (dislocation-) nucleation as well as of the wave-controlled 
growth of martensite
\cite{Kashchenko89,Vereshchagin89,Kashchenko90,Kashchenko91,Kash91,Vereshchagin91,Veresh91}. A calculus of the elastic fields of linear 
dislocations, being typical of the parent- $\gamma$- phase, have 
shown that even single mixed dislocations can play the role of nucleation 
centers (NC) of martensite crystals. NC-characteristics and macroscopic 
morphological features of martensite are (genetically) mutually linked up. 
In result there exist 24 different variants of matching patterns in the 
"NC-crystal".

The elastic field of dislocations disarranges the original lattice symmetry 
by selecting regions being most favorable for martensitic nucleation. Such a 
region features the shape of a perpendicular parallelepiped, its edges being 
oriented along the eigenvectors $\vec{\xi}_{i}$ of the 
strain tensor $\hat{\varepsilon}$, its eigenvalues $\varepsilon 
_{i}$ satisfying the following conditions:

\begin{displaymath}
\varepsilon_{1} > 0 ,\quad \varepsilon_{2} < 0, \quad \arrowvert 
\varepsilon_{3} \arrowvert < < \arrowvert  \varepsilon_{1,2}\arrowvert,
\end{displaymath}
thus ensuring the existence of slightly distorted surfaces (SDS) which are 
close to planes with normals

\begin{displaymath}
(\textbf{N}_{SDS})_{1,2} \:\Vert \:\vec{\xi}_{2}  \mp \vec{\xi}_{1}
\sqrt{\varepsilon_{1} / \arrowvert \varepsilon_{2} \arrowvert},\qquad 
\arrowvert \vec{\xi}_{1,2} \arrowvert  = 1. 
\end{displaymath}

Obviously, from a point of view of minimization of elastic distortion 
energy, phase-coupling is supported by weakly distorted (with 
$\varepsilon_{3} = 0$ invariants) planes. Thus it would be reasonable to 
expect that the normal of the habit-plane of the martensite crystal should 
match with one $\textbf{N}_{SDS}$.

In fact, among the $\textbf{N}_{SDS}$, there exist $\langle$ h k l $\rangle$ 
situated near $\langle  557 \rangle$, $\langle  225 \rangle$ ( for 60\r{ } - dislocations with 
lines $\langle 1\bar{1}0 \rangle$ ) and situated near $\langle  259 \rangle$, 
$\langle  3\: 10\: 15 \rangle$ ( for 30\r{} - dislocations with lines 
$\langle  1\bar{2}1 \rangle$), being evidence of certain differences among the 
NC of packet- vs explosive-martensite. 

Moreover, in the orientational relationship of the phase-lattice, there are 
included the slip-plane and the dislocation line, the latter one acting as a 
nucleation center, which suggests us to give preference to the 
Kurdjumov-Sachs- or Nishiyama-relationships, for various NC.

The question related to the orientation of macroscopic shear \textbf{S} will be 
resolved in conjunction with the choice of one of the two orientations of 
the normal $\textbf{N}_{SDS}$. For this aim, let us consider the notation of 
the distortional tensor of the elastic field, being represented as the sum 
of two diad products, and discriminate the part containing two addends

\begin{displaymath}
\textbf{S}_{1}  \cdot  \textbf{N}_{1} + \textbf{S}_{2}  
 \cdot  \textbf{N}_{2} ,\qquad \arrowvert \textbf{N}_{1,2} \arrowvert = 1.
\end{displaymath}

We recall that the diad product $\textbf{S} \cdot \textbf{N}$ 
defines a deformation with an invariant plane, where $\textbf{N}$ - normal of 
a plane and $\textbf{S}$ - vector characterizing the shape deformation. Further 
considering that austenite is metastable at the beginning of the martensitic 
transformation at $M_{S}$ - temperature, it is justified in the case of 
$\arrowvert \textbf{S}_{1} \arrowvert > \arrowvert \textbf{S}_{2} \arrowvert$, 
to surmise that the plane with the normal $\textbf{N}_{1}$ is distinguished, 
and that the anticipated orientation of macroscopic shear is close to 
$\textbf{S}_{1}$. And vice-versa, for $\arrowvert \textbf{S}_{2} \arrowvert >
\arrowvert \textbf{S}_{1} \arrowvert$, the components $\textbf{N}_{2}$ and 
$\textbf{S}_{2}$ will be discriminated, respectively. In the case of the 
straight $\gamma -\alpha$ - transformation, the results of this approach are 
in good accordance with experimental results.

With this approach, all macroscopic morphological characteristics of 
martensite attain a reasonable interpretation within the conceptual notion 
of nucleation at dislocations, where dislocations act as centers of forces 
disturbing the original lattice symmetry, their effect not 
being confined to the nuclear volume.

These findings match in detail with the ideas of the wave theory of growth, 
presupposing that the transformation starts with the emergence of an excited 
state with the shape of a parallelepiped, built up of vectors $\vec{\xi} 
_{i}$, its pairs of edges oscillating in opposed phase, 
thereby exciting controlling displacement waves orientated in the 
wave-normal $\textbf{n}_{1,2}$ close to $\vec{\xi}_{1,2}$. In the most simple 
approximation of the equations 

\begin{displaymath}
\textbf{n}_{1}= \vec{\xi}_{1},\quad \textbf{n}_{2} = \vec{\xi} 
_{2},
\end{displaymath}
the requirement of correspondence of $\textbf{N}_{WDS}$ with the wave-habit 
\eqref{5.3} delivers the following condition:

\begin{displaymath}
\varkappa = \frac{c_{2}}{c_{1}} \approx  \left[\frac{ \varepsilon_{1}}
{\arrowvert \varepsilon_{2}\arrowvert}\right] ^\frac{1}{2}
\end{displaymath}
which, if satisfied, ensures the possibility of a kinematic agreement of 
the wave description with the deformation description of the habit. 
Obviously, given the case that the ratio of tensile and compressive 
deformation in the wave-mode corresponds with $\varkappa^{2}$, then dynamic 
agreement will also be achieved. We further note that for 
the $\gamma - \alpha$ - transformation, which proceeds 
with increase of specific volume: $\varepsilon_{1} > \arrowvert 
\varepsilon_{2} \arrowvert$. Consequently, $c_{2} > c_{1}$, so 
that the tensile strain can be prescribed by the wave propagating 
with the smaller velocity $c_{1}$, whereas compressive strain can be 
prescribed by the wave propagating with the larger velocity $c_{2}$ (In the 
case of the $\alpha - \gamma$ - transformation, the 
situation will just be inverted). Specifically, the latter finding provides 
a physical criterion for the realistic choice of phase tuning of the waves 
controlling the transformation, as outlined in Chap. 5, as well as for the 
selection for the principal axes of tensile and compressive strain. Undoubtedly, 
the rules of deviation of the normal $\textbf{n}_{1,2}$ of the controlling 
waves from the axes of symmetry of an ideal cubic lattice are dictated by 
the symmetry of the elastic field of the NC in the nucleation stage, in 
conjunction with a hard mode of wave generation. It is further worth to note 
that the orientation of the axis $\vec{\xi}_{3}$ of the cylindrical nucleus 
of the $\alpha$ - phase can significantly deviate from the line of 
dislocations (most pronouncedly for the 30\r{} - dislocation line). 

Our specification of the nucleation stage, which also enabled us to define 
the orientation $\vec{\xi}_{3}$, besides lifetime, energy and 
spatial contour of the state of excitation, suggested possibilities for 
their physical simulation \cite{Letuchev90,Letuchev92}. The effect of laser impingement 
close to a linear trace, with a period of $\sim 2 \cdot 10^{-11}$ s 
and sufficient power density for atomic vaporization, could induce the 
onset of a $\gamma - \alpha$ - in Fe-31,5 Ni-single crystals, at a temperature 
some degrees above $M_{S}$, however only in cases where the orientation of the 
trace of impingement of laser radiation was close to the calculated orientation 
$\vec{\xi}_{3}$ on the flat surface of the specimen. Obviously, besides the 
envisaged precise measurement of the speed of growth of $\alpha$ - martensite, 
the already obtained results are fundamental for proving true the developed 
dynamical theory of the transformation. 

In order to get over to the dynamical description of the final stage, it 
would be useful, from our point of view, to represent the displacement waves 
by extremely small bundles of waves (with one of its transversal frontal 
dimensions of an order of magnitude of about one half wave-length), existing 
in the form of half-wave pulses (the dimensions of such pulses of order 
$\lambda / 2$ in their direction of propagation). In fact, given a 
level of deformation $\varepsilon  \sim  10^{-3}$, with $\lambda /2  
\sim  10^{-6}$ m, the utmost displacement associated with the wave-pulse 
would be $u_{max} \approx 1/4 \,\varepsilon \,\lambda$, i.e., of 
similar order of magnitude as the lattice-parameter. Consequently, such 
wave-pulses must destroy the continuity of the medium, by causing the 
following effects: Generation of structural defects; rotational 
distortions of the transforming lattice (being equivalent to the onset of 
"rotational modes" \cite{Panin82}), which re-establish the macroscopic continuity 
of the medium, as known from \cite{Umanskii78,Kurdjumov77,Panin82}, also leading to the emergence 
of a characteristic relief at the surface of the specimen subjected to a 
$\gamma -\alpha$ - transformation. Of course, all these processes, including 
the nucleation stage, are closely associated with the stage of growth and 
thus can only partly be put in relation with the stage of accommodation of 
coexisting phases. Nonetheless it is reasonable to classify as fundamental 
the identification of processes controlling the martensitic transformation, 
as well as a more detailed investigation of the microscopic mechanism of 
their development, within the frame of a comprehensive description of the 
growth mechanism of martensite. 

The identification of dislocation centers linked with martensite-nucleation 
opens up new possibilities for an interpretation of the laws and rules governing 
the formation of lattice-ensembles, by definition of the spectrum of new centers 
of nucleation, being immanent in the newly emerged $\alpha$ - phase crystal. 

Further progress of the general theory of martensitic $\gamma -\alpha$ 
- transformations will undoubtedly be associated with the 
following topics:
\begin{itemize}
\item[-]{Theoretical research on the dynamics of lattice-fluctuations including 
investigations on the effects of elastic dislocation fields on the 
nucleation stage of martensite. It can already be anticipated that the 
fluctuations leading to the above defined state of lattice-excitation in the 
form of an elongated parallelepiped, are most probable;}

\item[-]{Specification of the area in which the state of excitation is localized, 
within the field of dislocation (i.e. the shortest distance to the line of 
dislocation), as well as of its probable transversal dimension;}

\item[-]{Specification of the structure of the electronic spectrum in the proximity 
of the dislocation-nucleation center (DNC) and of the $B_{\gamma - \alpha}$
- region. Doing so, the following aspects have to be considered:
\begin{itemize}
\item[1)]{The potential coexistence of a mixture of electronic configurations of the 
iron-atoms (the basic pattern of calculus of a band-spectrum has already 
been proposed in \cite{Egorushkin84});}

\item[2)]{The effect of lattice-distortions on the shape of the S-surfaces as well as 
of energy-dispersion at the S-surfaces, in the proximity of the 
Fermi-surface.}
\end{itemize}}
\end{itemize}

Finally, we remark that S-surfaces resemble individual objects in 
quasi-momentum space, thus classifying electronic states with respect to 
their contribution to transport-processes. Furthermore, for investigations 
and interpretations of phenomena in transition-metal alloys, which can most 
conveniently be described on the basis of a first-order coherent-potential 
approximation \cite{Vediaev78}, it will be necessary to evaluate the possible existence 
of S-surfaces with large areas. Obviously, our proposed wave-model of growth 
control for an $\alpha$ - crystal will be universally applicable for 
most rapidly proceeding (1$^{st}$ - order) martensitic transformations, being 
related to a dislocation-nucleation mechanism. The threshold-deformation for 
such martensitic transformation comprises combined tensile and compressive 
strain, their main axes being oriented perpendicularly. For atoms in an 
intermediate phase region, the energy barrier can be overcome in a 
cooperative manner by a wave-mode. This way, the controlling 
displacement-waves imprint macroscopic characteristics in the 
martensite-crystals, however without determining the final positions of the 
atoms. The latter will mainly depend on the pattern of local interactions 
among the atoms. The atomic displacements towards their new 
equilibrium-positions can be initiated by waves of shorter wavelength (in 
relation to the control-waves), both in the pretransitional stage and in the 
growth-stage. Such an additional system of waves, in parallel with the 
accommodation-processes, must define the internal fine structure of the 
martensite plates.
\pagebreak
\begin{center}
Final conclusions
\end{center} 
As a final conclusion, it can be stated that the obtained results provide 
sufficient evidence of the onset of a new qualitative stage of research on 
the large field of $\gamma - \alpha$ - martensitic 
transformations, within the frame of the wave-approach, thus making 
possible, on principle, to establish and to describe the following 
fundamental chain of relationships of cause and effect: 
\begin{itemize}
\item{dislocation-nucleation-center}
\item{initial state of excitation}
\item{non-equilibrium conditions}
\item{particularities of the electronic structure}
\item{spectrum of the generated displacement-waves}
\item{observed morphological characteristics.}
\end{itemize}
It is apparent that just the growth-controlling displacement-waves carry the 
information about the elastic field created by the nucleation-center in the 
region of nucleation location, thereby ensuring a genetic correspondence of 
the characteristics of macroscopic martensite-crystals with those of the 
nucleation-center. It is also obvious that, within the developed dynamical 
theory of $\gamma - \alpha$ - martensitic transformation, the 
growth-controlling displacement-waves play the role of the previously 
missing link among micro- and macroscopic description.

Our general conclusion is that the new comprehensive model of growth of the 
martensite crystals in a wave regime is basically consistent with observations 
and phenomenological theories, now being complemented by contemporary 
dynamics of lattice, and also by thermodynamical and quantum theoretical 
aspects, missed or only partially considered in previous theories.

The author hopes that near-future research will confirm most of the 
predictions, enabling an extension of the insights of the wave-model of 
$\alpha$ - martensite-growth in ferrous alloys onto a variety of 
other solid-state martensitic transformations.}
\end{enumerate}

\pagebreak

\bibliographystyle{unsrt}
\addcontentsline{toc}{chapter}{Bibliography}
\bibliography{book}

\chapter*{Epilogue}
\addcontentsline{toc}{chapter}{Epilogue}

The author is sure that readers, who have reached up to the end of the 
monography, have understood that the mechanism controlling the spontaneous 
growth of martensite crystals (at cooling or heating) is connected with 
process of elastic waves generation by nonequilibrium 3d electrons (see also 
\cite{Kashchenko93,Kashchenko95}). It is self-evident that the content of the monography is not 
exhausted by exposition of this process as the severe fathoming of the 
martensitic transformation mechanism in iron alloys needs the knowledge of 
solid-state physics in full.

Really, the nucleation process exposition of martensite is guessing the 
knowledge of the elastic fields created by flaws (in particular by 
dislocations).

It is necessary to understand, on the one hand, features of an energy-band 
structure in transition metals, and on the other hand, the essence of the 
kinetic phenomenons in them to explain the generation mechanism of waves. 

It is necessary to take into account that not only pure metals must be 
discussed, but also alloys (in a wide range of concentrations and 
temperatures); not only paramagnetic state, but also the ferromagnetic 
state; not only linear, but also nonlinear waves with interior dynamic 
structure of a wave-front, etc. 

The explanations covering the physical essence of solved problems and 
distinctive role of the martensitic transformation $\gamma -\alpha$ in 
iron alloys among structural transformations in solids (as an example of 
phase transition with extremely high speed of growth) ensure worth of the 
given book for broad enough audience of specialists. 

The author has closely overlooked the monography once again and has 
convinced yourself that there is no need to make any essential changes in 
the text holding its topical nature (set of fine misprints have been 
eliminated at editing of translation).

It is worth noting that research directions marked in the monography have 
received the progress during last decade.
\begin{enumerate}
\item{ Dislocation nucleation centres were established for the martensitic 
crystals with all types of habit planes \cite{Letuchev95}. It has allowed to interpret the 
processes of forming of typical ensembles of the martensitic crystals (see 
for example \cite{Kashchenko94,Kashchenko1994,Kashchenko97}).}
\item{The formation of fine twin structure has been described as the 
consequence of the coordinated propagation of relatively short-wave and 
long-wave atomic displacements (see \cite{Kashchenko99,Kashchenko2000,
Chashchina2000}).}
\item{Calculations for the dislocation loops have been shown that their elastic 
fields (possessing the greater inhomogeneity in comparison with fields of 
the infinite rectilinear dislocations) create the such compact regions that 
the nucleation within them leads to observed distributions of habit planes.

Hence the distribution of habit planes gives the additional information for 
reconstruction of the nucleation process via morphological indications 
\cite{Letuchev98,Nefedov2000,Dzhemilev2000}.}
\item{The concept of the crystons was found convenient and constructive for 
description of the lattice losing a stability during propagation of the 
process controlling the martensite growth \cite{Kashchenko97}. This concept was originally 
used for interpretation of shear band formation (see for example
\cite{Kashchenko1996,Sokolova1998,Teplyakova1999,Semenovih2000,Semenovih2002}).}
\item{The cryston model has been extended to exposition of strain martensite. 

In this case the crystons (the shear carriers of superdislocation type) are 
direct carriers of threshold strain (see \cite{Semenovih2003,Kashchenko2003}).}
\end{enumerate}

Thus there is clear enough fathoming of physical mechanisms for all 
alternatives of the $\gamma -\alpha$ martensitic transformation in iron 
based alloys. 

It is significant that concepts of the heterogeneous nucleation and the 
strain controlling the crystal growth (the last is localized in the frontal 
region of the growing crystal) are the universal for the exposition of 
martensite crystal formation.

However the dynamic nature of controlling processes, as well as mechanisms 
of their energy support, essentially discriminate. 

For case of the cooling martensite this is the controlling wave process that 
is supported in the maser regime by nonequilibrium electrons due to the 
generation of energy in transforming phase.

For case of the strain martensite this is the processes of the crystons 
propagation that are supported in basic by energy of exterior stresses.

The author is sincerely grateful to mister U. Kayser-Herold and madam J. 
Gerlts. The translation that has been carried out by them, adequately 
transmits the content of the monography in Russian.

\renewcommand{\bibname}{Bibliography for epilogue}

\chapter*{}
\addcontentsline{toc}{chapter}{Summary}

\begin{center}
Kashchenko M.P."The wave model of martensite growth\\
for the $\gamma-\alpha$ transformation of iron-based alloys".\\
2005
\end{center}

\textbf{\huge Summary}
\\
\\
\\

Kashchenko Mikhail Petrovich, professor (Holder of Physics Chair, Urals State 
Forest Engineering University), doctor of physical-mathematical sciences. 
The sphere of his scientific inte­rests: physics of solids, high-excited 
states of solids, martensitic transformations, synergetics.

This book is the first monograph in the scientific literature, dedicated to 
the $\gamma-\alpha$ transfor­mation in iron-based alloys, in which the 
dynamical approach is used for the explanation of the martensite growth stage.

The rapid growth of a martensite crystal is considered as a self-organized 
process cont­rolled by the quasi-longitudinal lattice displacement waves (DW). 
The regime of the DW initial excitation is rigid. DW have the frequencies 
$\sim 10^{10} sec^{-1}$ from the hypersound band and the amplitudes providing 
the level of deformation $\sim 10^{-3}$. The conditions that are necessary for 
the generation of DW by non-equilibrium d-electrons are analyzed.

A wide range of questions (from peculiarities of the electronic spectrum to 
macroscopic morphological indicators), connected with the physical 
interpretation of the $\gamma-\alpha$ martensitic trans­formation in iron-based 
alloys, is discussed.

The short review of results having fundamental meaning for the creation of a 
physical model describing the martensite nucleation process is given in the 
monograph's conclusion. It is shown, that processes of the heterogeneous 
nucleation and wave growth have the genetic connection to the $\gamma-\alpha$ 
martensitic trans­formation.

\end{document}